\documentclass[openany]{book}
\usepackage[margin=1in]{geometry}

\usepackage{authblk}
\usepackage{amsmath}
\usepackage{graphicx}
\usepackage{epsfig}
\usepackage{epstopdf}
\usepackage{multirow}
\usepackage{overpic}
\usepackage{colortbl}
\usepackage{tabularx}
\usepackage{url}
\usepackage{multirow}
\usepackage{setspace}
\usepackage{amssymb}

\usepackage{bm}
\usepackage{color}
\usepackage{amsfonts,amsmath,amssymb}
\usepackage{bm}
\usepackage{titletoc}
\usepackage{makecell}
\usepackage[colorlinks,linkcolor=blue,anchorcolor=blue,citecolor=blue]{hyperref}
\usepackage{listings}
\hypersetup{
colorlinks=true,
allcolors=blue
}
\usepackage{float}
\usepackage[section]{placeins}
\usepackage{verbatim}
\usepackage{siunitx}
\usepackage[numbers,sort&compress]{natbib}
\usepackage{chapterbib} 
\usepackage{textcomp}

\usepackage{ifthen}
\newboolean{printBibInSubfiles}
\setboolean{printBibInSubfiles}{true} 
\def\bib{\ifthenelse{\boolean{printBibInSubfiles}}
        {\bibliographystyle{apsrev4-1}\bibliography{references}}
        {}
    }

\usepackage{subfiles}
\usepackage{booktabs}
\usepackage{subfigure, dcolumn}

\title{Volume I - Physics \& Detector }

\author[2,1,3]{Jing Chen}
\author[2,3]{Ji-Yuan Chen}
\author[12,13]{Jun-Feng Chen}
\author[2,3]{Xiang Chen}
\author[8,9]{Chang-Bo Fu}
\author[2,3]{Jun Guo}
\author[2,3]{\\Yi-Han Guo}
\author[1,2,3]{Kim Siang Khaw}
\author[2,3]{Jia-Lin Li}
\author[2,3]{Liang Li}
\author[1,2,3,6]{Shu Li}
\author[1,2,3]{Yu-ming Lin}
\author[1,2,3]{\\Dan-Ning Liu}
\author[1,2,3]{Kang Liu}
\author[1,2,3]{Kun Liu}
\author[1,2,3,14]{Qi-Bin Liu}
\author[4]{Zhi Liu}
\author[2,3]{Ze-Jia Lu}
\author[7]{Meng Lv}
\author[2,3]{\\Si-Yuan Song}
\author[1,2,3]{Tong Sun}
\author[2,3]{Jian-Nan Tang}
\author[17,4]{Wei-Shi Wan}
\author[5,4]{Dong Wang}
\author[8,9]{Xiao-Long Wang}
\author[1,2,3,15]{\\Yu-Feng Wang}
\author[1,2,3,10,11]{Zhen Wang}
\author[16]{Zi-Rui Wang}
\author[2,3]{Wei-Hao Wu}
\author[18,19,1]{Dao Xiang}
\author[2,1,3]{Hai-Jun Yang}
\author[1,2,3]{\\Lin Yang}
\author[2,3]{Yong Yang}
\author[1,2,3]{Dian Yu}
\author[1,2,3]{Rui Yuan}
\author[1,2,3]{Jun-Hua Zhang}
\author[2,3,14]{Yu-Lei Zhang}
\author[10,11]{\\Yun-Long Zhang}
\author[1,2,3]{Zhi-Yu Zhao}
\author[1,2,3]{Bai-Hong Zhou}
\author[2,3]{Chun-Xiang Zhu}
\author[1,2,3]{Xu-Liang Zhu}
\author[2,3]{Yi-Fan Zhu}

\affil[1]{\it Tsung-Dao Lee Institute, Shanghai Jiao Tong University, Shanghai 201210, China}
\affil[2]{\it Institute of Nuclear and Particle Physics, School of Physics and Astronomy, Shanghai Jiao Tong University, Shanghai 200240, China}
\affil[3]{\it Key Laboratory for Particle Astrophysics and Cosmology (MOE), Shanghai Key Laboratory for Particle Physics and Cosmology (SKLPPC), Shanghai 200240, China}
\affil[4]{\it Center for Transformative Sciences, ShanghaiTech University, Shanghai 201210, China}
\affil[5]{\it Shanghai Advanced Research Institute,Chinese Academy of Sciences, Shanghai 201210, China}
\affil[6]{\it Center for High Energy Physics, Peking University, Beijing 100871, China}
\affil[7]{\it School of Electronics, Information and Electrical Engineering, Shanghai Jiao Tong University, Shanghai 200240, China}
\affil[8]{\it Key Laboratory of Nuclear Physics and Ion-beam Application (MOE), Fudan University, Shanghai 200443, China}
\affil[9]{\it Institute of Modern Physics, Fudan University, Shanghai 200443, China}
\affil[10]{\it State Key Laboratory of Particle Detection and Electronics, University of Science and Technology of China, Hefei 230026, China}
\affil[11]{\it Department of Modern Physics, University of Science and Technology of China, Hefei 230026, China}
\affil[12]{\it Center of Materials Science and Optoelectronics Engineering, University of Chinese Academy of Science, Beijing, 100049 China}
\affil[13]{\it Shanghai Institute of Ceramics, Chinese Academy of Sciences, Shanghai, 201899 China}
\affil[14]{\it University of Washington, Seattle, WA 98195, United States of America}
\affil[15]{\it Deutsches Elektronen-Synchrotron (DESY), Hamburg, Germany}
\affil[16]{\it College of Literature, Science, and the Arts, University of Michigan, Ann Arbor, MI 48109-1040, United States of America}
\affil[17]{\it Quantum Science Center of Guangdong-HongKong-Macao Greater Bay Area}
\affil[18]{Key Laboratory for Laser Plasmas (MOE), School of Physics and Astronomy, Shanghai Jiao Tong University, Shanghai 200240, China}
\affil[19]{Zhangjiang Institute for Advanced Study, Shanghai Jiao Tong University, Shanghai 200240, China}
\begin{document}
\setboolean{printBibInSubfiles}{false}
% remember to use \bib in each subfile if you want to use bib!!

\input{title}
\maketitleLARGER

\setlength{\baselineskip}{0.5cm}
\tableofcontents
\newpage

\begin{chapter}{Physics}

\section{Introduction to Dark Matter}

Dark Matter (DM) has been widely regarded as an unknown physical mystery but with evidence from the astronomical observations and gravitational effects such as
the phenomena from galactic rotation curves, gravitational lensing, cosmic microwave background anisotropies, etc. Despite the physical interpretation of DM is yet to be discovered and verified,
the typical DM characteristics being Non-baryonic, massive, electrically neutral, gravitational and stable are broadly accepted by the physicists.
Many Beyond Standard Model (BSM) theories predict DM mechanisms, such as the weakly interacting massive particle (WIMP) being one of the popular Dark Matter Candidates.
Most of the BSM models predict the DM particles possibly produced through collider experiments interacting weakly with SM particles and pass invisibly through the particle detectors.
This would lead to common "Missing Transverse Energy" ($E_T^{miss}$) phenomena when a collision event does not balance in plane transverse to beam.

Over the past decades, collider based DM search is among those major approaches to search for DM, while direct detection approach making use of the nuclear recoils from DM-nuclei scatterings and indirect detection approach utilizing the products from DM annihilations are equally important.

Collider based DM search experiments make use of BSM predicted DM production mechanism in high-energy collisions, focusing on the $E_T^{miss}$ phenomena along with SM particles productions.
Many high energy collider experiments (e.g. the Large Hadron Collider (LHC), BESIII@BEPC-II, Belle-II@SuperKEKB, ...) naturally provide such search opportunity through either hadron-hadron collisions
or lepton-lepton collisions spanning wide mass range and examining many BSM contexts. Besides the general collider experiments, other accelerator based DM search experiments are equally motivated to
cover complementary search sensitivities, given the relatively lower center-of-mass energies compared to colliders.

When revisiting the DM search approaches, besides the regular searches aiming for detecting/observing DM candidate particles directly, one may also aim to look for DM mediators bridging
the SM particles and the DM particles. One of the widely used new physics context to describe such mediator is Dark Photon, which in the most simplified scenario introduces an extra $U(1)$ symmetry
predicting a new gauge field (X) and a corresponding new vector boson (i.e. Dark Photon $A^{\prime}$) and extending the SM $U(1)_{em}$ symmetry as $U(1)_{em}\times U(1)_X$.
Such theory context is naturally and easliy renormalizable and gauge invariant while the predicted experimental phenomena can be very straightforward. The free parameters in a simplified scenario can be
just the kinetic mixing constant ($\epsilon$) and mass ($m_{A^{\prime}}$).

\section{General Picture of Worldwide Research}

Over the past few decades, worldwide efforts are thoroughly and tremendously spent on exploring the mystery of the dark sector with not only the well-known direct search experiments (e.g. CDEX, LZ, PandaX, XENONnT, etc.) and the indirect search experiments in the space (e.g. DAMPE, HESS, IceCube, etc.), but also the general purpose collider experiments including both the energy frontier hadron collider experiments at LHC and the intensity frontier lepton collider experiments such as BES-III@BEPCII and Belle-II@SuperKEKB. As part of the accelerator based Dark Matter Search, collider experiment are in particular important and sensitive for the searches in the sub-GeV Dark Matter mass range as shown in Fig.~\ref{fig:ColliderSensitivity}.

\begin{figure}[h]
\centering
\includegraphics[width=0.8\linewidth]{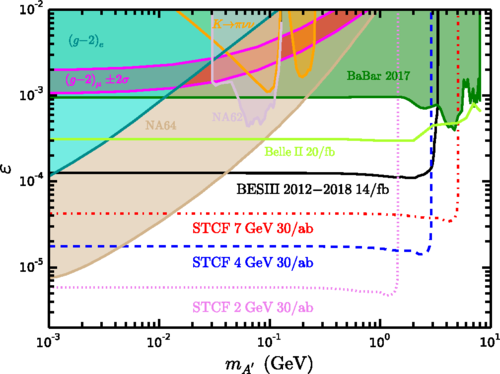}
\caption{\label{fig:ColliderSensitivity}~The expected 95\% C.L. exclusion limits on dark photon $A^\prime$ mixing parameter $\epsilon$ as a function of the mass $m_{A\prime}$ at present and future lepton colliders.~\cite{bib:DP-ColSensi}}
\end{figure}

When focusing on the well motivated dark photon searches representing the dark matter searches via probing the dark mediators,  One may easily anticipate that as more general purpose experiments of high energy colliders, the search sensitivities are in general limited due to the relatively higher collider center-of-mass energies designed for other energy frontier and intensity frontier physical requirements of precision measurements and other beyond Standard Model searches. To further explore the lower mass range of sub-X00 MeV, we would still live with accelerator based experiments with the emphasis on the fixed-target experiments. As shown in Fig.~\ref{fig:ColliderSensitivity}, the NA64 experiment~\cite{bib:NA64}, as one of the representatives for the fixed-target experiments, has the typically more competitive search sensitivities in MeV ~ 100 MeV mass range.

Many fixed-target experiments such as NA64 presently running (shown in Fig.~\ref{fig:NA64-sketch}) and LDMX~\cite{bib:LDMX} under proposal (shown in Fig.~\ref{fig:LDMX-Det}) typically search for dark photon via dark bremsstrahlungs through electron-on-target processes in electron beam running modes and in addition s-channel/t-channel annihilations in positron running modes. The effective center-of-mass energy of such experiments usually allows for sub-X00MeV dark photon probes. While the world 

\begin{figure}[h]
\centering
\includegraphics[width=0.8\linewidth]{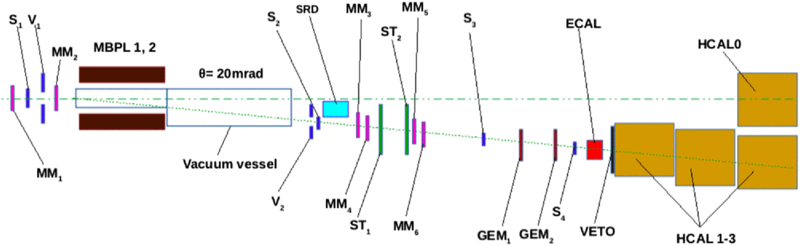}
\caption{\label{fig:NA64-sketch}~The schematic illustration of the NA64 experiment setup to search for dark photon $A^{\prime}$.~\cite{bib:NA64}}
\end{figure}

\begin{figure}[h]
\centering
\includegraphics[width=0.8\linewidth]{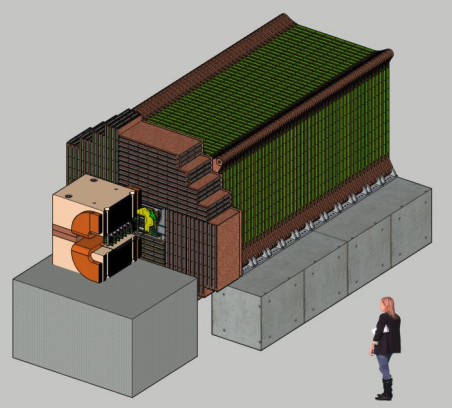}
\caption{\label{fig:LDMX-Det}~The prototype-like illustration of the LDMX experiment under proposal to search for dark photon $A\prime$ invisible decays.~\cite{bib:LDMX}}
\end{figure}

While there are many other fixed-target experiments worldwide such as DarkQuest~\cite{bib:DarkQuest} (as shown in Fig.~\ref{fig:DarkQuest}), DarkLight~\cite{bib:DarkLight} (as shown in Fig.~\ref{fig:DarkLight}) and DarkMESA~\cite{bib:DarkMESA} (as shown in Fig.~\ref{fig:DarkMESA}), the efforts to search for dark mediators to probe light dark matter candidate particles are broadly conducted. The Shanghai high repetition rate XFEL and extreme light facility (SHINE), presently under construction, offers the high repetition rate electron beam to probe the dark mediator and dark sector particles through an independent effort with dedicated setups to maximize the potential for future dark photon searches. Figure~\ref{fig:CompECAL} and~\ref{fig:CompHCAL} summarize the design and parametric comparisons between DarkSHINE calorimeter systems and NA64/LDMX experiments~\cite{bib:KSKhaw-ExpComp}. In order to achieve a better energy resolution for recoiled electron measurement with the ECAL, we choose the LYSO crystal as the baseline recipe for ECAL design with the z-depth up to ~40 $X_0$. To enhance the neutral hadron and muon background vetoing power aiming for better fake missing energy background rejection, we deploy the baseline design of HCAL as a scintilation sampling calorimeter with large volume so as to better contain the backgrounds, which will potentially in the future be revisited and further optimized so as to better fit the infrastructure restrictions such as the geometric dimensions and weighting limitations. Table~\ref{tab:CompTracker} summarizes the design and parateric comparisons of their tracking system. Both DarkSHINE and LDMX experiments employ precision silicon strip detectors to accurately measure the momentum of incident electrons and outgoing charged particles, allowing for the precise reconstruction of the missing momentum carried away by the dark photon. In contrast, the NA64-muon experiment uses trackers in conjunction with a calorimeter to reconstruct the energy of outgoing particles.

\begin{figure}[h]
\centering
\includegraphics[width=0.8\linewidth]{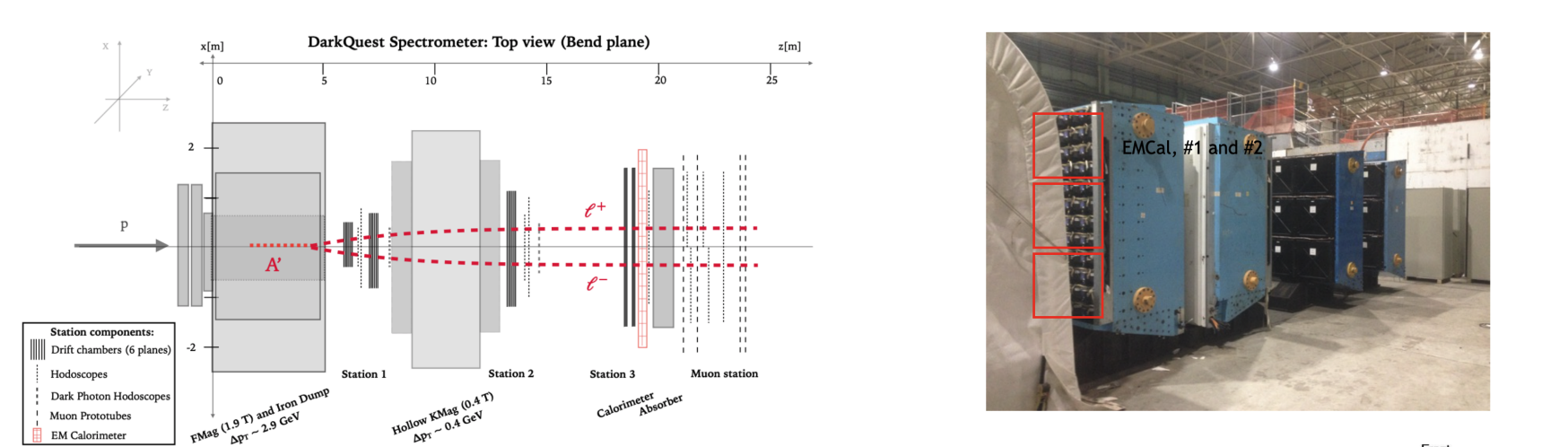}
\caption{\label{fig:DarkQuest}~The prototype-like illustration of the DarkQuest experiment under proposal to search for dark photon $A^{\prime}$.~\cite{bib:DarkQuest}}
\end{figure}

\begin{figure}[h]
\centering
\includegraphics[width=0.8\linewidth]{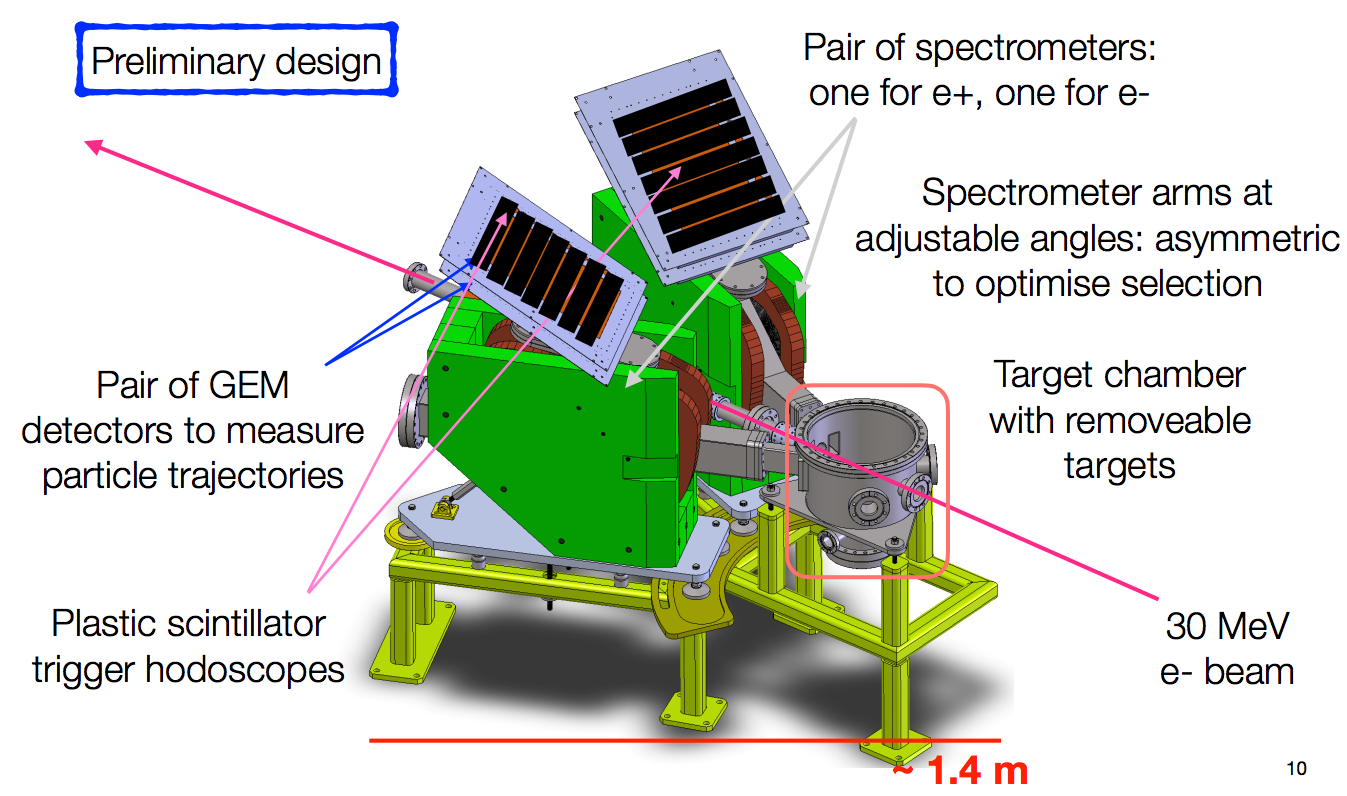}
\caption{\label{fig:DarkLight}~The prototype-like illustration of the DarkLight experiment under proposal to search for dark photon $A^{\prime}$.~\cite{bib:DarkLight}}
\end{figure}

\begin{figure}[h]
\centering
\includegraphics[width=0.8\linewidth]{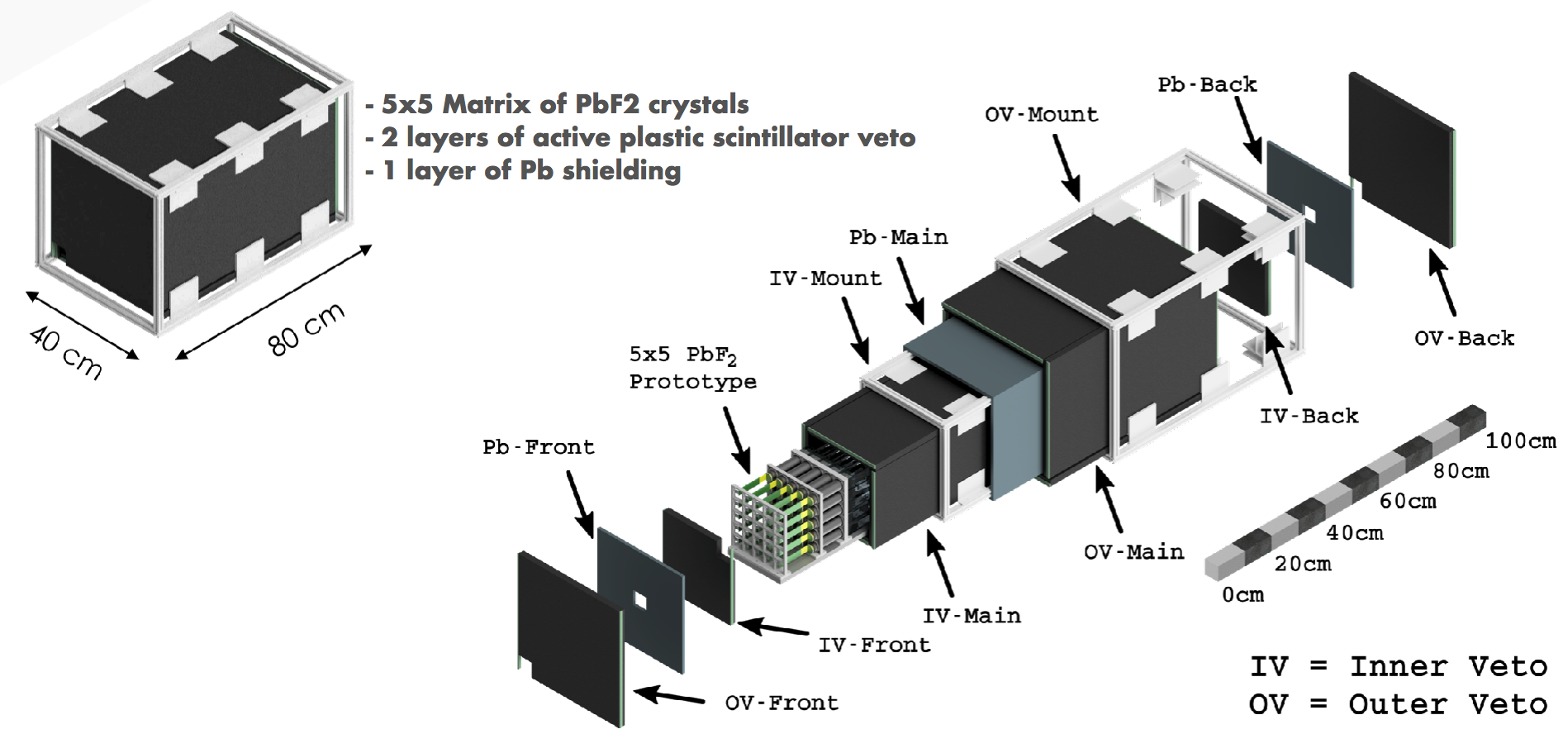}
\caption{\label{fig:DarkMESA}~The prototype-like illustration of the DarkMESA experiment under proposal to search for dark photon $A^{\prime}$.~\cite{bib:DarkMESA}}
\end{figure}

\begin{figure}[h]
\centering
\includegraphics[width=0.8\linewidth]{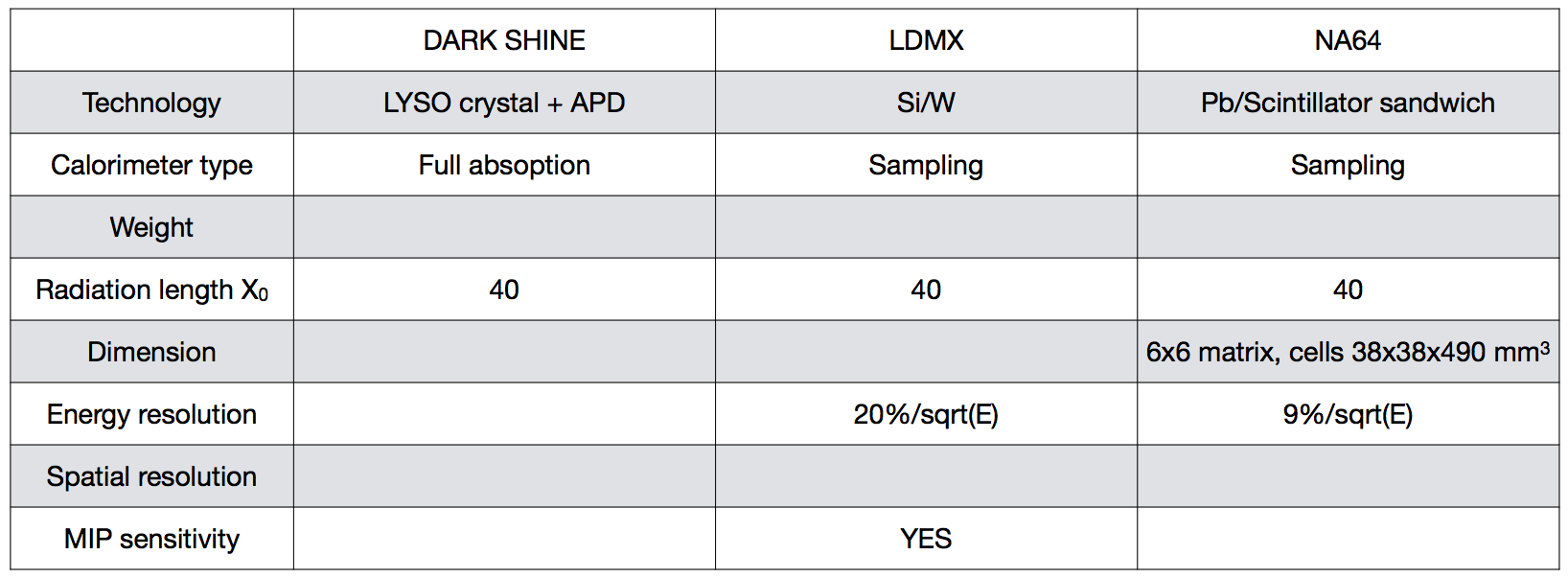}
\caption{\label{fig:CompECAL}~The Experimental setup comparison between DarkSHINE ECAL and NA64/LDMX.~\cite{bib:KSKhaw-ExpComp}}
\end{figure}

\begin{figure}[h]
\centering
\includegraphics[width=0.8\linewidth]{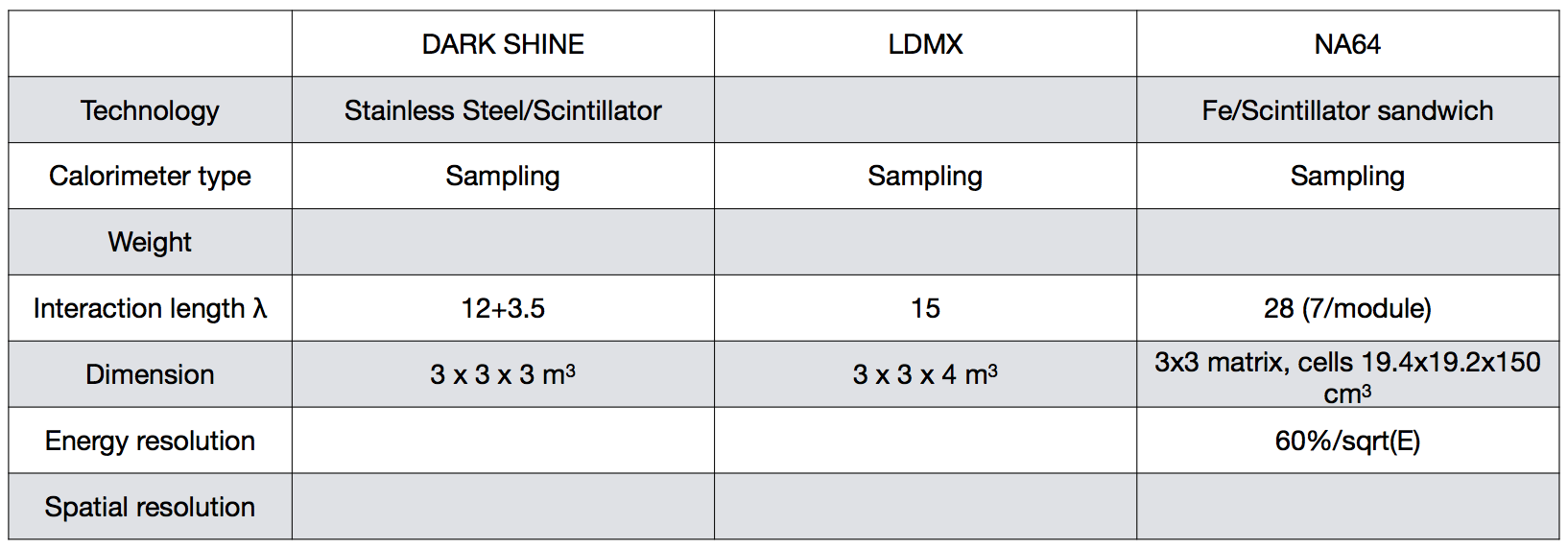}
\caption{\label{fig:CompHCAL}~The Experimental setup comparison between DarkSHINE HCAL and NA64/LDMX.~\cite{bib:KSKhaw-ExpComp}}
\end{figure}

\begin{table}[ht]
\centering
\begin{tabular}{|l|c|c|c|}
\hline
\textbf{Feature}                     & \textbf{DarkSHINE}                    & \textbf{LDMX}                     & \textbf{NA64-muon}                      \\ \hline
\textbf{Tagging Tracker Type}         & Silicon strips                                & Silicon strips                           & Micromegas + Straw-tubes         \\ \hline
\textbf{Pitch}                        & $30 \sim 50 \mu m$                                  & 30 \textmu m                              & Micromegas + Straw-tubes  \\ \hline
\textbf{Number of Layers}             & 7 layers                                      & 7 layers                                 & 4 Micromegas + 2 Straw-tubes                                       \\ \hline
\textbf{Active area}      & $4\times 10 cm^2$                                  & $4\times 10 cm^2$                               & $20\times 20 cm^2$                           \\ \hline
%\textbf{Resolution (Horizontal)}      & 10 \textmu m                                  & 6 \textmu m                               & 100 \textmu m                           \\ \hline
%\textbf{Resolution (Vertical)}        & 60 \textmu m                                  & 60 \textmu m                              & -                                    \\ \hline
%\textbf{Timing Resolution}            & 50 ps                                            & -                                        & 15 ns (Micromegas) / 5 ns (Straw-tubes)   \\ \hline
\textbf{Recoil Tracker Type}          & Silicon strips                                & Silicon strips                           & Micromegas + GEM  \\ \hline
\textbf{Pitch (Recoil Tracker)}       & $30\sim50 \mu m$                                  & 30 \textmu m                              & Micromegas + Straw-tubes   \\ \hline
\textbf{Number of Layers (Recoil)}    & 6 layers                                      & 6 layers                                 & 3 Micromegas + 2 Straw-tubes + 4 GEM   \\ \hline
\textbf{Active area}      & $4\times 10 cm^2$                                  & $4\times 10 cm^2$                               & $20\times 20 cm^2$  \\ \hline
%\textbf{Resolution (Horizontal)}      & 10 \textmu m                                  & 6 \textmu m                               & 115 \textmu m                            \\ \hline
%\textbf{Resolution (Vertical)}        & 60 \textmu m                                  & 60 \textmu m                              & -                       \\ \hline
\end{tabular}
\caption{\label{tab:CompTracker}Summary of tracker designs for DarkSHINE, LDMX~\cite{bib:LDMX}, and NA64-muon~\cite{NA64:2024nwj} experiments.}
\end{table}

\section{DarkSHINE Physics Program}

As inspired by the global efforts to disclose the Dark Matter mystery in general, the proposed DarkSHINE experiment initiative aims to introduce an independent search for the dark matter interaction's vector portal mediator: Dark Photon. Generically, Figure~\ref{fig:DarkPhotonProduction} and~\ref{fig:DarkPhotonDecays} show the four leading production modes of Dark Photon and the major decay products.

The Dark Photon ($A'$) couples to the electromagnetic current via kinetic mixing between the SM hypercharge and the $A'$ field strength tensor. The anticipated coupling strength is factorized by $ \varepsilon $, the kinetic mixing term relative to that of the photon by a factor. Such mechanism manages to provide a portal through which dark photons can interact with SM.
The Lagrangian of such kinetic mixing mechanism can be presented as follows: $$L = L_{sm}+\varepsilon F^{\mu \nu}F'_{\mu \nu}+\frac{1}{4}F'^{\mu \nu}F'_{\mu \nu}+m_{A'}^{2}A'^{\mu}A'_{\mu}, $$
where $m_{A'}$ represents the dark photon mass, $A'_{\mu}$ represents the dark photon field, and $F'_{\mu \nu}$ represents the field strength tensor. The value of $ \varepsilon $ ranges from $10^{-8}$ to $10^{-2}$ for each mass point to be examined. The relative values between $m_{A'}$ and $2m_{\chi}$ drives the Dark Photon decay modes whether to the SM particles or the dark sector products. In the simplified Dark Photon theory context, three free parameters (the kinetic mixing parameter $\varepsilon$, the dark photon mass $m_{A'}$, and the decay branching ratio of the dark photon into invisible dark sector), are usually considered to be complete enough to describe the Dark Photon characteristics.
In the baseline design of DarkSHINE experiment, we focuses on the scenario of [$m_{A'}$, $\varepsilon$] parameter space to be surveyed and characterized in the experimental phenomena, assuming the decay branching ratio of dark photons to DM as 100\% for further simplified treatment.

Figure~\ref{fig:DarkPhotonProduction} shows four leading production modes of Dark Photon: bremsstrahlung through electron-on-target events ($eZ\rightarrow eZA'$ and $pZ\rightarrow pZA'$), annihilation at $e^{+}e^{-}$ collider ($e^{+}e^{-}\rightarrow A'\gamma$), meson decays (e.g., $\pi^{0}\rightarrow A'\gamma$ or $\eta\rightarrow A'\gamma$ for dark photons with $m_{A'} < m_{\pi,\eta}$), and Drell--Yan process ($q\bar{q}\rightarrow A'$).
The bremsstrahlung is taken as the benchmark process for Dark Photon production for Dark SHINE, with the cross section values varying as a function of $m_{A'}$ and $\varepsilon$:
%\textcolor{blue}{$$\frac{d\sigma}{dx_e} = 4\alpha^3\varepsilon^2\xi\sqrt{1-\frac{m^2_{A'}}{E^2_e}}\frac{1-x_e+\frac{x^2_e}{3}}{m^2_{A'}\frac{1-x_e}{x_e}+m^2_ex_e},$$}
$$\frac{d\sigma}{dx_e} = 4\alpha^3\varepsilon^2\xi\sqrt{1-\frac{m^2_{A'}}{E^2_e}}\frac{1-x_e+\frac{x^2_e}{3}}{m^2_{A'}\frac{1-x_e}{x_e}+m^2_ex_e},$$
where $\varepsilon$ is the kinetic mixing parameter mentioned above and $x_e = E_{A'}/E_e$ is the fraction of the incoming electron's energy carried by the dark photon. The effective flux of photons $\xi$ is given by $$\xi(E_e,m_{A'},Z,A) = \int^{t_{\rm max}}_{\rm t_{\rm min}}dt\frac{t-t_{\rm min}}{t^2}G_{2}(t),$$
where $t_{\rm min} = (m^2_{A'}/2E_e)^2$, $t_{\rm max} = m^2_{A'}$, and the electric form factor $G_2(t)$ consists of elastic and an inelastic contribution, both depend on the atomic number Z and mass A.

The dark photon decays into DM (i.e., ``invisible decay'') if it is kinetically allowed, i.e., $m_{A'} > 2m_{\chi}$. Otherwise, the dark photon will decay into visible SM final states. Figure~\ref{fig:DarkPhotonDecays} shows the leading Feynman diagrams for the corresponding processes. The partial widths of dark photon invisible decay used in this study are given by the following:
%\textcolor{blue}{$$ \Gamma(A'\rightarrow\chi\bar{\chi}) = \frac{1}{3}\alpha_D m_{A'}\sqrt{1-\frac{4m^2_{\chi}}{m^2_{A'}}}(1+\frac{2m^2_{\chi}}{m^2_{A'}}), $$}
$$ \Gamma(A'\rightarrow\chi\bar{\chi}) = \frac{1}{3}\alpha_D m_{A'}\sqrt{1-\frac{4m^2_{\chi}}{m^2_{A'}}}(1+\frac{2m^2_{\chi}}{m^2_{A'}}), $$
where $\alpha_D$ is the dark sector fine structure constant. Assuming $\varepsilon = 1$, Figure~\ref{fig:cross-section} shows the cross section of a dark photon as a function of $m_{A'}$ for an incident electron with an energy of 8~GeV, according to the designed beam energies of the SHINE facility.
To maximize the searched sensitivities, the DarkSHINE focuses on searching for Dark Photon Decays into invisibles (i.e. the dark matter candidate particles with masses lighter than the Dark Photon itself). This is a hypothetically pre-assumption based on the facts that we anticipate the production of Dark Photon is a rare process as long as it has never been discovered experimentally so that the kinetic mixing parameter strength would not be a large factor. Despite it is a pre-assumption, it still introduces a natural scenario that is coherent with all the known/observed phenomena throughout the worldwide experimental frontiers.

\begin{figure}[H]
\centering
\subfigure{
\includegraphics[width=3.5 cm]{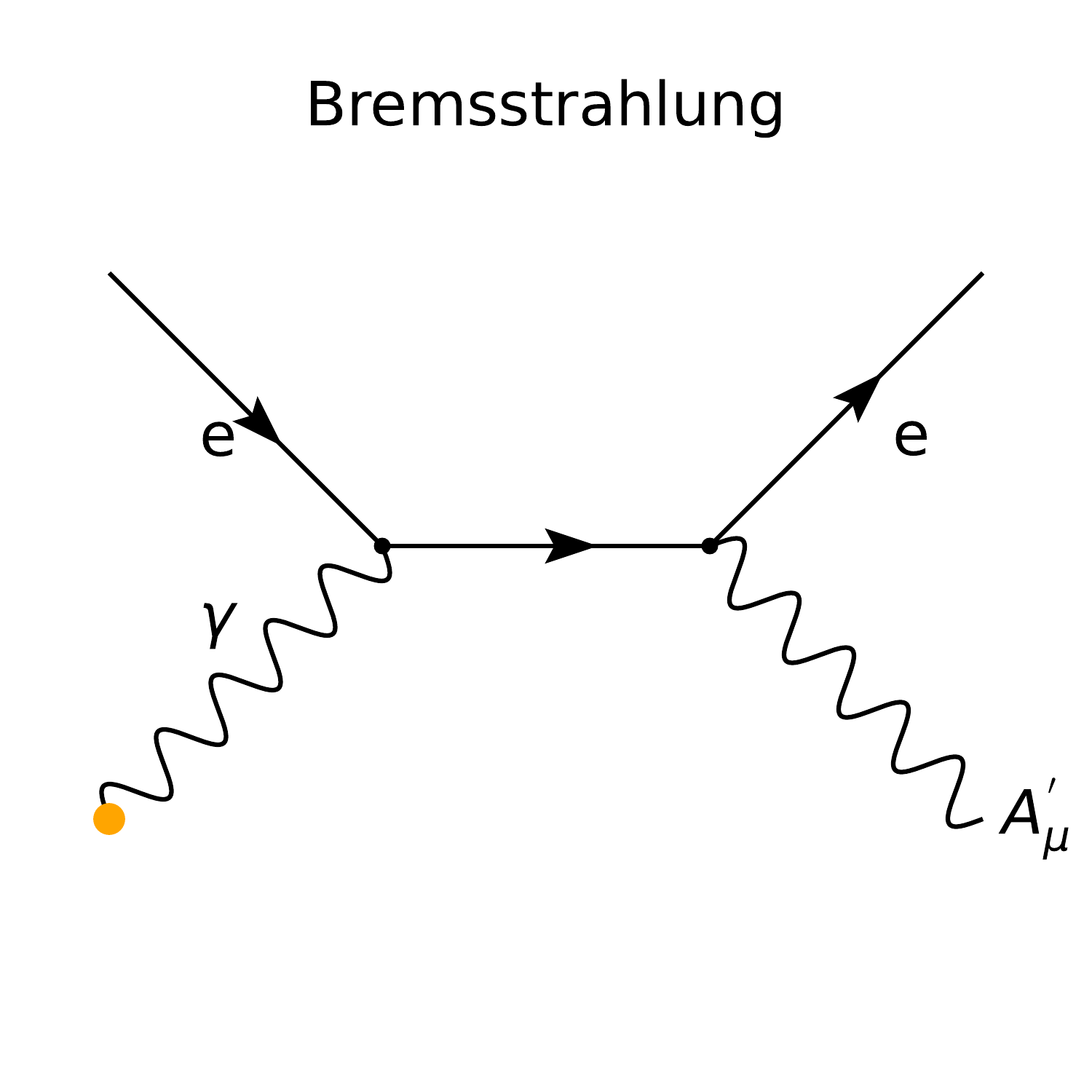}
}
\subfigure{
\includegraphics[width=3.5 cm]{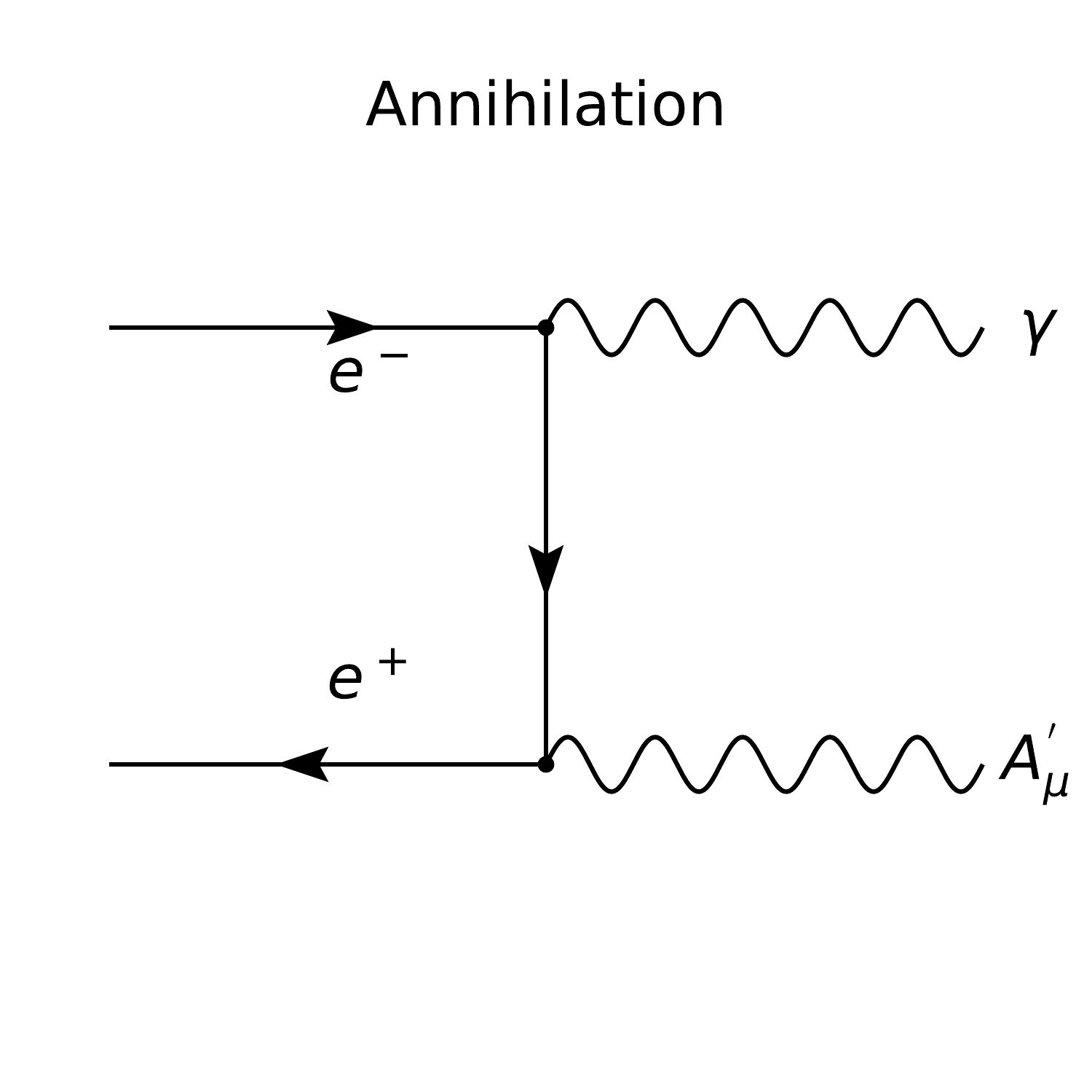}
}
\\
\subfigure{
\includegraphics[width=3.5 cm]{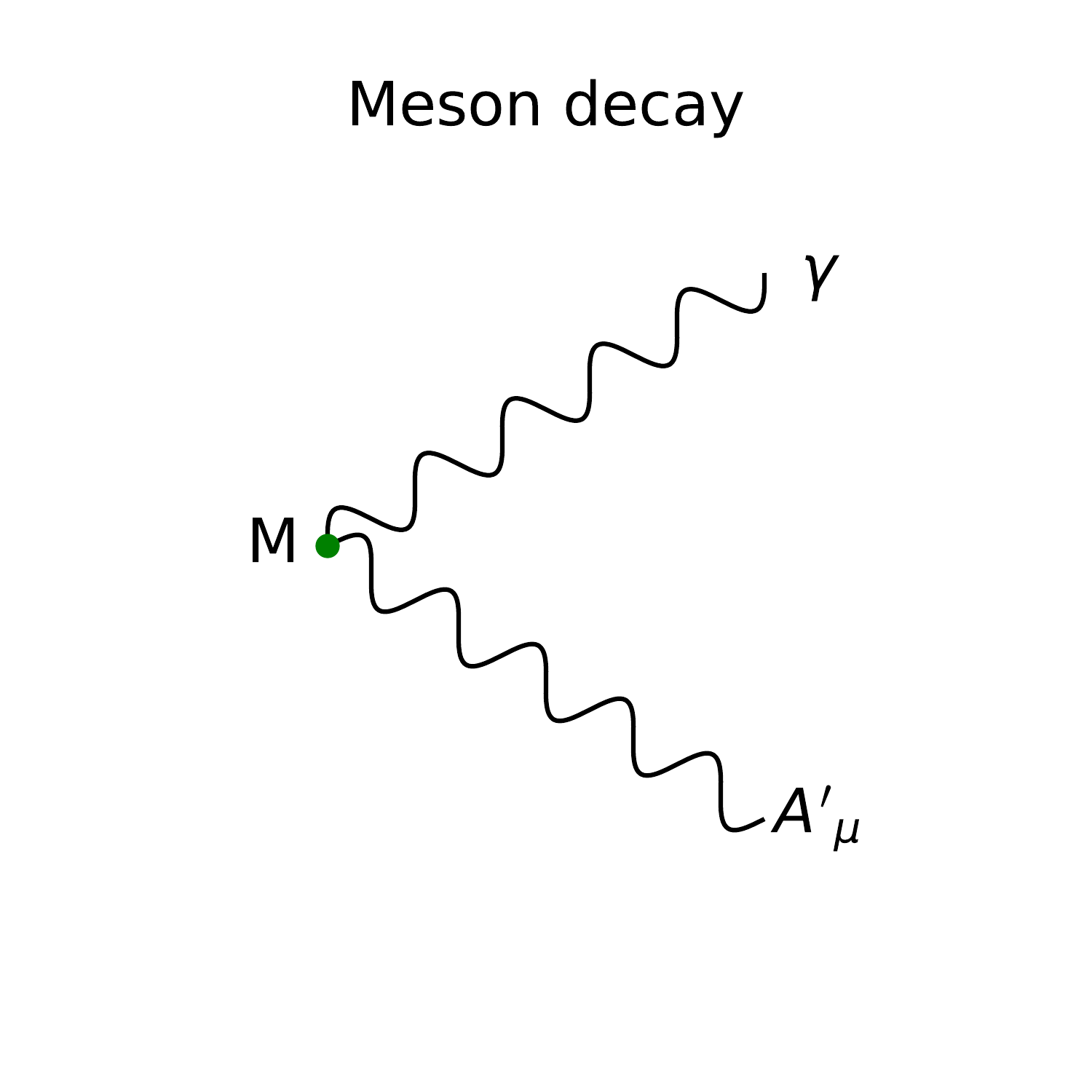}
}
\subfigure{
\includegraphics[width=3.5 cm]{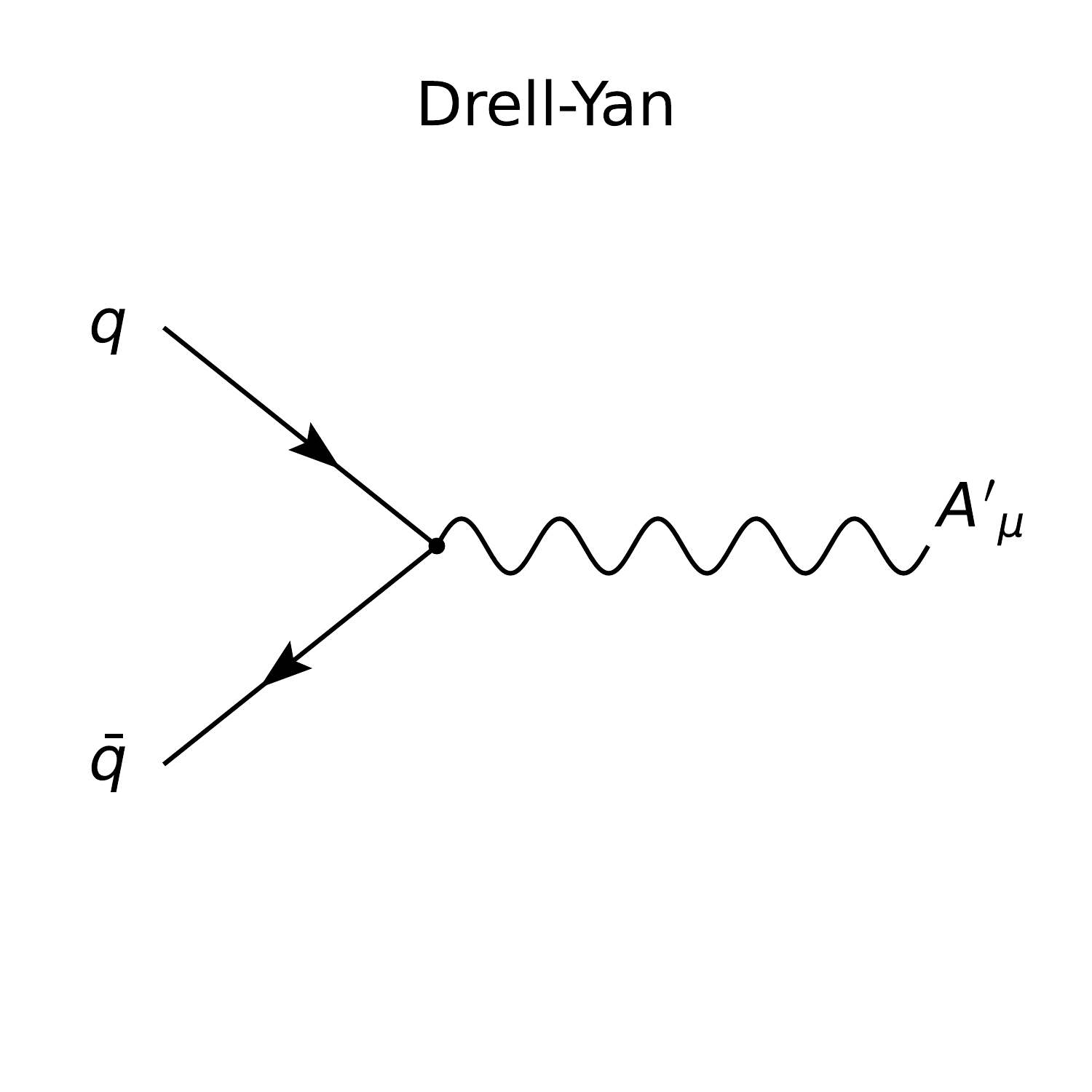}
}
\caption{(Color online) Production of dark photons: bremsstrahlung, annihilation, meson decay, and Drell-Yan.~\cite{Fabbrichesi:2020wbt}}
\label{fig:DarkPhotonProduction}
\end{figure}

\begin{figure}[H]
\centering
\subfigure{
\includegraphics[width=3.5 cm]{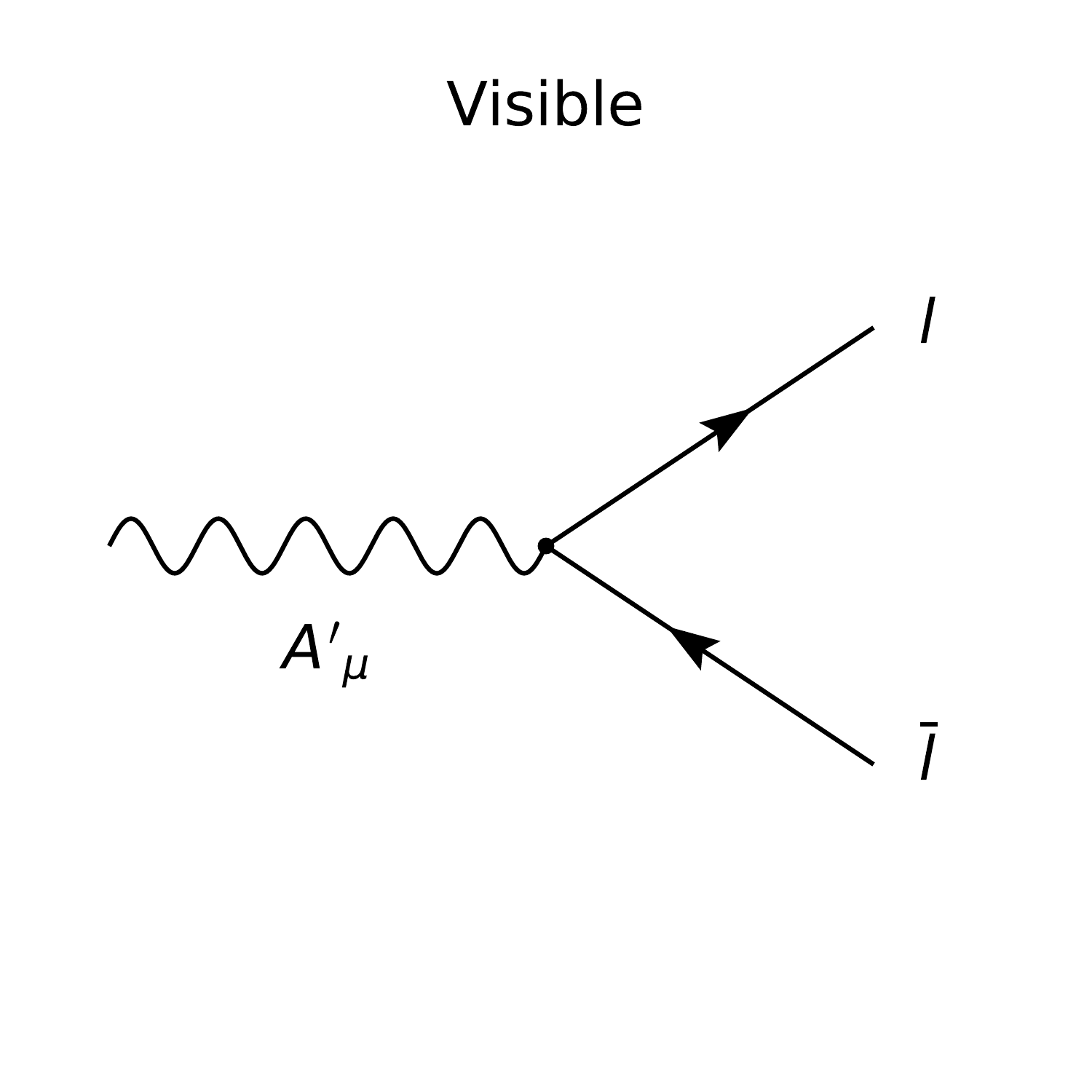}
}
\subfigure{
\includegraphics[width=3.5 cm]{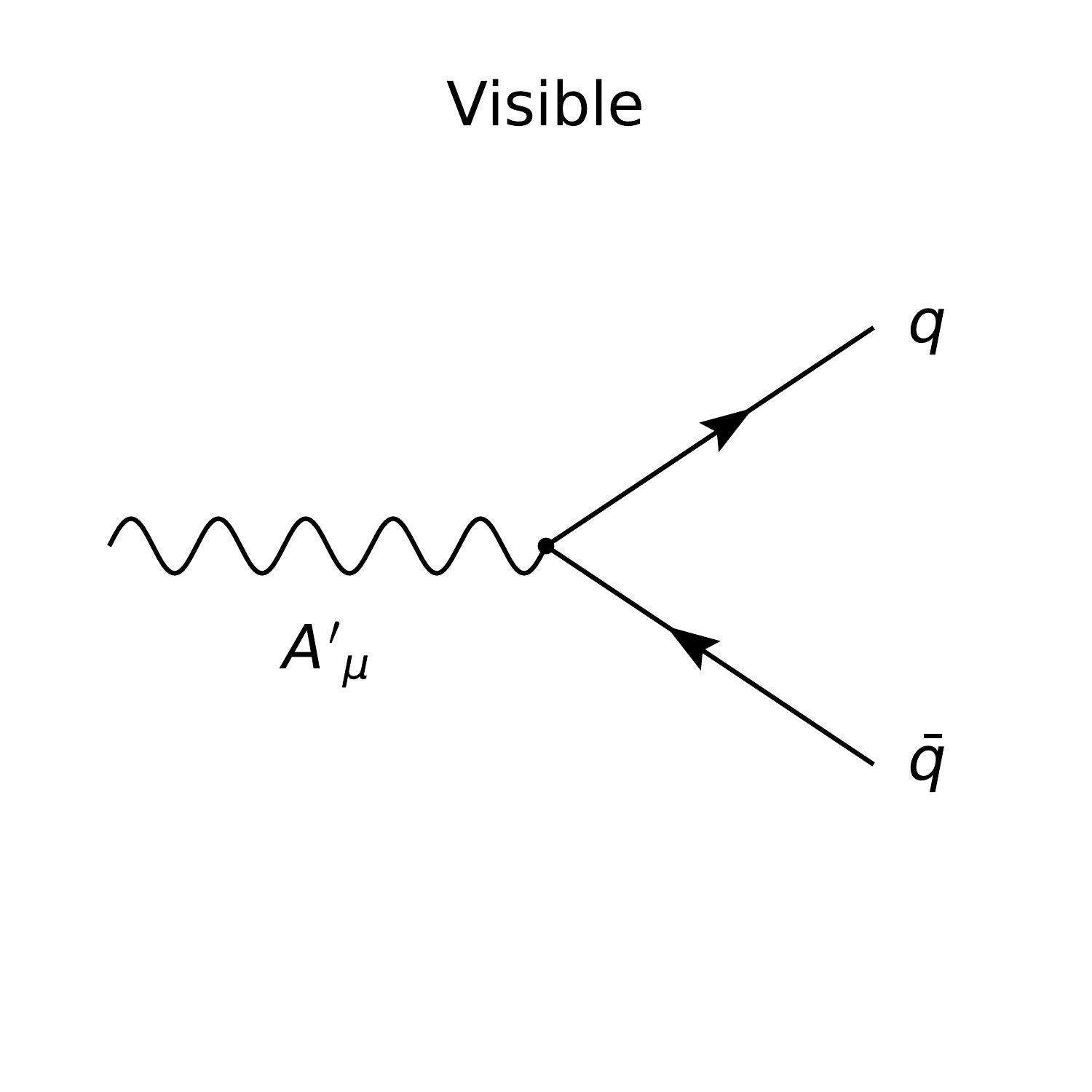}
}
\\
\subfigure{
\includegraphics[width=3.5 cm]{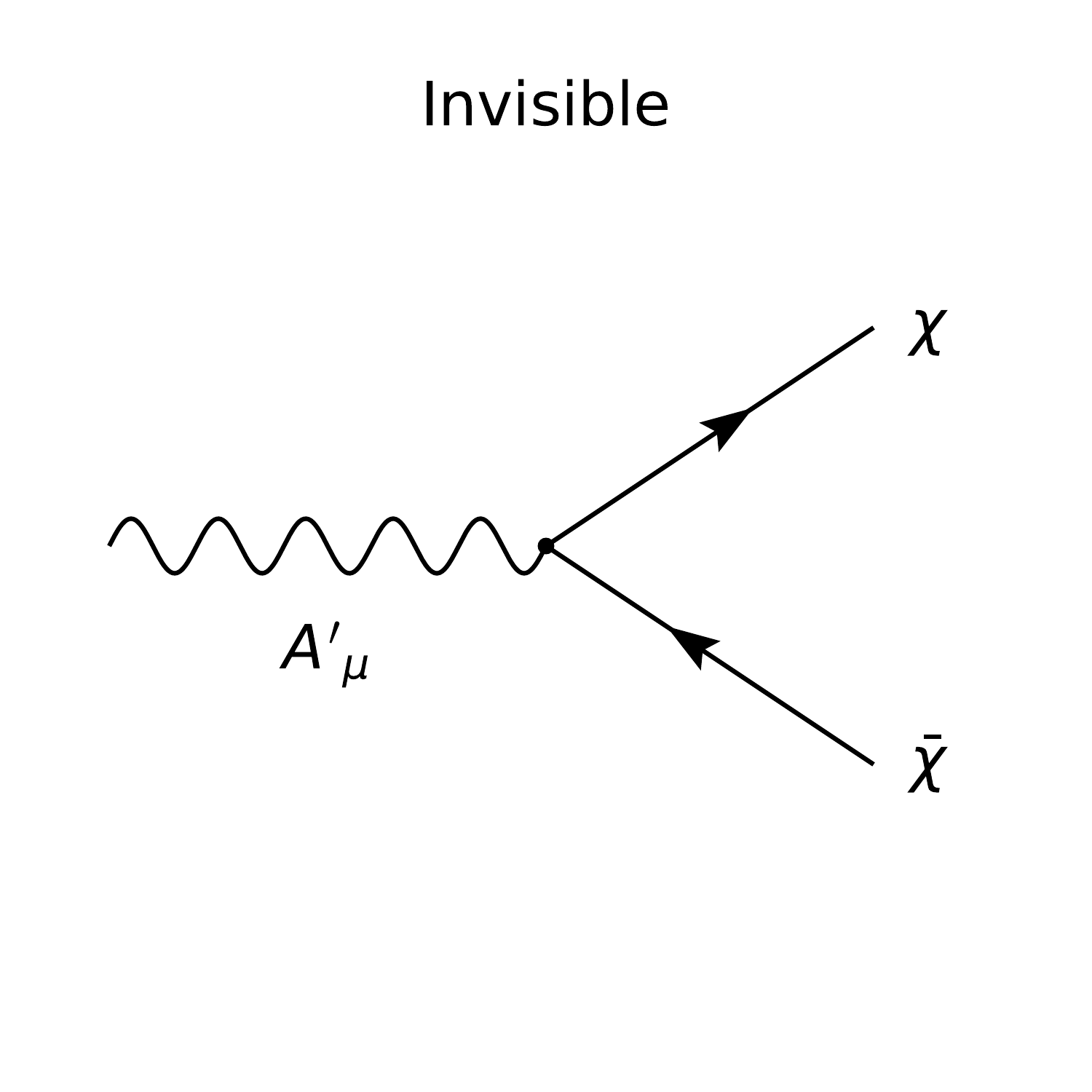}
}
\caption{Decay of the massive dark photon into visible (SM leptons or hadrons) and invisible (DM) modes.~\cite{Fabbrichesi:2020wbt}}
\label{fig:DarkPhotonDecays}
\end{figure}

\bib

\end{chapter}   

\begin{chapter}{Detectors}

\section{Introduction to DarkSHINE Beamline}

\begin{figure*}[h]
\centering
\includegraphics[width=15.0 cm]{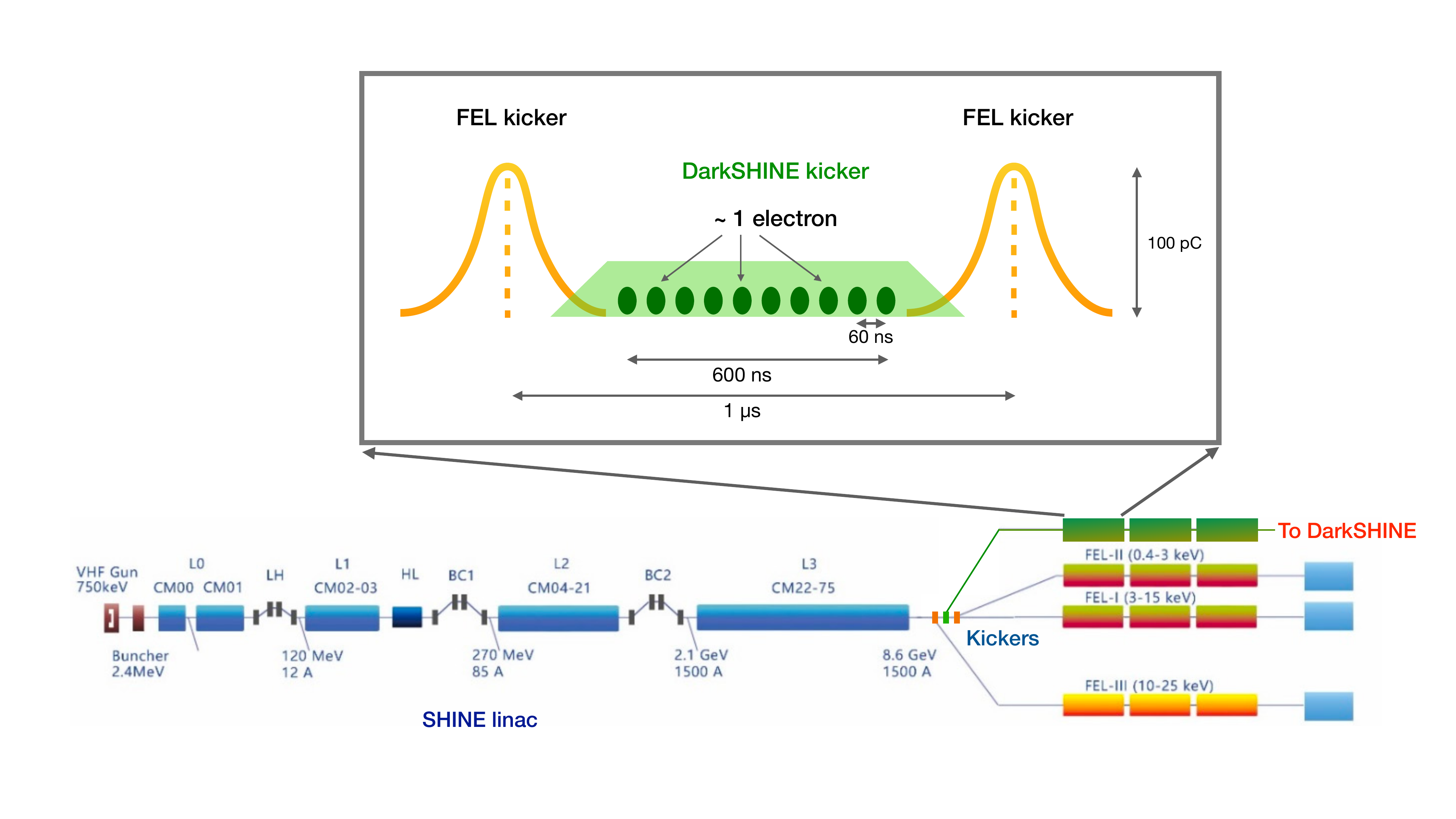}
\caption{(Color online) Illustration of the SHINE linac~\cite{Zhao:2018lcl}, the FEL kicker system, and the DarkSHINE kicker~\cite{Nosochkov:2017xoc}. There are $1.3\times10^{9}$ buckets per second provided by the 1.3~GHz microwave of the accelerator. For every 1300 buckets, 100~pC of electrons are placed into one bucket to be accelerated (corresponding to 1~MHz). For the DarkSHINE experiment, an additional laser with lower energy is used to produce a single electron bunch. For each empty bucket, one electron will then be put in. The accelerated electrons will be distributed by different kicker systems, i.e., FEL-I, II, III, and DarkSHINE~(green).}
\label{fig:kicker}    
\end{figure*}

As a newly proposed fixed-target experiment initiative, DarkSHINE utilizes the high repetition rate single electron beam, to be explored and deployed by SHINE facility, so as to search for Dark Photon invisible decay signals through the hypothetical dark bremsstrahlung processes of electron-on-target (EOT) events. In order to maximize the statistical power accumulating sufficient number of electron-on-target events within a reasonably short period of machine time, a conceptual beamline design is proposed with technical feasibility elaborated in Figure~\ref{fig:kicker}. The high repetition rate at 1~MHz will lead to a competitive amount of Dataset at the order of $10^{13}$ EOT per year, substantiating a high intensity experiment to probe the Dark Photon mediating the interactions to the dark portal.

\section{Overview of Detector System}

The global design of the DarkSHINE detector system is illustrated in Figure~\ref{fig:detectorsketch}, which consists of the magnetic and tracking system, the tungsten target, the electromagetic calorimeter (ECAL) and the hadronic calorimeter (HCAL). The dimensions depicted are intended to facilitate the visualization of sub-detector system designs and do not correspond rigorously to their actual proportions. For accurate sizes, please see the mechanical system~\ref{sec:machanical}. The magnetic field and tracking system is introduced in Sec~\ref{sec:magnet} and \ref{sec:tracker}, which aims to detect and reconstruct precisely the trajectory of the incident/recoiled electrons. The electrons hit the tungsten target and are recoiled to be detected by the recoil tracker and the ECAL system, which are the key subsystems to give the precise information of reconstructed recoiled electron momentum and recoiled electron energy. The ECAL system~\ref{sec:ECAL} is designed to be a homogeneous crystal calorimeter with the LYSO\cite{1239590} being the crystal materials providing competitive electromagnetic resolution with high light yield and rapid scintillation decay time, so that the recoild electron energies could be precisely reconstructed and measured. The HCAL system~\ref{sec:HCAL} is a sampling scintillation calorimeter comprised of sandwiched plastic scintillators and iron absorbers while the sensitive layer consisting of plastic scintillator strips. HCAL plays a very important role to compensate the energy losses after incident electrons are recoiled with radiation by-products, in particular those inducing muons and neutral hadrons which come from bremsstrahlungs, photon-nulcear interactions and electron-nuclear interactions but easily fake into the missing momenta and missing energies. With the combined working systems of all these sub-detectors, DarkSHINE are ensured to be sensitive to precise momentum and energy measurement as well as missing momentum and energy detections. Consequentially, the Dark Photon invisible decay signals with missing momenta/energies will be hunted against the enormous SM backgrounds.

\begin{figure}[H]
\centering
\includegraphics[width=8.0 cm]{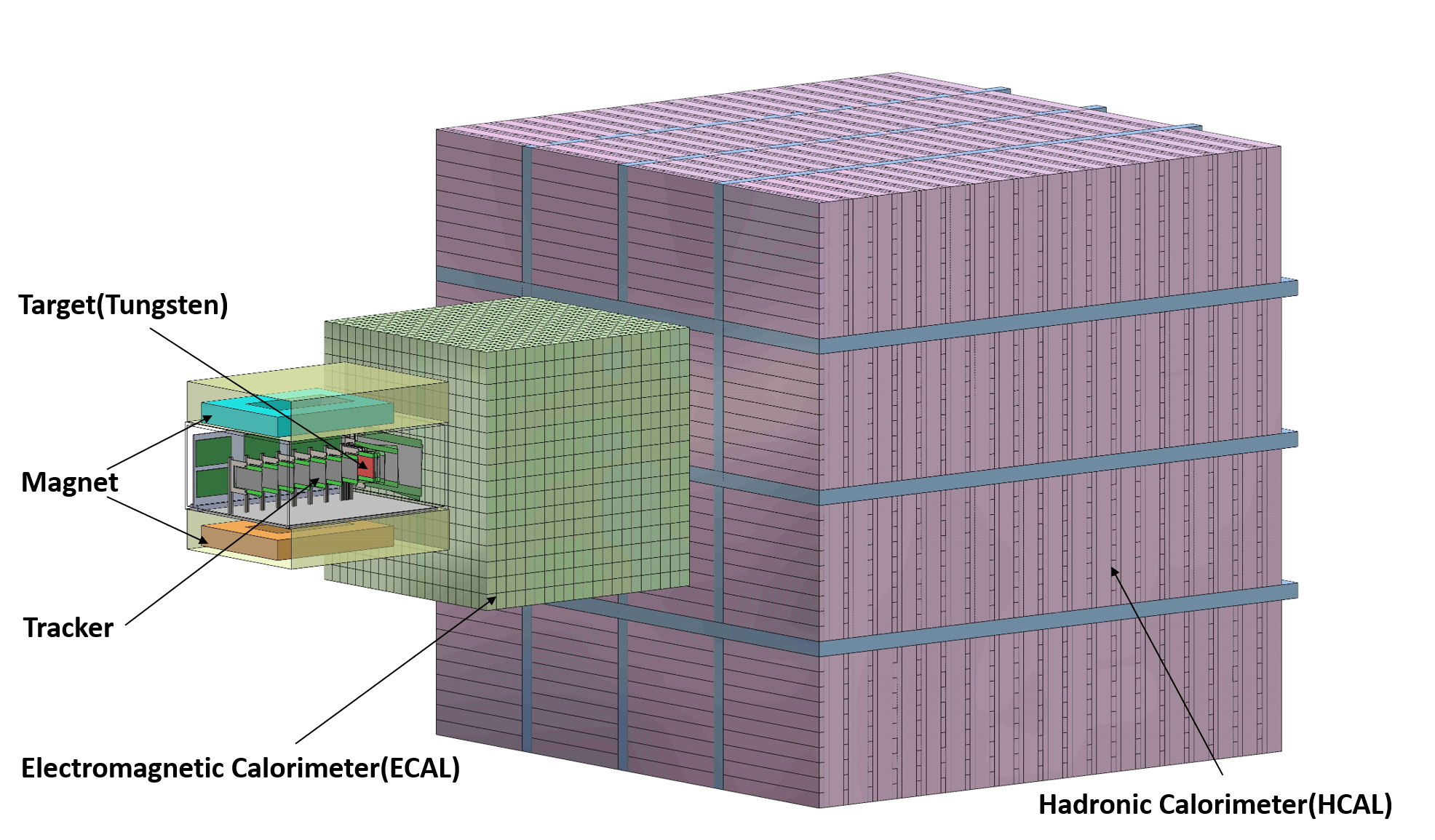}
\caption{(Color online) The detector sketch picture. Along the electron incident direction from left to right in the picture, the yellow and cyan square material is the dipole magnet. The tagging tracker is placed at the center of it. The recoil tracker is located at the edge of the magnet (1.5~T). The target is caught in the middle. ECAL is placed after the recoil tracker, followed by HCAL.}
\label{fig:detectorsketch}
\end{figure}

\section{Magnetic Field System}
\label{sec:magnet}

The design of the magnets for this experiment plays a critical role in achieving the desired magnetic fields for both the tagging and recoil trackers. For the tagging tracker, the key requirement is to maintain a uniform of 1.5 T magnetic field. This uniformity ensures that incoming electrons are bent in a predictable manner, allowing for precise reconstruction of their momentum. In contrast, the recoil tracker requires in-homogeneous magnetic field, where the magnetic field gradually reduces from 1.5 T. This non-uniformity serves a dual purpose: bending the recoil electrons while simultaneously keeping the acceptance for low-momentum electrons which typically in the range of 30 to 50 MeV. This tailored design optimizes the tracking performance for electrons across different energy ranges, ensuring accuracy in momentum reconstruction and data collection.

To achieve these magnetic field configurations, square Helmholtz coils positioned outside the tagging tracker are an optimal choice. These coils are capable of generating a uniform magnetic field within the tagging tracker while simultaneously creating a non-uniform field at the edges, which benefits the recoil tracker. As illustrated in Figure ~\ref{fig:MagneticStructure}, this configuration ensures that the necessary field uniformity is maintained for the tagging tracker, while the desired field gradient is achieved for the recoil tracker. The magnetic field map for this setup has been simulated, and the results are presented in Figure ~\ref{fig:MagenticFieldSimulation}, further validating the effectiveness of the design in meeting the experimental requirements. Typically, the magnetic field reduces to half at the end of recoil tracker, and then gradually drops to zero within the Calorimeter.

\begin{figure}[h]
\centering
\includegraphics[width=0.50\linewidth]{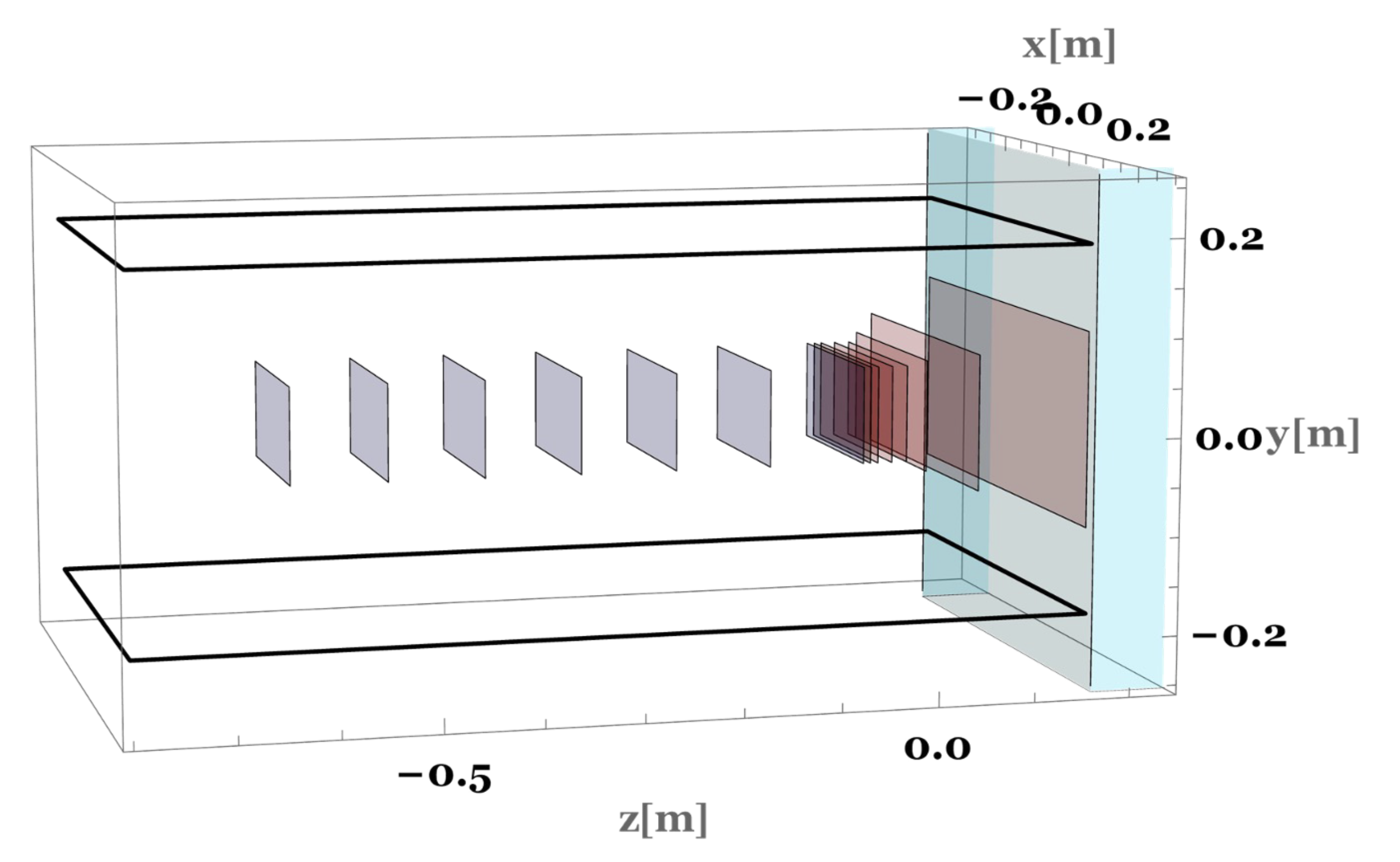}
\includegraphics[width=0.35\linewidth]{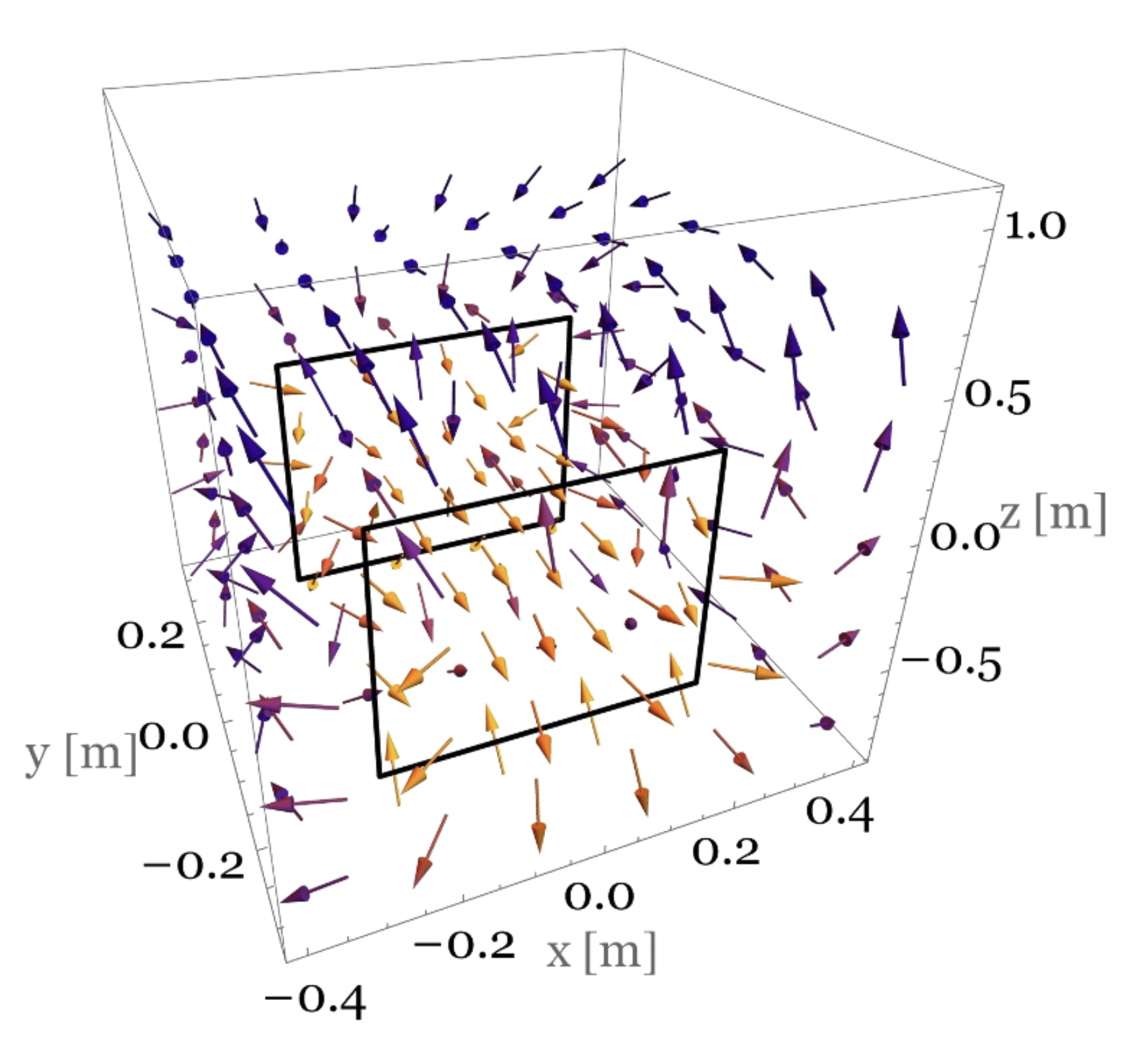}
\caption{\label{fig:MagneticStructure}. The left plot shows a skeleton of DarkSHINE tracker. The right plot shows the vector of Magnetic field provided by square Helmholtz coils.}
\end{figure}

\begin{figure}[h]
\centering
\includegraphics[width=0.52\linewidth]{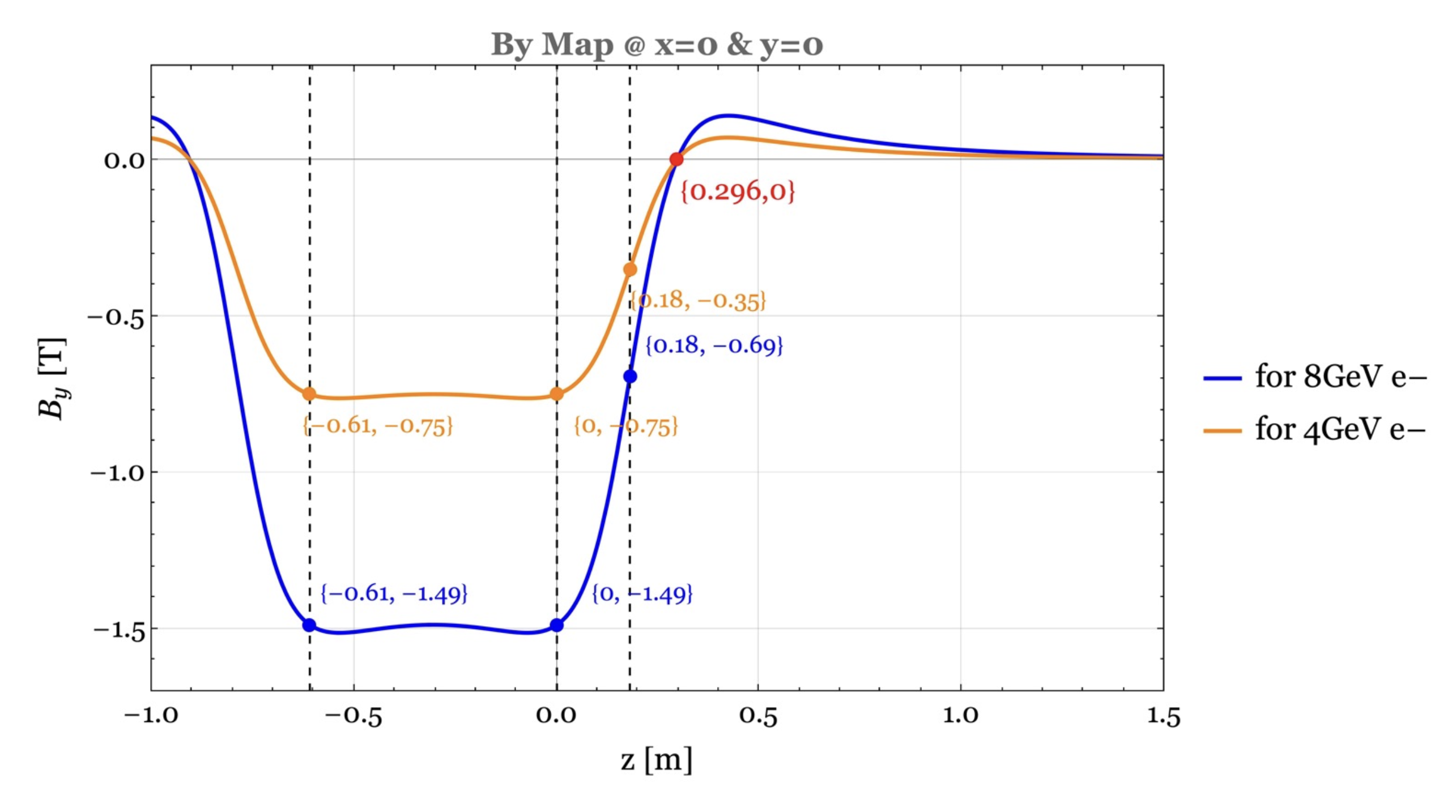}
\includegraphics[width=0.35\linewidth]{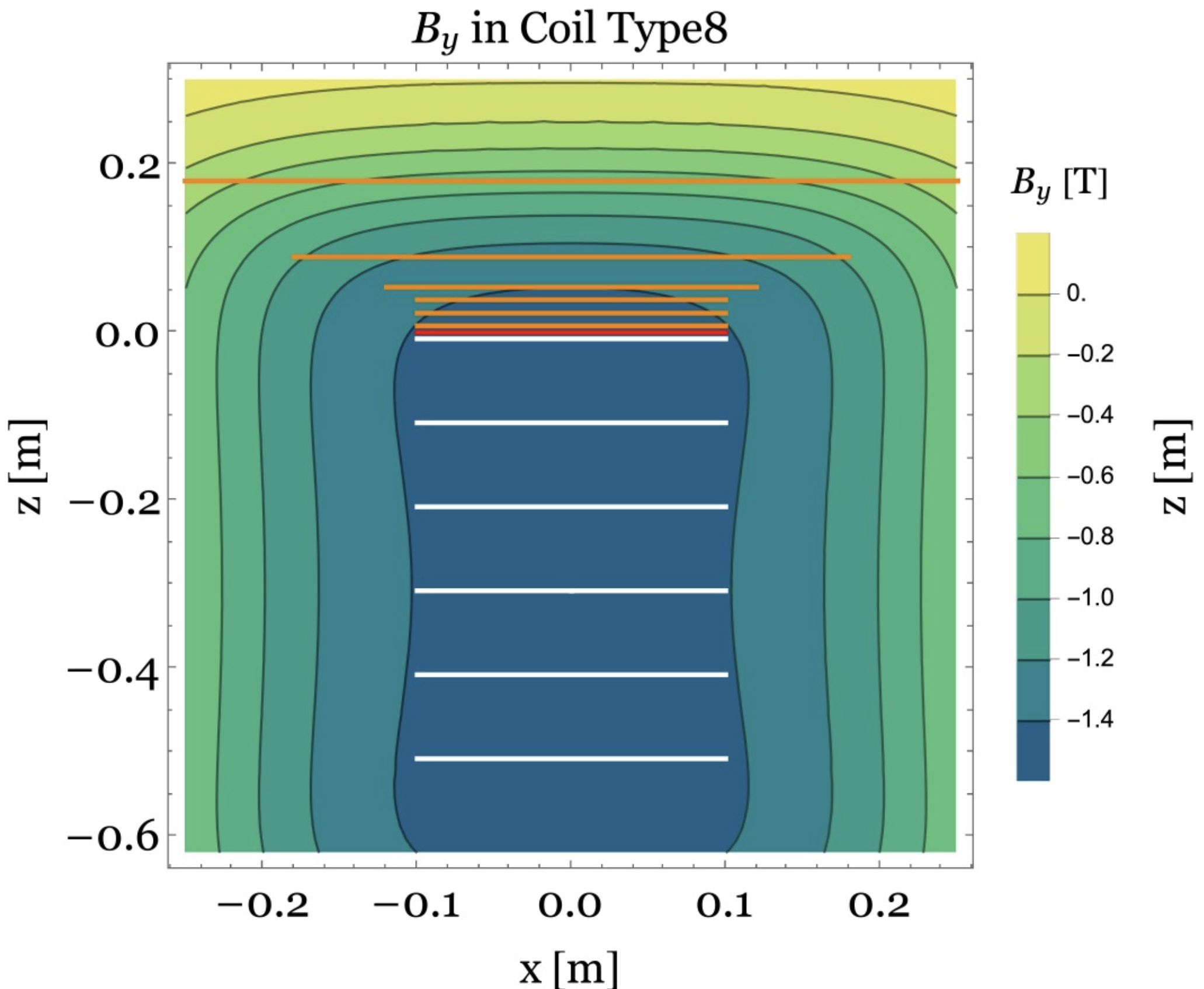}
\caption{\label{fig:MagenticFieldSimulation}The left plot shows magnetic field as a function distance along beam direction ($z$). The right plot shows 2D map of magnetic field in $x-y$ plane.}
\end{figure}

\section{Tracking System}
\label{sec:tracker}

In the DarkSHINE experiment, the primary goal is to search for the invisible decay of dark photons. The signature of missing momentum can be reconstructed from vector difference of momentum from incident electron and recoil electron. Thus, detecting such rare events requires a highly precise and efficient tracking system capable of identifying individual particle trajectories with exceptional spatial and timing resolution. Since DarkSHINE involves a 4 GeV electron beam impinging on a target, precise tracking of the outgoing particles is essential to suppress background events and isolate the subtle signatures of dark photon interactions. The tracking system must not only provide accurate positional information to reconstruct particle paths but also deliver precise timing measurements to distinguish between potential signal and background events in the challenging experimental environment. These requirements motivate the use of advanced detector technologies like AC-LGAD, which offer high granularity, excellent time resolution, and robust performance in high-energy physics experiments.

The tracking system for the DarkSHINE experiment must meet stringent performance criteria to ensure the successful detection of dark photon signals. One of the key requirements is achieving a position resolution better than 10 micrometers, which is crucial for accurately reconstructing the trajectory of particles emerging from electron-target interactions. This high spatial resolution is necessary to distinguish signal events from background noise, especially when dealing with low cross-section processes like dark photon production. Additionally, the recoil tracking system must be capable of reconstructing electrons with momenta ranging from as low as 30 MeV to as high as 2 GeV. This wide momentum range ensures that both low-energy recoils, which are critical for detecting soft electron emissions, and higher-energy particles, relevant for background discrimination, can be efficiently tracked. Meeting these requirements is essential for the overall success of the experiment, as it ensures that all relevant particle trajectories are captured and accurately reconstructed in the analysis.

Sketch of the DarkSHINE tracking system is shown in Figure~\ref{fig:TrackerSketch} left plot. It includes tagging tracker (electron reconstruction before target) and recoil tracker (charged particle reconstruction after target). Figure~\ref{fig:TrackerSketch} right plot shows AC-LGAD strips sensor. Key parameters of tagging and recoil tracker are provided in Table~\ref{tab:TrackerKeyParameters}.
\begin{figure}[h]
\centering
\includegraphics[width=0.50\linewidth]{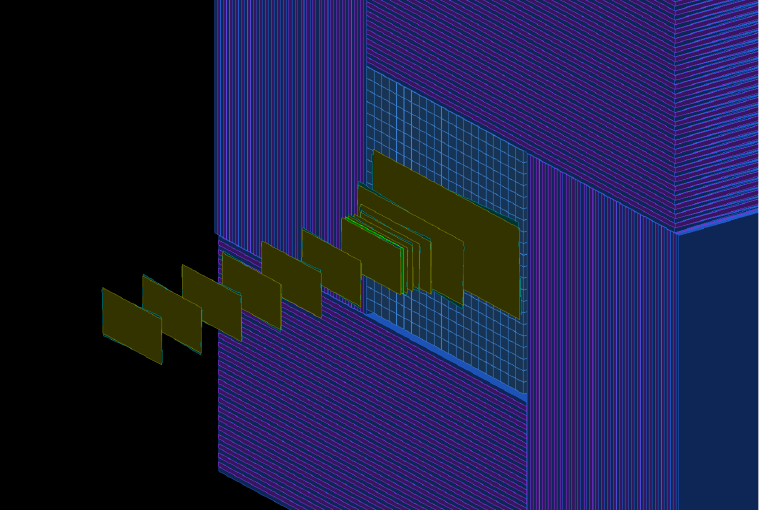}
\includegraphics[width=0.35\linewidth]{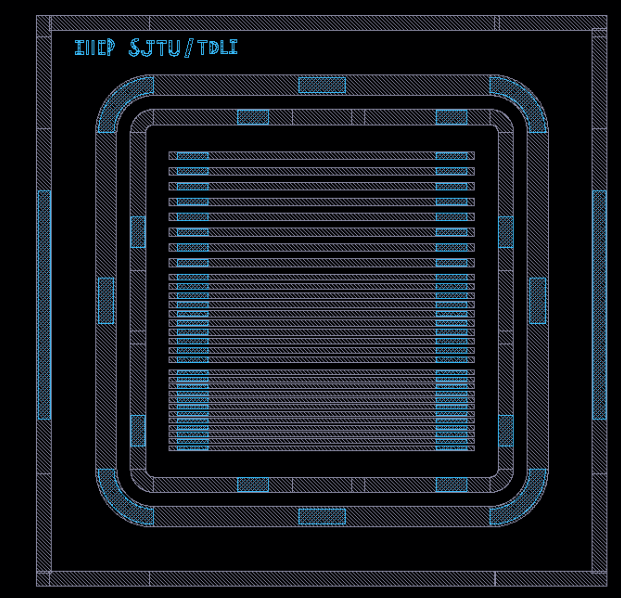}
\caption{\label{fig:TrackerSketch}~Left: Sketch of
the DarkSHINE tracking system. From left to right: tagging tracker (with seven layers of tracking module), the tungsten target, and the recoil tracker (with six layers of tracking module). For each layer, two strip sensors are shown in the sketch, placed at a small angle (100 mrad). Right: Design of the AC-LGAD strips sensor.}
\end{figure}

\begin{table*}[t]
\footnotesize
\caption{\label{tab:TrackerKeyParameters} Key parameters of Tracker design.}
\tabcolsep 14pt %space between two columns. 
\begin{tabular*}{0.974\textwidth}{c|c|ccc|c|c}
\toprule  
\centering
Node  & \multicolumn{1}{r|}{Centre (mm)} & \multicolumn{3}{c|}{Size (mm)} & Arrangement & Comments \\
    & \multicolumn{1}{c|}{z} & \multicolumn{1}{c}{x} & \multicolumn{1}{c}{y} & \multicolumn{1}{c|}{z} & & \\ \hline
    Tagging Tracker & -307.783 & 200   & 400   & 600.216 & 7 layers & \multicolumn{1}{p{8em}}{Second layer rotation: 0.1 rad} \\ \hline
Target & 0     & 100   & 200   & 0.35 &  & \\ \hline
Recoil Tracker & 94.032 & 500   & 800   & 172.714 & 6 layers & \multicolumn{1}{p{8em}}{Second layer rotation: 0.1 rad} \\
\bottomrule
\end{tabular*}
\end{table*}

\subsection{Tagging Tracker}

For the DarkSHINE experiment, the tagging tracker must deliver a spatial resolution of 10 micrometers to accurately track the trajectories of 4 GeV incoming electrons. This high-resolution capability is essential for distinguishing between signal and background events with the required precision.

The number of layers in the tagging tracker has been optimized based on the trade-off between incoming electron acceptance efficiency and momentum resolution. As illustrated in Figure~\ref{fig:TrackerTaggingNoLayer}, increasing the number of tagging tracker layers increases the momentum resolution for incoming electrons but simultaneously reduces the acceptance efficiency. To achieve an optimal balance between these parameters, a configuration of 7 layers has been selected for the tagging tracker. This configuration ensures an acceptance efficiency exceeding 99.5\% while maintaining a momentum resolution better than 2\%.

\begin{figure}[h]
\centering
\includegraphics[width=0.45\linewidth]{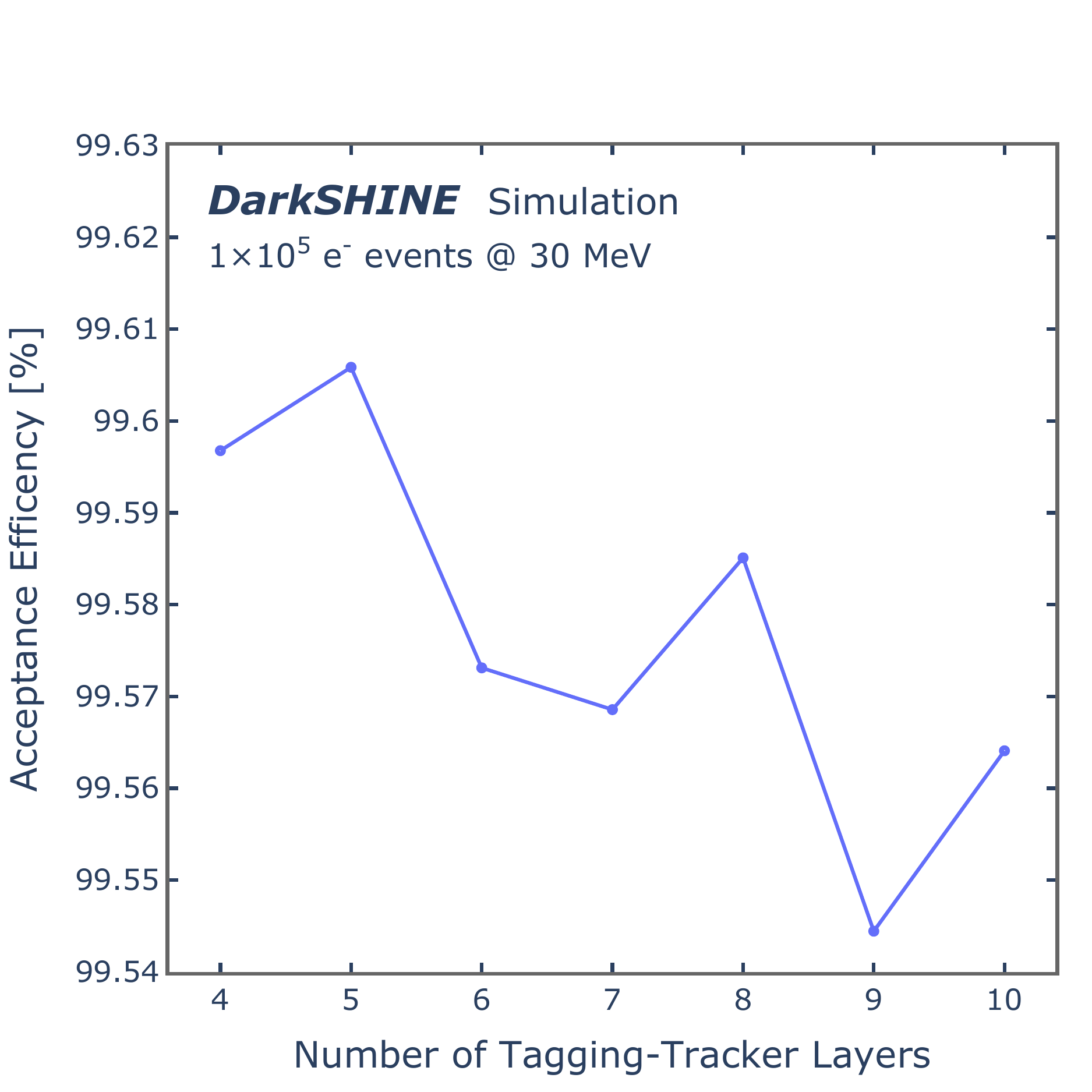}
\includegraphics[width=0.45\linewidth]{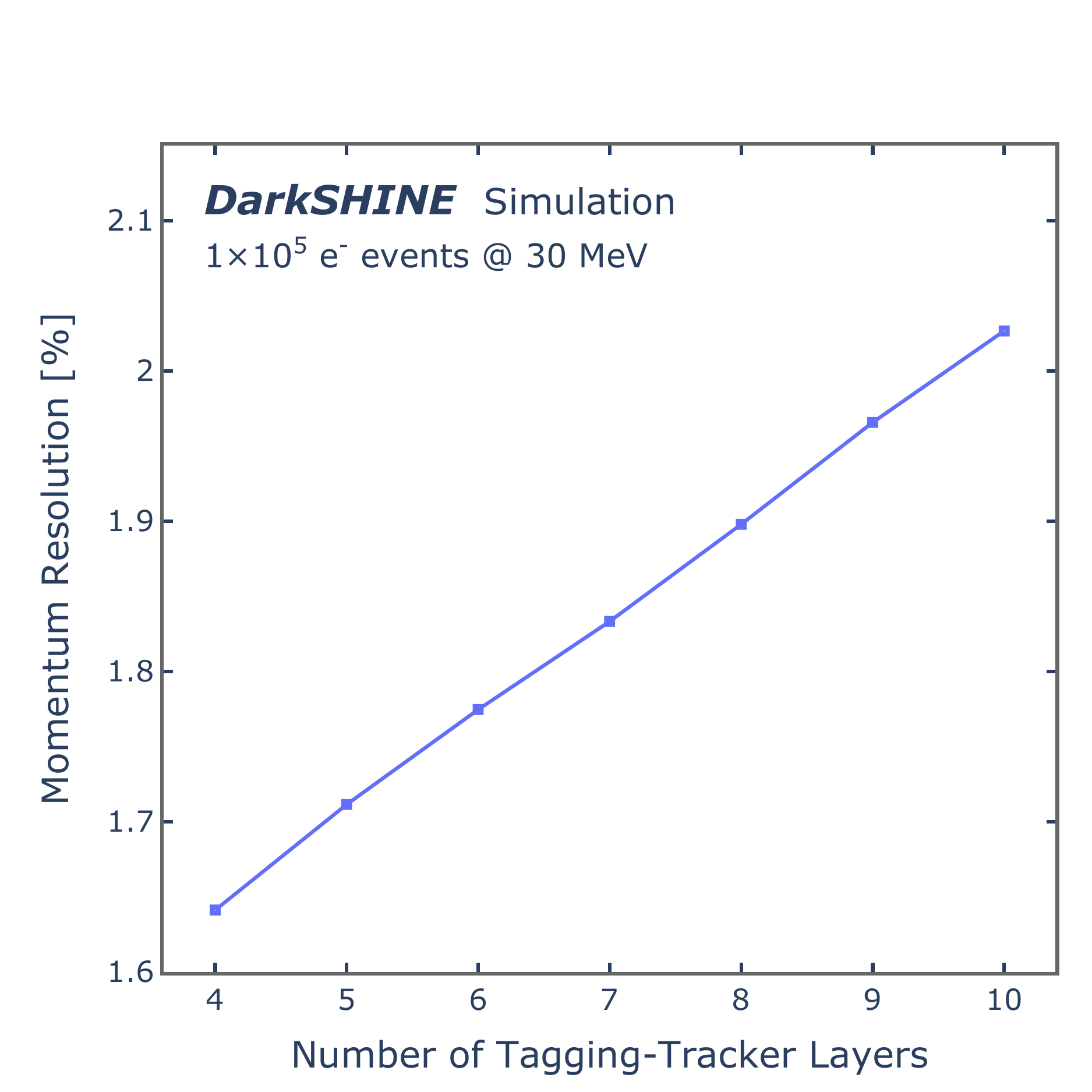}
\caption{\label{fig:TrackerTaggingNoLayer}~As a function of number of tagging tracker layers, the left plot shows acceptance efficiency, and the right plot shows momentum reconstruction resolution.}
\end{figure}

\subsection{Recoil Tracker}

The recoil tracker is designed to precisely reconstruct the trajectories of charged particles produced by the interaction of the incoming electron with the target. For the signal process, the primary outcomes are electrons with momenta typically ranging from 30 MeV to 2 GeV. In contrast, for background processes, the resulting charged particles are predominantly hadrons such as $\mu^{\pm}$, $\pi^{\pm}$/$K^{\pm}$ etc. The ability to accurately track and differentiate these particles is crucial for isolating potential dark photon signals from background events. The challenge is to achieving high acceptance efficiency for wide momentum range of the recoil electrons while also distinguishing them from the heavier background particles.

\begin{figure}[h]
\centering
\includegraphics[width=0.41\linewidth]{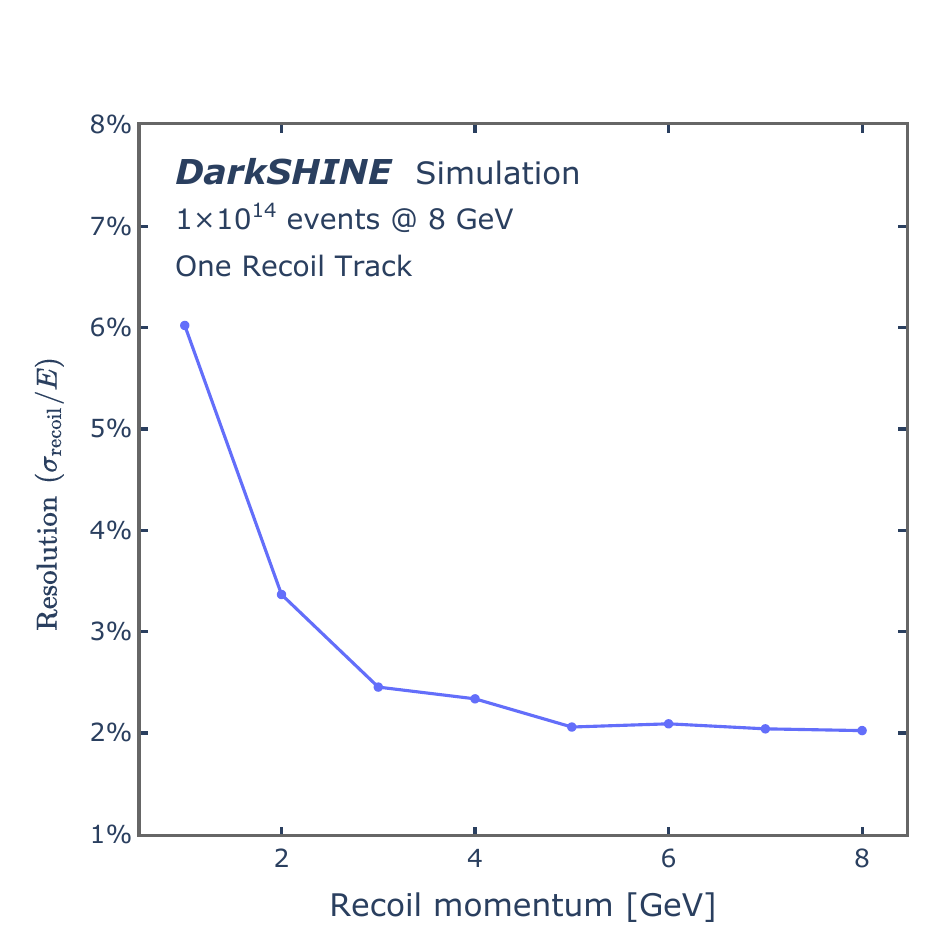}
\includegraphics[width=0.50\linewidth]{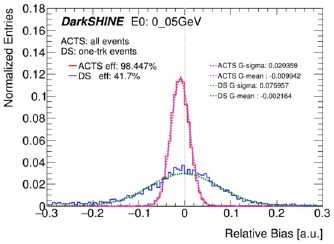}
\caption{\label{fig:TrackerRecoilMomentumReso}~Left: Recoil tracker momentum reconstruction resolution. Right: Comparison of resolution from two reconstruction algorithms, i.e.: ACTs and preliminary Kalman filtering method developed for the DarkSHINE experiment.}
\end{figure}

The experiment utilizes a non-uniform magnetic field in the recoil tracker, along with six layers of silicon strip detectors, optimized in both position and size. As shown in Figure~\ref{fig:TrackerRecoilMomentumReso}, the left plot illustrates the momentum reconstruction resolution as a function of the recoil electron momentum, while the right plot compares the momentum resolution achieved by two different algorithms. For the following study of this paper, the ACTs algorithm has been adopted due to its superior performance.

\begin{figure}[h]
\centering
\includegraphics[width=0.40\linewidth]{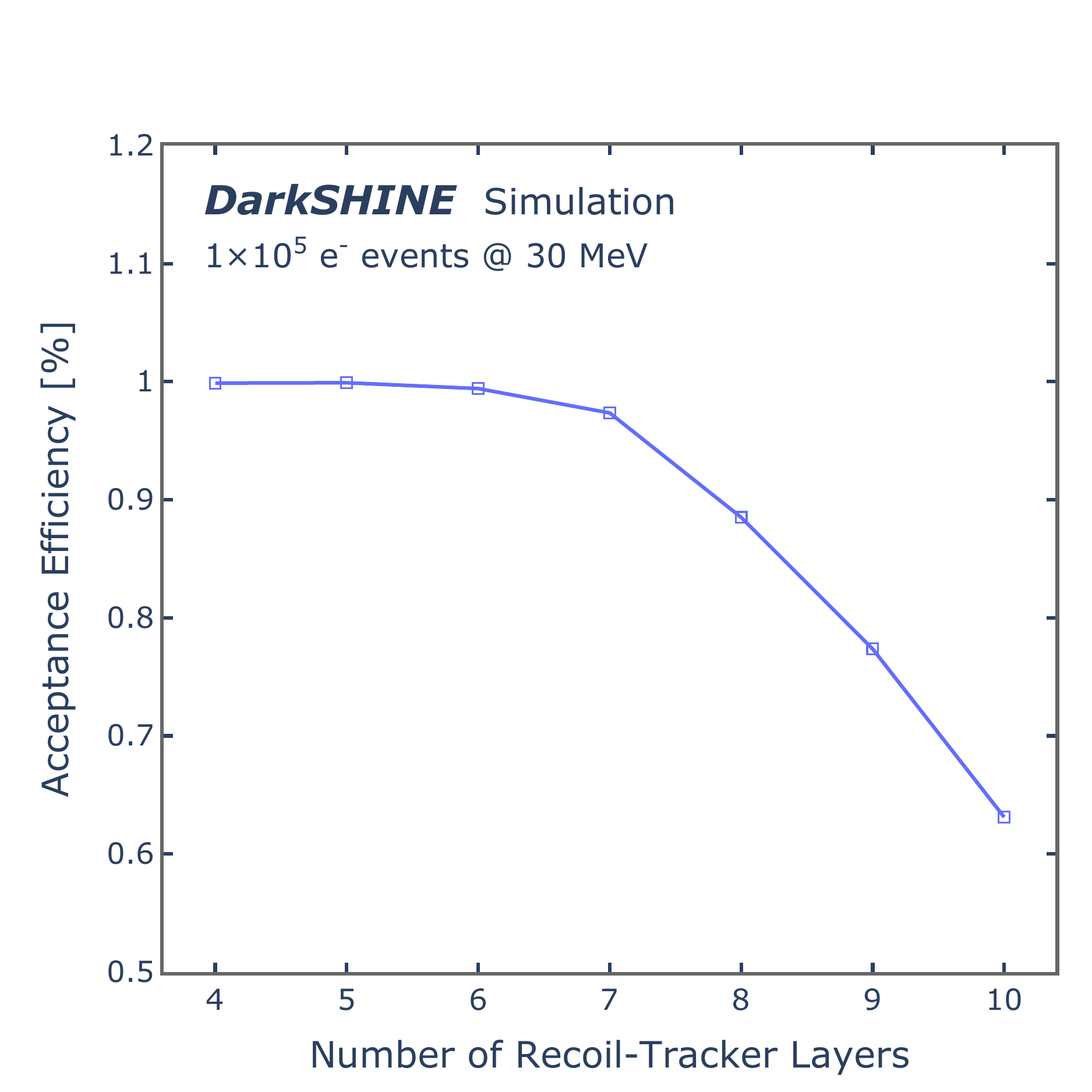}
\includegraphics[width=0.40\linewidth]{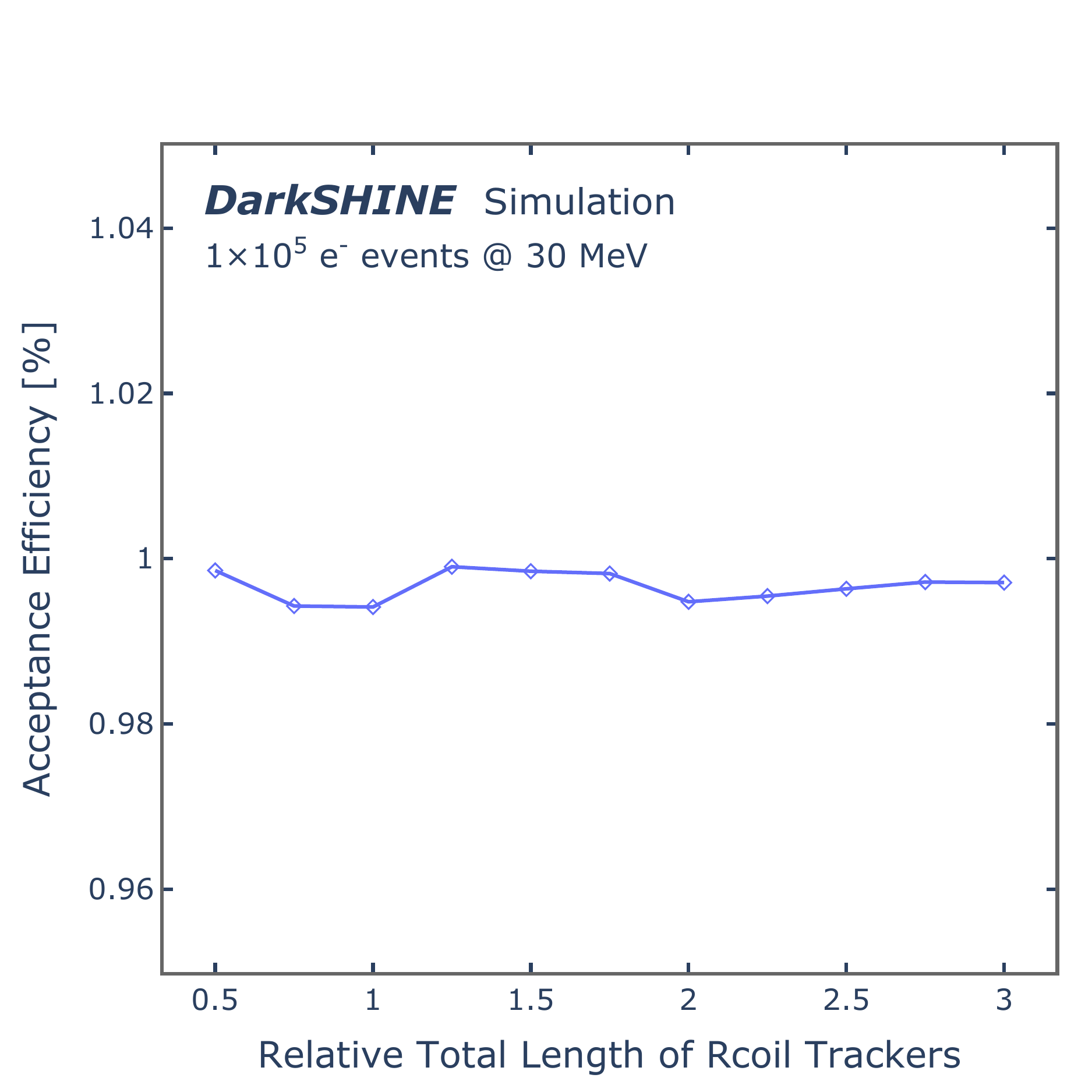}
\caption{\label{fig:TrackerRecoilOpt}~Left: Recoil tracker acceptance efficiency for 30 MeV electrons vs number of layers. Right: Recoil tracker acceptance efficiency for 30 MeV electrons vs total length (scaling to baseline design) of the recoil tracker.}
\end{figure}

\subsection{HCAL Expected performance}
\label{sec:HCALPerformance}

As stated in Section~\ref{sec:HCALIntroduction}, the HCAL must satisfy a veto inefficiency requirement of less than $10^{-5}$ for high-energy neutrons and less than $10^{-3}$ for low-energy neutrons, respectively. Figure~\ref{fig:baseline1VetoIneff} presents the evaluation of veto inefficiency by injecting neutrons with different energy. 

Events that are not rejected meet the following criterion.

\begin{itemize}
    \item total energy reconstructed in HCAL, $\mathrm{E_{HCAL}^{total}}<30$~MeV;
    \item maximum cell energy in HCAL, $\mathrm{E_{HCAL}^{MaxCell}}<0.1$~MeV.
\end{itemize}

This result includes multiple points representing incident neutron energies ranging from 100~MeV to 3~GeV, encompassing the energy spectrum depicted in Fig~\ref{fig:HCALneutron_energy}. To avoid conflicting with the logarithmic Y-axis, when there are no events remaining after the rejection, the veto inefficiency is set to $10^{-6}$ and only the bottom half of the error bar is displayed. This adjustment accounts for the fact that veto inefficiency cannot be precisely calculated within a range between $10^{-6}$ and 0 due to limitations imposed by sample statistics. It is evident that this design fulfills the physics requirements, with the veto performance primarily dependent on the front section for low-energy neutrons.

\begin{figure}[htb]
\centering
\includegraphics
  [width=0.5\hsize] {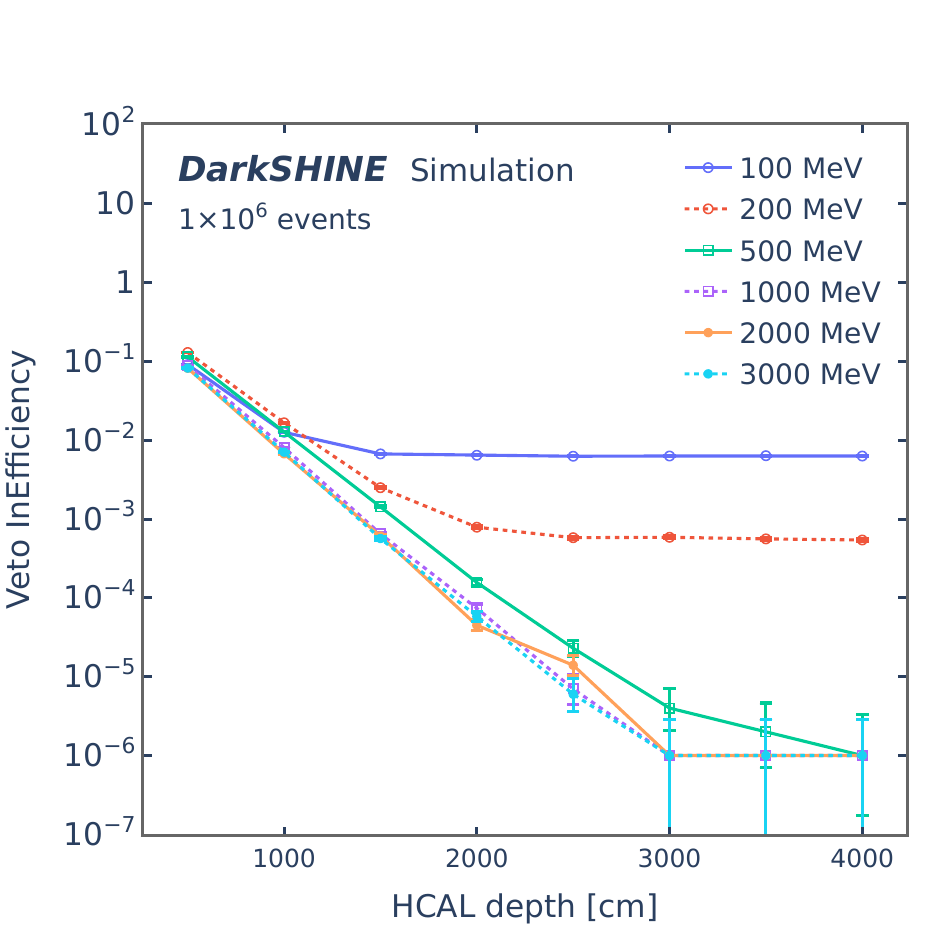}
\caption{The veto inefficiency of neutrons is investigated as a function of the depth of the HCAL, with each curve representing different incident energy choices. The x-axis denotes the HCAL depth, encompassing both sensitive and absorber layers.}
\label{fig:baseline1VetoIneff}
\end{figure}

As a sampling detector, only part of events' energy deposits, and in order to understand and validate the design, a scale factor for calibration is studied to re-scale the energy in HCAL back to the particle initial energy. The sampling efficiency of neutrons with different energies are shown in Fig~\ref{fig:HCAL_cali}. 

The curve exhibits good linearity and has been fitted to determine the scale factor. This study encompasses several particle types, most of which demonstrate a similar trend with approximately equal fitted slopes. While muons do yield different factors, it is important to note that not all muons are deposited, and applying two distinct factors across the entire HCAL would be impractical. Consequently, since particle identification is not required for the HCAL, the factors calculated using hadrons are ultimately employed for energy calibration.

\begin{figure}[htb]
\centering
\includegraphics
  [width=0.6\hsize]
  {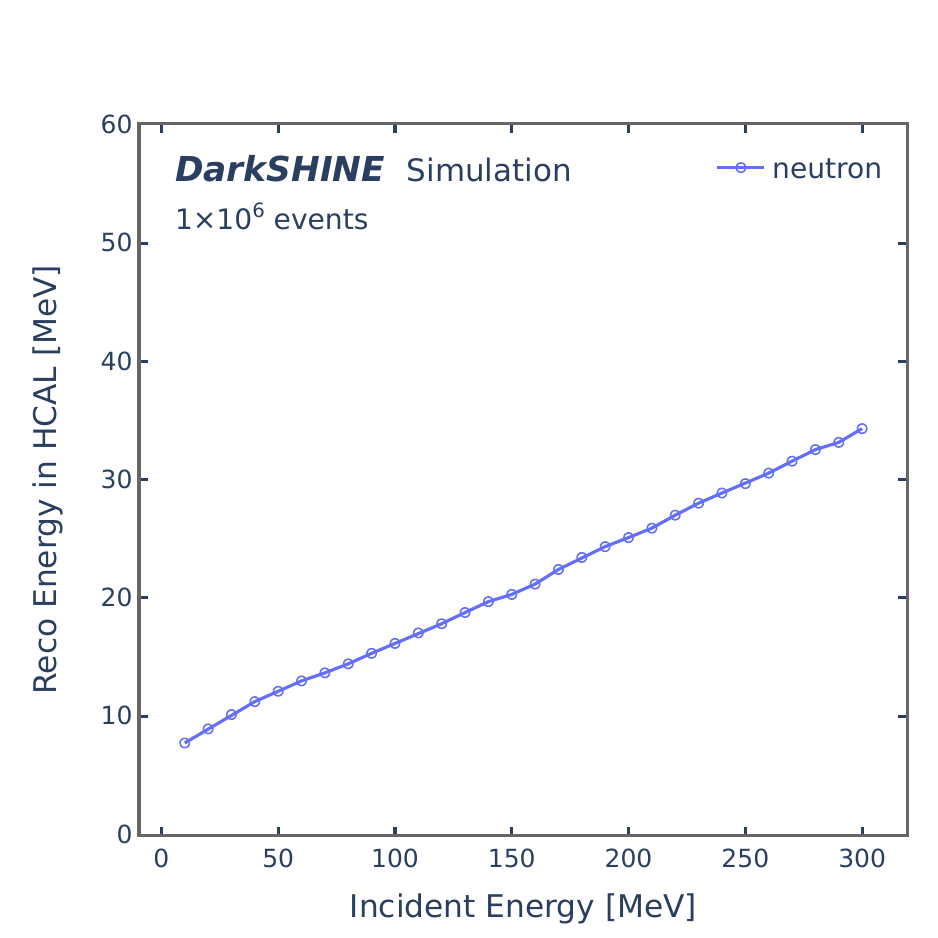}
\caption{The total energy deposited in HCAL as the function of incident neutron energy. Good linearity is observed. Calibration factor is obtained by fitting this curve.}
\label{fig:HCAL_cali}
\end{figure}

The scale factor obtained can be applied to the HCAL energy deposition of events, resulting in a multiplication of the raw energy. Consequently, 2D plots depicted in Fig~\ref{fig:2DHCALECAL_Inc} are generated, clearly illustrating that the sum of ECAL total energy and calibrated HCAL energy is approximately 8~GeV, equivalent to the incident electron energy. 

\begin{figure}[htb]
\centering
\includegraphics
  [width=0.6\hsize]
  {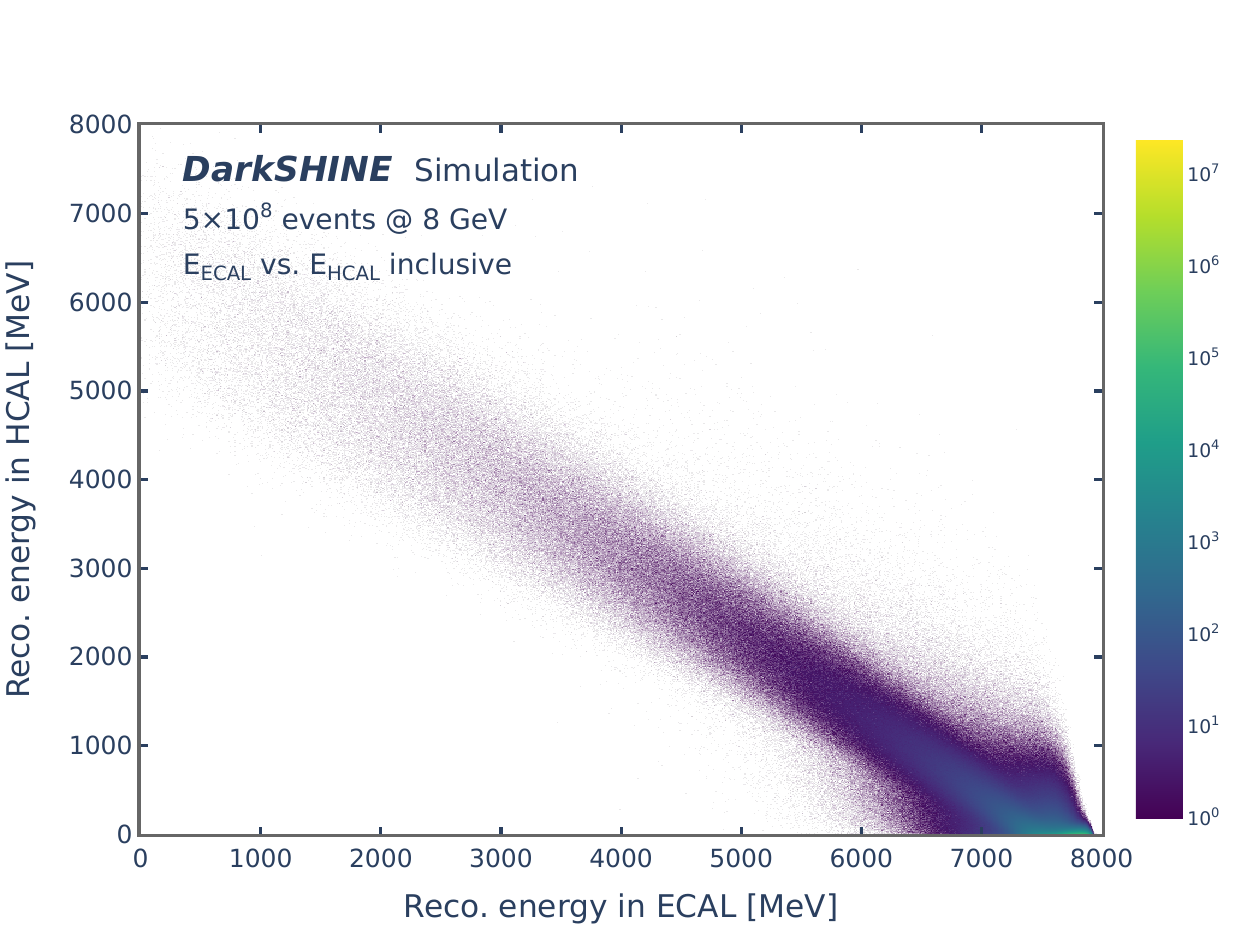}
\caption{ECAL vs. HCAL total energy after calibration. Sum of $E_{ECAL}$ and $E_{HCAL}$ is approximately 8~GeV, which is the incident energy.}
\label{fig:2DHCALECAL_Inc}
\end{figure}

\clearpage

\subsection{ECAL Electronics}
\label{sec:ECALElectronics}

The DarkSHINE ECAL readout electronics system is designed to meet the requirements of high energy precision and fast event rates. The system is responsible for processing signals from silicon photomultipliers (SiPMs) that detect the scintillation light produced by the LYSO crystals in the ECAL. To achieve high precision within a wide dynamic range and high event rate, the electronics system includes a pre-amplifier board, ADC board, and FPGA board, as shown in Figure~\ref{fig:ECALEle_amp}, \ref{fig:ECALEle_ADC} and \ref{fig:ECALEle_FPGA}. Each component is optimized for low-noise, high-speed operation.~\cite{guo2024}

\subsubsection{Pre-amplifier Board}

The SiPM signals are fast-rising (5 ns) and low-amplitude current pulses that need to be amplified and converted into voltage signals. The pre-amplifier board employs a trans-impedance amplifier (TIA) for this conversion. Unlike charge-sensitive amplifiers (CSA), which may cause waveform distortion, TIA offers better performance for this application. The TIA bandwidth is calculated using the formula:

\begin{equation}
    F_{-3dB} = \frac{\text{GBP}}{2\pi \cdot R_F \cdot C_S}
\end{equation}

where GBP is the gain-bandwidth product of the amplifier (in this case, 4.0 GHz), $R_F$ is the trans-impedance gain resistor (1,000 $\Omega$), and $C_S$ is the parasitic capacitance of the SiPM. The resulting bandwidth is approximately 150 MHz, which prevents signal distortion during amplification.

To ensure the system can handle the high event rate, the output signal must be shaped to a small width. A TIA-CR-RC2 filter circuit is used for this purpose. The CR stage filters out low-frequency components, narrowing the trailing edge, while the two RC stages reduce high-frequency noise and smooth the waveform. This design results in a more Gaussian-like signal with an improved signal-to-noise ratio (SNR).

\begin{figure}[h]
    \centering 
    \subfigure[]{
    \includegraphics[width=0.8\textwidth]{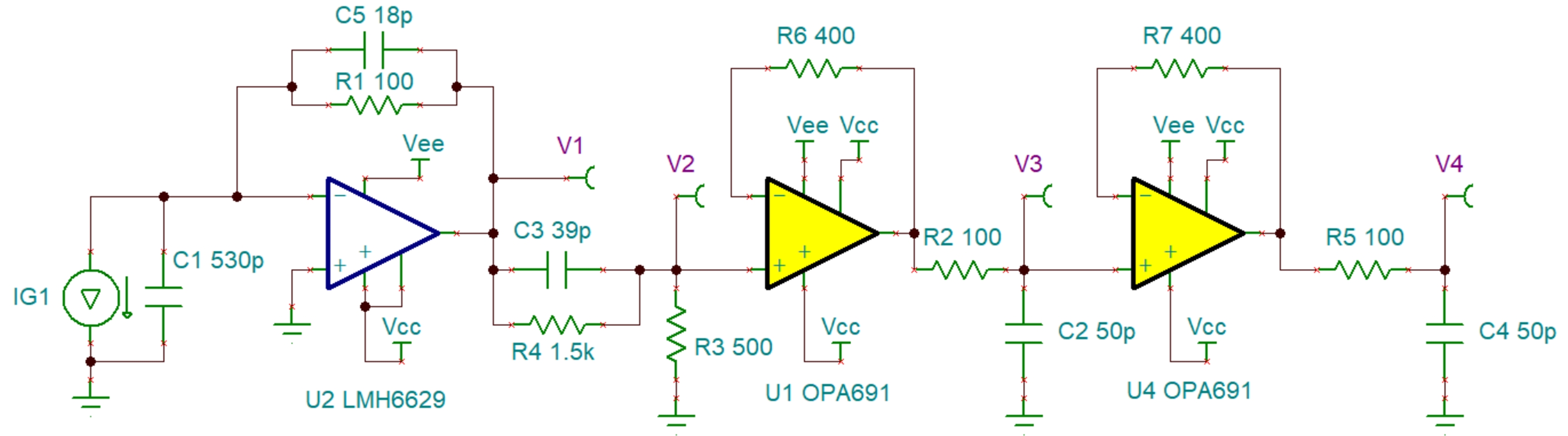}}
    \subfigure[]{
    \includegraphics[width=0.8\textwidth]{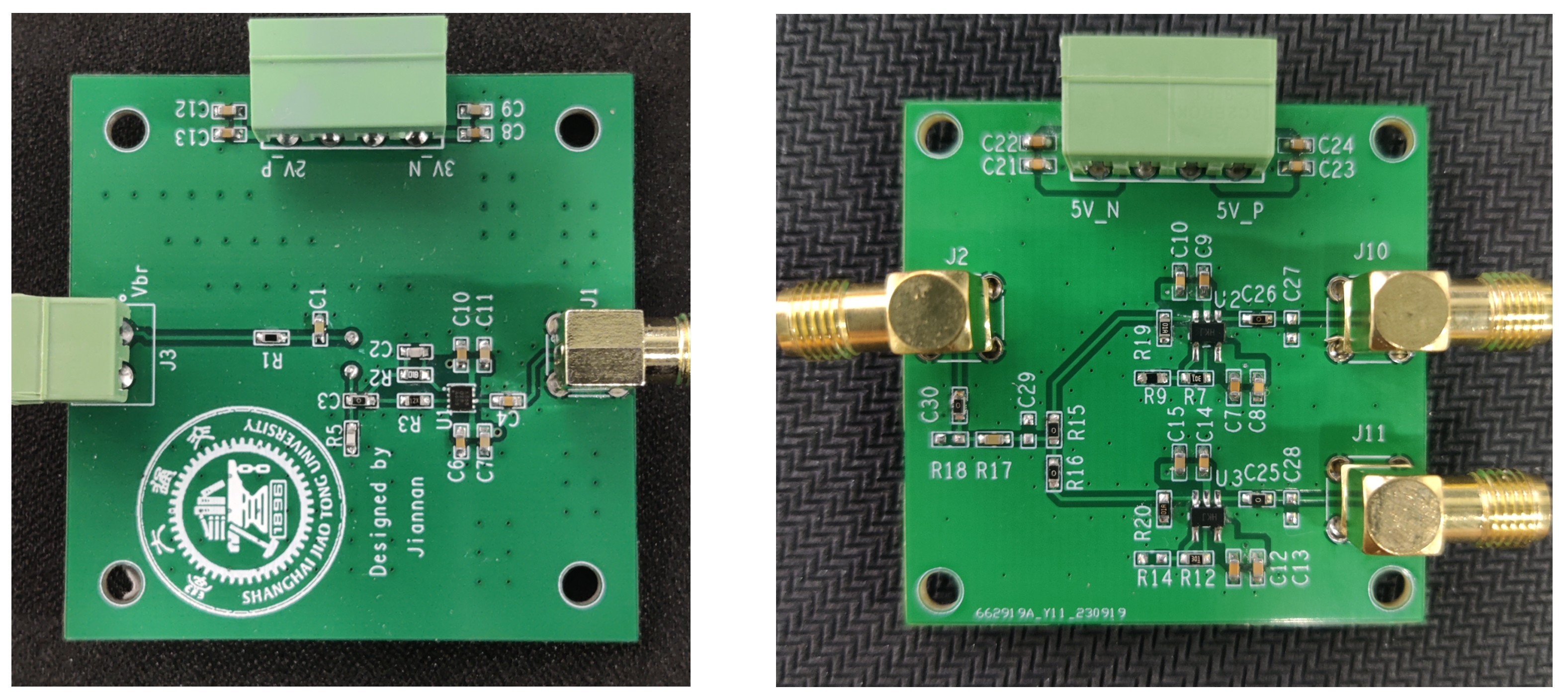}}
    \caption{\label{fig:ECALEle_amp}~The schematic and photograph of the pre-amplifier board.~\cite{guo2024}}
\end{figure}

To handle the wide dynamic range (10 dB), a dual-gain amplifier is implemented. For input energies below 40 MeV, a high gain (20x) setting is applied, enhancing the signal-to-noise ratio. For energies above this threshold, the gain is reduced (1x) to prevent saturation of the ADC input.

\subsubsection{ADC Board}

After amplification and shaping, the signals are digitized by a dual-channel, 14-bit, 1 GSPS ADC (AD9680). This high-speed ADC ensures precise waveform digitization, which is crucial for accurate energy reconstruction. The ADC board is connected to the FPGA board using the JESD204B high-speed serial interface, capable of transmitting data at 10 Gbps per channel.

The ADC board features LTC6409 single-ended to differential drivers that convert the analog signal for digitization by the ADC. The digitized data is then sent to the FPGA for further processing.

\begin{figure}[h]
    \centering 
    \subfigure[]{
    \includegraphics[width=0.45\textwidth]{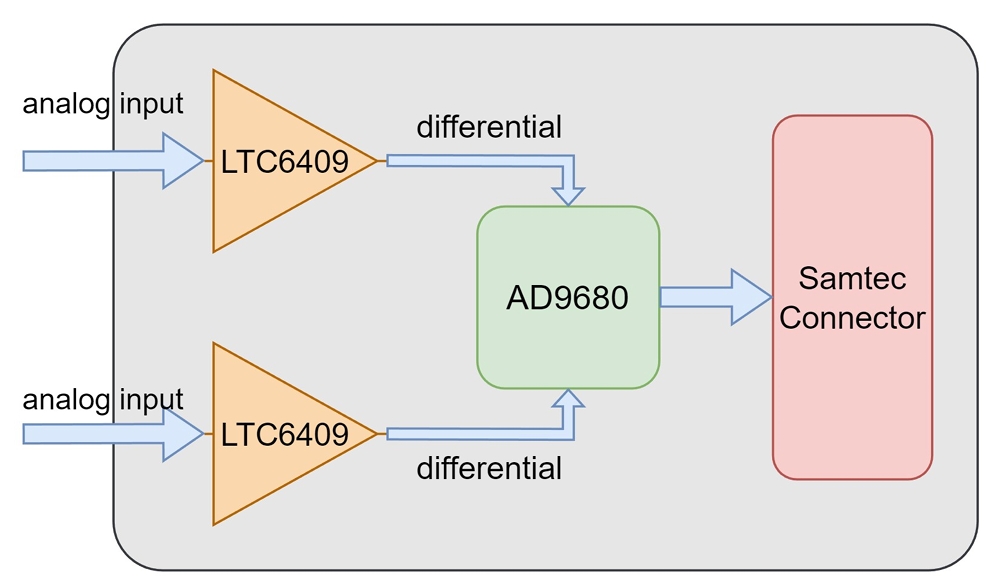}}
    \subfigure[]{
    \includegraphics[width=0.45\textwidth]{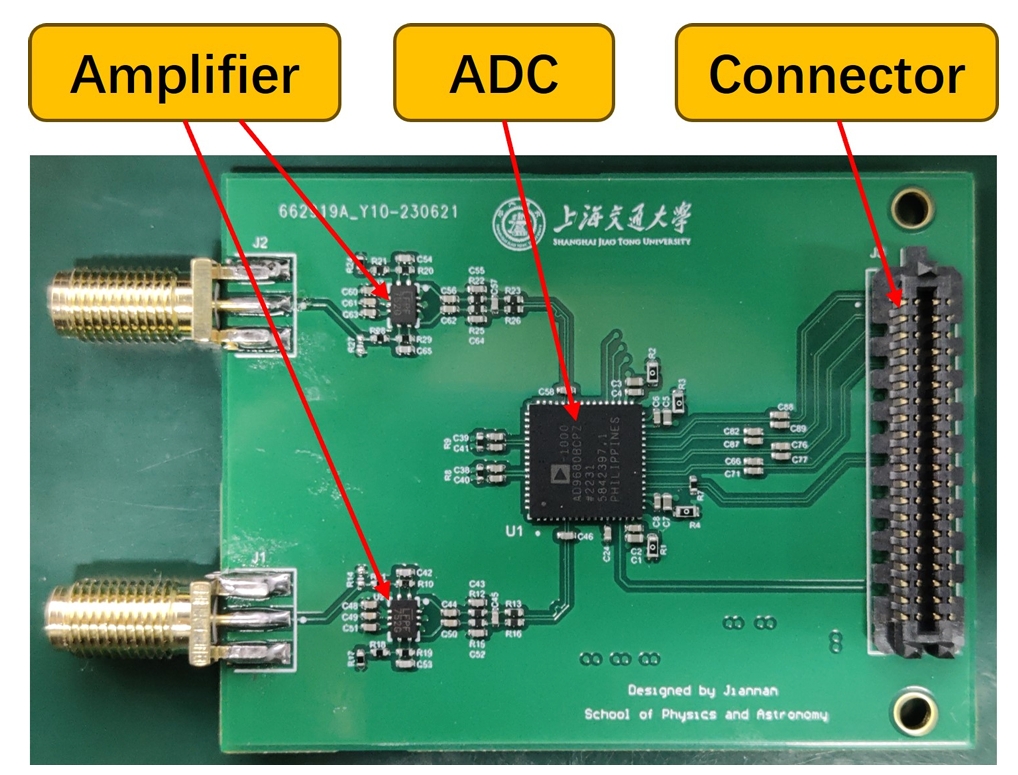}}
    \caption{\label{fig:ECALEle_ADC}~The block diagram and photograph of the ADC board.~\cite{guo2024}}
\end{figure}

\subsubsection{FPGA Board}

The FPGA board is responsible for processing the digitized signals from the ADC. Based on the Xilinx Kintex-7 series FPGA (XC7K420T-FFG901), it provides high-speed data transmission and storage capabilities. The board supports up to 28 GTX channels for high-speed serial data transmission, with each channel capable of handling data rates of up to 12.5 Gbps. The JESD204B protocol ensures the synchronization of data across multiple ADC channels.

A ring buffer is implemented in the FPGA to temporarily store the digitized waveforms. The buffer operates as a circular memory, where older data is replaced by newer data once the buffer is full. A trigger system is also integrated into the FPGA, allowing the system to capture waveform data whenever a specific threshold is crossed. This threshold, along with other parameters such as trigger offset and waveform length, can be configured via the data acquisition software.

The FPGA also manages the clock distribution and power supply for the ADC board. A 200 MHz system clock is provided by a SIT9121 oscillator, while the AD9528 clock distribution IC handles clock synchronization between the ADC and FPGA.

\begin{figure}[h]
    \centering
    \includegraphics[width=0.6\linewidth]{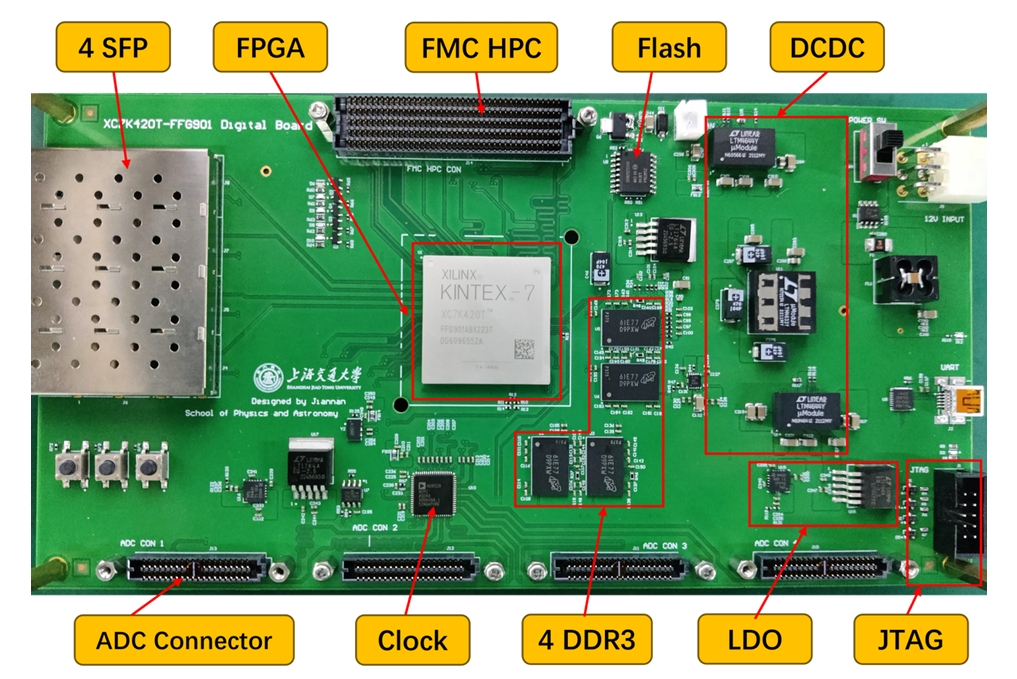}
    \caption{\label{fig:ECALEle_FPGA}~The FPGA board for the DarkSHINE ECAL readout electronics.}
\end{figure}

\subsubsection{Data Acquisition System}

\begin{figure}[h]
    \centering 
    \subfigure[]{
    \includegraphics[width=0.45\textwidth]{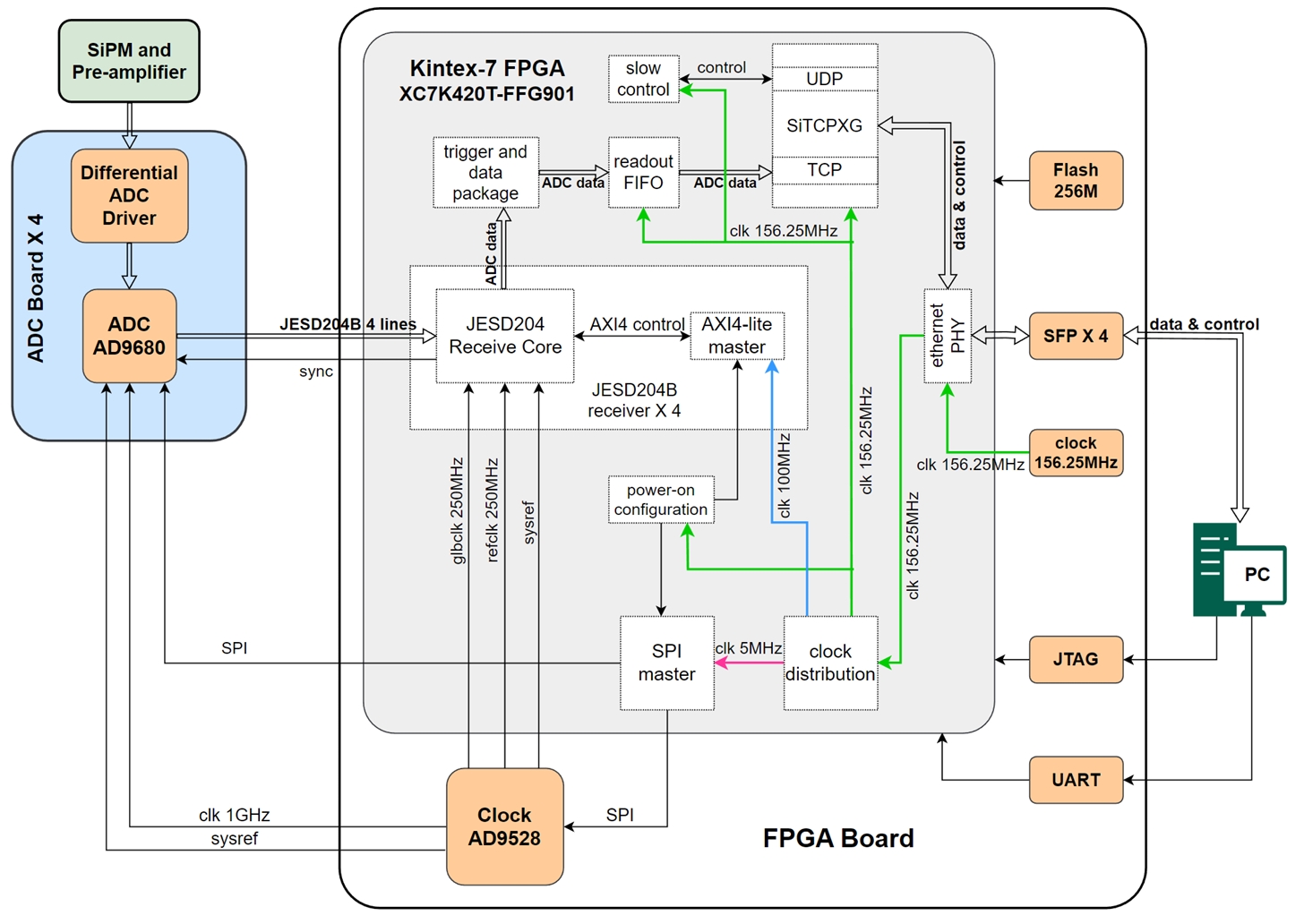}}
    \subfigure[]{
    \includegraphics[width=0.45\textwidth]{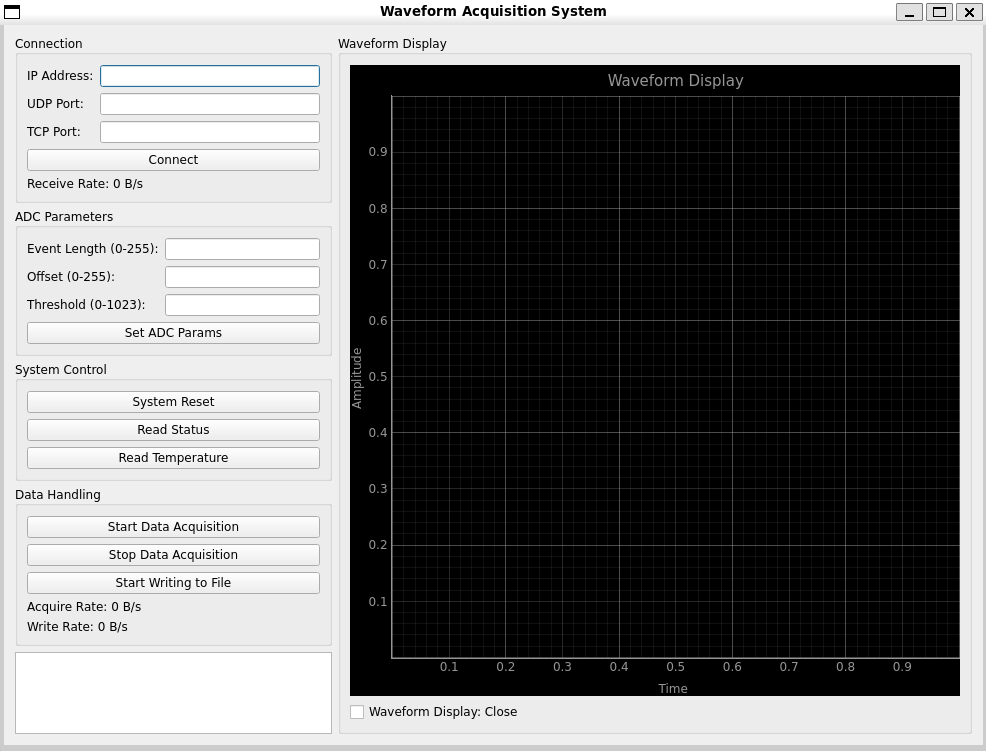}}
    \caption{\label{fig:ECALEle_DAQ}~(a) The block diagram and photograph of the DAQ system.~\cite{guo2024} (b)User interface of the DAQ software for DarkSHINE ECAL.}
\end{figure}

The data acquisition (DAQ) system (Figure~\ref{fig:ECALEle_DAQ} (a)) is responsible for managing data flow between the FPGA and the PC. The firmware running on the FPGA handles the reception of high-speed serial data from the ADCs, triggers data capture, and formats the output data for transmission via Ethernet.

A DAQ software was developed to provide a graphical user interface (GUI) for configuring the system, viewing real-time waveforms, and storing data (Figure~\ref{fig:ECALEle_DAQ} (b)). It communicates with the FPGA over Ethernet using the SiTCP protocol, which employs TCP for data transmission and UDP for slow control. The software allows users to configure various parameters such as the trigger threshold, event length, and channel selection. It also provides monitoring functions to display the system status, including sample rate, synchronization, and link status.

\section{Target System}\label{sec:target Sys}

%\textcolor{red}{Target (with brief mentioning of comparisons)}

DarkSHINE is an electron-on-target experiment with 1~MHz single electron beam at 8~GeV. After the first run, DarkSHINE can collect $3\times 10^{14}$~electron-on-target. The expected signal yield $N_{sig}$ can be described by the following equation: 
\begin{equation}
\label{eq:signalYields}
N_{sig}=\sigma_{A^{\prime}} \times \rho \times 0.1 X_{0} \times L \times N_{A} / M_{target} \times 10^{-36} \times \varepsilon^{2}   
\end{equation}

where $\sigma_{A^{\prime}}$ is the production cross section of a dark photon with mass $m_{A^{\prime}}$, $\rho$ is the density of the target, $0.1X_{0}$ represents the thickness of the target, $X_{0}$ is radiation length of the target material, $L=3\times10^{14}$~(EOTs) is the number of events available, $N_{A}$ is the Avogadro constant, $M_{target}$ is the atomic mass of the target, and $\varepsilon$ is the kinetic mixing parameter. 

Among the parameters in equation~\ref{eq:signalYields}, signal cross section, the radiation length of the target material, and the atomic mass of the target will affect the final signal yield. The key parameters of different target materials\cite{Workman:2022ynf} are summarized in Table~\ref{tab:TargetInfos}. The cross section ratios and signal yield ratios with different target materials divided by tungsten target case as a function of $m_{A^{\prime}}$ are compared in the left and right parts of Figure~\ref{fig:xsection1}. As shown in the right part of Figure~\ref{fig:xsection1}, signal yields ratio between lead (Pb) and tungsten (W) target is close to 1. The signal yield ratios increase in the high-mass region of dark photon with the aluminum (Al) and beryllium (Be) target. 

\begin{figure}[h]
\centering
\includegraphics[width=0.45\linewidth]{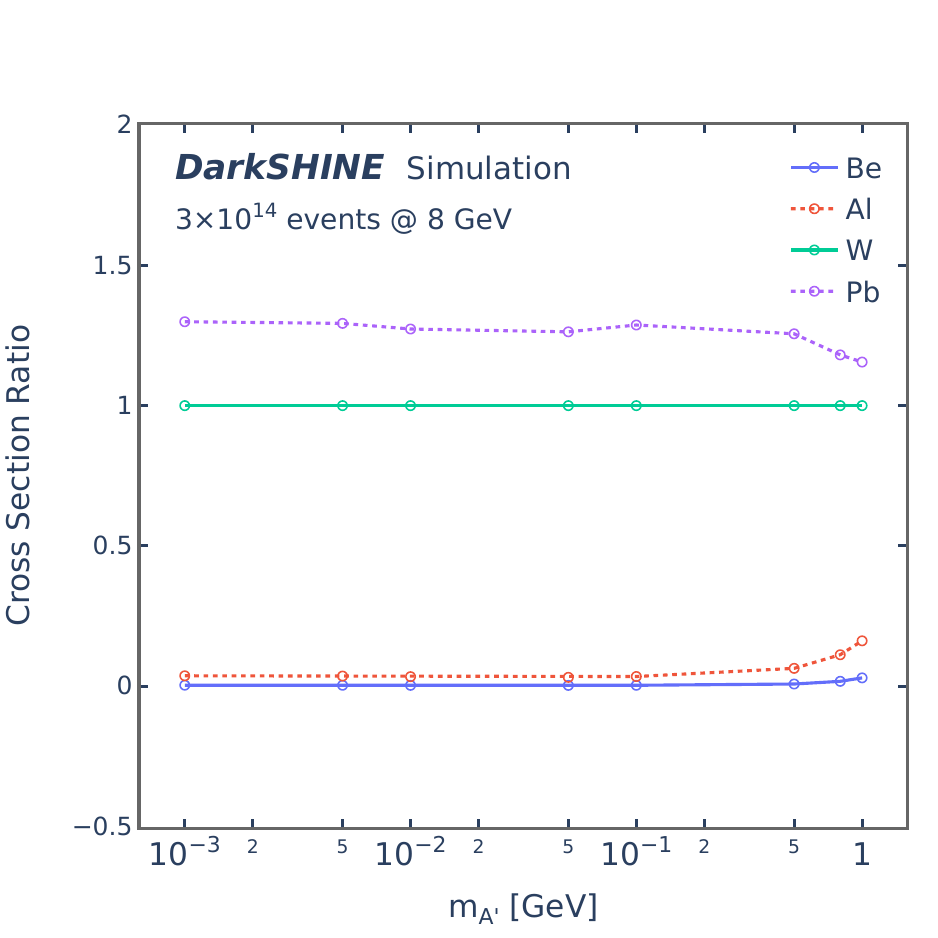}
\includegraphics[width=0.45\linewidth]{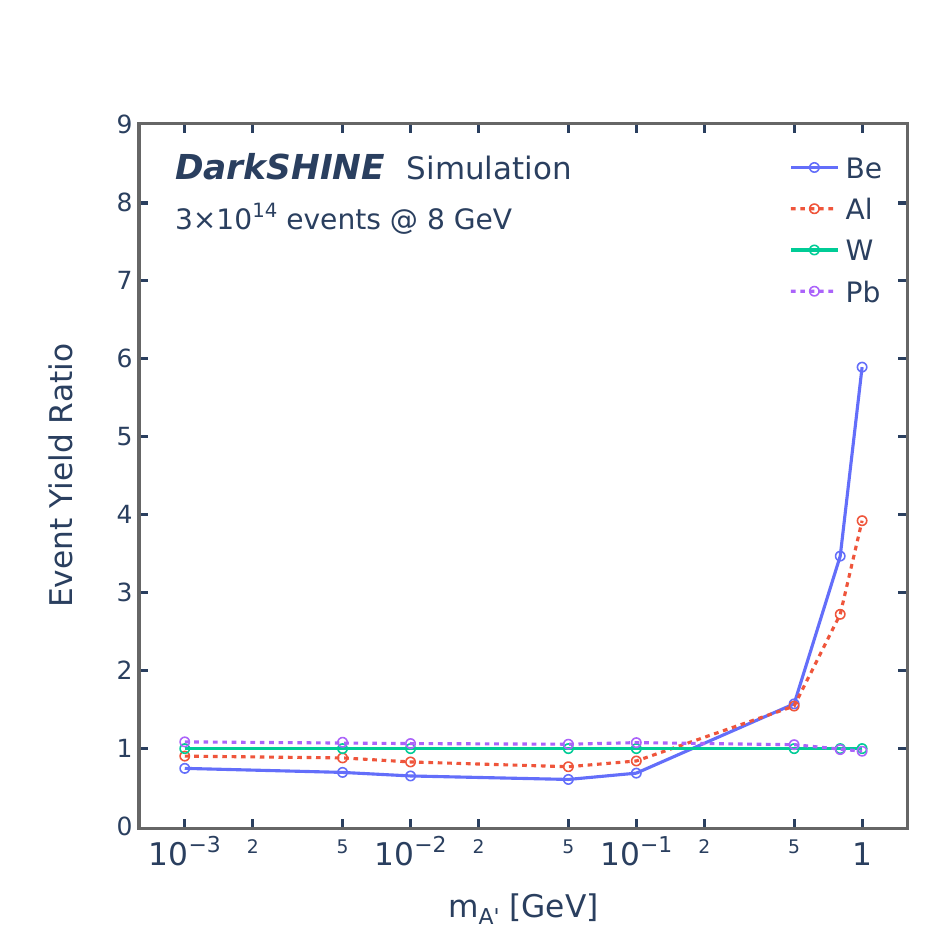}
\caption{\label{fig:xsection1}~Left: Cross section ratios derived from CalcHep as a function of $m_{A^{\prime}}$. Right: Signal yield ratios as a function of $m_{A^{\prime}}$ with $3\times 10^{14}$~EOT, $\varepsilon$=1.}
\end{figure}

In most cases, the electrons in the beam pass through the target without any interaction. However, due to the presence of sufficient material within the target, electrons have potential to occur other processes as they pass through. 

A small fraction of electrons generate additional photons via hard bremsstrahlung process. These bremsstrahlung photons may either contribute to an electromagnetic shower within the ECAL or end up in conversion into hadrons or lepton pairs, which can occur at both target or ECAL. These backgrounds are illustrated in Fig~\ref{fig:flowbackground}. In addition to photon conversion, these bremsstrahlung photons can also undergo interactions with materials in the target and ECAL, leading to photon-nuclear reactions that result in hadron production. Moreover, hadronic interactions of the incident beam electron with the target and ECAL are also important backgrounds. Although the branch ratios of these processes (electron-nuclear, photon-nuclear, $\gamma \to \mu\mu$) are relatively small, they cannot be ignored due to the high repetition rate of electron-on-target events, and effectively suppressing these events is crucial for achieving low background expectations.

What's more, there exist irreducible physics backgrounds that encompass neutrino processes. The processes involved the production of high energy neutrinos are Moller $+$ Charged-Current Exchange (e.g. $e^{-}e^{-} \to e^{-}e^{-}$, $e^{-}p \to \nu n$), Neutrino Pair Production (e.g. $e N \to e \nu \Bar{\nu} N$), Bremsstrahlung $+$ Charged-Current Exchange (a low energy electromagnetic shower initiated by soft bremsstrahlung followed by a charged-current quasi-elastic reaction) and Charged-Current Exchange with Exclusive $\pi^{0}$ Final State (e.g. $e^{-} p \to \nu n \pi^{0}$)~\cite{PhysRevD.91.094026}. Among the four processes described above, Moller $+$ Charged-Current Exchange (Moller + CCQE) and Charged-Current Exchange with Exclusive $\pi^{0}$ Final State (CCQE with $\pi^{0}$) are correlated with the target material.

%\subsection{Target materials}

Rare processes yields are proportional to the target radiation length, density of the target, and cross sections ($N_{rare} \thicksim \rho \times X_{0}\times \sigma$). The cross sections of rare processes per atom are shown in left part of Figure~\ref{fig:xsection2}. Rare processes cross sections per atom are similar in Pb and W. While in Al and Be target, the cross sections per atom are much smaller. As shown in the right part of Figure~\ref{fig:xsection2}, considering the density of material, the event yields of rare processes are similar in Pb and W target. For $\gamma \to \mu \mu$, the event yields in Al and Be target are also similar. However, electron-nuclear and photon-nuclear processes event yields are larger in Al and Be target. Corresponding to W target, the electron-nuclear event yields are 2.5 and 5.3 times larger in Al target and Be target. The ratio of photon-nuclear event yields between Al/Be and W are 4 and 11.7. 

Dark photon search is based on the $E_T^{miss}$ phenomena. Therefore, real missing energy background is an important component of background. The event number of Moller $+$ Charged-Current Exchange scales with $M_{target}/(Z(Z+1))^{2}$, where $M_{target}$ is the atomic mass and $Z$ is atomic number. The event number of Charged-Current Exchange and Charged-Current Exchange with Exclusive $\pi^{0}$ Final State scaling as $M_{target}/Z(Z+1)$. Table~\ref{tab:realMissEBackground} summarizes the event yields of Moller $+$ Charged-Current Exchange and Charged-Current Exchange with Exclusive $\pi^{0}$ Final State per $3\times 10^{14}$ EOT with $E_{beam}=10$~GeV. The background yields of these two processes are small and similar in the W and Pb targets. In the Al target or Be target, these yields are much larger. 

Although signal yields are much larger in high-mass region of dark photon with Al target and Be target, the irreducible real missing energy background processes are quit large in Al and Be cases. As shown in equation~\ref{eq:upperlimit}, larger irreducible background lowers the sensitivity. Therefore, Al target and Be target are not used in DarkSHINE. 

For W and Pb, signal and background yields are similar. Pb has a relatively high density ($11.35~g/cm^{3}$) and relatively low melting point ($327.5^\circ C$). W has a higher density ($19.3~g/cm^{3}$)  and  a very high melting point ($3422^\circ C$), allowing it to remain stable under high temperatures. W's high melting point allows it to perform well in high-temperature environments, while Pb's deficiencies in this area make it more vulnerable to damage and failure under similar conditions. Therefore, DarkSHINE chooses a 350 micron W target. 

%~\cite{bib:}
%As shown in Figure～\ref{fig:xsection1} and Figure~\ref{fig:xsection1}, 
%\textcolor{ref}{(the figure need to be add)}.

\begin{table*}[t]
\footnotesize
\caption{\label{tab:TargetInfos} Key parameters of target material.}
\tabcolsep 10pt %space between two columns. 
\begin{tabular*}{0.974\textwidth}{c|c|c|c|c}
\toprule  
\centering
Material & Atomic Number $Z$ & Atomic Mass $M_{target}$ & Radiation Length $X_{0}$ [$cm$] & Density $\rho$ [$g/cm^{3}$] \\ \hline 
W        & 74            & 184         & 0.35                          & 19.30 \\ \hline
Pb       & 82            & 207         & 0.56                          & 11.35 \\ \hline
Al       & 13            & 27          & 8.90                         & 2.70  \\ \hline
Be       & 4             & 9           & 35.28                         & 1.85  \\ 
\bottomrule
\end{tabular*}
\end{table*}

\begin{table*}
\footnotesize
\caption{\label{tab:realMissEBackground} Real missing energy background event yields with different target material.}
\tabcolsep 21pt %space between two columns. 
\begin{tabular*}{0.974\textwidth}{c|c|c|c|c}
\toprule
\centering
Events per $3\time 10^{14}$ EOT & W & Pb & Al & Be \\ \hline
Moller + CCQE & $3.6\times 10^{-4}$ & $3.0\times 10^{-4}$ & $8.6\times 10^{-3}$ & $7.3\times 10^{-2}$\\ \hline
CCQE with $\pi{0}$ & $0.3\sim 0.6$ & $0.28\sim 0.55$ & $1.3\sim 2.7$ & $4.1 \sim 8.2$ \\
\bottomrule
\end{tabular*}
\end{table*}

\begin{figure}[h]
\centering
\includegraphics[width=0.45\linewidth]{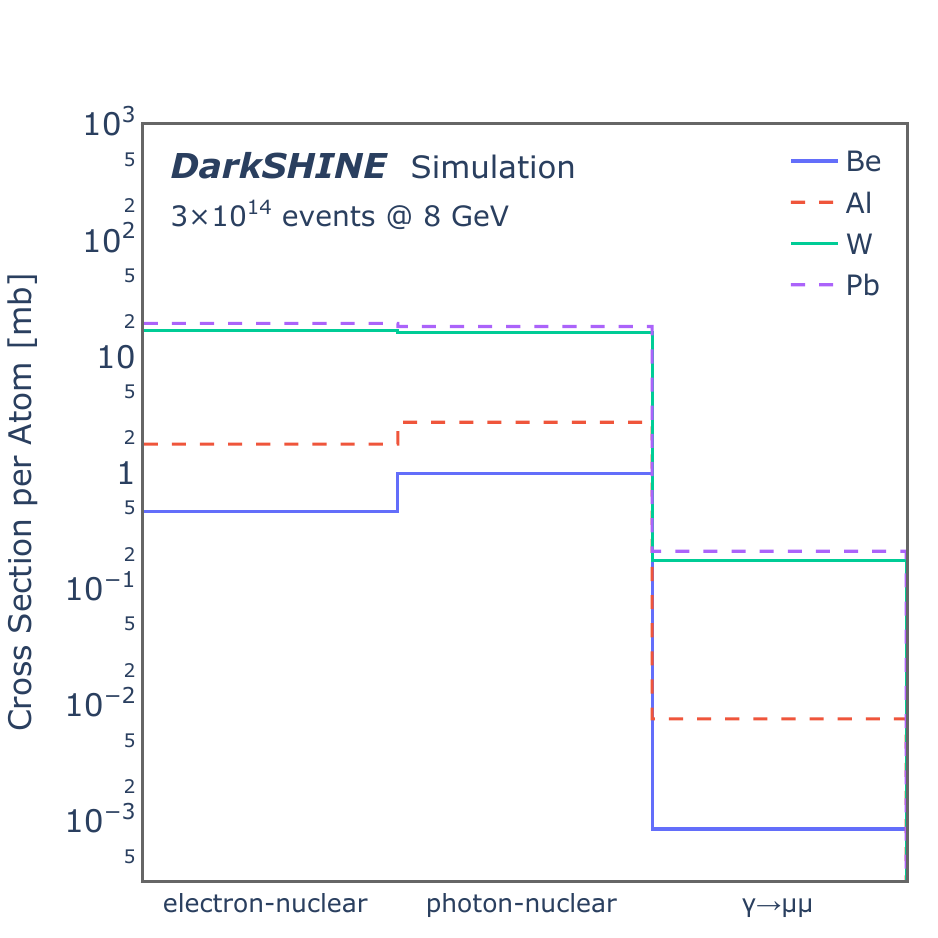}
\includegraphics[width=0.45\linewidth]{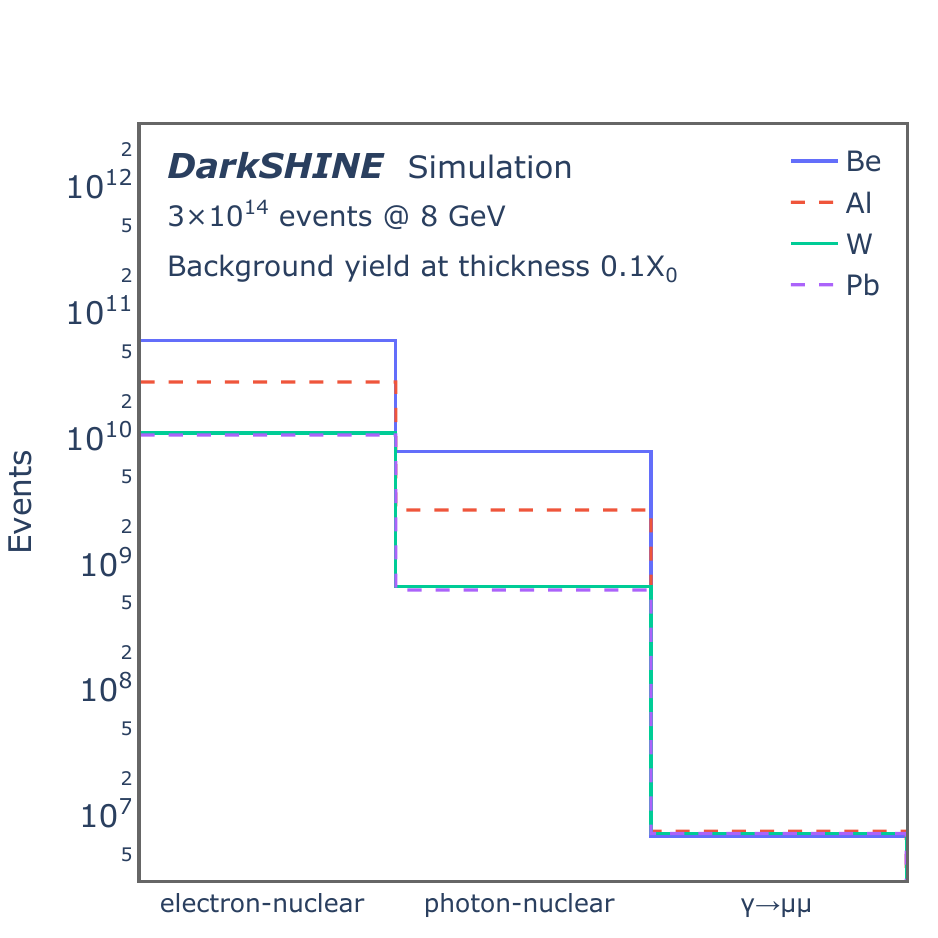}
\caption{\label{fig:xsection2}~Left: Rare processes cross sections per atom with different target material. Right: Rare processes background yields with $3\times 10^{14}$~EOT, $0.1X_{0}$ targets.}
\end{figure}

\section{Electromagnetic Calorimeter System}
\label{sec:ECAL}

\subsection{Introduction}
\label{sec:HCALIntroduction}

The role of the HCAL is crucial in rejection of those rare background processes mentioned in Sec~\ref{sec:target Sys}, and a schematic diagram is provided in Figure~\ref{fig:flowbackground}. In addition to processes initiated by bremsstrahlung photons, electron interactions with the materials of ECAL and target can also result in nucleon production and yield final states similar to those produced by photon-nuclear processes.

These rare processes, such as photon-nuclear (PN) and electron-nuclear (EN) reactions, can be further categorized based on whether they occur in the target or ECAL. Thus, they are referred to as PN-target, PN-ECAL, EN-target, and EN-ECAL, respectively.

The conversion of photons into electron pairs can be effectively addressed by applying a direct $\mathrm{E_{ECAL}}$ cut. However, the challenges arise when dealing with conversions into muon pairs and hadron pairs. Muons, being minimum ionizing particles (MIP), pass through the ECAL, thereby diminishing the efficacy of the $\mathrm{E_{ECAL}}$ cut. While the ECAL offers potential advantages in terms of providing additional information such as tracks and topology within its domain, it is crucial to emphasize that HCAL remains paramount and provides straightforward information without necessitating complex reconstruction algorithms. Similar considerations apply to final states involving charged hadron pairs, where combining information from both ECAL and HCAL may lead to exclusionary outcomes. However, for neutral hadrons that do not decay within the ECAL volume, discrimination power heavily relies on HCAL.

Compared to neutral hadrons, muons are easier to detect due to their significant energy deposition in multiple layers of scintillators. The primary concern lies in the HCAL's ability to detect neutral hadrons, as the veto power of these particles becomes a crucial function and design consideration. Table~\ref{tab:HCALbackground_process} provides a summary of the proportions of the most frequently generated particles from these rare processes. Given that neutrons constitute the largest proportion and protons can be excluded through a combination of tracker and ECAL information, neutrons are utilized for identifying the veto power of the HCAL.

\begin{table}[!htb]
\caption{Particle types and frequencies from electron-nuclear and photon-nuclear process, neutrons are predominant.
\label{tab:HCALbackground_process}}
\centering
\vspace{0.3cm}
\begin{tabular}{l|l|l|l|l}
\toprule
Process         & Neutron & Proton  & Pion    & Kaon   \\
\midrule
Electron\_Nuclear & 73.42\% & 21.52\% & 4.64\%  & 0.42\% \\
Photon\_Nuclear   & 64.95\% & 18.56\% & 14.43\% & 2.06\% \\
\bottomrule
\end{tabular}
\end{table}

The HCAL employs energy deposition cuts to reject events, and the veto efficiency varies for neutrons of different energies. Notably, while the single-neutron event has an identical veto efficiency as that of the neutron, a multi-neutron event's efficiency is equivalent to at least one of these neutrons' veto efficiencies. This performance is evaluated by calculating the ratio between the number of events or neutrons not being vetoed and the total number, referred to as \textbf{veto inefficiency}. 

The energy distribution and number of neutrons in prediction events are studied and are illustrated in Figure~\ref{fig:HCALneutron_energy}. As discussed before, the ECAL of DarkSHINE absorbs all photon and electron energy, providing the total deposited energy quantity. The variable $\mathrm{E_{ECAL}}$ can effectively discriminate against numerous background events, as the majority of background events tend to exhibit higher values of $\mathrm{E_{ECAL}}$ compared to the signal. To specifically focus on events that cannot be rejected by other sub-detectors but rely on the rejection power of HCAL, only events satisfying the cut $\mathrm{E_{ECAL}} < 2.5$~GeV~\cite{Chen:2022liu} are presented. 

This result involves simulating 1$\times$10$^{8}$ electrons hitting the target. Considering that the ECAL energy loss is approximately 5.5 GeV, even if there is a neutron with an energy of around 2.5 GeV in the event, which accounts for roughly half of the energy loss, it can still be inferred that the event comprises other components with similar energies. Consequently, events involving a single high-energy neutron make a significant contribution to the overall energy loss, while the remaining components lack sufficient energy for veto and are rare. To assess the quality of the optimization result, neutrons around 2~GeV are considered as energetic particles. 

\begin{figure}[htb]
\centering
\includegraphics
  [width=0.6\hsize]
  {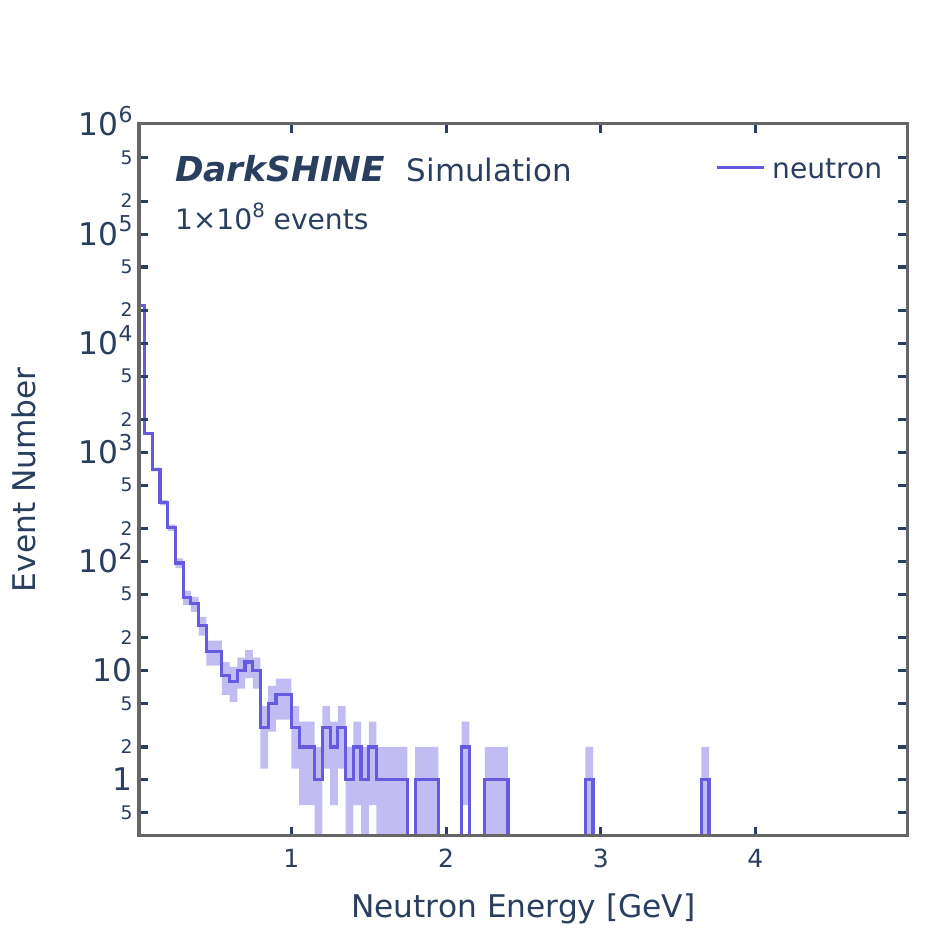}
\caption{Neutron energy distribution after applying cut on ECAL energy to request  $\mathrm{E_{ECAL}} < 2.5$~GeV. The result shows that very few neutrons with energy greater than 2.5~GeV is left after ECAL cut. Which means under assumptions of 1$\times$10$^{14}$ EOT, one can expect there to be at most 1$\times$10$^{6}$ level energetic neutrons.}
\label{fig:HCALneutron_energy}
\end{figure}

Given that only a few or fewer neutrons survive in this phase space, it is expected that around 1$\times$10$^{6}$ energetic neutrons will be generated with the conditions of 1$\times$10$^{14}$ EOT, which is consistent with the predicted number to be collected within one year~\cite{Chen:2022liu}. Consequently, a veto inefficiency $<$ 10$^{-5}$ is chosen as the performance benchmark for high-energy neutron rejection, capable of reducing energetic neutrons to the unit level. Conversely, in the absence of high-energy neutrons and presence of solely low-energy neutrons, it is implausible for these particles to be the sole particles detected in the event; otherwise, the ECAL would have recorded an energy deposition closer to 8~GeV. In such a scenario, a veto inefficiency $<$ 10$^{-3}$ would suffice to achieve equivalent rejection power if multiple neutrons are present. Considering the aforementioned estimates and constraints regarding sample generation time and storage, a sample size of 1$\times$10$^{6}$ is employed for investigating the performance of HCAL.

\subsection{Software configuration}
\label{sec:ECALSoftware}

\subsubsection{Simulation}
\label{sec:ECALSimulation}

The Monte Carlo simulation of ECAL is performed with the DarkSHINE Software (DSS) software framework based on GEANT4\cite{GEANT4:2002zbu}. The full size detector including crystal cube, wrapper, SiPM sensor, and supporting materials. Other parts in Dark SHINE experiments are simulated together such as Tracker and HCAL, in order to provide the seeding reference for ECAL reconstruction and basic analysis analysis cut to define the proper phase space for ECAL analysis.

The tungsten target are used only to simulate the inclusive and rare process, denoting as "full simulation", where 8GeV electron beam hits the target as the experiment proposed and secondary particles go through all the detector parts. Besides, to reduce the time of simulation and focus on the ECAL response, "ECAL-only" simulation is performed as well where particles with pre-defined type, energy and direction incident after the target to mimic the secondary particles in inclusive events. Only those particles with large energy, accepted direction entering ECAL and of physics interest are simulated and it highlights the ECAL performance. In the following the ECAL-only simuation is used for the performance evaluation of ECAL and ECAL-unit while the full simulation used to produce the events for the physics analysis.

% Detailed geometry description. Figure~\ref{fig:ECALGeo} and Table~\ref{tab:ECALPara}. 
% Why we need a ECAL with such a large size. (Do we need the same optimization on ECAL size as baseline v1.6?)\

% \begin{figure}[h]
%     \centering
%     \includegraphics[width=0.3\linewidth]{Detectors/Figures/ECAL/ECAL_1.0.png}
%     \caption{\label{fig:ECALGeo} ECAL geometry.}
% \end{figure}

% \begin{table}[h]
% \centering
% \fontsize{9}{13}\selectfont
%     \begin{tabular}{cccccc}
%         \toprule
%         \makecell[c]{Total size} &\makecell[c]{Crystal number}  &\makecell[c]{Crystal size} &\makecell[c]{Sensor}  &\makecell[c]{Supporting material} &\makecell[c]{Others}\\
%         \midrule
%         - & - & - & - & - & - \\
%         \bottomrule
%     \end{tabular}
%     \caption{\label{tab:ECALPara} Parameters in ECAL simulation.}
% \end{table}

\subsubsection{Digitization}
\label{sec:ECALDigitization}
The accurate simulation of response of crystal-based detector requires precise description of optical process and digitization effect. The GEANT4 optical simulation is used in ECAL only simulation of single crystal unit together with hit-based digitization. 
The performance of single unit is then analyzed and the energy smearing parameters are extracted by fitting the smearing formula, as shown in \ref{eq:ECAL_smearing}

\begin{equation}
\label{eq:ECAL_smearing}
\frac{\sigma}{E}=\frac{A}{\sqrt{E}}\oplus B\oplus \frac{C}{E}
\end{equation}

The parameters, $A$, $B$ and $C$, are then used in the full size ECAL detector and full simulation, to mimic the realistic effect of detector as well as efficiently simulating thousands of units and saving the computing. The energy deposition of each crystal unit is collected in the full-size simulation and it is denoted as "truth" information before optical and digitization effect applied. Then a Gaussian smearing is applied per unit with center value at the truth energy deposition and sigma at the smearing parameters mention before to add the realistic detector into full simulation with affordable computing. Four setups of crystal unit are studied in full simulation, which considers the different properties such as wrapper material and crystal unit performance and are validated by the single unit measurement, and in all the analysis mentioned in this paper without further notice, the setup-1 is used for a realistic setup. The full setup of the four sets of crystal unit are shown in Table~\ref{tab:ECAL_smearing_par} with corresponding smearing parameter extracted in Table~\ref{tab:ECAL_smearing}.

% Geant4 simulation: full simulation with target and all detectors,  ECAL-only simulation.

% Digitization in baseline v1.0. Figure~\ref{fig:ECALDigi_v1.0}

% \begin{figure}[h]
%     \centering
%     \includegraphics[width=0.3\linewidth]{Detectors/Figures/ECAL/Digitization_v1.0.png}
%     \caption{\label{fig:ECALDigi_v1.0} ECAL geometry.}
% \end{figure}

\begin{table}[h]
\footnotesize
\fontsize{8}{12}\selectfont
\caption{\label{tab:ECAL_smearing_par} Setup of the ECAL crystal unit}
\centering
\begin{tabular}{c c c c c c}
\hline
                     & Cube Dimensions     & Wrapper               & SiPM Size ($mm^2$) & Coupling*QE (\%)    & Yield/MeV          \\ \hline
R90\_LYSO             & 2.5$\times$2.5$\times$4 cm & Ref=90\% (ref.)      & 9                 & 20\%               & 30000 (LYSO)       \\ 
R10\_LYSO             & 2.5$\times$2.5$\times$4 cm & Ref=10\% (abs.)      & 9                 & 20\%               & 30000 (LYSO)       \\ 
R90\_S9\_PWO4         & 2.5$\times$2.5$\times$4 cm & Ref=90\% (ref.)      & 9                 & 20\%               & 200 (PWO)          \\ 
R90\_S36\_PWO4        & 2.5$\times$2.5$\times$4 cm & Ref=90\% (ref.)      & 36                & 20\%               & 200 (PWO)          \\ \hline
\end{tabular}
\end{table} 

\begin{table}[h]
\footnotesize
\fontsize{8}{12}\selectfont
\caption{\label{tab:ECAL_smearing} Smearing parameters of ECAL crystal unit simulation }
\tabcolsep 16pt %space between two columns. 
\centering
\begin{tabular}{cccc}
\hline
                      & $A/\sqrt{\text{GeV}}$ & $B$ & $C/\text{MeV}$ \\ \hline
R90\_LYSO             & 1.00\%                & 0.00\% & 0.0000         \\
R10\_LYSO             & 6.69\%                & 0.00\% & 0.0851         \\
R90\_S9\_PWO4         & 4.26\%                & 0.70\% & 0.0001         \\
R90\_S36\_PWO4        & 2.32\%                & 0.17\% & 0.7051         \\ \hline
\end{tabular}
\end{table} 

\subsubsection{Reconstruction}
\label{sec:ECALReconstruction}

A dedicated reconstruction algorithm based on shower topology was developed to provide additional information beyond the total ECAL energy, enhancing the rejection of rare background events. Table~\ref{tab:ECAL_shower_variable} presents several reconstructed variables along with their definitions. The effectiveness of reconstruction is directly linked to the granularity of the ECAL --- finer granularity provides more detailed shower information. Therefore, we maintained a moderate granularity in the ECAL, aiming to leverage reconstruction information for improving signal selection. Section~\ref{sec:ECALReconstructedVariables} shows some preliminary results of reconstruction, demonstrating the potential of using reconstructed variables in multivariate analysis to suppress rare backgrounds.

% The total energy deposition in ECAL is reconstructed by summing up all the energy in each unit, after smearing. Shower variables are defined as Table~\ref{tab:ECAL_shower_variable} and studied to reduce the background processes, significantly benefited from the high granularity of ECAL design. The performance of signal/background with shower variables are shown in the following Section~\ref{sec:ECALReconstructionEfficiency}.

\begin{table}[h]
\footnotesize
\fontsize{8}{12}\selectfont
\caption{\label{tab:ECAL_shower_variable}~Several reconstructed variables of the ECAL.}
\tabcolsep 16pt %space between two columns. 
\centering
\begin{tabular}{cc}
\hline
Reconstructed Variable & Definition \\ \hline
$E_\mathrm{ECAL}^\mathrm{total}$   & Total energy deposition in ECAL\\  
$N_\mathrm{hits}$            & Number of fired cells (hits) with non-zero energy deposition \\ 
$E_\mathrm{mean}$          & Average energy deposition of the hits, $E_\mathrm{ECAL}^\mathrm{total}$ /$N_\mathrm{hits}$  \\ 
Hit layers & Number of layers with at least one hit \\  
Shower density   & Average number of hits in 3 $\times$ 3 $\times$ 3 cells around every fired one \\ 
Shower layer & Number of layers with energy deposition and $\mathrm{XWidth}, \mathrm{YWidth} \ge 5 \ \mathrm{cm}$\\ \hline
\end{tabular}
\end{table}

\subsection{Radiation damage}
\label{sec:ECALRadiationDamage}

Given the high-energy and high-frequency beam environment, the ECAL, particularly its central region, is subjected to significant radiation dose that may degrade its performance. Therefore, the radiation damage to the ECAL must be evaluated, focusing on crystal damage primarily from ionizing energy loss and silicon sensor damage from non-ionizing energy loss. Simulations were conducted using Geant4 to estimate the radiation damage to crystals and silicon sensors under $3\times10^{14}$ electrons-on-target events. 

\begin{figure}[htbp]
\centering
    \subfigure[]{
    \includegraphics[width=0.45\textwidth]{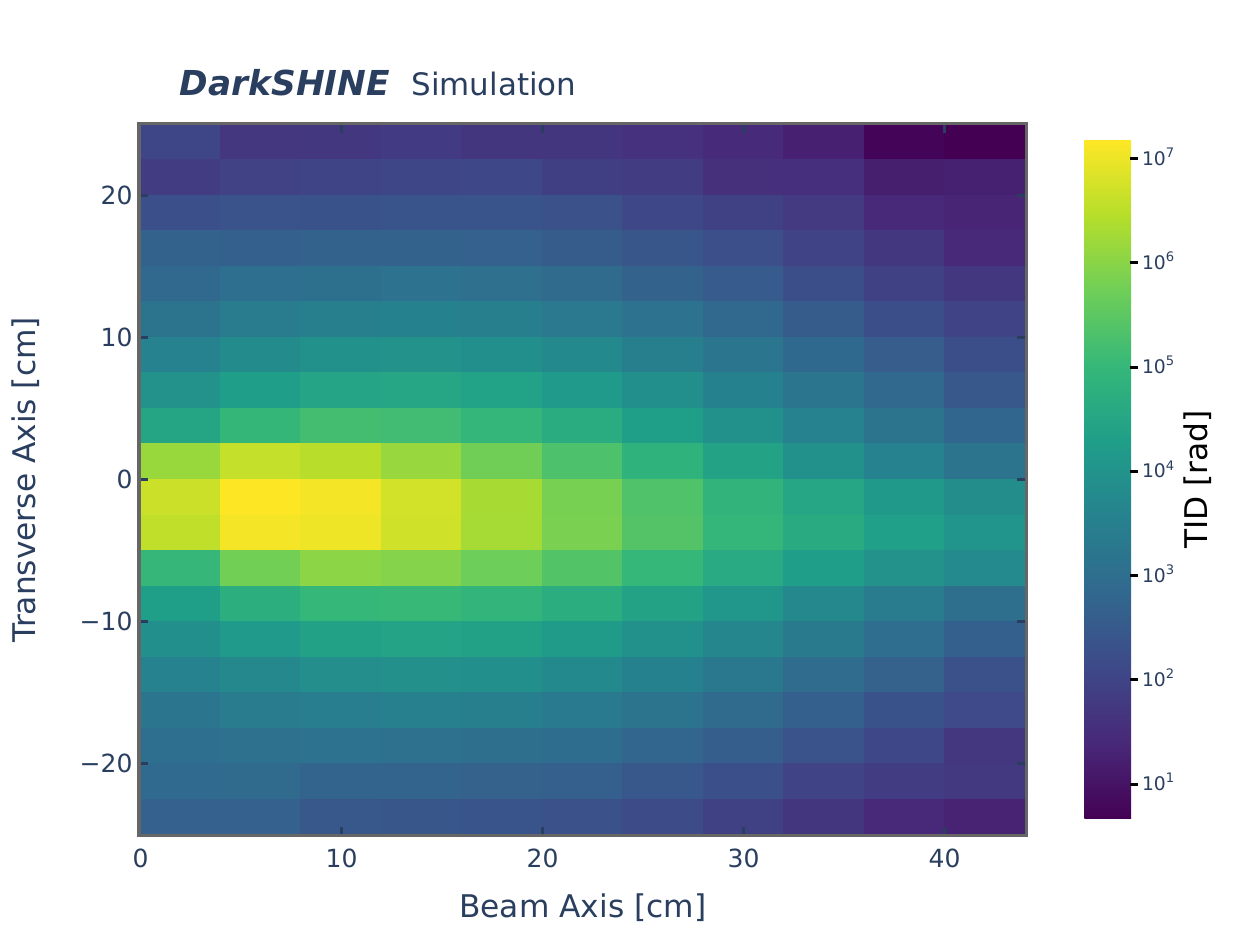}}
    \subfigure[]{
    \includegraphics[width=0.45\textwidth]{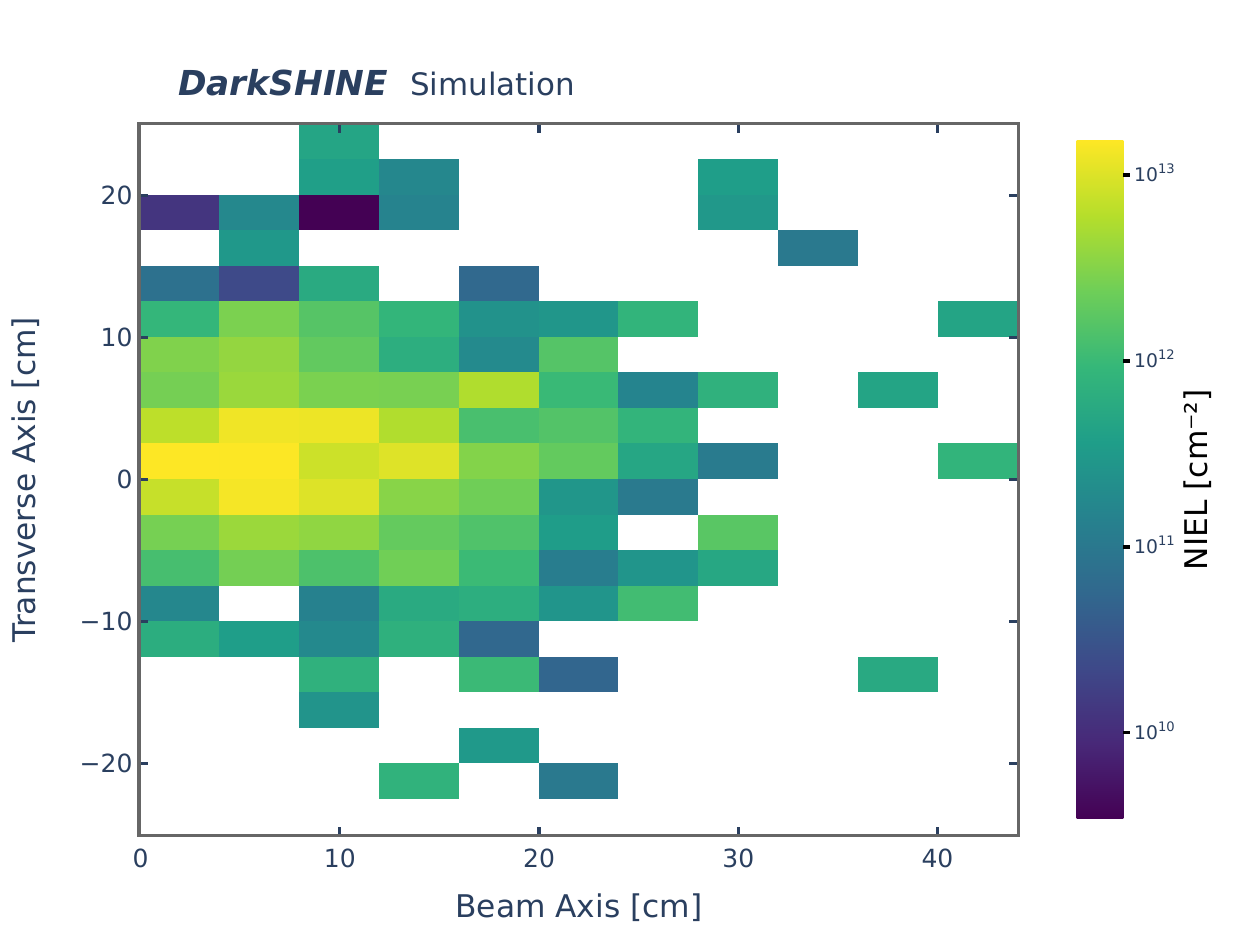}}
    \caption{\label{fig:RadiationDamage}~Distribution of radiation damage in the ECAL region under $3\times10^{14}$ electrons-on-target events. The two distributions illustrate the radiation damage along the ECAL symmetry plane in beam direction. In figures, the horizontal axis represents the beam direction, while the vertical axis corresponds to the transverse directions. Each segment indicates the position of a crystal. (a) Total Ionizing Dose (TID) absorbed by crystals, with a maximum value of $10^7$ rad. (b) Non-Ionizing Energy Loss (NIEL) in silicon sensors, expressed as the equivalent 1 MeV neutron flux, with a maximum value of $10^{13}$ per square centimeter.~\cite{zhao2024DSECAL}}
\end{figure}

Radiation damage to crystals mainly results from ionizing energy loss of incident particles, which can be evaluated by the Total Ionizing Dose (TID). The TID in crystal is defined as the ionizing energy absorbed per unit mass of crystal. The average ionizing energy loss absorbed by each crystal in the ECAL per event was calculated based on one million electrons-on-target events. This average value was then multiplied by $3\times10^{14}$ to estimate the TID for each crystal. For the crystal absorb the maximum dose, the value of TID is about $10^7$ rad (Figure~\ref{fig:RadiationDamage}(a)). Most inorganic scintillators commonly used in high-energy physics detectors, like CsI, BGO, and PWO, lose significant light yield after such a dose. However, LYSO exhibits only a small reduction in light yield\cite{Zhu:2019ihr}, meeting the radiation resistance requirements for the DarkSHINE ECAL.

Radiation damage to silicon sensors mainly results from Non-Ionizing Energy Loss (NIEL), typically expressed using the equivalent 1 MeV neutron flux. To obtain the flux on each sensor, the average NIEL per event for each sensor was first calculated through simulation, referred to as $E_1$. Then, the average NIEL for a 1 MeV neutron passing through a single sensor was simulated, referred to as $E_2$. Finally, the ratio $E_1/E_2$ provides the equivalent 1 MeV neutron flux for each sensor. In the DarkSHINE ECAL, the equivalent 1 MeV neutron flux on silicon sensors in the most heavily irradiated area is about $10^{13}$ per square centimeter (Figure~\ref{fig:RadiationDamage}(b)). This significant radiation could cause the dark current of general sensors to increase by several orders of magnitude, rendering them unusable. Therefore, silicon sensors with excellent radiation resistance are required for our experiment.\cite{ULYANOV2020164203, SANCHEZMAJOS2009506, PREGHENELLA2023167661}

\subsection{R\&D activities}

\subsubsection{Studies on SiPMs}
\label{sec:ECALSiPMTest}

SiPM is a type of photon sensor composed of an array of avalanche photodiodes (APDs) operating in Geiger mode above the breakdown voltage, allow a single photon to initiate a self-sustaining avalanche. Noted for their high gain, superior timing resolution, low voltage requirements, and insensitivity to magnetic fields, SiPMs are extensively utilized in diverse fields, including medical imaging (notably in PET scans), LIDAR systems, astrophysical research, and high-energy physics. 

In DarkSHINE ECAL, the SiPM should have a large dynamic range to match the high light output of LYSO crystal scintillator. The dynamic range of SiPM is inherently linked to its pixel pitch and the number of pixels it contains. Saturation occurs when the number of received photons exceeds the available pixels, becoming more pronounced under high photon influx. When measuring scintillation light, a higher pixel density in the SiPM makes it less prone to saturation. Therefore, we selected SiPMs with small pixel sizes for testing, such as the HAMAMATSU S14160-3010PS~\cite{S14160-3010PS} with 10 $\mu$m pixels and the NDL EQR06-11-3030D-S~\cite{EQR06} with 6 $\mu$m pixels.

\begin{figure}[htbp]
\centering
    \subfigure[]{
    \includegraphics[width=0.26\textwidth]{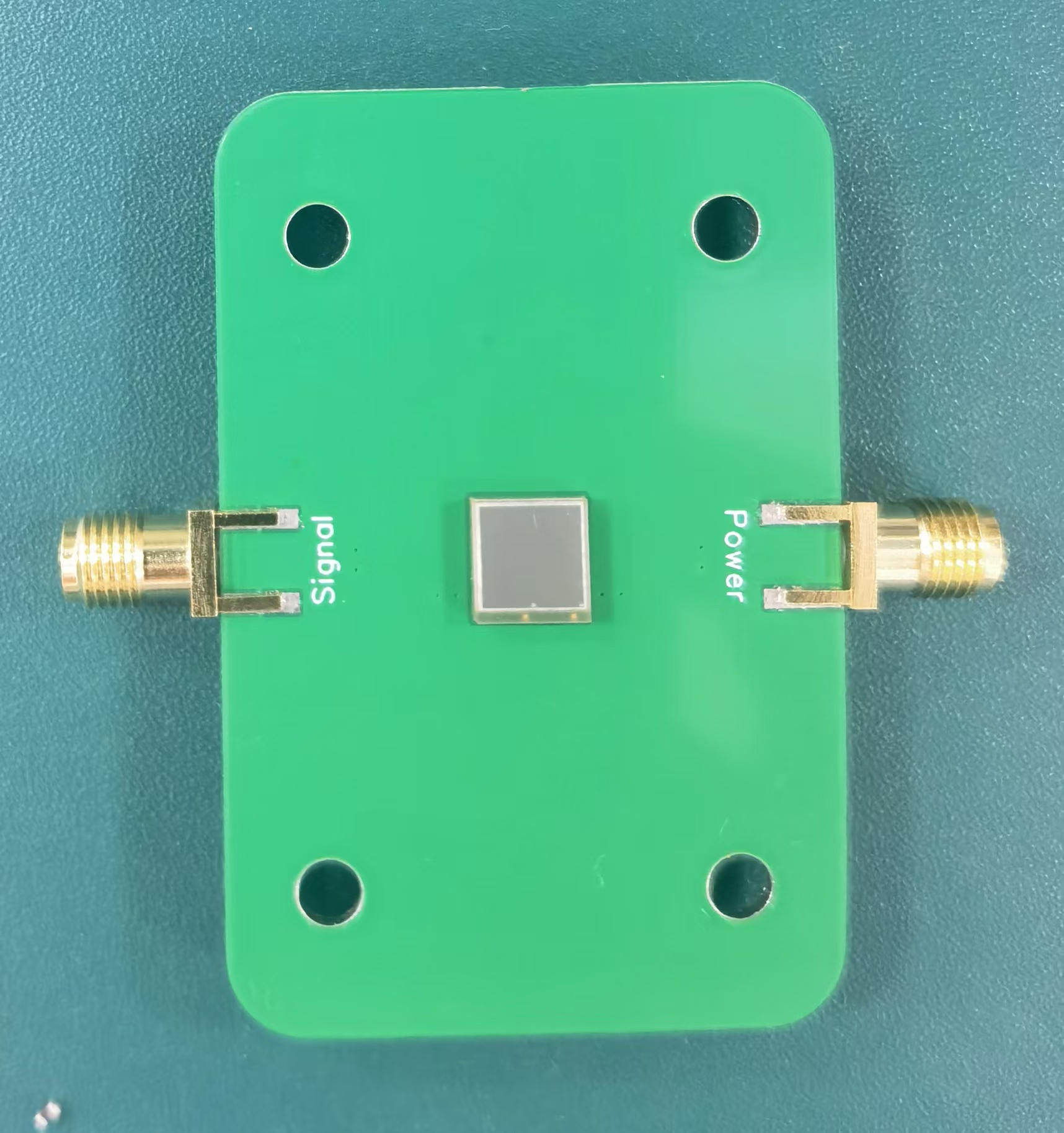}}
    \subfigure[]{
    \includegraphics[width=0.38\textwidth]{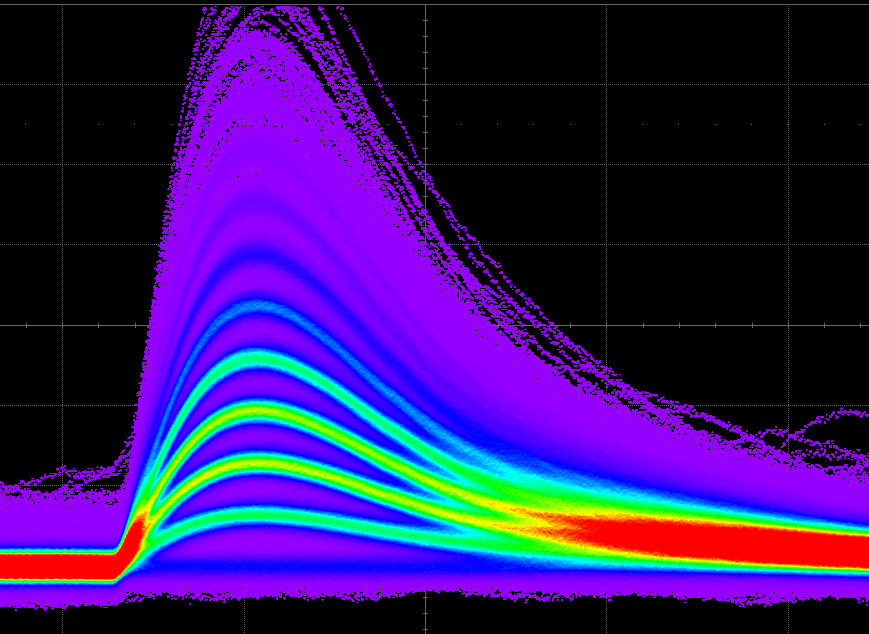}}
    \subfigure[]{
    \includegraphics[width=0.32\textwidth]{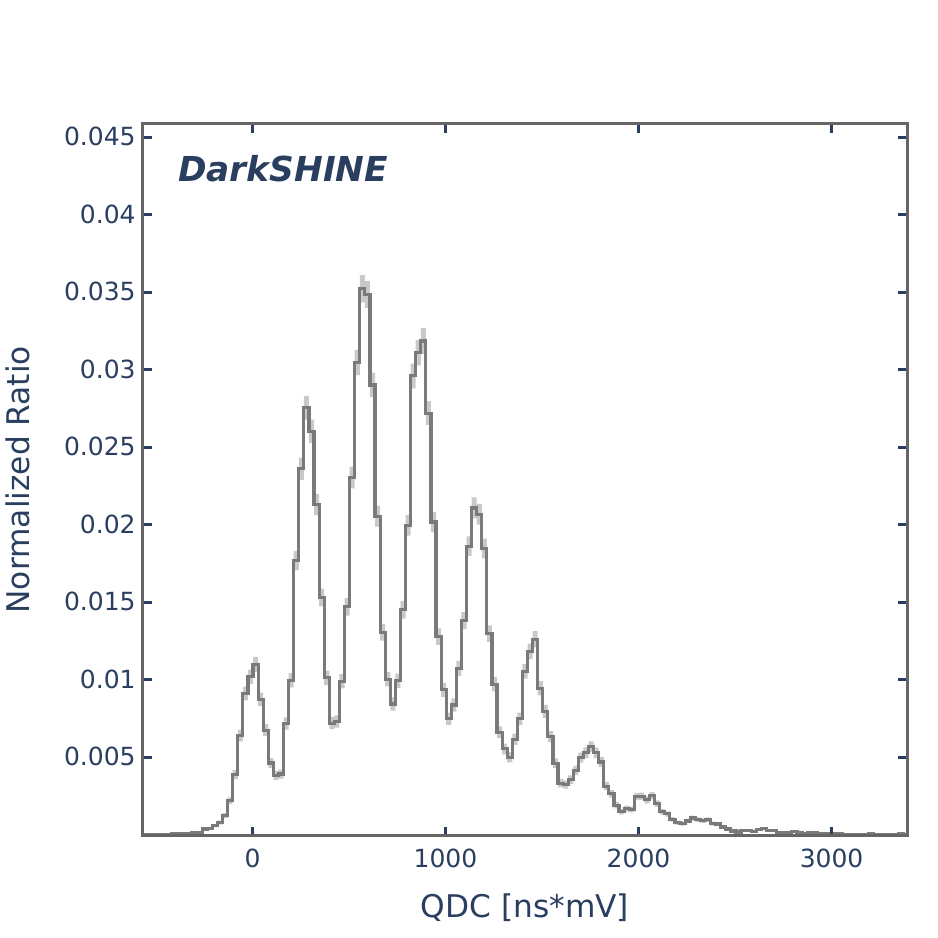}}
\caption{\label{fig:SiPMGain}~(a) SiPM (HAMAMATSU S13360-6025PE~\cite{S13360-6025PE}) soldered on the front-end electronics board. (b) LED signals detected by SiPM. (c) QDC Spectra corresponding to several photoelectrons response of SiPM.~\cite{zhao2024dynamic}}
\end{figure}

The gain of a SiPM is defined as the number of charge carriers (electrons) produced per photoelectrons, it determines the signal magnitude generated by the SiPM when detecting a unit quantity of photoelectrons. Since various factors, including temperature and bias voltage, can influence SiPM gain, calibration for the gain of SiPM is necessary. During the calibration, a LED was used as the light source. By adjusting the bias voltage of the LED, the SiPM can detect only a few photoelectron signals (Figure~\ref{fig:SiPMGain} (a) and (b)). The integrated Charge-to-Digital-Converter (QDC) spectra in Figure~\ref{fig:SiPMGain} (c) shows a clear discrimination for signals with different numbers of photoelectrons. The number of QDC per photoelectron, which is proportional to SiPM gain, can be obtained with a multi-Gaussian fitting applied on the QDC spectra. Table~\ref{tab:SiPMGain} lists the measured gains of three SiPMs.

\begin{table}[h]
\fontsize{8}{12}\selectfont
\centering
\caption{\label{tab:SiPMGain}~Calibrated QDC-to-p.e. ratios of SiPMs.~\cite{zhao2024dynamic}}
    \begin{tabular}{ccc}
        \toprule
        \makecell[c]{S13360-6025PE} &\makecell[c]{S14160-3010PS}  &\makecell[c]{EQR06 11-3030D-S}\\ 
        \midrule
        6.42 ns$\cdot$mV/p.e. & 1.69 ns$\cdot$mV/p.e. & 0.9 ns$\cdot$mV/p.e. \\
        \bottomrule
    \end{tabular}
\end{table}

The SiPM dark count refers to noise of the device in the absence of incident light. Dark count result from thermal noise, spontaneous emission, or other processes that trigger the SiPM's detection mechanism, producing false signals. Dark count is an intrinsic property of SiPMs and contribute to the overall noise in the system. The rate at which these dark counts occur is referred to as the Dark Count Rate (DCR). According to our measurements, the DCR of the SiPM is on the order of hundreds to thousands of Hertz. However, due to the short scintillation decay time of the LYSO crystal, combined with the fast shaping time of the electronics, the dark count contribution of the SiPM is minimal compared to the light output of LYSO and can be considered negligible.

\begin{figure}[htbp]
\centering
    \subfigure[]{
    \includegraphics[width=0.46\textwidth]{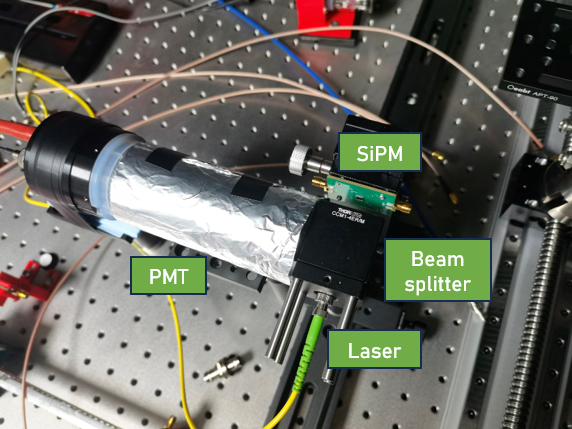}}
    \subfigure[]{
    \includegraphics[width=0.4\textwidth]{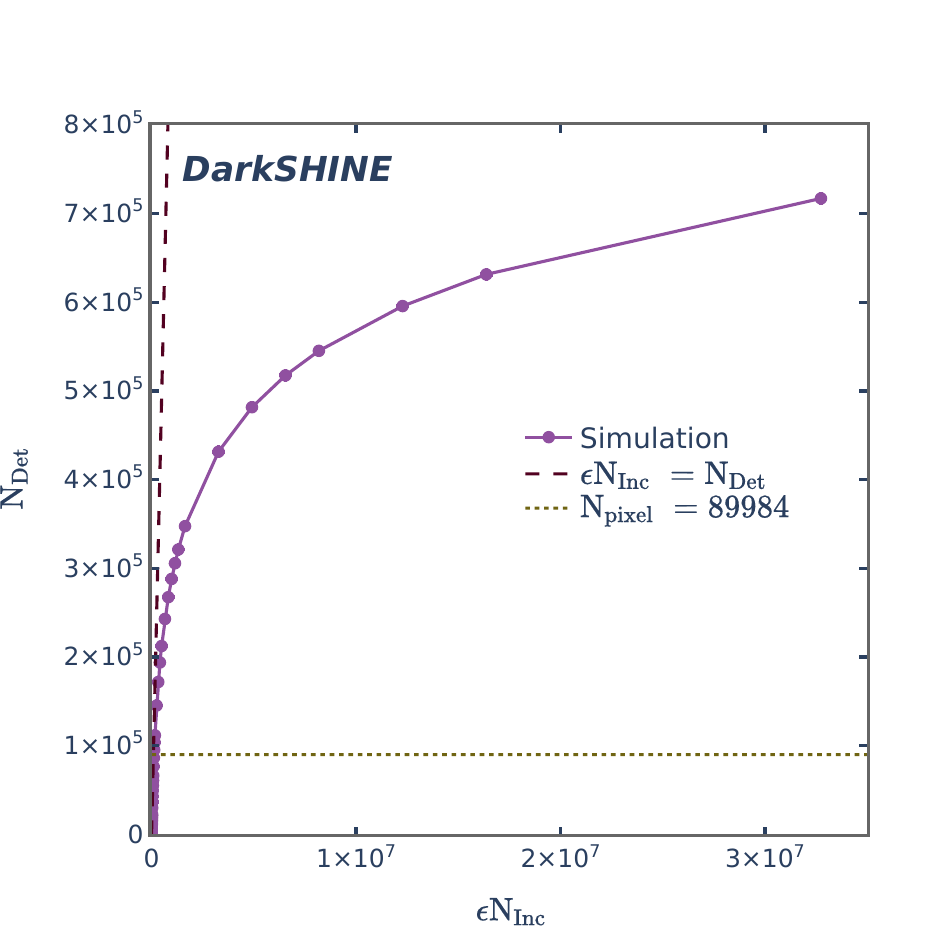}}
\caption{\label{fig:SiPMDynamic}~(a) Setup of an experiment to measure the SiPM intrinsic dynamic range under pico-second laser. (b) SiPM response under pico-second laser. The red points represent the experimental measurement results, the blue dashed line represents the nominal pixel counts of the SiPM, and the purple dashed line has a slope of one.~\cite{zhao2024dynamic}}
\end{figure}

A dedicated experiment was designed to measure the intrinsic dynamic range of the SiPM under a laser source. In this setup, a pico-second laser serves as the light source, with a PMT acting as an auxiliary calibration device. The laser pulse width is within 45 ps to ensure that the SiPM pixels are not re-illuminated. As shown in Figure~\ref{fig:SiPMDynamic} (a), the laser beam is split by a beam splitter into two parts: one directed at the SiPM under test, and the other received by the PMT. The PMT is maintained in a linear response mode by applying varying bias voltages according to the light intensity. In Figure~\ref{fig:SiPMDynamic} (b), the horizontal axis represents the effective photon count, which corresponds to the calibrated PMT response and can be used to estimate the number of photoelectrons the SiPM should detect without saturation effects. Initially, the SiPM output increases almost linearly but gradually deviates from linearity as the effective photon count rises, eventually reaching a saturation region. For the measured SiPM, nonlinearity begins to appear at roughly 10\% of the total pixel count. The maximum output under laser illumination is close to, but slightly less than, SiPM's nominal pixel count, due to manufacturing variations. Additionally, we also built a monte carlo model to simulate the nonlinear response of the SiPM.~\cite{zhao2024dynamic}

Due to the high repetition rate of the electron beam, the SiPMs in the ECAL will experience significant radiation damage, as discussed in Section~\ref{sec:ECALRadiationDamage}. A potential method for addressing this radiation damage in the DarkSHINE experiment is the in-situ current annealing technique proposed in~\cite{Tsang:2016cmc, Cordelli:2018kgh, GU2023168381}, which does not require the disassembly of the SiPM or the use of additional heating devices.

\subsubsection{Studies on crystal Scintillators}
\label{sec:ECALScinLightTest}

The LYSO crystal scintillator demonstrates excellent performance in terms of scintillation decay time, radiation resistance, and light yield~\cite{Kalinnikov:2023coj, Zhu:2019ihr}. Some experiments to evaluate the performance of LYSO produced by CETC were conducted. 

\begin{figure}[h]
\centering
    \subfigure[]{
    \includegraphics[width=0.35\textwidth]{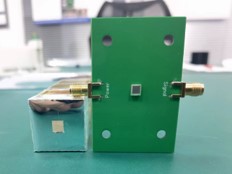}}
    \subfigure[]{
    \includegraphics[width=0.45\textwidth]{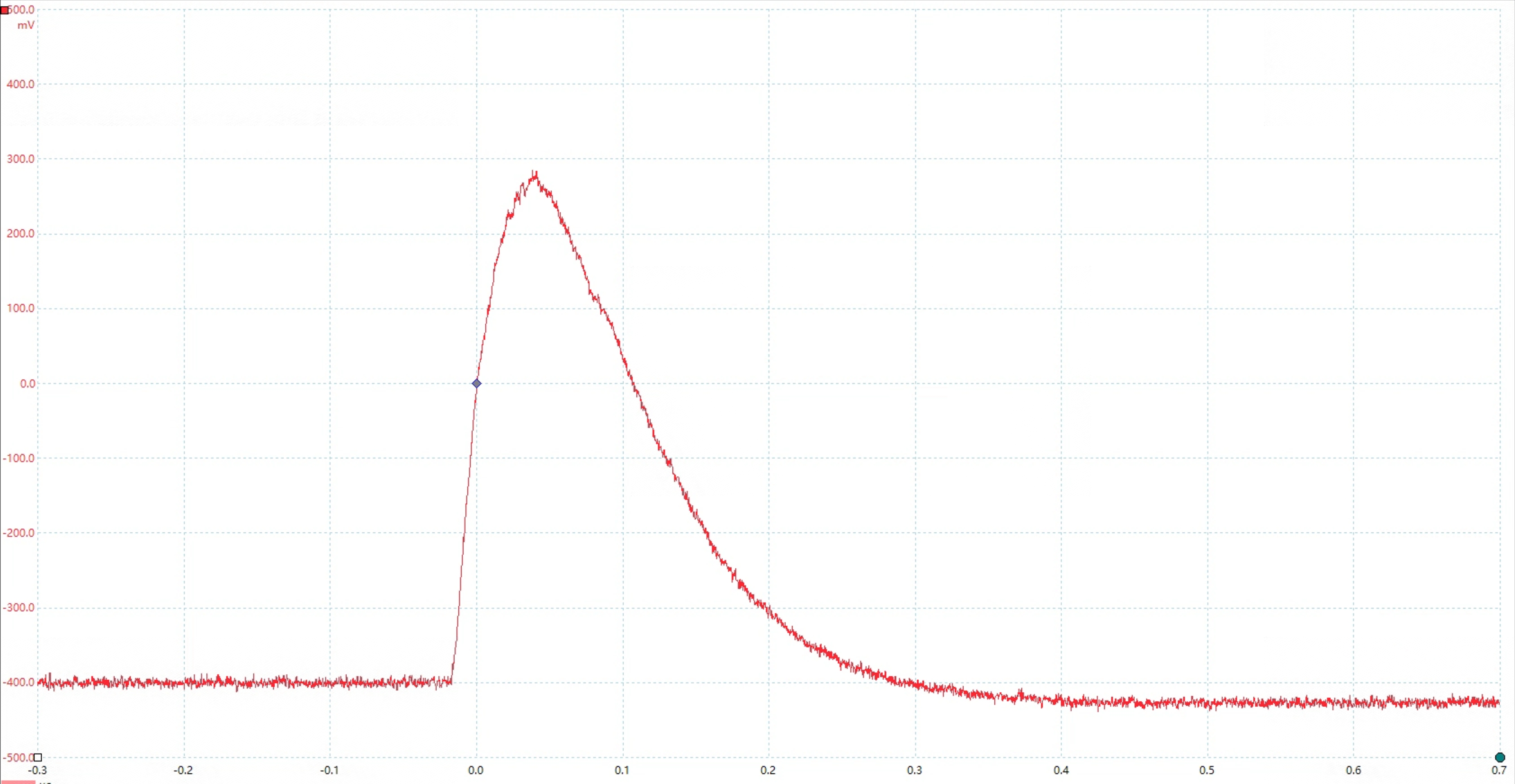}}
\caption{\label{fig:LYSOWaveform}~(a) LYSO-SiPM detection unit. (b) Signal waveform output by LYSO-SiPM unit.}
\end{figure}

As a fixed-target experiment operating at a high repetition rate of 1 MHz, channels in ECAL will encounter an extremely high event rate up to 1 MHz, which means that the ECAL crystal units need to have a response time of less than 1 $\mu$s. The scintillation decay time of LYSO crystals is approximately 40 ns. The signal width for a single channel in the ECAL is around 200 ns (Figure~\ref{fig:LYSOWaveform}). Therefore, the LYSO crystal ECAL has a sufficiently fast response time to prevent event pile-up effects in DarkSHINE experiment.

% (Figure~\ref{fig:LYSOWaveform} (a)), which may result in event pileup. The event rate of channel is defined as the trigger rate when the channel energy exceeds 2 MeV. Minimizing the signal width is crucial for mitigating the pileup effect. The LYSO crystal, with its short scintillation decay time of approximately 40 ns, offers a potential solution. Combined with fast shaping electronics, the signal width for a single channel in the ECAL can be constrained to under 200 ns. As shown in (Figure~\ref{fig:LYSOWaveform} (b), the LYSO-SiPM detection unit consists of a LYSO crystal wrapped in a reflective film of ESR, with one open face for coupling to the SiPM. Figure~\ref{fig:LYSOWaveform} (c) shows a typical waveform of the signal measured by the detection unit with a LYSO crystal with dimensions of 2.5$\times$2.5$\times$4 cm$^3$. The overall width of the waveform is generally below 200 ns, characterized by a rapid rising edge and a slower falling edge. These features are governed by the scintillation time of the crystal, the detector's geometry, and the electronics shaping time. A fitting formula was applied to analyze similar waveforms, as shown in Figure~\ref{fig:LYSOWaveform} (c), yielding rise and fall time constants of 10 ns and 50 ns, respectively.

\begin{figure}[h]
    \centering  %图片全局居中
    \subfigure[]{
    \includegraphics[width=0.46\textwidth]{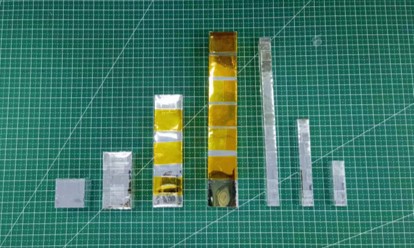}}
    \subfigure[]{
    \includegraphics[width=0.42\textwidth]{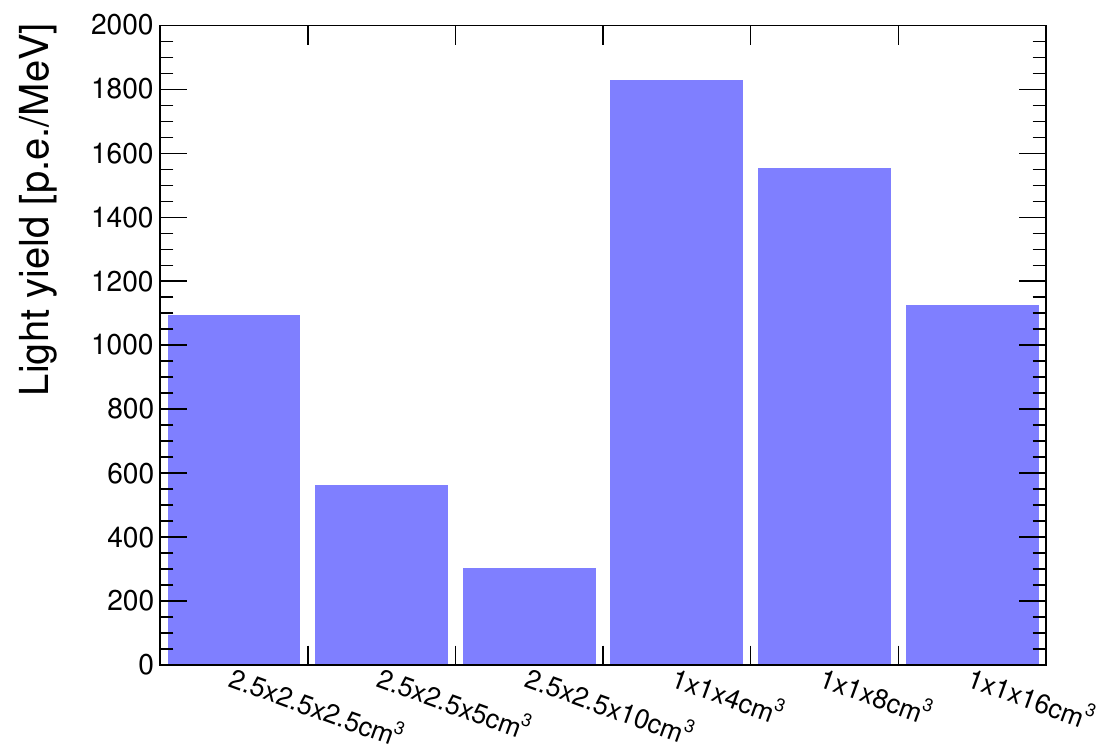}}
    \caption{\label{fig:LYSOLY}~(a) LYSO scintillators with different sizes. (b) Light yield of LYSO scintillator with different sizes.}
\end{figure}

The light yield of LYSO-SiPM detection units with various scintillator sizes was measured with a radioactive source, and is presented in Figure~\ref{fig:LYSOLY}. The LYSO crystals are wrapped in ESR reflective film and coupled to a SiPM at one open end. The same SiPM of NDL EQR06 11-3030D-S was used for testing LYSO crystals of different sizes. Figure~\ref{fig:LYSOLY} (b) shows that shorter crystal lengths and smaller cross-sectional areas result in higher effective light yield, as the light travels a shorter distance within the crystal. These measured light yield results provide an important reference for our detector simulation and digitization.

\begin{figure}[h]
    \centering  %图片全局居中
    \subfigure[]{
    \includegraphics[width=0.23\textwidth]{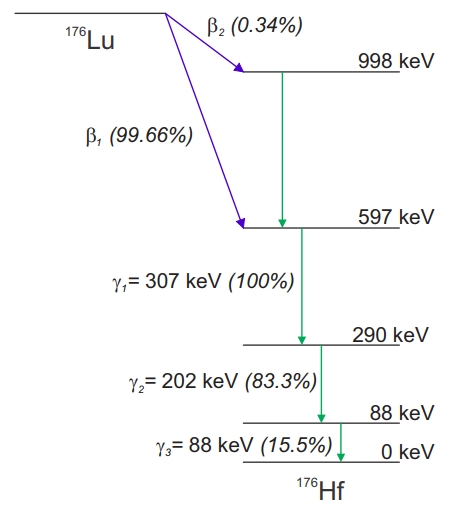}}
    \subfigure[]{
    \includegraphics[width=0.35\textwidth]{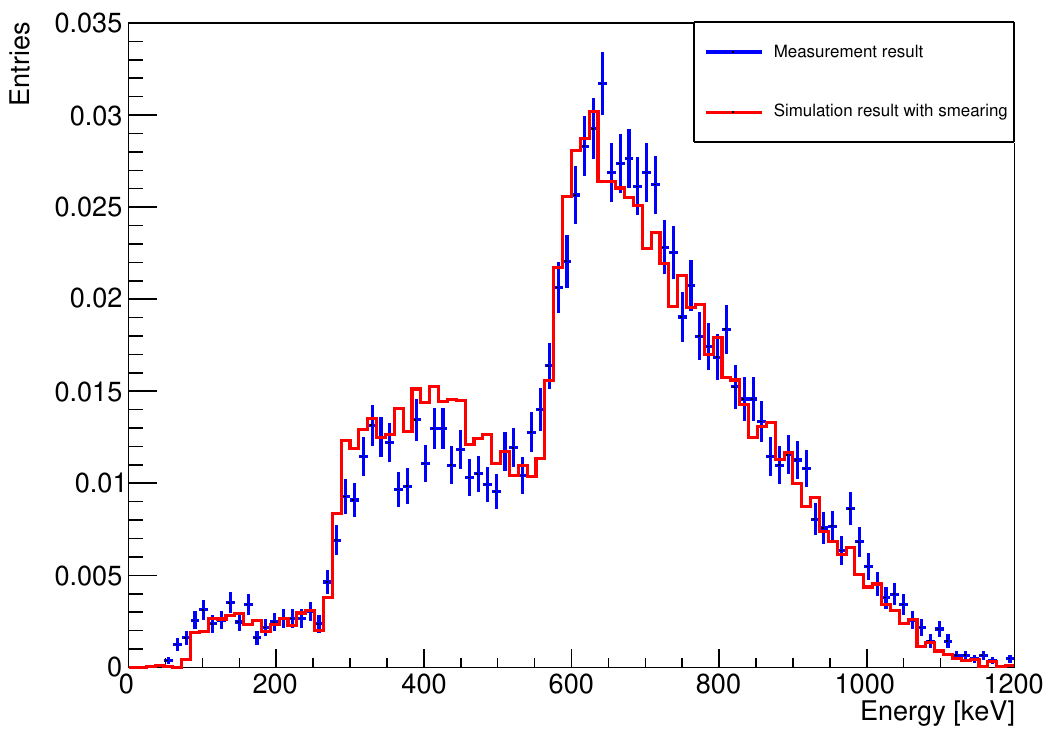}}
    \subfigure[]{
    \includegraphics[width=0.37\textwidth]{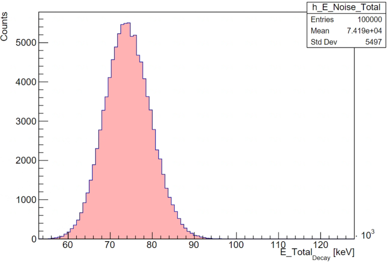}}
    \caption{\label{fig:LYSOIR}~(a) Decay scheme of Lutetium-176.~\cite{LYSOIR} (b) Intrinsic radioactive background spectrum of a 2.5$\times$2.5$\times$2.5 cm$^3$ LYSO crystal scintillator. (c) The average ECAL noise energy introduced by LYSO intrinsic radiation across all channels without energy threshold.}
\end{figure}

The LYSO scintillator contains intrinsic radioactive background originating from the decay of Lutetium-176 into Hafnium-176, which involves one beta decay followed by three consecutive gamma decays (Figure~\ref{fig:LYSOIR} (a)).~\cite{LYSOIR} The energy spectrum of this background is illustrated in Figure~\ref{fig:LYSOIR} (b), where the blue points represent experimental data, and the red histogram corresponds to simulation results, showing good agreement. With a half-life of 10$^{10}$ years, Lutetium-176 contributes approximately 240k Bq of radioactive activity in a single 2.5$\times$2.5$\times$4 cm$^3$ LYSO crystal. Without applying an energy threshold to channels, the contribution of this intrinsic background to the total energy measured by the ECAL is around 74 MeV, as shown in Figure~\ref{fig:LYSOIR} (c). Such level of background has a negligible impact on our dark photon search, and intrinsic noise can be further suppressed by setting appropriate channel thresholds. However, given its well-defined energy spectrum and long half-life, the background radiation can also be used to monitor and calibrate the response of each detector channel, improving overall measurement precision.

\subsubsection{Beam test on a crystal module}
\label{sec:CrystalModule}

To investigate the detector's response to high-energy particles and explore detector modularization techniques, we developed a compact four-channel crystal module, which comprises four short LYSO crystal bars and front-end electronics board integrated with four SiPMs (Figure~\ref{fig:Beamtest} (a)). A high-energy electron beam tests, with energies ranging from 1 to 5 GeV, were conducted on the crystal module at DESY TB-22, in 2023. The crystal module was placed inside a dark box, with the beam particles triggered by two 1 cm$^3$ plastic scintillators before reaching one end of the crystal along its longitudinal direction(Figure~\ref{fig:Beamtest} (b)).

\begin{figure}[h]
    \centering  %图片全局居中
    \subfigure[]{
    \includegraphics[width=0.38\textwidth]{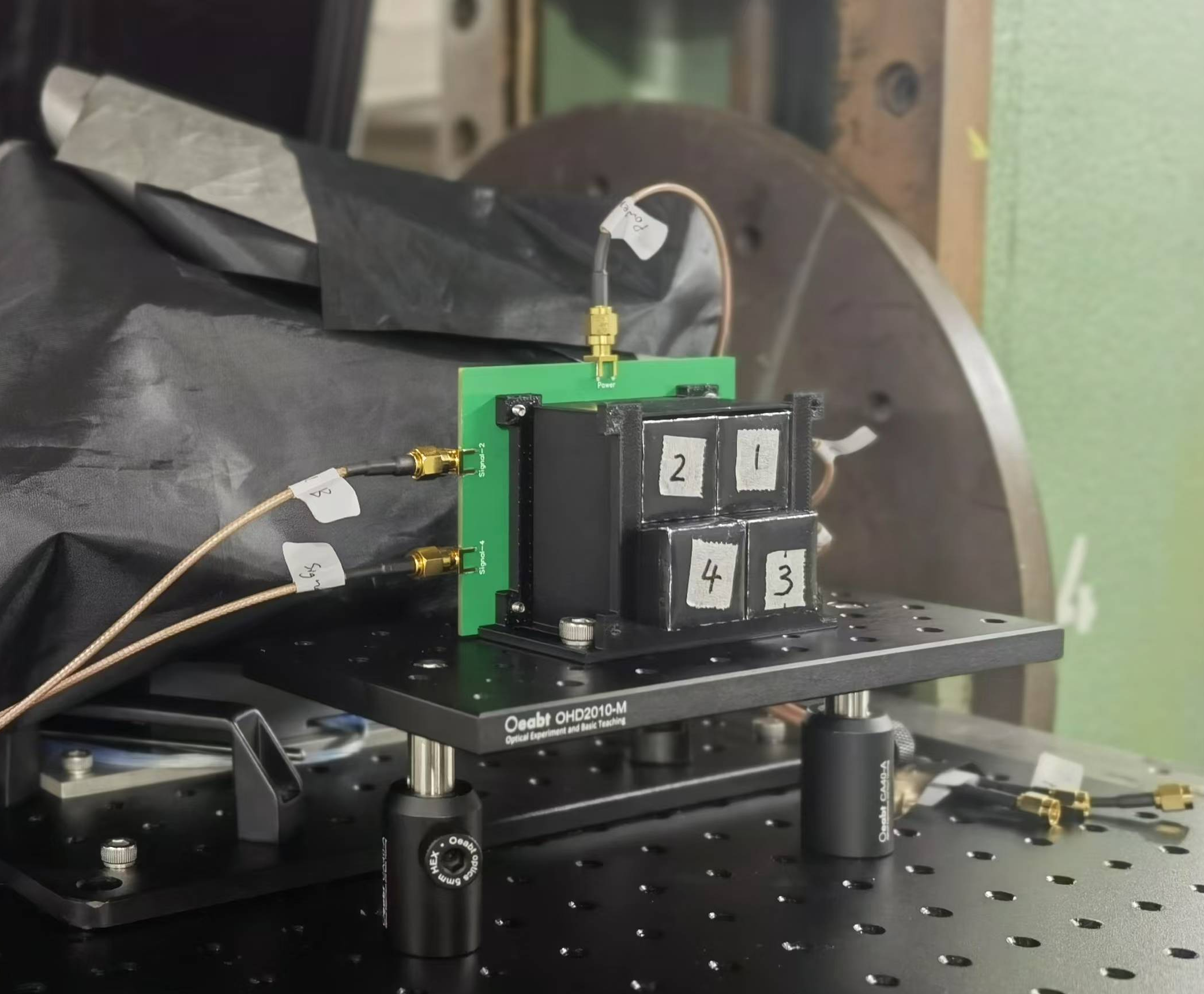}}
    \subfigure[]{
    \includegraphics[width=0.59\textwidth]{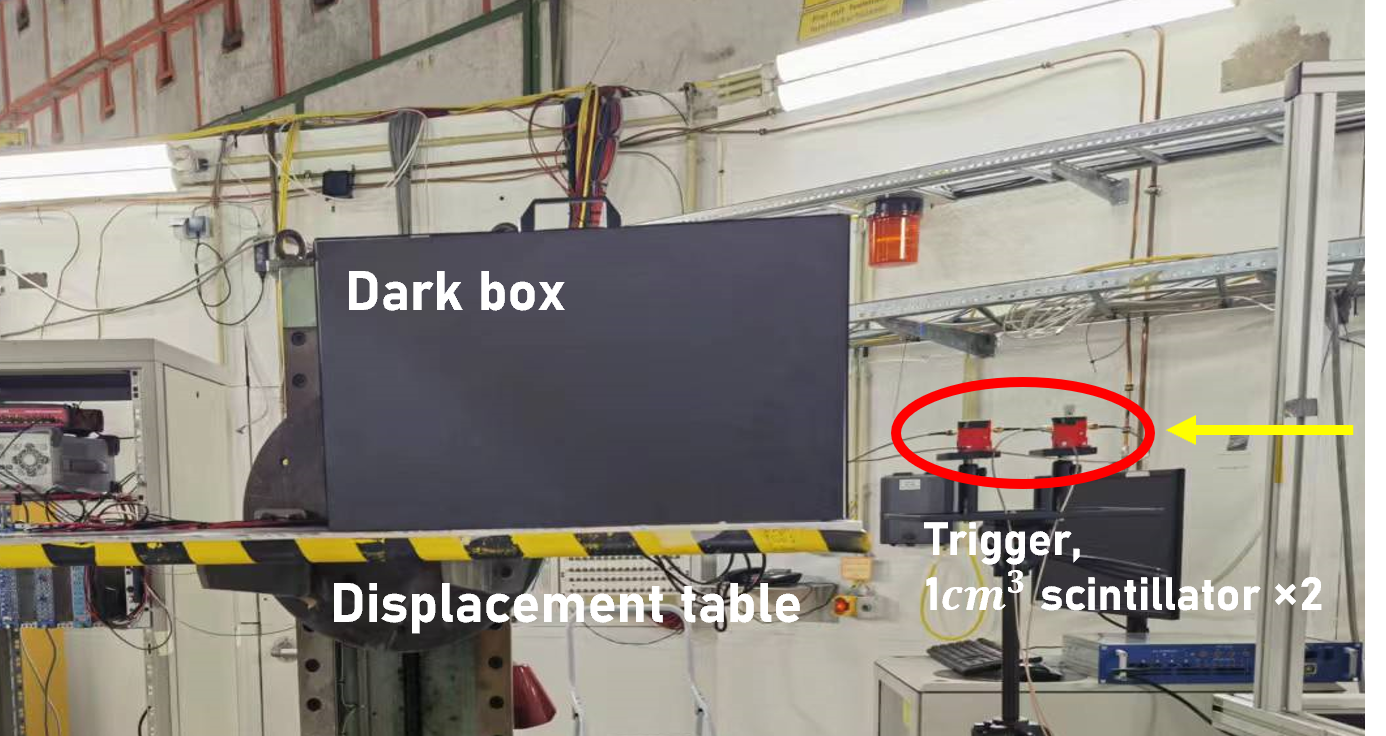}}
    \caption{\label{fig:Beamtest}~(a) Configuration of a four-channel crystal module. (b) Experimental setup of beam test on crystal module at DESY TB-22.}
\end{figure}

\begin{table}[h]
\fontsize{8}{12}\selectfont
\caption{~Beam, crystal, and SiPM used in the beam test.}
    \centering
    \begin{small}
    \begin{tabular}{cccc}
    \hline
    \textbf{Beam article} & \textbf{Beam energy} & \textbf{LYSO crystal} & \textbf{SiPM}\\ 
    \hline
     e$^-$ & 1, 2, 3, 4, 5 GeV& \makecell[c]{2.5$\times$2.5$\times$4cm$^3$ $\times$2,\\2.5$\times$2.5$\times$5cm$^3$ $\times$2}  & HPK S14160-3010PS \\
    \hline
    \end{tabular}
    \end{small}
    \label{tab:TableBeam}
\end{table}

\begin{figure}[h]
\centering
    \subfigure[]{
    \includegraphics[width=0.32\textwidth]{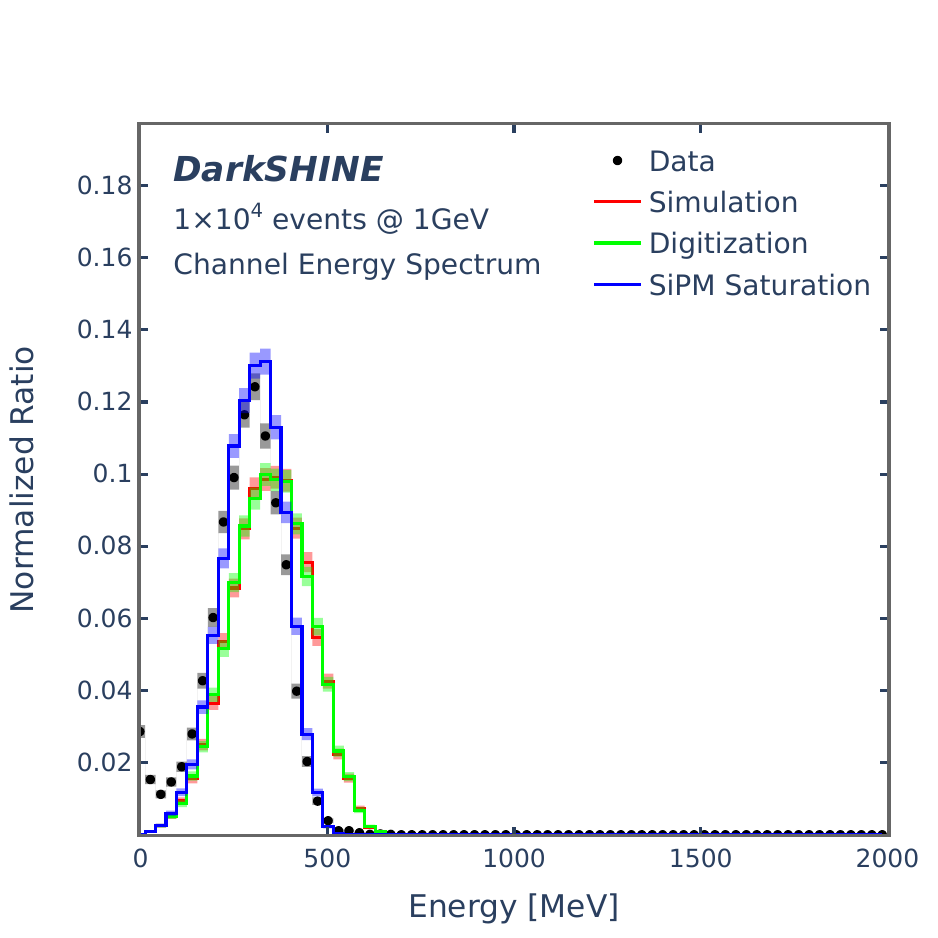}}
    \subfigure[]{
    \includegraphics[width=0.32\textwidth]{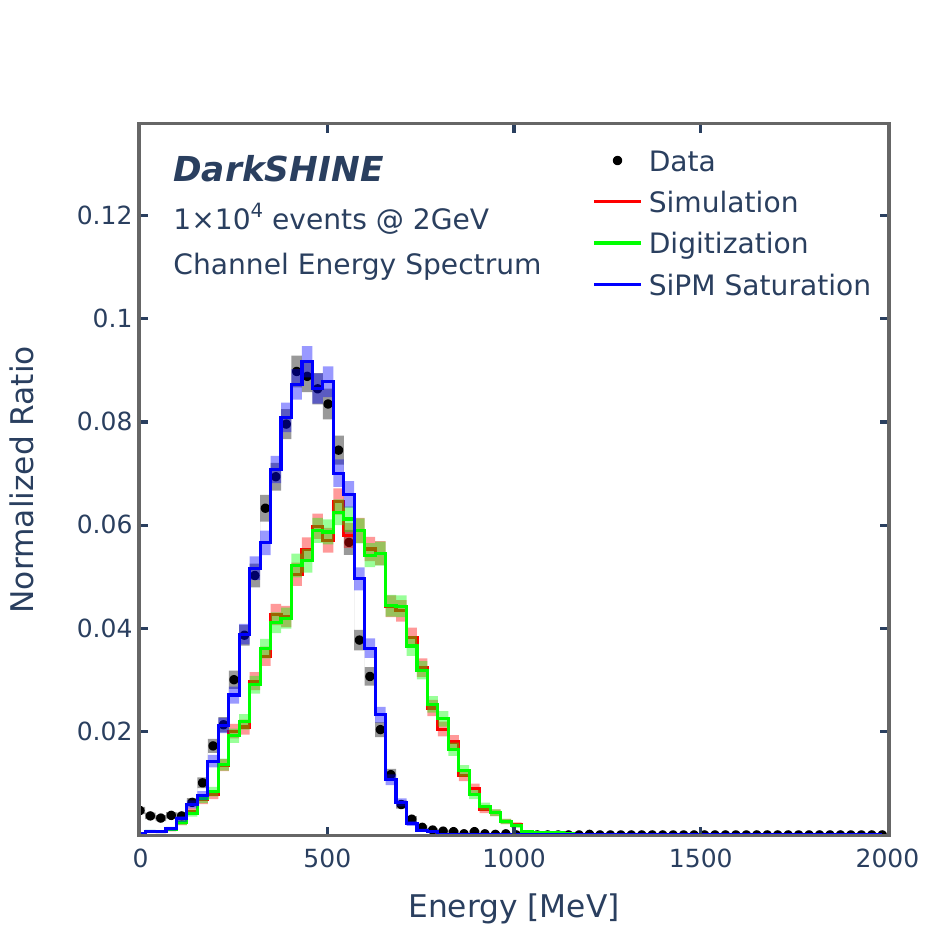}}
    \subfigure[]{
    \includegraphics[width=0.32\textwidth]{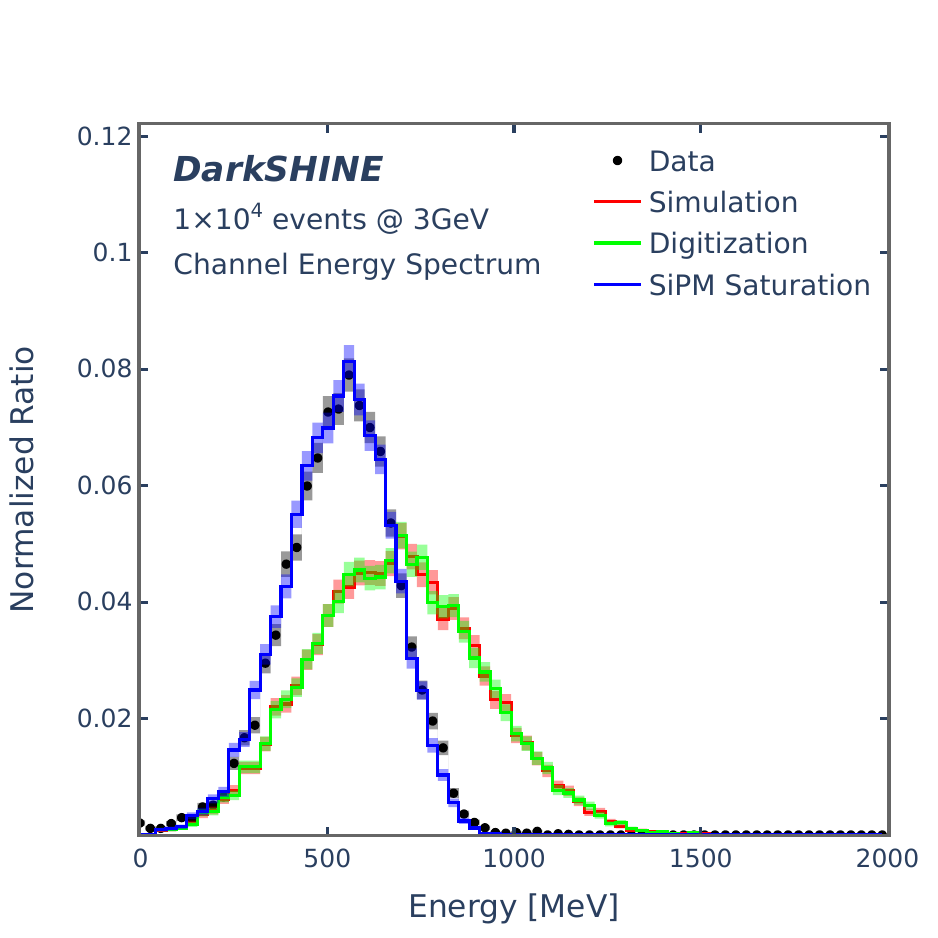}}
    \subfigure[]{
    \includegraphics[width=0.32\textwidth]{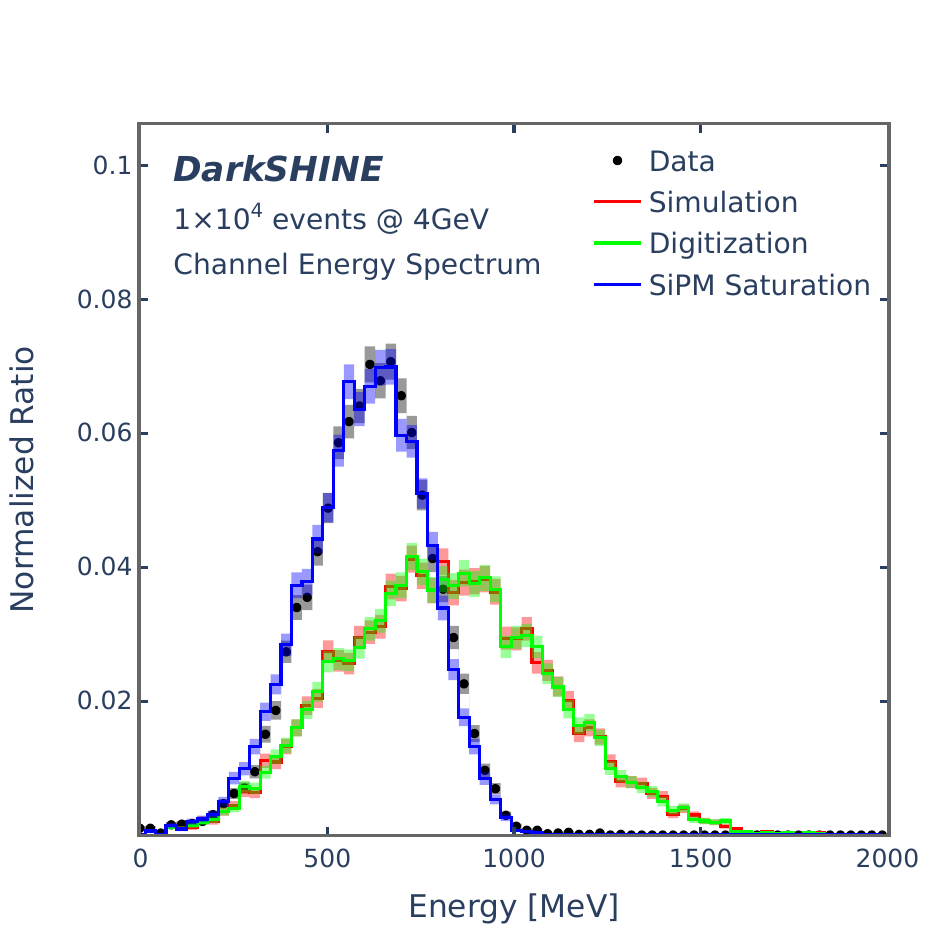}}
    \subfigure[]{
    \includegraphics[width=0.32\textwidth]{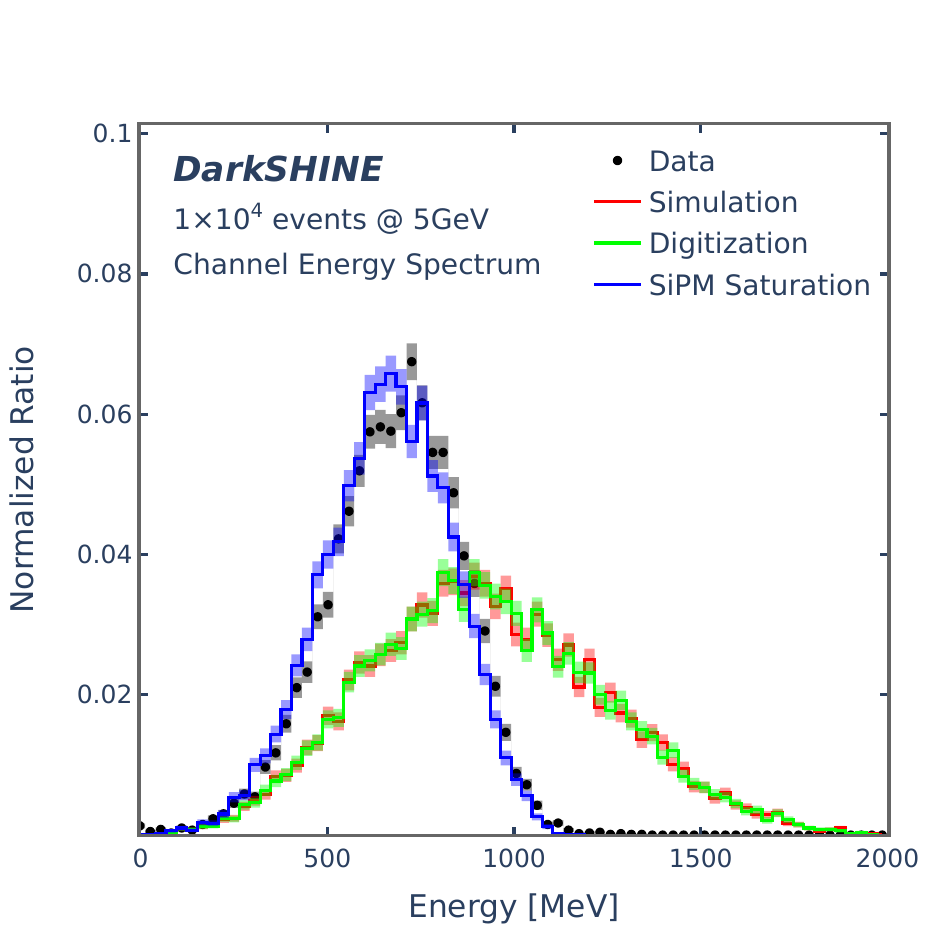}}
\caption{\label{fig:Beam_ChannelEnergy}~The energy spectrum of the crystal(2.5$\times$2.5$\times$5cm$^3$) unit's response to 1-5 GeV electrons.}
\end{figure}

In the test, the incident particles were directed to hit the center of the end of a single crystal, and the energy spectrum obtained from measuring 1-5 GeV electrons using this single channel was analyzed. Additionally, related simulations were conducted to better understand the detector’s behavior. The results of both the data and simulations are shown in Figure~\ref{fig:Beam_ChannelEnergy}. The black dots in the figures represent the measured results. The measured energy was found to be significantly lower than the actual energy of the incident particles, with the deviation increasing as the incident energy increased. This was attributed to the limited volume of the single crystal, which can absorb only a small portion of the energy. As the particle energy increases, more energy is lost due to leakage.

For the simulation results, the difference between the ideal energy deposition (Simulation) and the digitized energy deposition (Digitization) was minimal, as the fluctuations in energy deposition within the single crystal were much larger than those caused by the energy resolution of a single channel. However, the impact of the SiPM saturation effect on energy measurement was significant. Without considering the SiPM saturation effect, the deviation between the simulation and measured results increased with the incident particle energy. After incorporating the SiPM saturation effect, the simulation results and measured results were found to align closely. Since LYSO crystals have a high light yield, with a single channel’s effective light yield approximately 150 photoelectrons per MeV, saturation of the SiPM occurs when large energy deposits are measured. However, the results shown in Figure~\ref{fig:Beam_ChannelEnergy} suggest that the simulation and SiPM response model are able to describe the detector’s behavior accurately, and the SiPM saturation effect can be corrected with the model. This has significant reference value for the design of future real detectors.~\cite{zhao2024dynamic}

\subsubsection{Studies on readout electronics}
\label{sec:ElectronicsTest}

\begin{figure}[h!]
    \centering
    \includegraphics[width=0.6\textwidth]{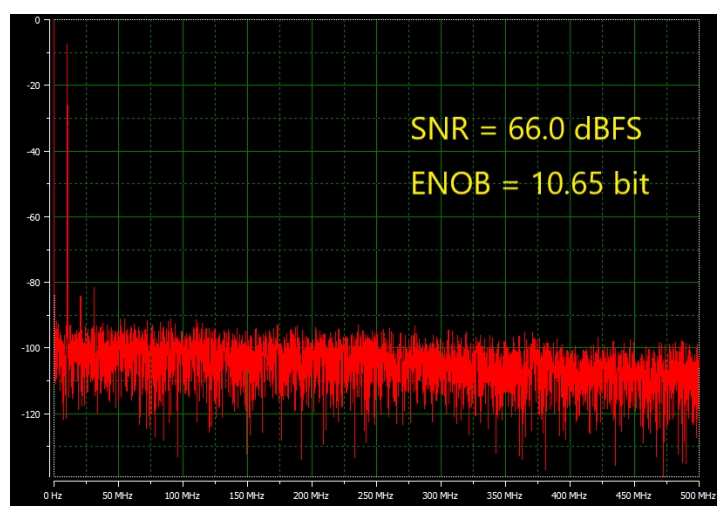}
    \caption{\label{fig:ECALEle_SNR}~The frequency spectrum of the ADC output with the output of 10.3 MHz sine wave.~\cite{guo2024}}
\end{figure}

The readout system was evaluated using a 10.3 MHz sine wave from a signal generator (Figure~\ref{fig:ECALEle_SNR}). The fast Fourier transform (FFT) was applied to convert the ADC output from the time domain to the frequency domain, allowing the system's performance to be analyzed. The test results showed a signal-to-noise ratio (SNR) of 66 dBFS and an effective number of bits (ENOB) of 10.6 bits.~\cite{guo2024}

% The DarkSHINE experiment has a high demand on event rate. Figure~\ref{fig:Completeness} (a) shows, at a repetition rate of 1 MHz, the event rate of channels in the central region of ECAL can already reach 1 MHz. The event rate of channel is defined as the events ratio with channel energy exceeds 2 MeV. Therefore, to meet the measurement requirement of 1 MHz repetition rate, the ECAL readout electronics should be able to handle an event rate of at least 1 MHz. 

% \begin{figure}[h]
%     \centering  %图片全局居中
%     \subfigure[]{
%     \includegraphics[width=0.45\textwidth]{Detectors/Figures/ECAL/Completeness_SingleChannel.pdf}}
%     \subfigure[]{
%     \includegraphics[width=0.45\textwidth]{Detectors/Figures/ECAL/EDis_SamplingRate.pdf}}
%     \caption{\label{fig:Completeness}~(a) Signal completeness of readout electronics for single channels at 1 MHz repetition rate. (b) Energy spectrum measured by a single LYSO-SiPM detection unit with 1 GeV incident electrons. Different numbers of samples per event were achieved by keeping the same integration window for the channel but lowering the sampling rate. The data was acquired by an oscilloscope.}
% \end{figure}

The DarkSHINE experiment has a high demand on event rate. Crystals in the central region of the ECAL have a high energy deposition in almost every event. Therefore, to meet the measurement requirement of 1 MHz repetition rate, the ECAL readout electronics should be able to handle an event rate of at least 1 MHz.

\begin{figure}[h]
    \centering  
    \subfigure[]{
    \includegraphics[width=0.45\textwidth]{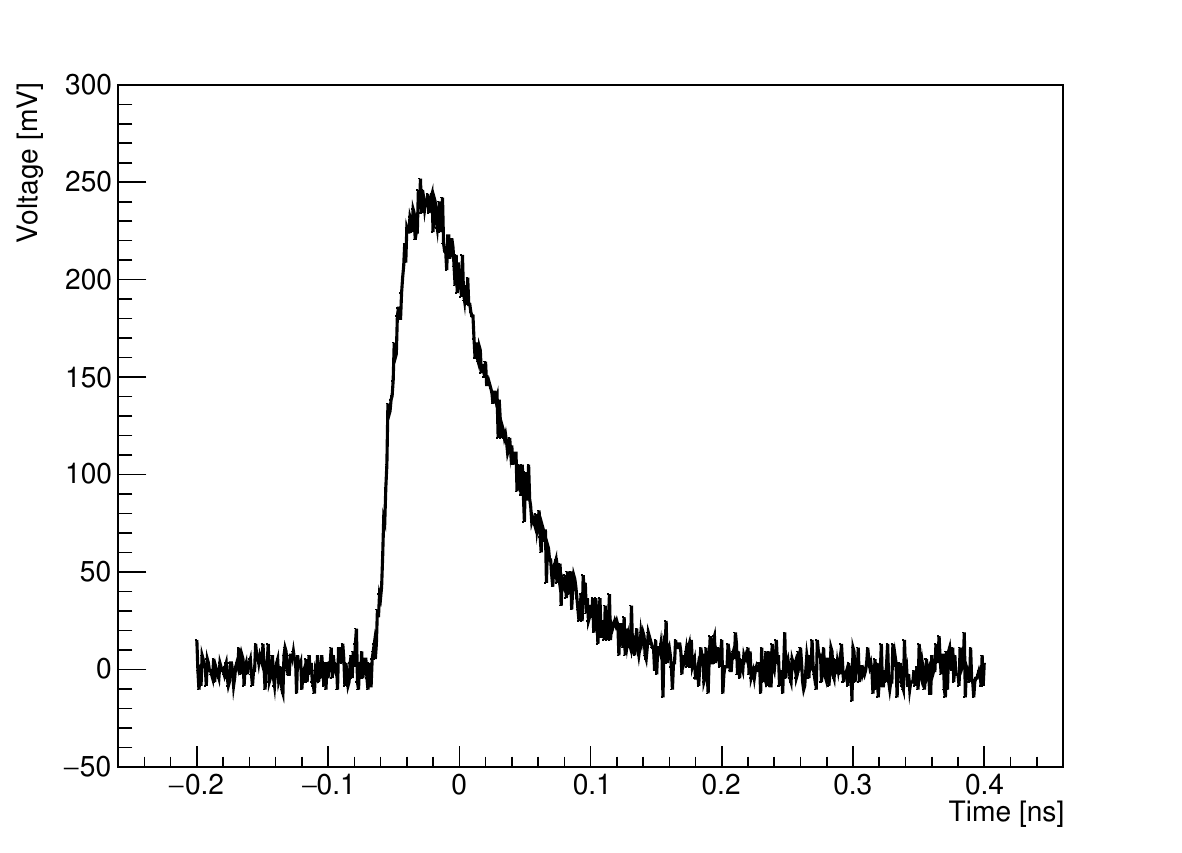}}
    \subfigure[]{
    \includegraphics[width=0.45\textwidth]{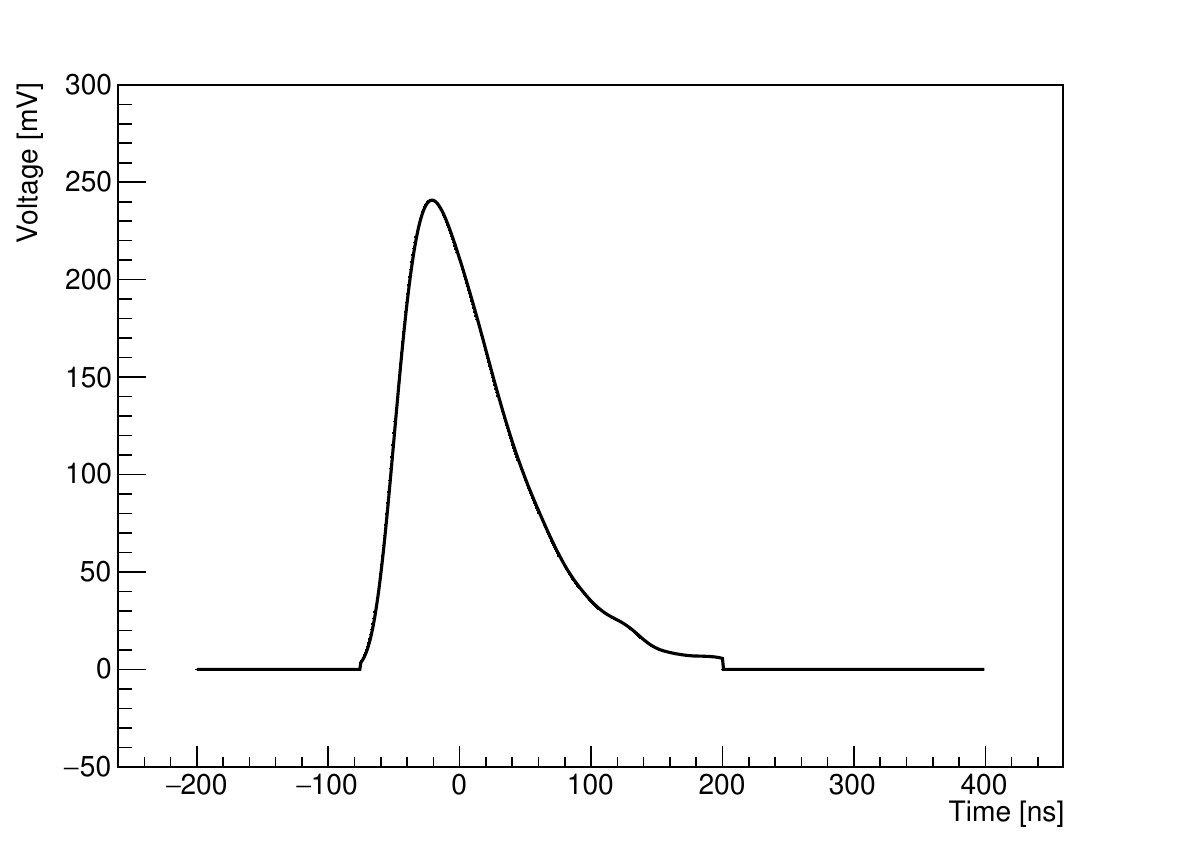}}
    \caption{\label{fig:WaveformGen}~(a) Signal waveform of the LYSO-SiPM detection unit under a real particle beam. (b) A 30 MHz low-pass filter was applied to the waveform in (a), and the pedestal was set to zero.}
\end{figure}

\begin{figure}[h]
    \centering  
    \subfigure[]{
    \includegraphics[width=0.45\textwidth]{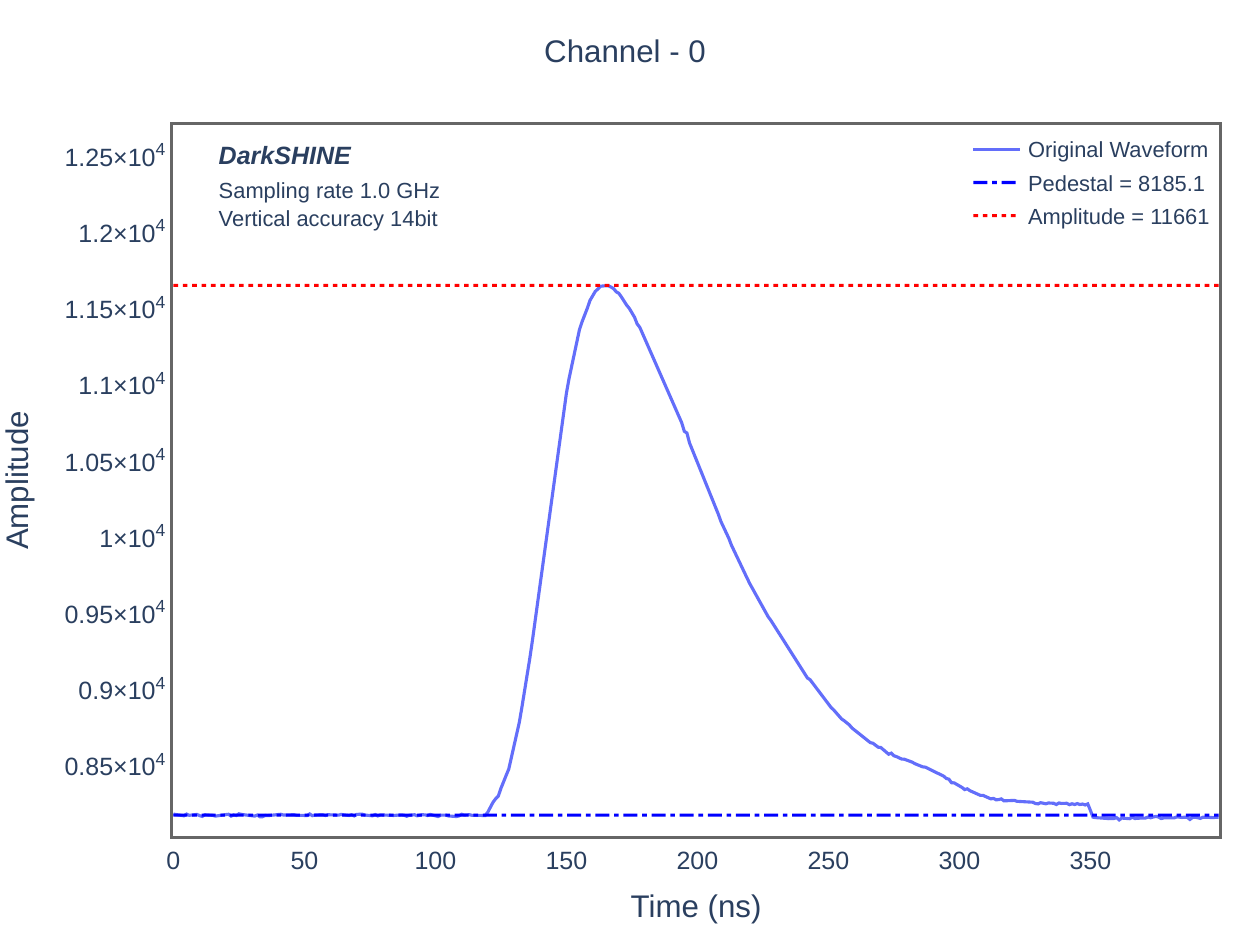}}
    \subfigure[]{
    \includegraphics[width=0.45\textwidth]{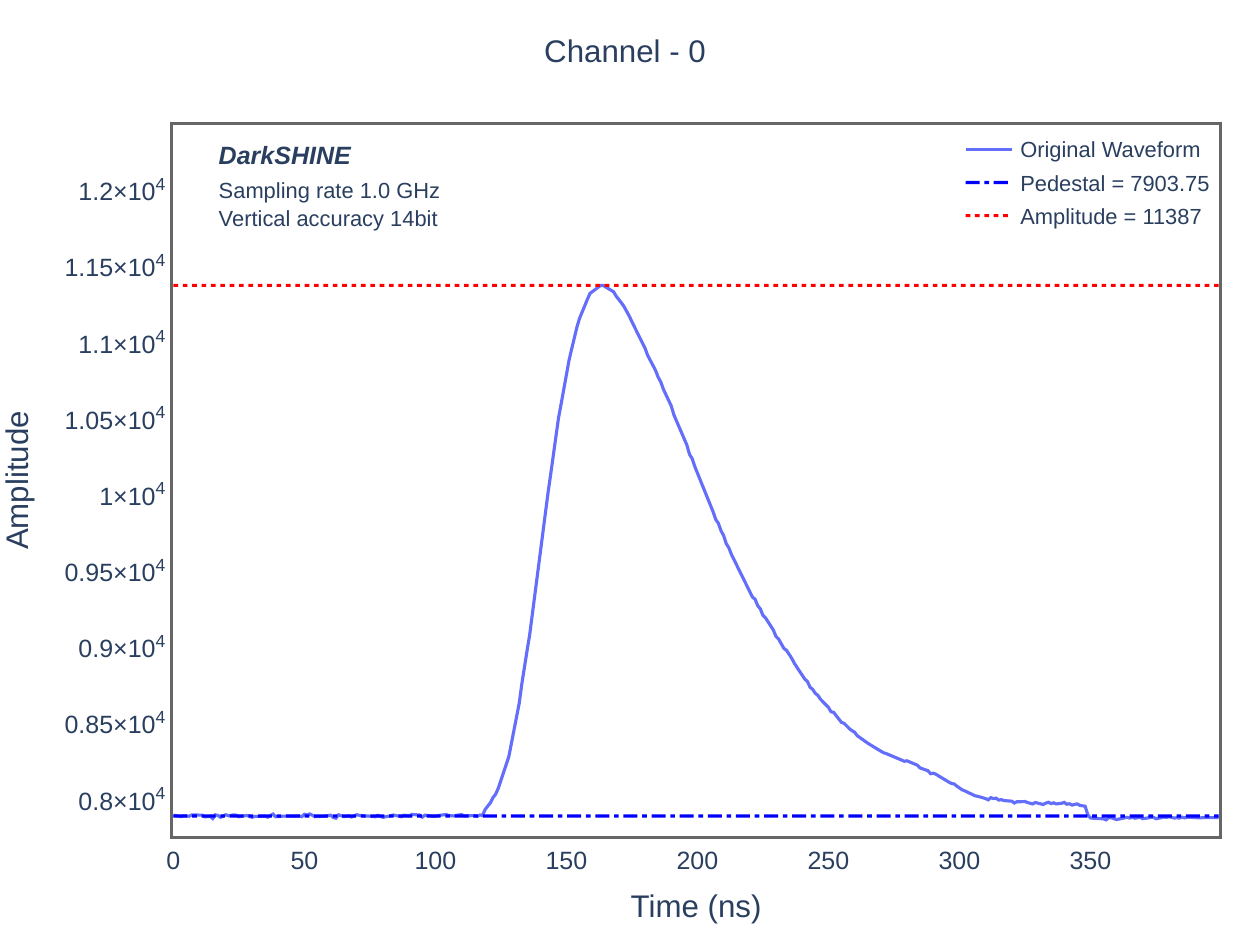}}
    \caption{\label{fig:WaveformDet}~(a) The waveform measured by Channel-0 at a 1 kHz repetition rate. (b) The waveform measured by Channel-0 at a 1 MHz repetition rate.}
\end{figure}

We designed a specific experiment to measure the response of the readout electronics prototype to input signal waveforms at different repetition frequencies. First, an arbitrary waveform generator was used to simulate the signal waveform of the LYSO-SiPM detection unit under a real particle beam (Figure~\ref{fig:WaveformGen} (a)). To facilitate the evaluation of the measurement accuracy of the electronics prototype, a 30 MHz low-pass filter was applied to the waveform shown in Figure~\ref{fig:WaveformGen} (a), and the amplitude of the pedestal region was set to zero, resulting in the waveform shown in Figure~\ref{fig:WaveformGen} (b). The waveform generated by the signal generator, as shown in Figure~\ref{fig:WaveformGen} (b), was then input into the electronics prototype, and the actual measured waveform is shown in Figure~\ref{fig:WaveformDet}. It can be observed that, for different repetition frequencies, our electronics prototype demonstrates great waveform digitization performance.

% \begin{figure}[h]
%     \centering
%     \includegraphics[width=0.9\textwidth]{Detectors/Figures/ECAL/ILA_1MHz.png}
%     \caption{\label{fig:ILA}~ILA waveform.}
% \end{figure}

\begin{figure}[h]
    \centering  
    \subfigure[]{
    \includegraphics[width=0.45\textwidth]{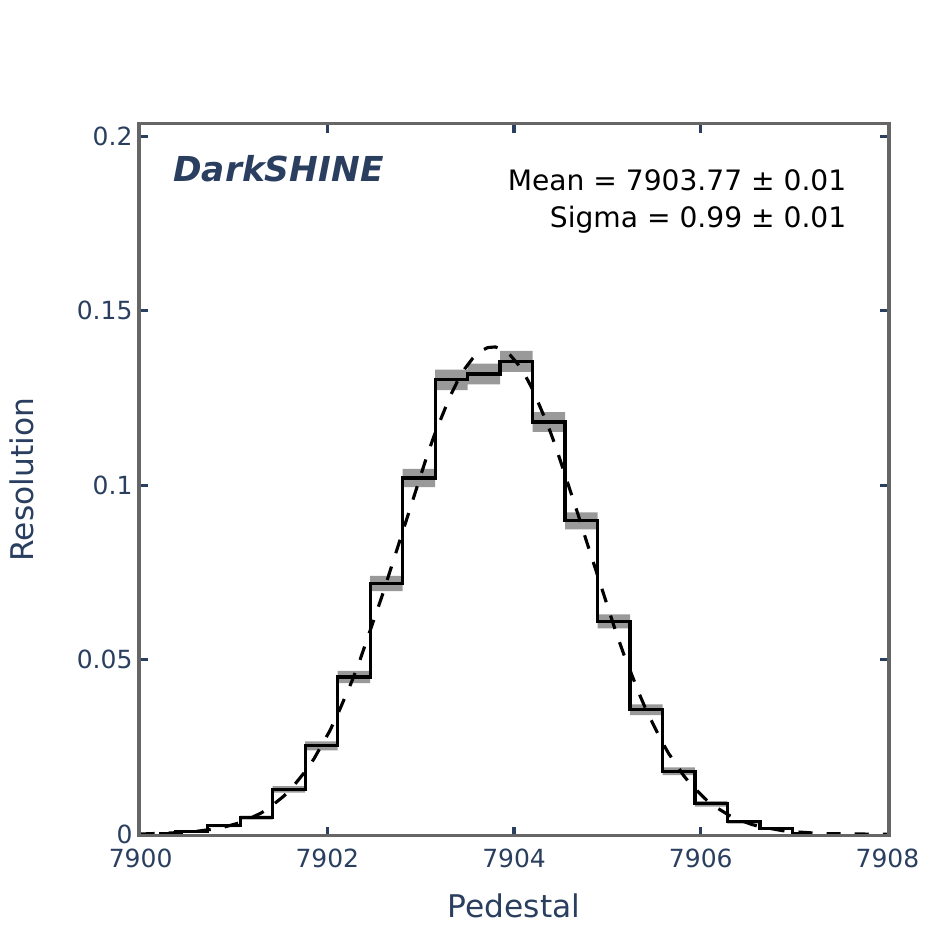}}
    \subfigure[]{
    \includegraphics[width=0.45\textwidth]{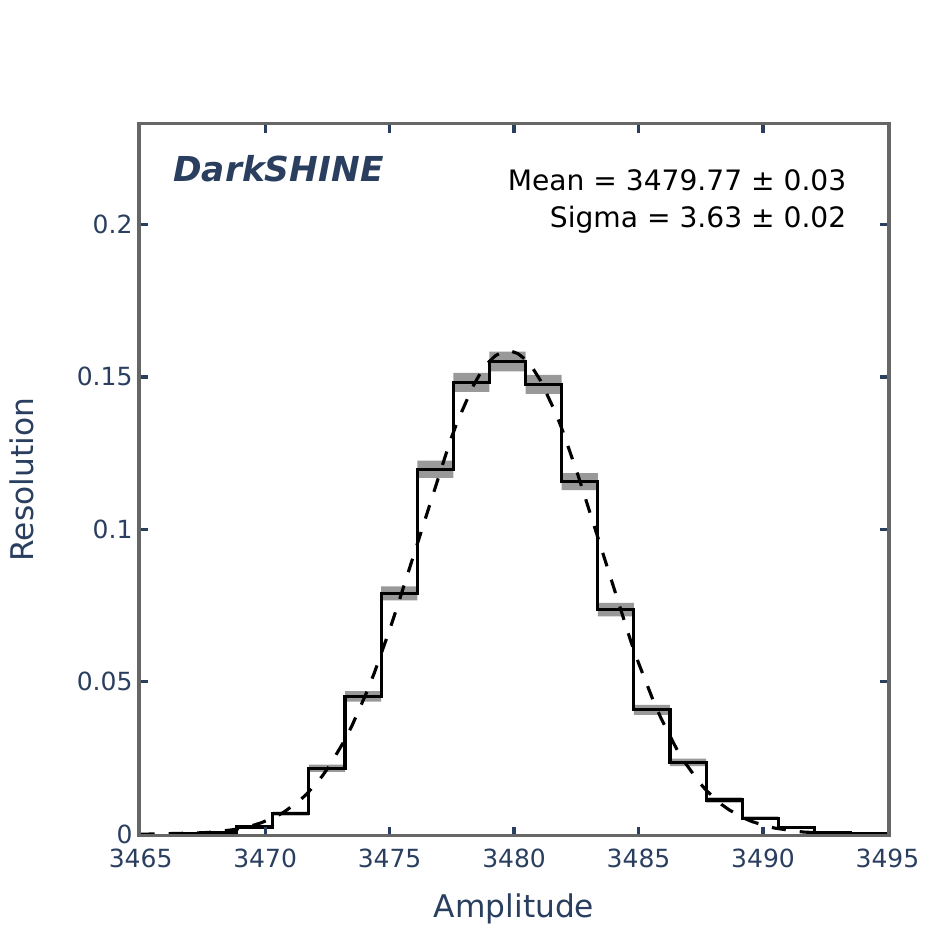}}
    \caption{\label{fig:PedestalAmplitude}~(a) Pedestal distribution measured by Channel-0 at a 1 MHz repetition rate. (b) Amplitude distribution measured by Channel-0 at a 1 MHz repetition rate.}
\end{figure}

By analyzing the pedestal and amplitude of a large dataset, the distribution shown in Figure~\ref{fig:PedestalAmplitude} was obtained. A Gaussian fit was applied to the distribution, yielding the mean and standard deviation. The analysis results for pedestal and amplitude at different repetition frequencies are summarized in Table~\ref{tab:PedestalAmplitude_tab}. Although the pedestal of the electronics prototype exhibits slight drift at 1 kHz and 1 MHz event rates, the amplitude remains nearly constant. In practical measurements, the effect of pedestal drift can be mitigated through pedestal subtraction. This indicates that the electronics prototype is capable of signal readout at an event rate of 1 MHz.

\begin{table}[h]
\fontsize{8}{12}\selectfont
\caption{\label{tab:PedestalAmplitude_tab}~Pedestals and amplitudes of the measured waveforms at different repetition rates.}
    \centering
    \begin{small}
    \begin{tabular}{cccc}
    \hline
    \textbf{Repetition rate} & \textbf{Channel} & \textbf{Pedestal} & \textbf{Amplitude}\\ 
    \hline
     1 MHz & 0 & 7903.77$\pm$0.01  & 3479.77$\pm$0.03 \\
     1 kHz & 0 & 8185.63$\pm$0.01  & 3479.57$\pm$0.05 \\
     1 MHz & 1 & 7924.41$\pm$0.01  & 3483.32$\pm$0.03 \\
     1 kHz & 1 & 8205.55$\pm$0.01  & 3481.50$\pm$0.05 \\
    \hline
    \end{tabular}
    \end{small}
\end{table}

Preliminary tests on the signal completeness of the readout electronics prototype were also conducted, with an ADC sampling rate of 1 GS/s. A sine wave with 1 MHz repetition rate was generated and fed into an individual channel. The completeness of the signal under different data transmission volumes was studied by varying the number of sampling points. The completeness of signal is defined as the number of acquired events divided by the difference between the maximum and minimum trigger IDs. The results of completeness are shown in Figure~\ref{fig:Completeness} (a). As the number of samples per event increased from 8 to 180, the signal completeness gradually decreased but remained above 99\%, due to the limitation of bandwidth. When the number of samples per event was less than 120, there was virtually no event loss.

\begin{figure}[h]
    \centering
    \includegraphics[width=0.6\textwidth]{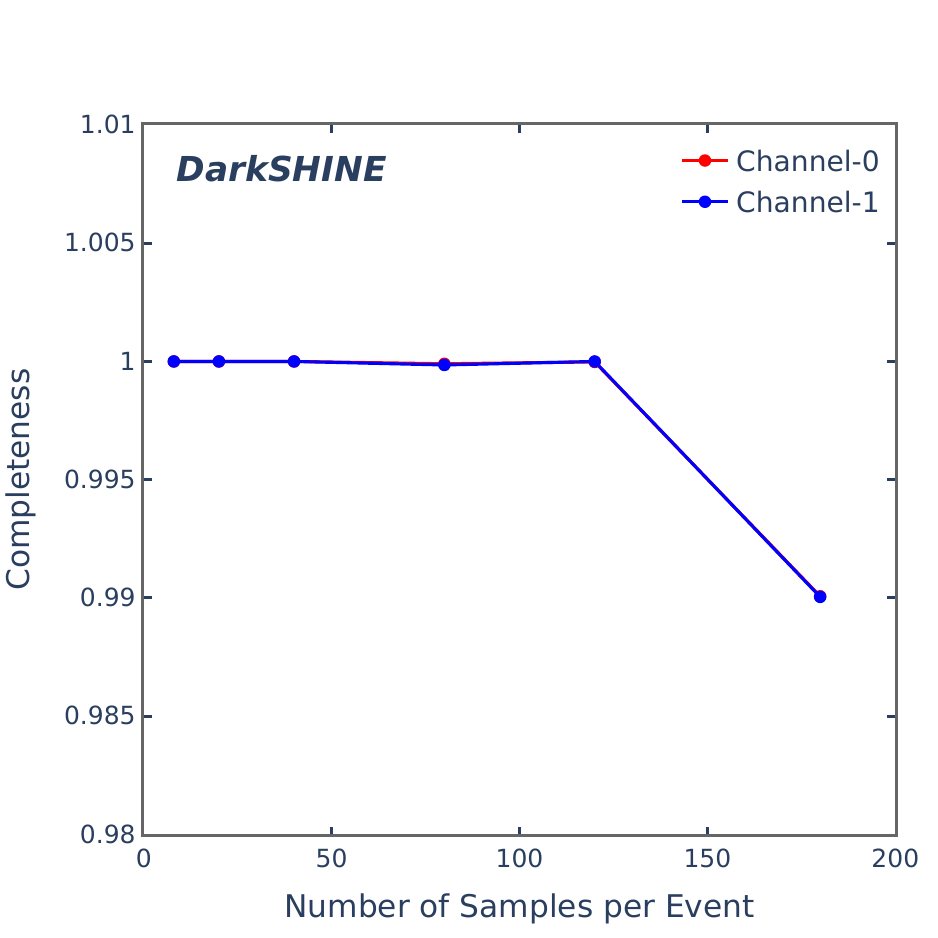}
    \caption{\label{fig:Completeness}~Signal completeness of readout electronics prototype for single channels at 1 MHz repetition rate.}
\end{figure}

\begin{figure}[h]
\centering
    \subfigure[]{
    \includegraphics[width=0.32\textwidth]{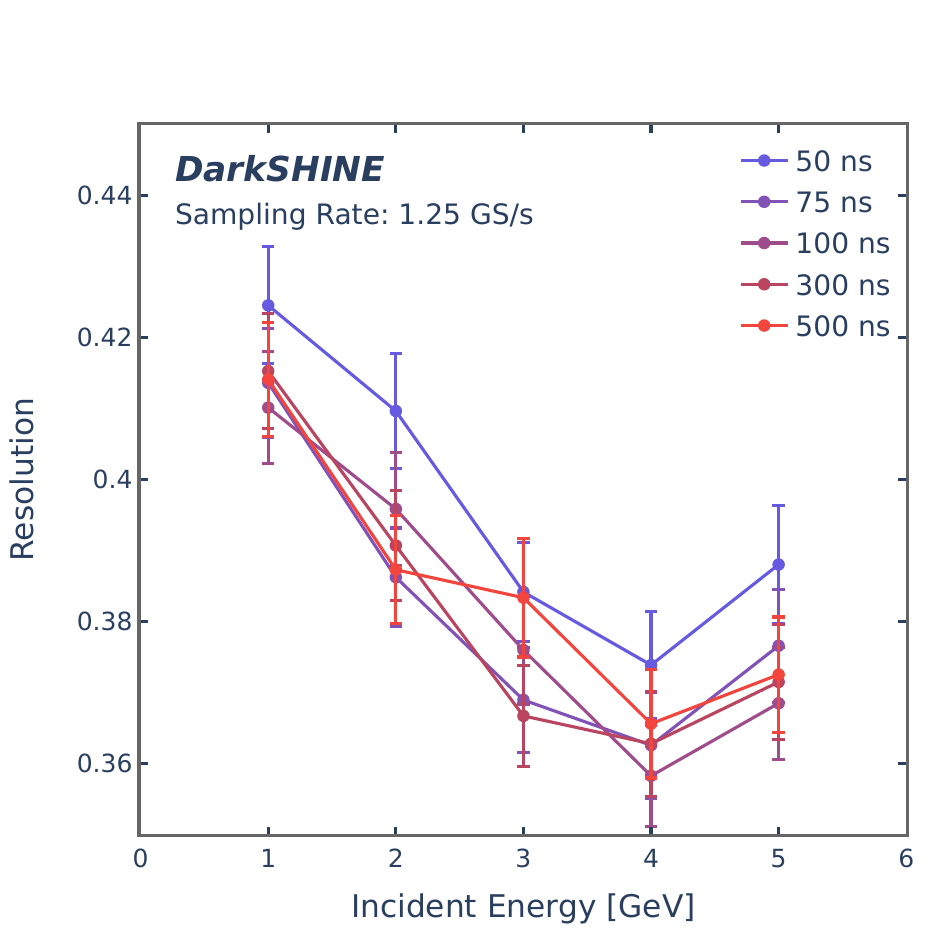}}
    \subfigure[]{
    \includegraphics[width=0.32\textwidth]{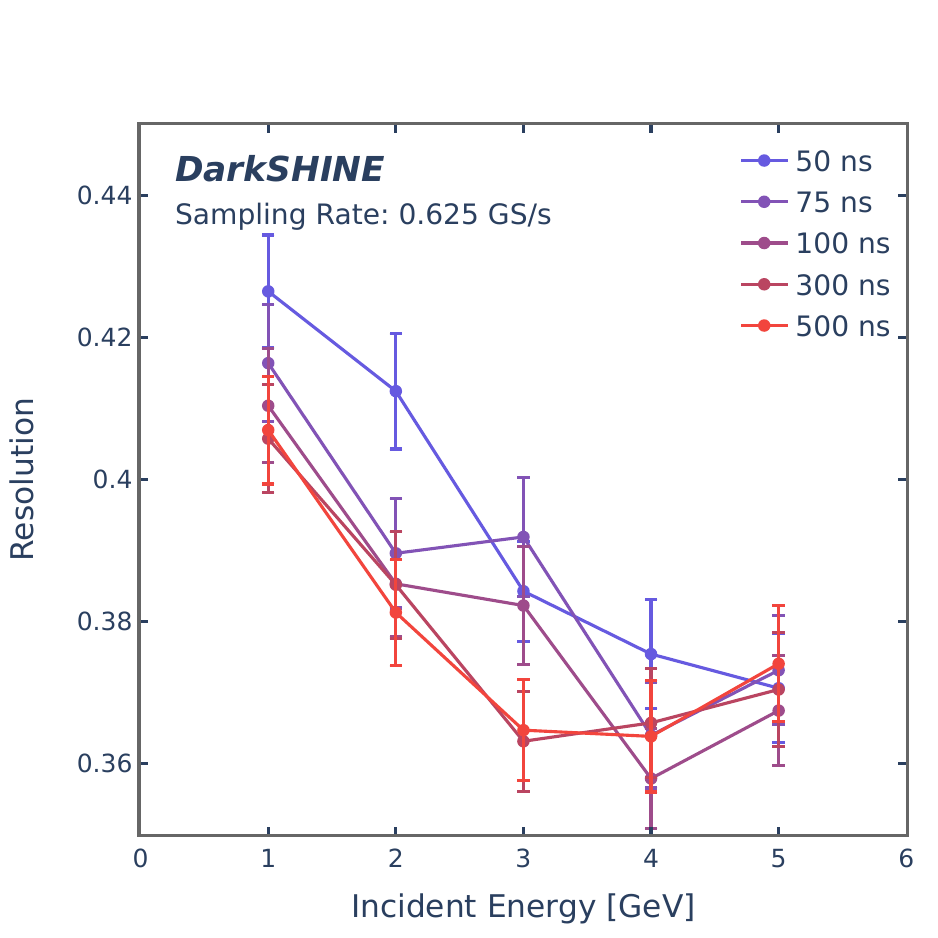}}
    \subfigure[]{
    \includegraphics[width=0.32\textwidth]{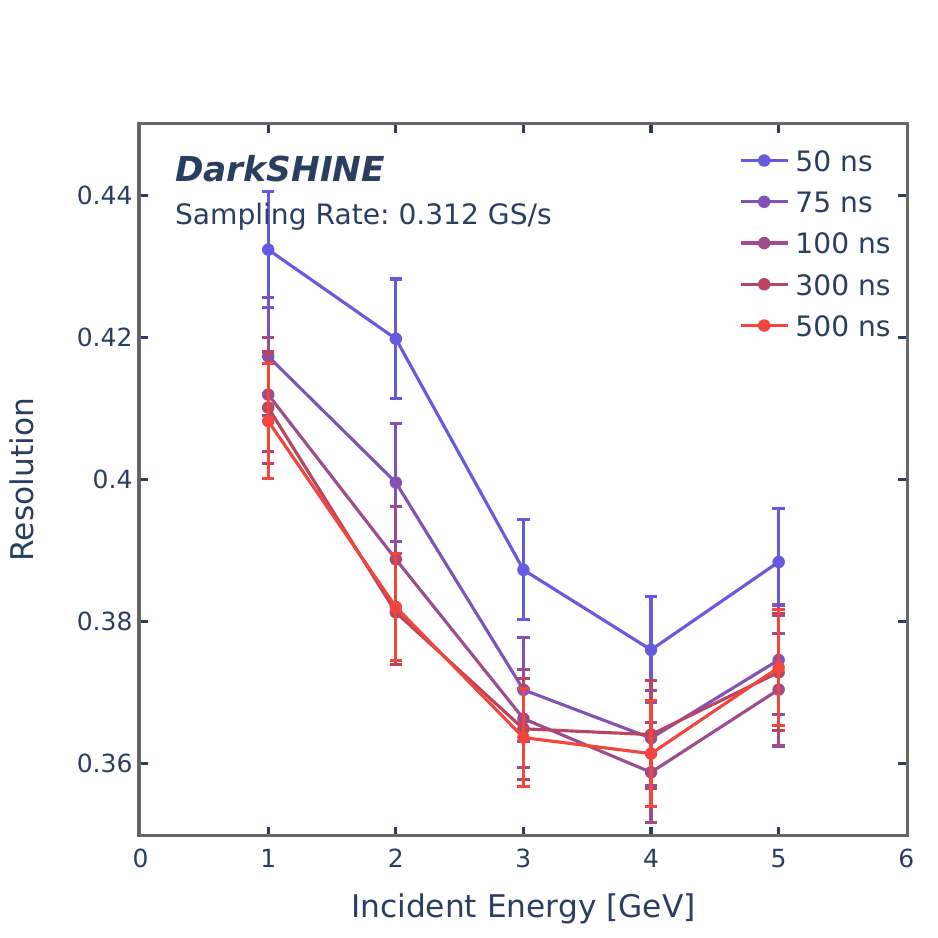}}
    \subfigure[]{
    \includegraphics[width=0.32\textwidth]{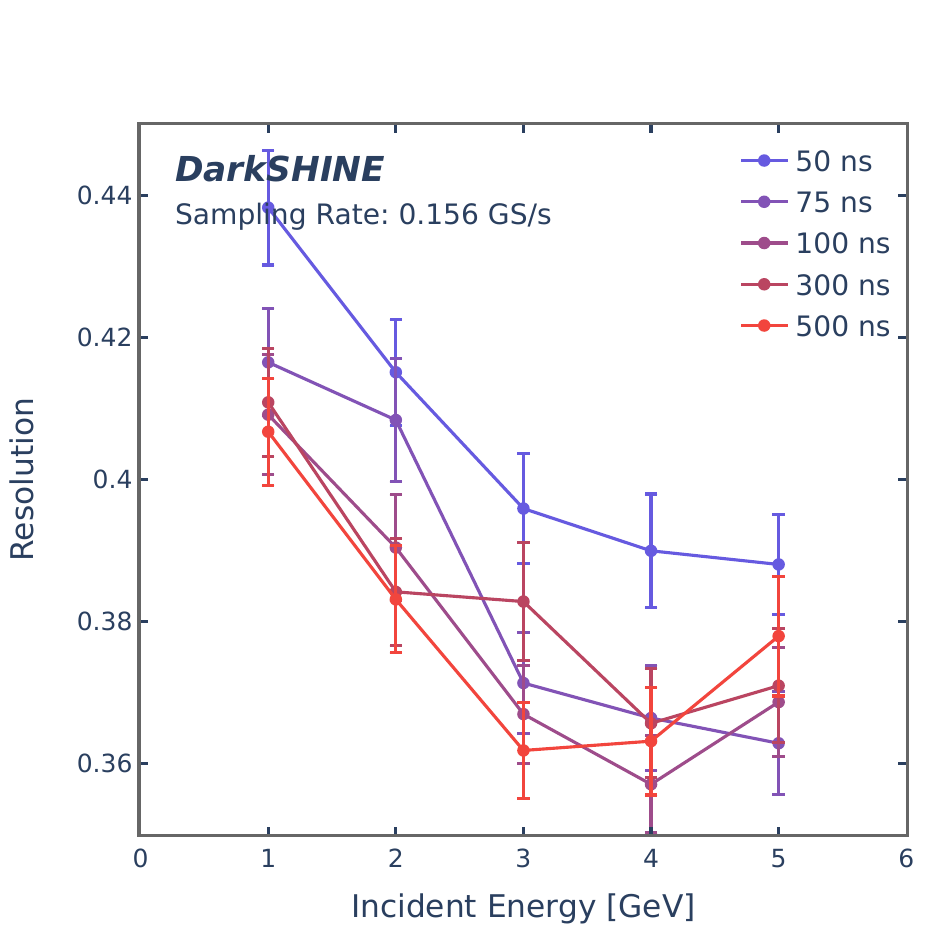}}
    \subfigure[]{
    \includegraphics[width=0.32\textwidth]{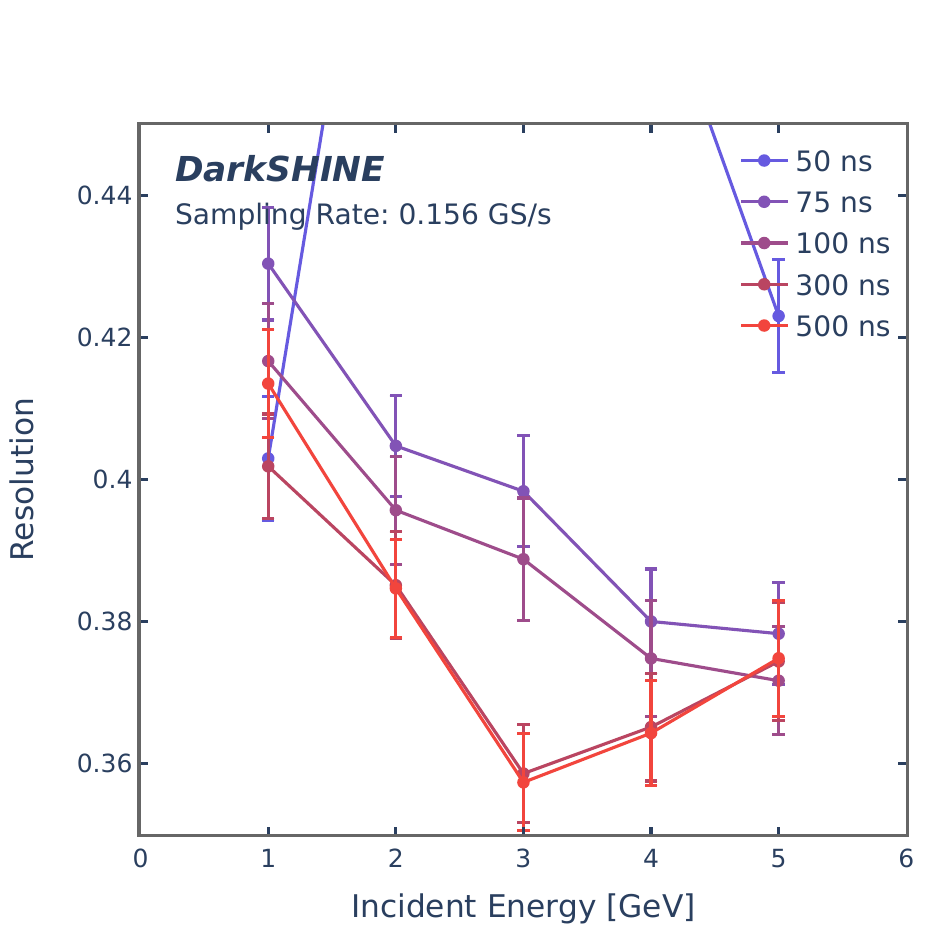}}
\caption{\label{fig:Electronics_Res}~Energy resolution of one LYSO-SiPM detection unit for 1–5 GeV electron beams under varying sampling rates and event lengths.}
\end{figure}

The effects of the sampling rate and event length of the readout electronics system on energy measurements were studied using the beam test(Figure~\ref{fig:Beamtest}) results of crystal units. Figure~\ref{fig:Electronics_Res} presents the energy resolution of one LYSO-SiPM detection unit for 1–5 GeV electron beams under varying sampling rates and event lengths. It is observed that when the sampling rate is at least 0.156 GS/s and the event length is at least 100 ns, the energy resolution shows minimal variation. This implies that, to minimize the impact of the readout electronics on the detector's resolution, the readout electronics should meet the minimum requirements of a sampling rate of at least 0.156 GS/s and an event length of at least 100 ns, corresponding to a minimum of 16 sampling points per event. Based on the single-channel signal completeness results shown in Figure~\ref{fig:Completeness}, the current electronics design is adequate for testing a small number of channels. For future detectors with several thousand channels, an online triggering system is proposed to filter out the majority of background events and conserve data bandwidth. Additionally, the resolution shown in the Figure~\ref{fig:Electronics_Res} is much worse than the digitized single-channel resolution listed in the Table~\ref{tab:ECAL_smearing}. This is because the limited crystal size results in significant fluctuations in the energy deposition of high-energy particles within the crystal. Consequently, the ratio of the measured energy spectrum's standard deviation to its mean value is much larger than the energy resolution of the detector.

% \subsubsection{Cosmic ray test}
% \label{sec:ECALMCT}

% Cosmic rays, serving as a natural source of particles, are useful for measuring the detector’s MIP response. In the laboratory, we performed cosmic ray tests on a multi-channel crystal module to assess the MIP response. All channels exhibited a clear MIP energy distribution with low noise levels. After calibrating the collected cosmic ray energy using the scintillator's intrinsic radioactive background, the calibration results showed good agreement with the simulation results.

\section{Hadronic Calorimeter System}
\label{sec:HCAL}

The HCAL is a sampling calorimeter composed of alternating layers of iron absorbers and plastic scintillator sensitive layers along the beam line direction. It serves as a \textbf{veto system} to exclude events containing muon pairs or neutral hadrons, such as neutrons, which often exhibit similar behavior to signal processes in the ECAL and pose challenges for effective vetoing. The capability to reject these events directly determines whether DarkSHINE can truly achieve its goal of being a low-background experiment with high sensitivity in searching for dark photons.

In this section, we present a baseline design of DarkSHINE HCAL and its performance as documented in ref~\cite{Chen:2022liu}. Furthermore, we provide an overview of the sensitive unit set along with its performance test results. Lastly, Section~\ref{sec:HCALFuture} offers a concise introduction to future optimization prospects.

\subsection{HCAL Conceptual Design}
\label{sec:HCALConceptual}

The HCAL is comprised of layers of plastic scintillators and iron absorbers, with the sensitive layer consisting of plastic scintillator strips. The main HCAL has a transverse size of 4 $\times$ 4 m$^2$ and is divided into sixteen horizontal modules measuring 100 $\times$ 100 cm$^2$ each. This design ensures that the length of an individual plastic scintillator is not excessively long. The scintillation strip size on the main hcal is 100 $\times$ 5 $\times$ 1 cm$^3$, and each layer on each module contains 20 scintillators. 

The schematic of the HCAL geometry is illustrated in Fig~\ref{fig:HCALSimuFigure}, drawn using the DarkSHINE simulation framework. The front surface is immediately followed by the ECAL, which is a sensitive layer.

\begin{figure}[htb]
\centering
\includegraphics
  [width=0.55\hsize]
  {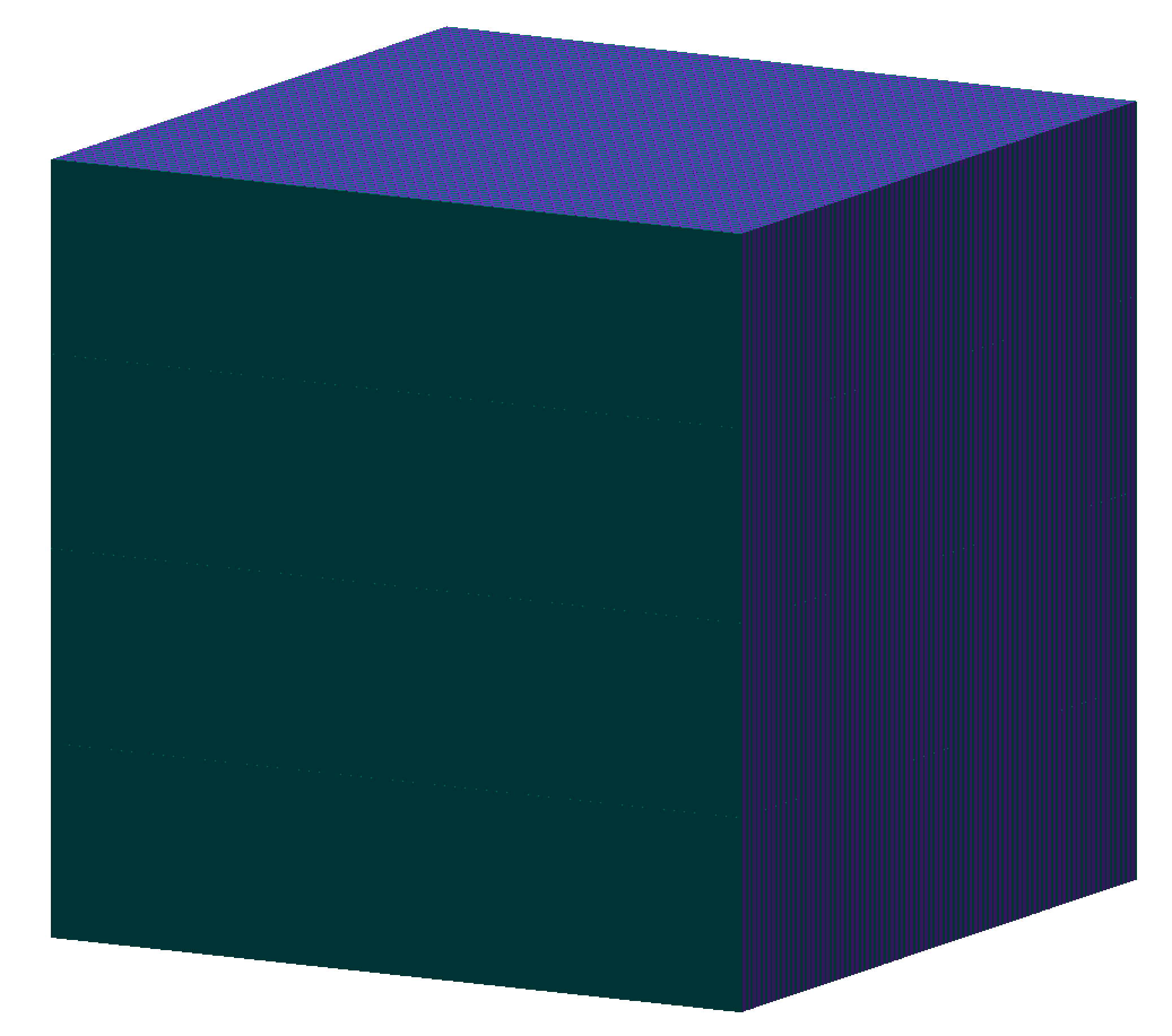}
\caption{HCAL design, a cuboid formed by the sequential arrangement of iron absorbers and plastic scintillator strips.}
\label{fig:HCALSimuFigure}
\end{figure}

There are two sensitive layers between the two absorbers, and the scintillator strips in the two adjacent sensitive layers are oriented differently, along the x or y direction, respectively. The arrangement of the two layers is as shown in Fig~\ref{fig:HCALFeSc}. The main HCAL consists of alternating layers of plastic scintillator and iron, with a total length slightly longer than 400~cm~(including the wrapper of scintillator) in the z direction. There are 80 layers absorber, each with a thickness of 3~cm, and 162 layers of plastic scintillator, each with a thickness of 1~cm. 

\begin{figure}[htb]
\centering
\includegraphics
  [width=0.55\hsize]
  {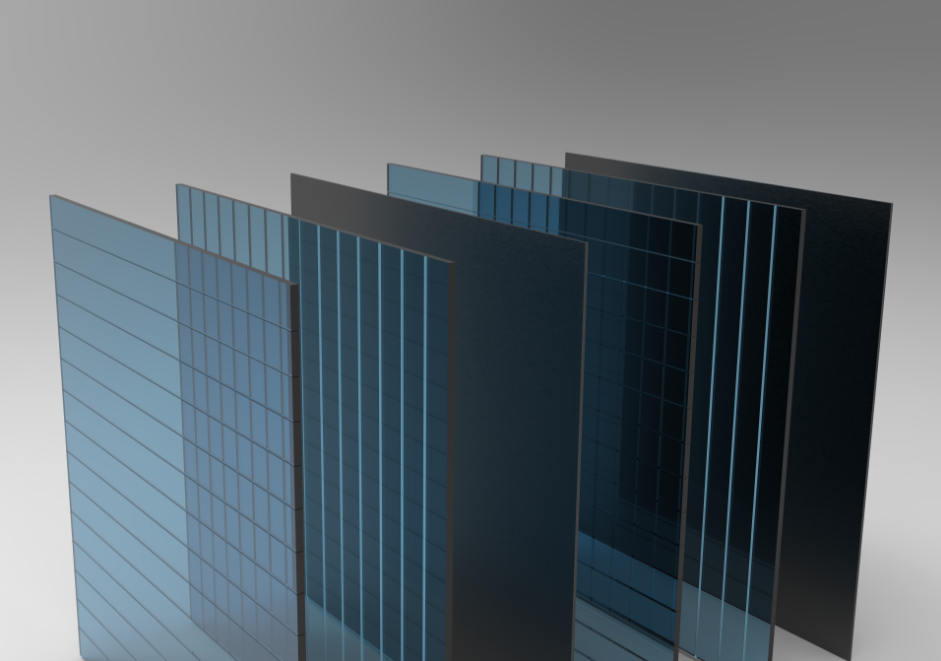}
\caption{Arrangement of scintillators and absorbers is such that the translucent layers consist of scintillators, with adjacent layers having long sides of their respective scintillator bars perpendicular to each other. The opaque layers represent absorbers.}
\label{fig:HCALFeSc}
\end{figure}

The design of HCAL considered the situation that high energy neutral hadrons need enough interaction length, which leads to the deep design in Z-zaxis. On the other hand, the low energy particles may have a bigger $\eta$, which request a bigger transverse size to provide enough acceptance. The ratio of particles go in to the different range of HCAL is shown in table~\ref{tab:n_region_ratio}.

\begin{table}[!htb]
\caption{Ratio of neutrons inside the range of 50$\times$50~cm$^2$, 150$\times$150~cm$^2$, and 400$\times$400~cm$^2$ regions of the HCAL, respectively. No other cuts were added.}
\label{tab:n_region_ratio}
\centering
\begin{tabular}{|c|c|c|c|}
\hline
Process & 50$\times$50~cm$^2$ & 150$\times$150~cm$^2$ & 400$\times$400~cm$^2$\\ \hline
EN\_ECAL & 0.204 & 0.556 & 0.804 \\ \hline
EN\_target & 0.185 & 0.59 & 0.835 \\ \hline
PN\_ECAL & 0.241 & 0.595 & 0.823 \\ \hline
PN\_target & 0.219 & 0.594 & 0.828 \\ \hline

\end{tabular}
\end{table}

As mentioned at the beginning of this section, each module consists of absorber layers and scintillator layers. The absorber layer is a complete iron plate, while the scintillator layers consist of scintillator strips. The plastic scintillator strip is wrapped in a reflective film, grooved on the surface, and put into wavelength shift fibers, which collect the photons inside the scintillator and transmit it to the SiPM. This design ensures that the readout part of the HCAL is concentrated on the side of the module and does not appear inside the HCAL. 

The rejection of particles/events and be implemented on the energy calculated after calibration, or directly on the number of photons collected. Since reconstruction of events is not required in HCAL the two cut conditions are equivalent.

\subsection{Electronic}
\label{sec:HCALElectronic}

The electronic system for HCAL includes the SiPM supporting board, the pre-amplifier, and the readout system, which can be directly inherited from the system described in Section~\ref{sec:ECALElectronics}. It is designed to accommodate a high repetition frequency of events. Conversely, unlike ECAL's stringent event reconstruction requirements and high event rates, HCAL does not demand such levels. In principle, under normal background events, the full-absorption ECAL will not leak energy to HCAL except for those rare processes mentioned in the previous section. As depicted in Figure~\ref{fig:ECALStaggered}(b), events with leakage are three orders of magnitude less frequent than those entering into ECAL; this implies that if the event rate in ECAL is 1~MHz, HCAL only needs to handle no more than 10~kHz events.

The current hardware test of HCAL utilizes an alternative electronic readout system, originally designed for pandaX-4T~\cite{He:2021sbc}, which fulfills our requirements. This section provides an introduction to this system.

The electronic readout is a 6U VME-standard module, as shown in Figure~\ref{fig:HCALElectronic1}. This electronic readout is a 500 MS/s waveform digitizer designed to support both external-trigger and trigger-less readout. Considering that a complete waveform is about 200~ns, there are at least 100 points in each waveform, and this number is far beyond what we need. As studied in cosmic ray test mentioned in Section~\ref{sec:HCALUnit}, a waveform with the time interval between points is 0.4~ns shows similar resolution in test results to a waveform with that time intercal equal to 8~ns.

Furthermore, the flexible selection of the sampling interval enables us to effectively transmit all data within a specific bandwidth range, thereby mitigating any potential packet loss. With 100 samples collected per event and each sample occupying two Bytes, at a sampling rate of 10~kHz, the resulting data volume for one channel amounts to 2~MB/s. Considering that eight channels operate simultaneously, the total data transmission is 16~MB/s, which remains well below the board's bandwidth capacity of 90 MB/s.

The digitizer combines two analog-to-digital converters (ADCs) with a Field Programmable Gate Array (FPGA). Each input single-ended analog signal is converted to a differential pair through a differential amplifier. The gain is set to be 1.6. After the amplifier, the signal is attenuated by resistors and a low-pass filter before entering into the ADCs. The overall amplification is about 1.25. In the ADC, four input signals are simultaneously sampled and digitized with a sampling rate of 500 MS/s and a 14-bit resolution. The dynamic range is set to be 2.16 Vpp. The digital data are transferred to the FPGA through the high-speed JESD204B serialized interface.~\cite{He:2021sbc}

\begin{figure}[htb]
\centering
\includegraphics
  [width=0.6\hsize]
  {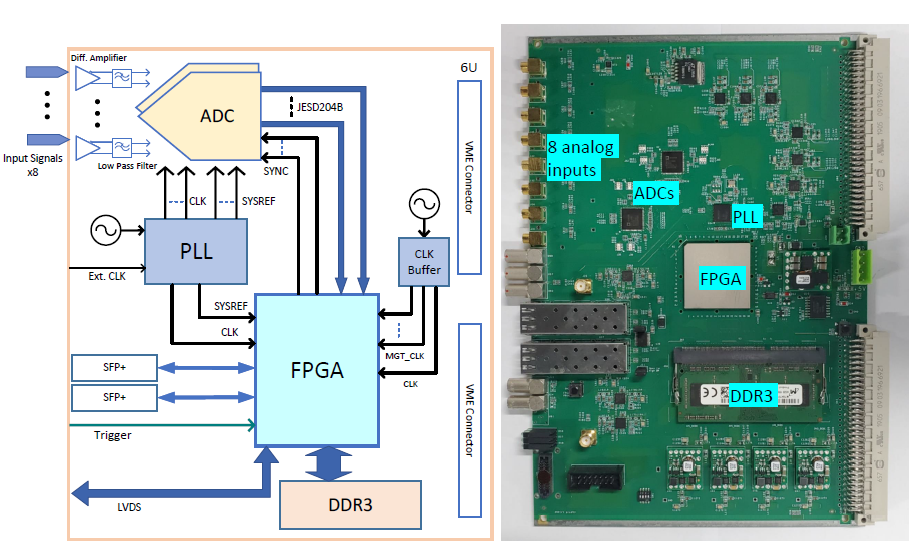}
\caption{The schematic drawing (left) and the photo (right) of the new digitizer. The digitizer is a 16-layer PCB, hosting 8 differential amplifiers (ADI LTC6409), 8 low-pass filters (Mini-Circuits DLFCN-290+), two ADCs (ADI AD9694), one FPGA (Xilinx XC7K325T), one PLL (TI LMK04610), one DDR3 memory module (Micron 4GB MT8KTF51264HZ-1G9P1) and other parts.~\cite{He:2021sbc}}
\label{fig:HCALElectronic1}
\end{figure}

In the entire HCAL, more than ten thousands of readout channels should work as a whole, in a multi-digitizer system with the clock fanout module, the channel-to-channel synchronization within single board and between two boards is measured to be better than 0.2 ns~\cite{He:2021sbc}, This performance is good enough for the entire HCAL multi-channel readout.

\subsection{HCAL Sensitive Unit and Performance Test}
\label{sec:HCALUnit}

As mentioned in Sec~\ref{sec:HCALConceptual}, the sensitive layer is consisted of plastic scintillator~\cite{Luo:2023inu} strips, and each scintillator strip has grooves on its surface to put wavelength shift fibers. One side of the fibers are wrapped with reflective adhesive, and the other side is polished and couple to SiPM. Finally, the scintillator is wrapped with ESR reflecting film. These materials are shown in Fig~\ref{fig:HCALScWLSSiPM}. An experiment platform friendly size of scintillator is picked, each strip is 75 $\times$ 5 $\times$ 1 cm$^3$, and has three grooves on it (an alternative 2 grooves design is also prepared for test). WLS fiber has a $d = 1 mm$, three of the fibers will couple to one SiPM. 

\begin{figure*}[htb]
\centering
\includegraphics
  [width=0.45\hsize]
  {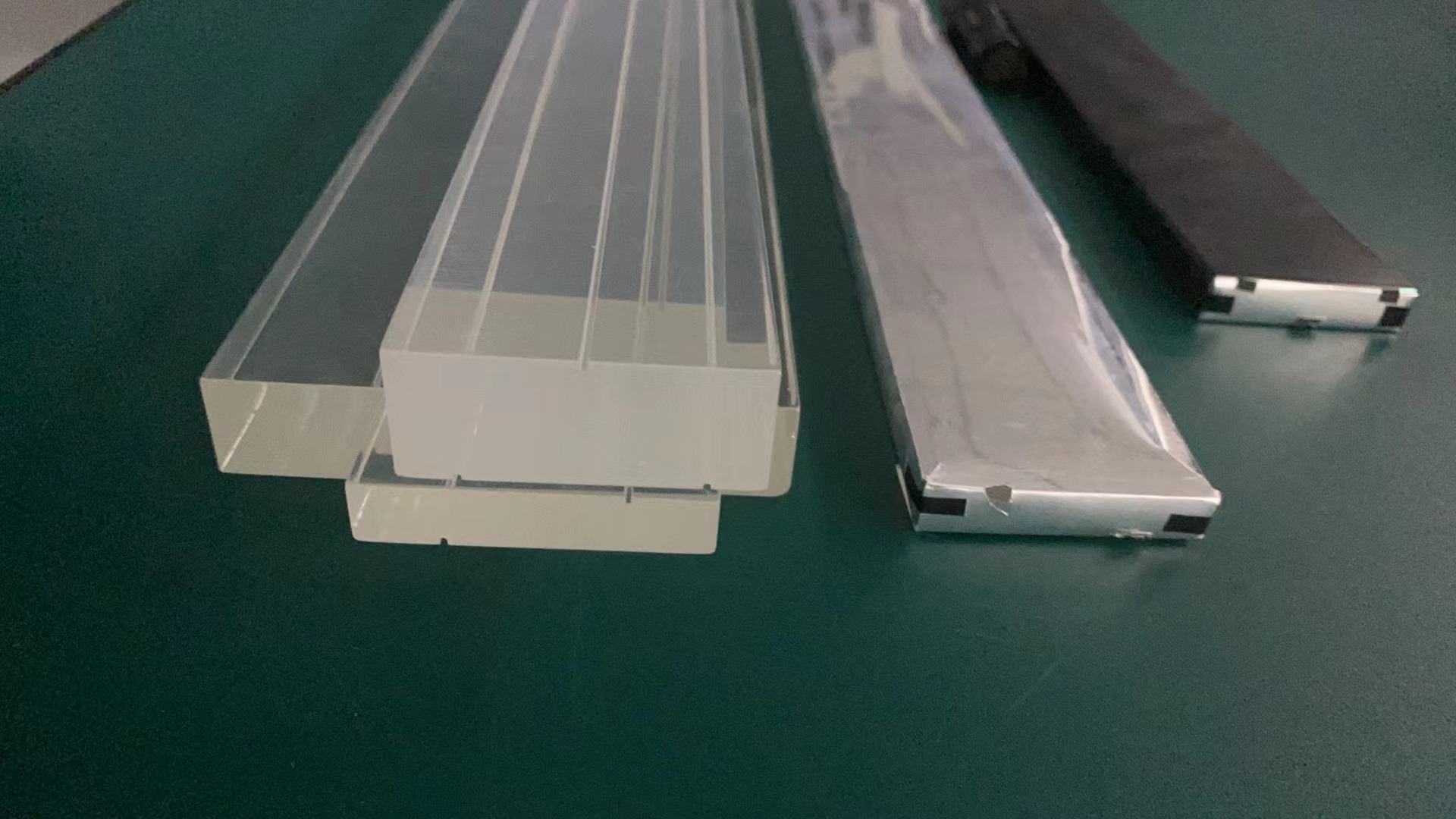}
\includegraphics
  [width=0.45\hsize]
  {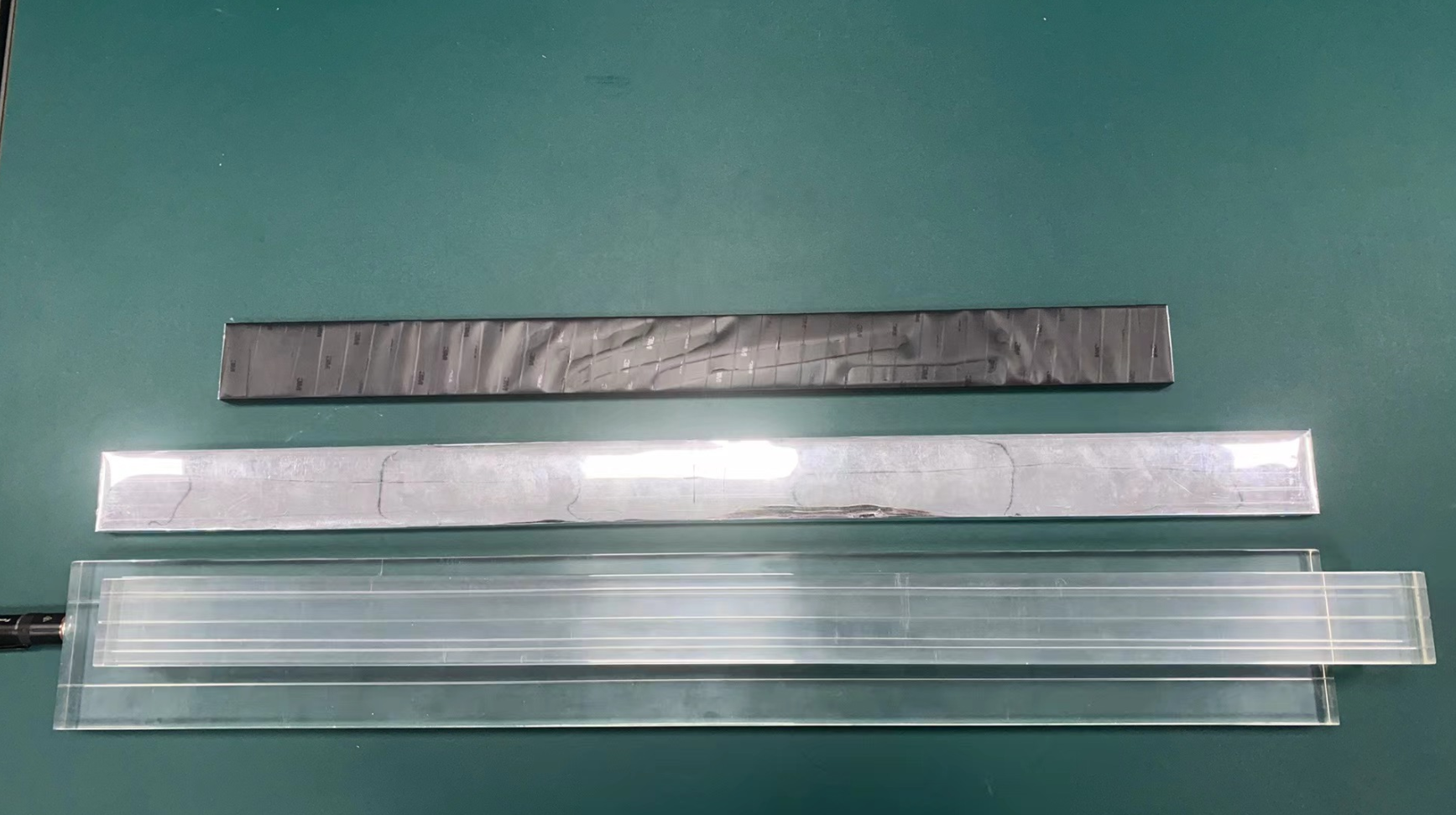}
\includegraphics
  [width=0.45\hsize]
  {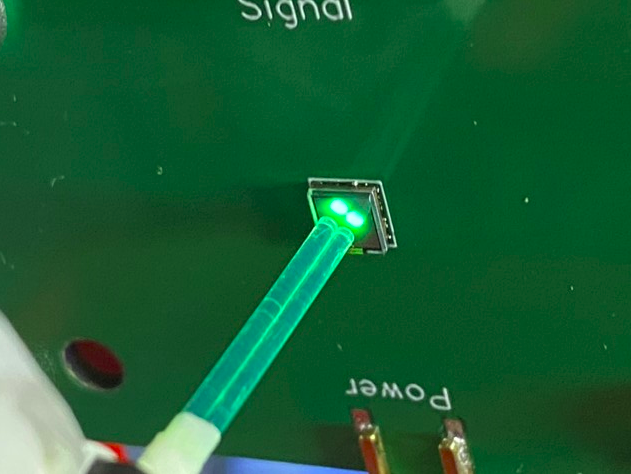}
\includegraphics
  [width=0.45\hsize]
  {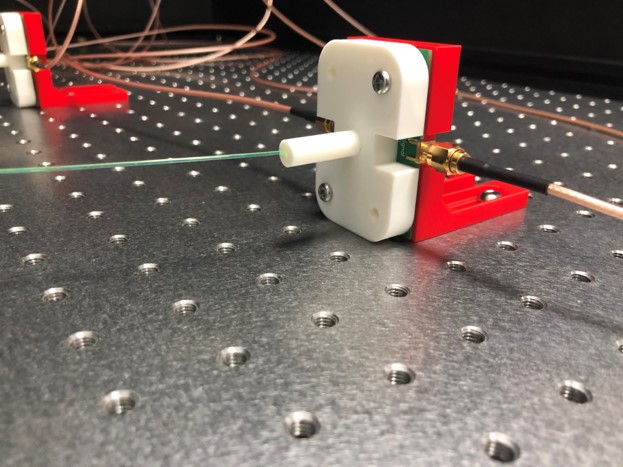}
\caption{Scintillators, wavelength shift fiber, and it's coupling to SiPM. In the top plots, the upper side of scintillator features two grooves, while the third groove is located on the opposite side. In the bottom-left plot, fibers from the same scintillator are coupled to a single SiPM. In practical applications, a collimating structure fabricated using 3D printing technology will be employed to cover the SiPM board and provide shading.}
\label{fig:HCALScWLSSiPM}
\end{figure*}

Before the testing of Scintillator, the property of several different types of SiPM~\cite{S14160-3015PS,S13360-3025CS,S13360-3050CS,EQR06} is also studied same as in ECAL, since there will be three fibers coupling into the same SiPM, the size of SiPM in HCAL is different from those in ECAL. Several SiPM including the S13360-3050PS SiPM from HAMAMATSU with 3 $\times$ 3 mm$^2$ size are tested to study their dark current rates, as shown in Fig~\ref{fig:HCALDCR}. S13360-3050PS is selected due to the consideration of both DCR and gain,

\begin{figure}[htb]
\centering
\includegraphics
  [width=0.45\hsize]
  {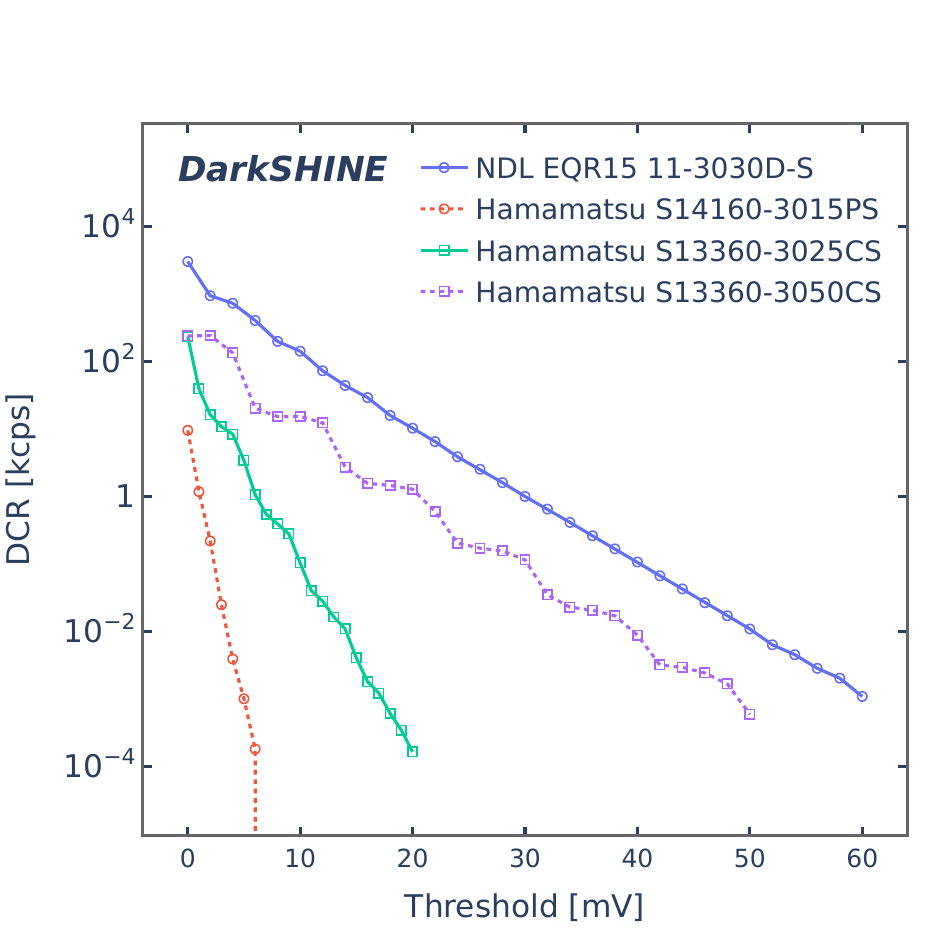}
\caption{Dark current rate of SiPMs used in HCAL unit test. The choice of s13360-3050PS can obtain a balance in the gain coefficient and noise level}
\label{fig:HCALDCR}
\end{figure}

The noise level of this type of SiPM is 6 P.E., which leads to a cut threshold of 8 P.E. in the usage of HCAL information. The cell which has a photon yields greater than 8 P.E. can be considered to have collected energy from the events. 

Additionally, the plastic scintillator has been tested in gamma spectrometer to ensure the background noise of material will not affect the cut threshold. Result is illustrated in right plot of figure~\ref{fig:PStest}. The blue line is when there is no material, and the red line is when there is material. Conclusion can be obtained, the plastic scintillator don't have self-background which will affect the target particle rejection.

\begin{figure*}[htb]
\centering
\includegraphics
  [width=0.45\hsize]
  {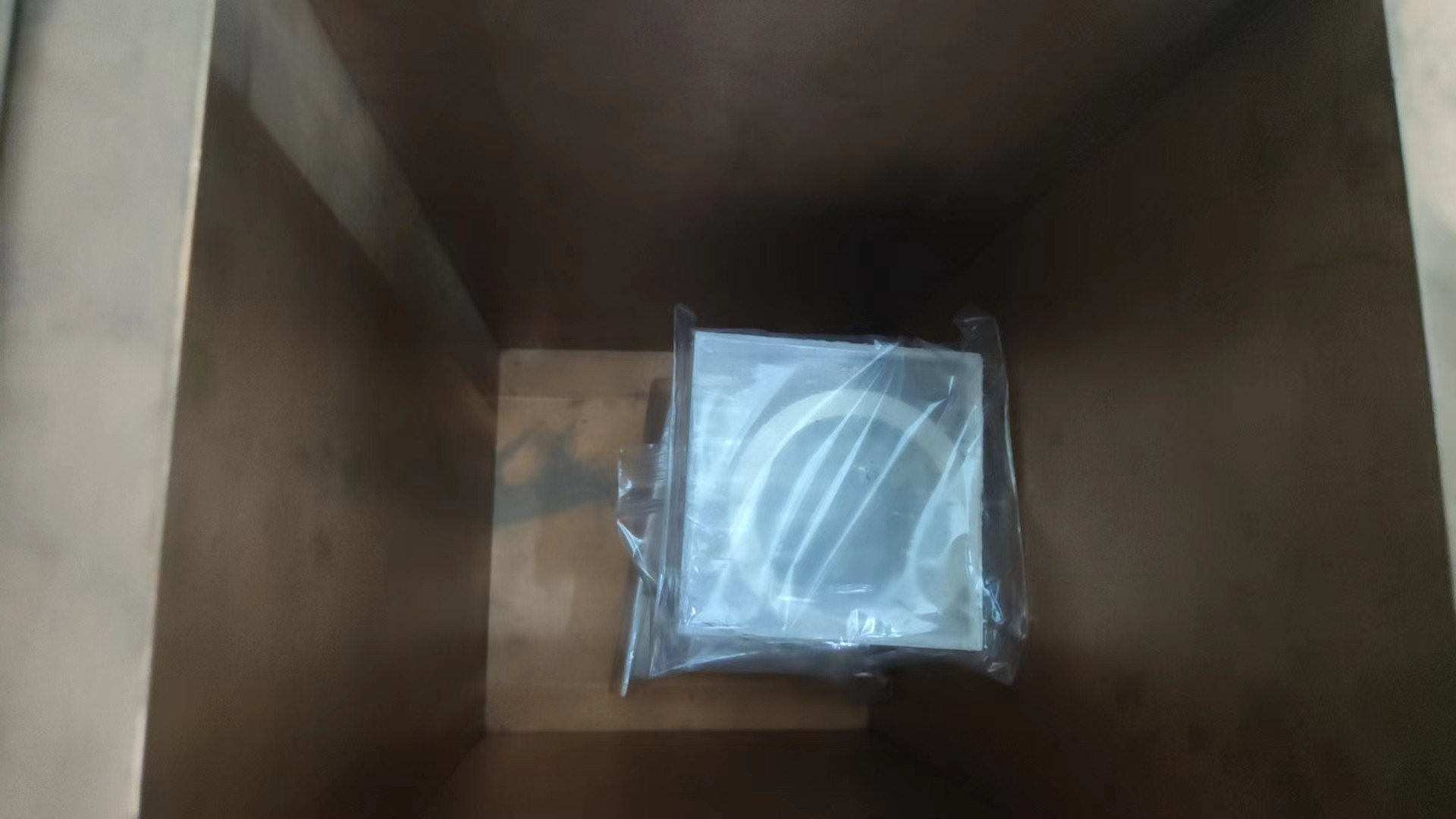}
\includegraphics
  [width=0.45\hsize]
  {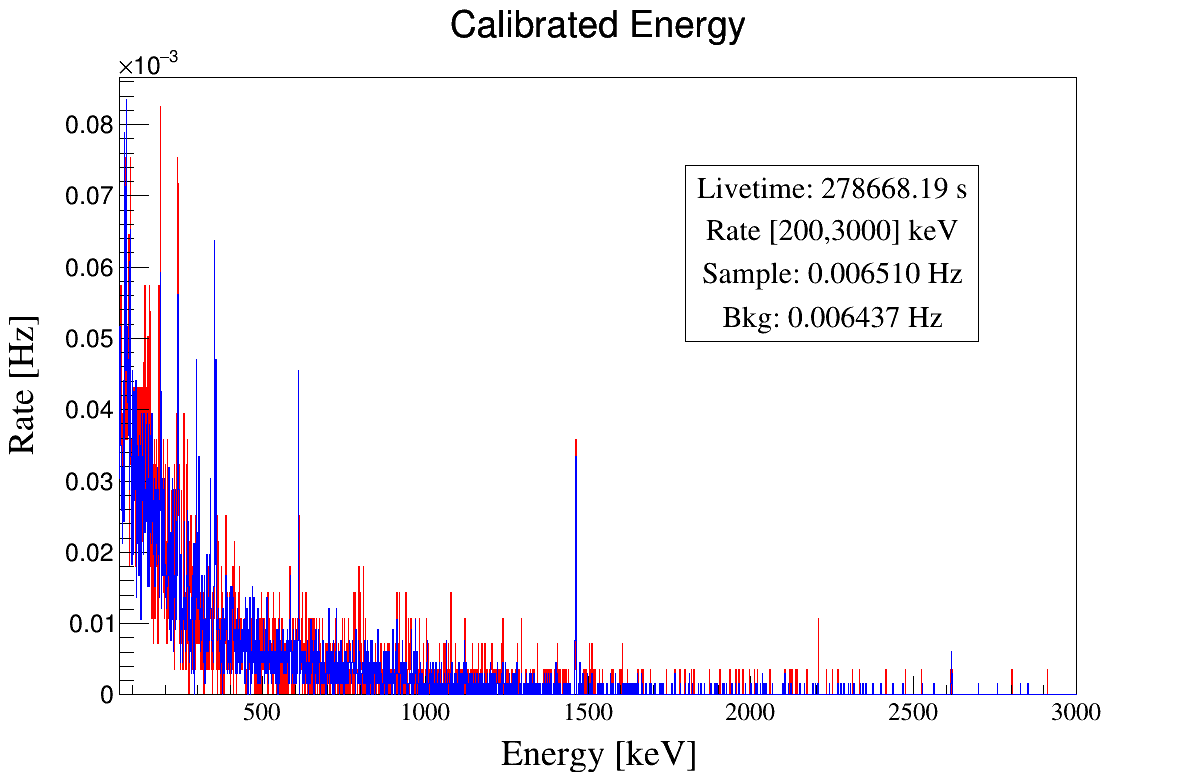}
\caption{Left is the photo of plastic scintillator inside the gamma spectrometer, right is the test result.}
\label{fig:PStest}
\end{figure*}

In order to establish the correct calibration between photon yields and deposited energy, a cosmic ray test is implemented. The experimental setup is shown in Fig~\ref{fig:HCALCRTest}. Two small scintillator pieces are used as a coincident trigger system, meaning that a signal will only be recorded if both pieces detect a signal above the threshold. The longer one, located between the two small pieces, is the sample to be tested. In order to understand the relationship between the photon yield and several freedom degree of the scintillator design, such as the thickness and the number of fibers, different types are measured. Results are collected in Fig~\ref{fig:HCALPY}.

\begin{figure*}[htb]
\centering
\includegraphics
  [width=0.45\hsize]
  {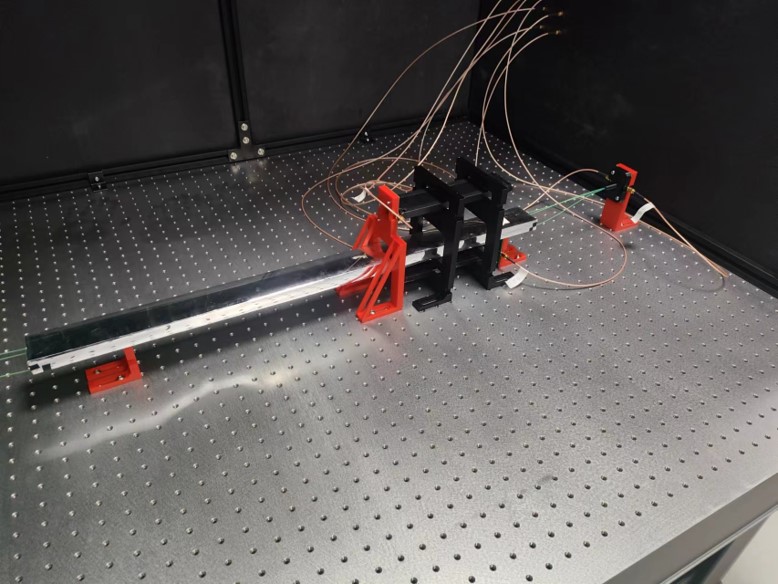}
\includegraphics
  [width=0.45\hsize]
  {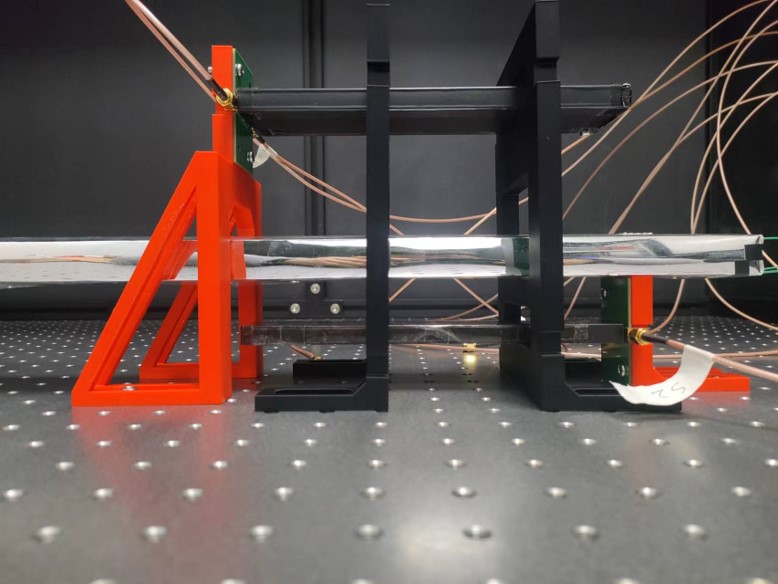}
\caption{Cosmic ray test plateform.}
\label{fig:HCALCRTest}
\end{figure*}

The simulation results and test results demonstrate a good consistency. Increasing the number of fibers leads to higher photon yield, particularly within the range of 1 to 3 fibers where linearity is observed. A thicker scintillator has the ability to capture more energy from muons; however, it exhibits a lower photon yield per MeV compared to thinner scintillators. Additionally, we conducted tests on scintillators provided by EJ company which utilize polyvinyl toluene for comparison with those from HND and HTX that employ polystyrene. The EJ scintillator exhibited better performance, although the HND scintillator also demonstrated satisfactory results. Considering a design with three grooves and a thickness of 1 cm, an 8 P.E threshold corresponds to approximately 0.12 MeV.

\begin{figure}[htb]
\centering
\includegraphics
  [width=0.45\hsize]
  {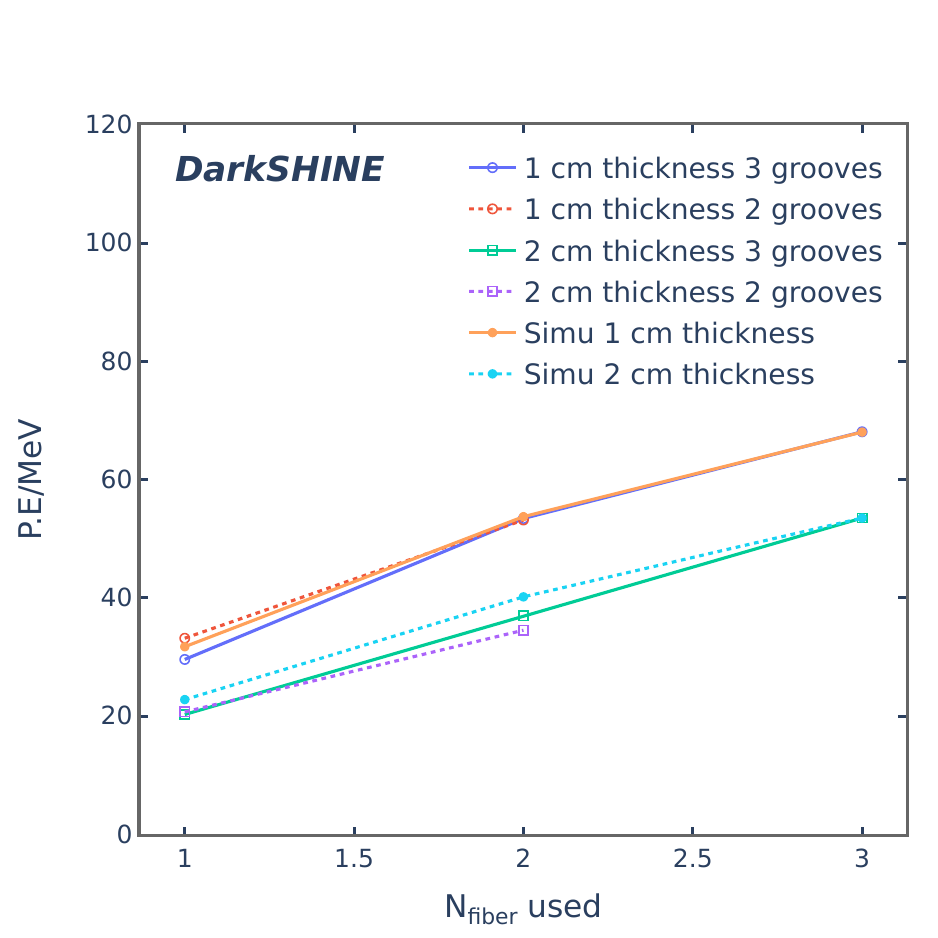}
\includegraphics
  [width=0.45\hsize]
  {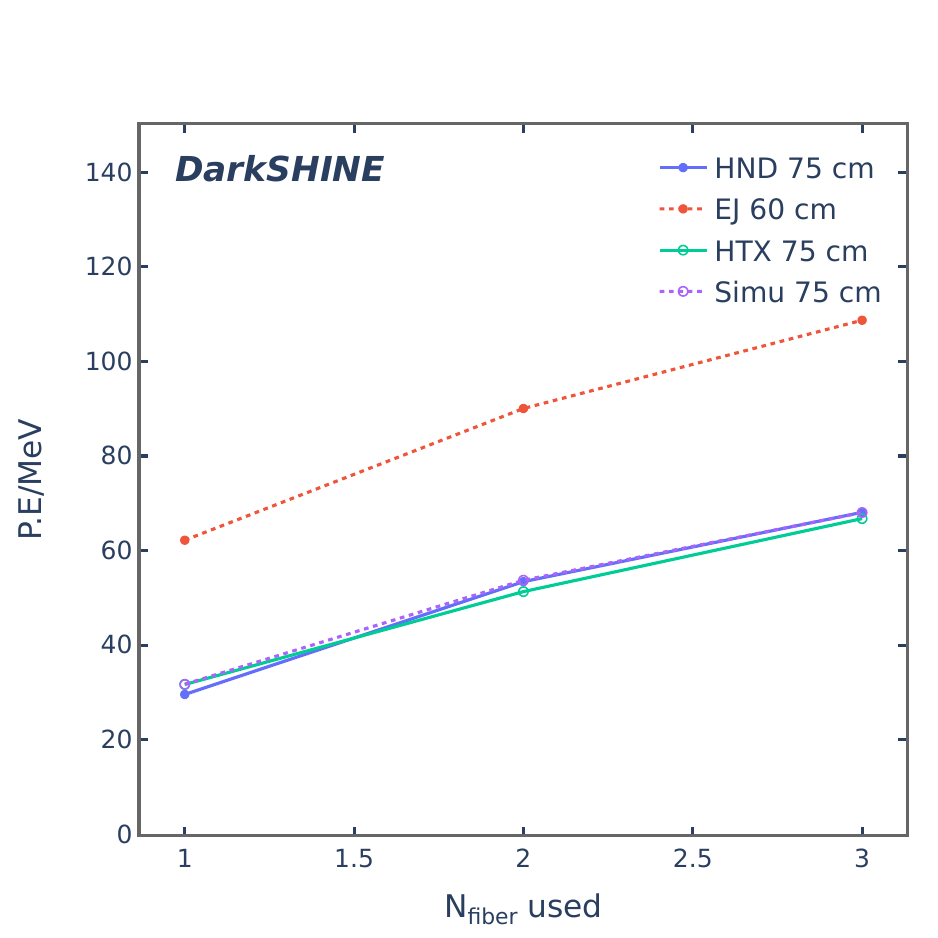}
\caption{Photon yields test results. Y-axis is the photon yield per MeV energy deposited, X-axis is the number of fibers put into the scintillator.}
\label{fig:HCALPY}
\end{figure}

\section{Mechanical System}\label{sec:machanical}

\subsection{Constrains and Requirements}

The DarkSHINE detector will be installed in the existing experimental hall. Therefore the DarkSHINE mechanical integration design is constrained by the dimensions of the experimental hall and the weight capacity of the floor. Additionally, requirements for the mechanical integration design are as follows:

1. Provide all necessary support and adjustments for all sub-detectors to ensure they are positioned within specified tolerances.

2. Allow convenient access to sub-detectors for maintenance, such as access to the tracker when the sensor is broken.

3. Provide adequate routes and space for cables and cooling pipes.

\subsection{Beamline and Detector Interface}

The DarkSHINE beamline includes a beampipe that terminates at a thin vacuum window positioned just upstream, in front of the tracker detector. The tracker system is housed in a support box. The square Helmholtz coils are designed to be positioned above and below the tracker system's support box. The entire DarkSHINE detector setup, as shown in Figure~\ref{fig:overviewDetector}, is designed with an integrated support and rail system to facilitate efficient operation and maintenance. To support the square Helmholtz coils, a specialized mounting structure is used. This structure consists of a rigid, non-magnetic frame to prevent interference with the generated magnetic field. Adjustable brackets secure the coils firmly, maintaining their parallel alignment and fixed distance to ensure optimal field uniformity. For easy maintenance, the support system is modular, allowing for quick access to the coils and tracker system when adjustments are required. The ECAL and magnet support frames are mounted on a shared rail system, allowing both components to be smoothly moved along the rails for maintenance or alignment adjustments. The HCAL, located directly behind the ECAL and magnet, is installed on an independent support frame that is separate from the rail system. This dedicated frame provides stability and precise alignment for the HCAL without interfering with the rail-based mobility of the magnet and ECAL. 

\begin{figure}[h]
    \centering
    \includegraphics[width=0.6\linewidth]{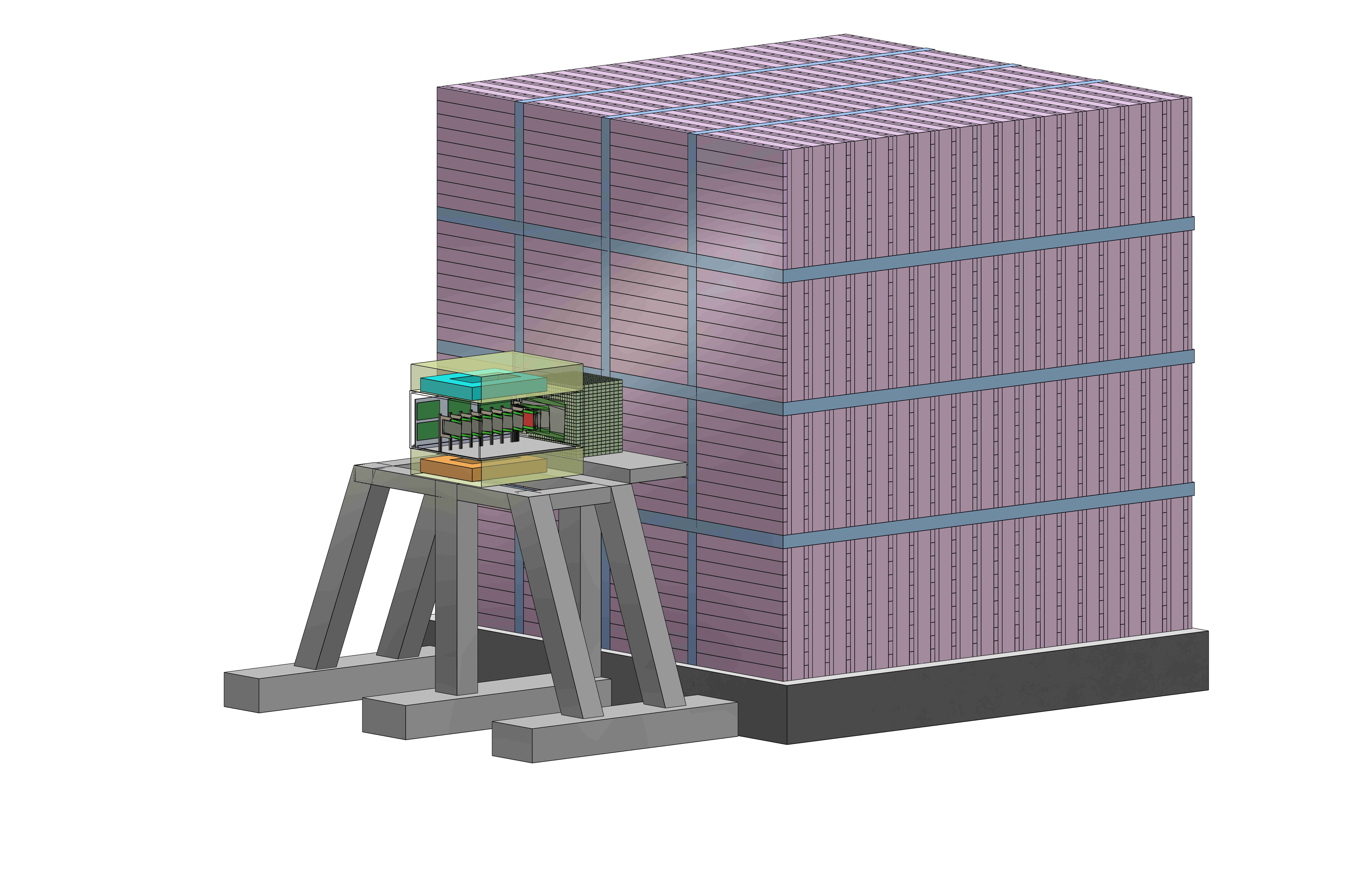}
    \caption{An overview of DarkSHINE detector mechanical design}
    \label{fig:overviewDetector}
\end{figure}

\subsection{Support Structure of Tracker System}

Tracking system consists of tagging tracker and recoil tracker, which share the same support structures and data acquisition hardware, as shown in Figure~\ref{fig:overviewTracker}. The main mechanical structure of this system is a horizontally oriented aluminum plate and each layer of tracker is mounted onto the plate. To provide cooling, a copper tube through which coolant flows is pressed into a machined groove in the plate. Starting from the upstream end of the magnet, the plate is placed inside a support box that is aligned and secured within the magnet bore. Another similar plate is placed into the support box on the positron side, housing the Front End Boards (FEBs) responsible for distributing power and control signals from the DAQ, as well as digitizing raw data from the modules for transmission to the external DAQ. The cooling lines of the tracker supports and the FEB support are routed to a cooling manifold at the upstream end of the magnet. In addition, the target is interposed onto the same plate between the last layer of the tagging tracker and the first layer of the recoil tracker.

\begin{figure}[h]
    \centering
    \includegraphics[width=0.5\linewidth]{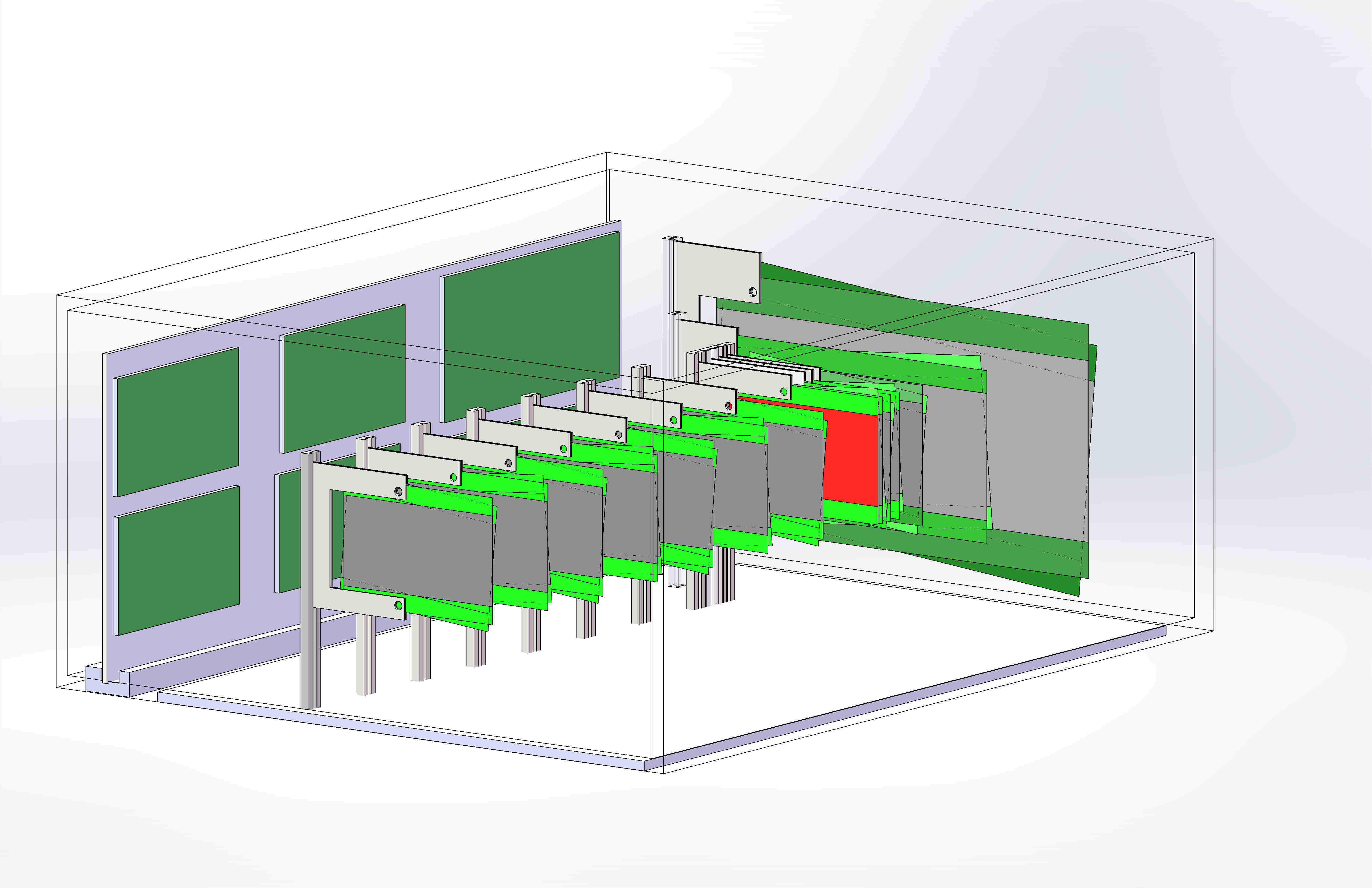}
    \caption{An overview of the tracking systems and target mechanical design}
    \label{fig:overviewTracker}
\end{figure}

\subsection{Support for Electromagnetic Calorimeter System}

The entire ECAL is mounted on a robust support frame that sits on the same rails as tracker system, allowing the entire assembly to be moved along the track for maintenance access. The ECAL is positioned directly behind the tracker system, as shown in Figure~\ref{fig:ECAL_support}. The baseline layout of it consists of 21$\times$21$\times$11 LYSO crystals. Each crystal is coupled with a SiPM for readout, which is mounted on a dedicated printed circuit board (PCB). Behind the PCB, a cooling plate is positioned to maintain the necessary thermal conditions. The cooling system of the ECAL should be arranged reasonably to ensure stable heat dissipation.

\begin{figure}[h]
    \centering
    \includegraphics[width=0.6\linewidth]{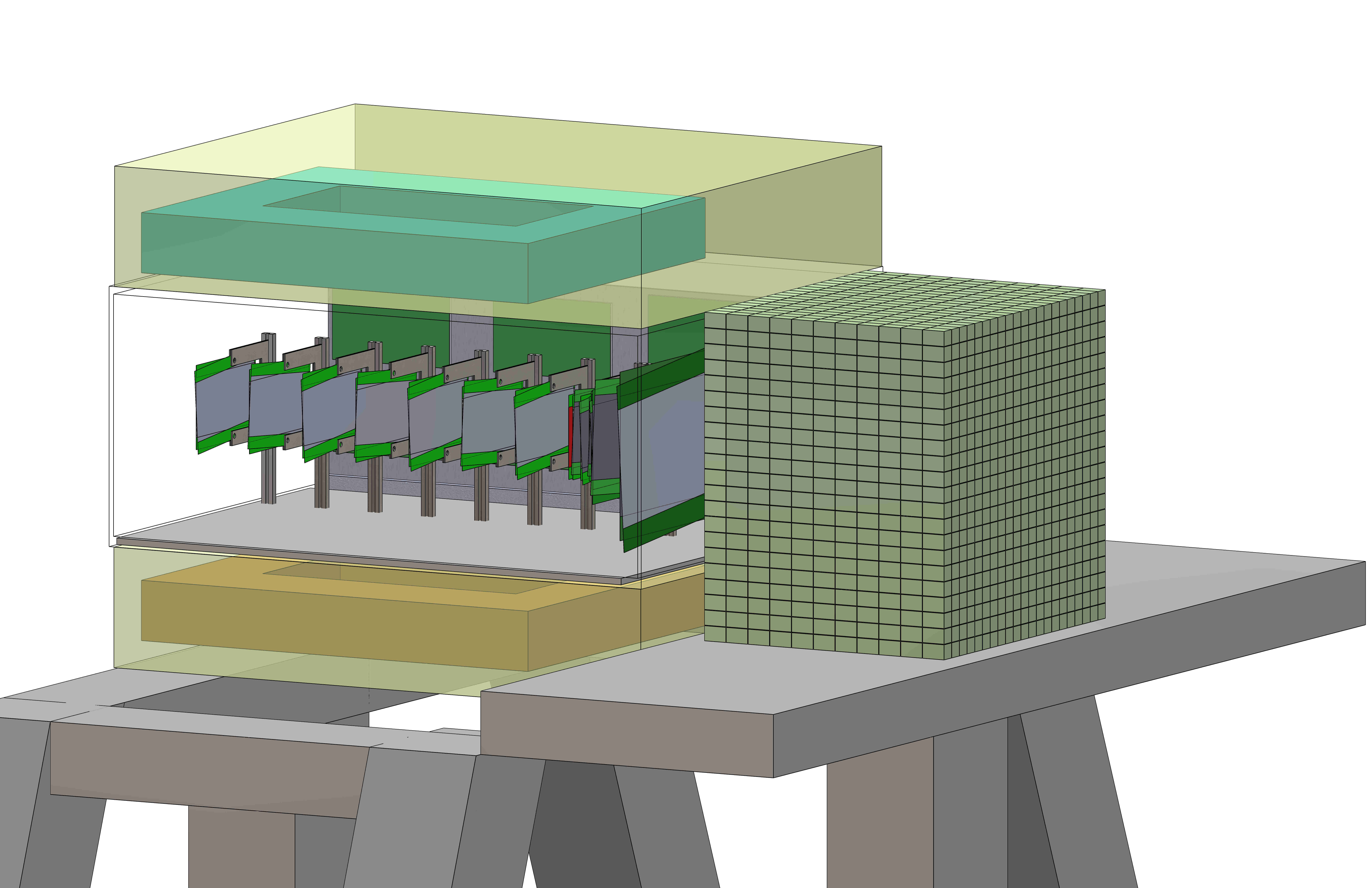}
    \caption{Support structure of ECAL along with tracking system, target and magnets}
    \label{fig:ECAL_support}
\end{figure}

\subsection{Support for Hadronic Calorimeter System}

The design of the HCAL is introduced in Section~\ref{sec:HCALConceptual}. As the heaviest sub-detector in the DarkSHINE detector, it requires robust support. To evenly distribute its weight, the HCAL is placed on a support plate. The HCAL consists of 16 modules, each connected with supporting frames to ensure stability and facilitate maintenance, as is shown in Figure~\ref{fig:overviewHCAL}. To minimize deformations due to gravity, the mechanical support must provide adequate stiffness. In the redesign, the supporting platform under the HCAL will use a material optimized for both strength and weight distribution, enhancing overall stability. In addition, The readout modules are mounted on dedicated brackets attached to the HCAL support frame. These brackets are designed for modularity, allowing quick removal and replacement during maintenance. Additionally, the placement of readout electronics should take into account several key factors. The brackets should be made from high thermal conductivity materials, with integrated ventilation slots to enhance heat dissipation. Thoughtful cable management should be implemented to reduce signal interference and ensure clean, organized routing. These considerations will improve the stability, performance, and maintainability of the HCAL system.

\begin{figure}[htb]
    \centering
    \includegraphics[width=0.5\linewidth]{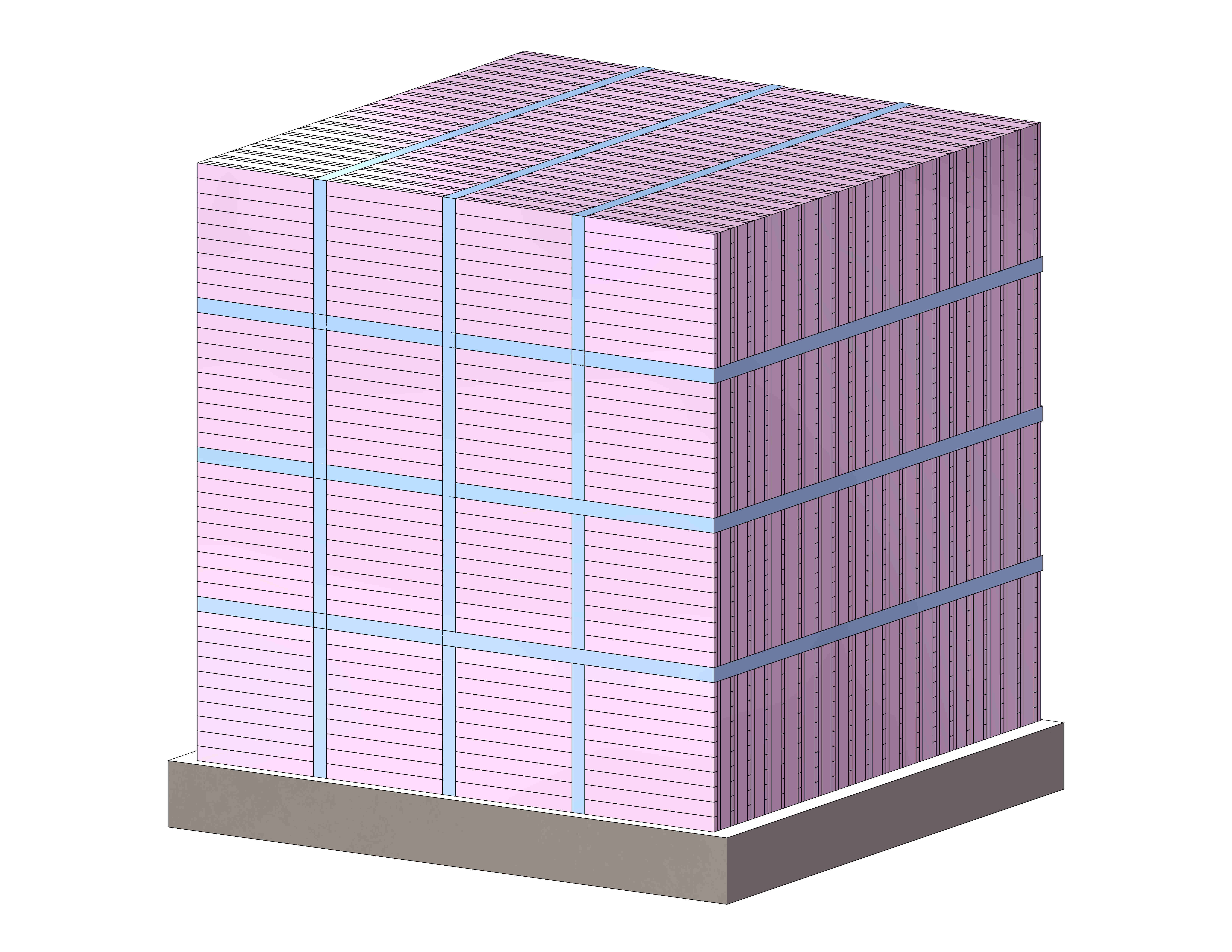}
    \caption{An overview of HCAL mechanical design}
    \label{fig:overviewHCAL}
\end{figure}

\end{chapter}

\begin{chapter}{Signal and Background}

\section{Simulation Software}
\label{sec:simulation}

DarkSHINE Software integrates detector simulation, digitization, event reconstruction and data analysis into one package. The event data is a self-defined C++ class stored in the ROOT~\cite{rene_brun_2019_3895860} file, containing all the necessary event information for analysis. There are two main parts, \textbf{D}arkSHINE \textbf{Simu}lation program, \textbf{D}arkSHINE Reconstruction and \textbf{Ana}lysis program:

\textbf{DSimu} is responsible for the detector simulation. Similar to all other High Energy Physics (HEP) experiments, the physics simulation is based on Geant4~\cite{Agostinelli:2002hh}, which is a C++ toolkit for the simulation of the passage of particles through matter. The basic idea of simulation is to build a self-defined detector in program with explicitly defined materials.

For physics simulation, we are using the default Geant4 physics list \verb|FTFP_BERT|. To enhance the gamma physics, we added a physics class including \verb|G4GammaConversionToMuons| process. When a particle, for example, electron passes through the detector materials, it will deposit energy inside the material and scatter with the atomic nucleus. Geant4 will help to simulate the whole physics process including standard model electromagnetic physics, hadron-nucleus, nucleus-nucleus scattering, intranuclear cascade, and de-excitation process, with energy deposition information, including deposition type (EM / hadronic) recorded in the event data. 

The initial input particle can be generated by 2 methods: particle gun and external generator. The particle gun is the default method for generating primary particles, which can be easily configured. It is commonly used to simulate Standard Model (SM) process. In our case, the dark photon is beyond Standard Model, so it's better to generate the BSM process using external generator (CalcHEP~\cite{Belyaev:2012qa}) and then input the result of differential cross section of final state energy ratio and scatter angle into the simulation program \textbf{DSimu}.  With the initial particle entering the detector, it will produce tons of secondaries and deposit energy in the material. The deposited energy (as mentioned before) and those secondary particles are recorded in the event data as the truth hit collection and truth particle collections, which is known only in simulation. 

In reality, it's impossible to know the truth energy deposition and secondary particles. Capturing optical photons is the main method to detect. When the particle passes through materials, it will not only deposit energy, but also emit optical photon through scintillation and Cerenkov radiation. With optical photon detected, next step is to convert the optical signal, which is called digitization. Digitized hit collection is used in real analysis instead of truth information.

For detector geometry used in simulation, we considered the sensitive detector material, wrapper material, electronics and supporting structures. The experiment setup is shown in Figure~\ref{fig:dsimugeometry}, consists of tagging tracker, target, recoil tracker, ECAL and HCAL arranged in a line. Non-uniform magnetic field along y-axis are added in the tagging tracker and the recoil tracker region.

\begin{figure}[htbp]
    \centering
    \includegraphics[width=0.75\textwidth]{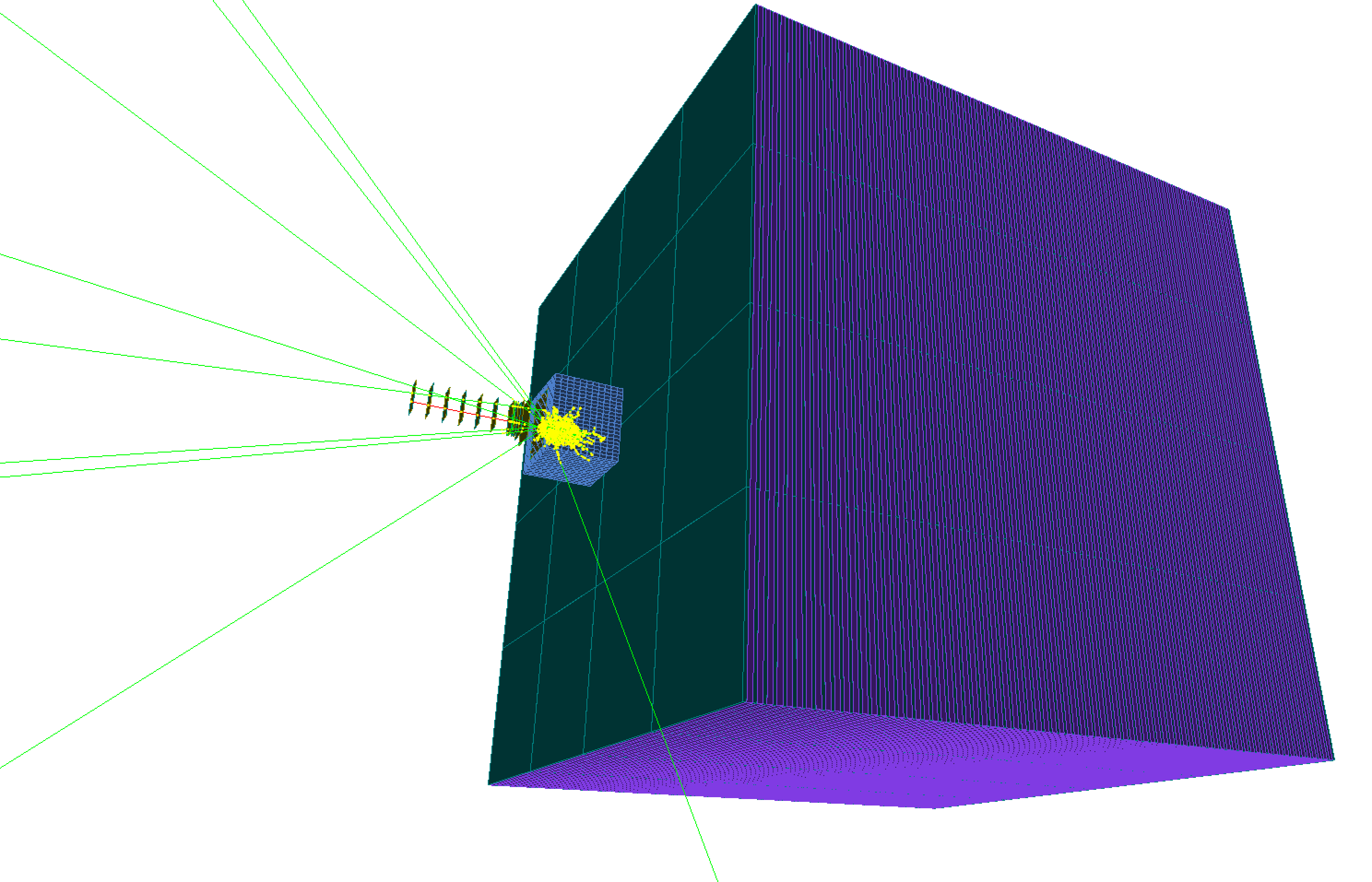}
    \caption{Overview of default geometry used in the DarkSHINE simulation, and an example background event display. The electron (red line) is passing through tagging tracker, target and recoil tracker, with non-uniform magnetic field in this region. Then the recoil electron initiates an EM shower inside the ECAL region. The green lines represent tracks of neutral particles, and yellow dots are track step points.}
    \label{fig:dsimugeometry}
\end{figure}

\begin{table}[!ht]
    \centering
    \begin{tabular}{lllll}
    \toprule
        Subsystem & Materials & Logical Vol. & Physical Vol. & Vol. Depth \\
        \midrule
        Tagging Tracker & 2 & 43 & 94753 & 3 \\ 
        Target & 1 & 1 & 1 & 0 \\ 
        Recoil Tracker & 2 & 37 & 114749 & 3 \\ 
        ECAL & 4 & 5 & 14653 & 3 \\ 
        Side HCAL & 6 & 31 & 624 & 2 \\ 
        HCAL & 7 & 96 & 27234 & 4 \\
        \midrule
        Dark SHINE Total & 9 & 214 & 252015 & 5 \\
        \bottomrule
    \end{tabular}
    \caption{Quantities of materials, logical volumes, physical volumes, and volume depth of the DarkSHINE detector in simulation.}
\end{table}

Biasing is a set of techniques to enhance the event of interest in the interested geometry region, thus provide large acceleration factors for rare events simulation. Biasing is used to accelerate our signal production and important background process production. For signal production, the signal cross-section on the target is enlarged by a factor of $10^{21}$. The gamma to mu pair, photon-nuclear and electron-nuclear samples are also prepared with biasing.

In summary, \textbf{DSimu} simulates the entire process and outputs event data, including truth particle collections, truth hit collections, and optionally, optical photon information.

\textbf{DAna} is a framework for the analysis and reconstruction tools. It requires the output ROOT file (Geometry, Magnetic field map and event data) from DSimu. The first thing to do after simulation is reconstruction. Reconstruction is to rebuild the whole event only based on the known information. The rebuild process can be divided into several sub processes, for example, calculating the transverse momentum of charged particles according to the hits on trackers, clustering calorimeter hits and particle identification (PID). The digitized hit collection will be converted to reconstructed particle collection after reconstruction. The latter one is used in analysis such as event selection.

In summary, the workflow of Dark SHINE Software is simple and obvious, which is summarized in Figure~\ref{fig:DSS_Workflow}. Firstly, running \textbf{DSimu} produces the simulation results with and other truth information. Then using \textbf{DAna} to reconstruct the event gives the output file for final analysis. In addition, we have a \textbf{D}arkSHINE Event \textbf{Dis}play program.

\begin{figure}[htb]
\centering
\includegraphics
  [width=0.9\hsize]
  {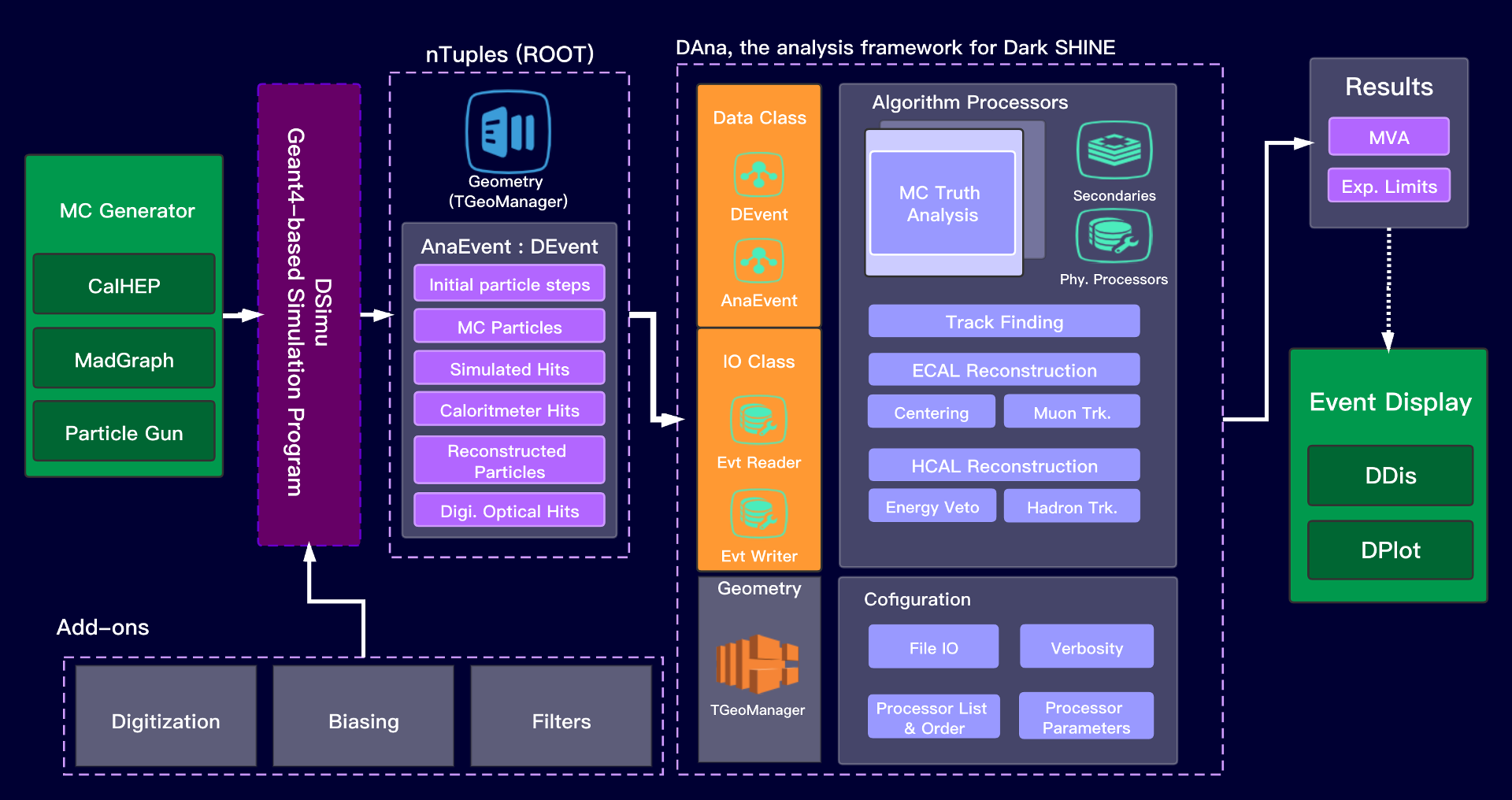}
\medskip
\caption{Workflow for the whole Dark SHINE Software package. \textbf{D}arkSHINE \textbf{Simu}lation program is a Geant4-based program. For initial particles generation, it uses Geant4 particle gun (standard model events), or CalcHEP (signal events). The output ROOT file includes event data, geometry and magent information.  \textbf{D}arkSHINE Reconstruction and \textbf{Ana}lysis program read the ROOT file, run digitization and reconstruction algorithms, and output the result ROOT file for physics analysis.}
\label{fig:DSS_Workflow}
\end{figure}

    \section{Signal Efficiency and Background Rejection}\label{sec:SiganlAndBackground}

This section briefly introduces the signal and background production and related studies based on the DarkShine experiment. 

The simulation of the dark photon emission is done by the CalcHEP generator~\cite{Belyaev_2013} and simulated in Geant4~\cite{Agostinelli:2002hh}. The input beam during simulation is fixed at 8 GeV. The parameters scanned in the CaclcHEP are the dark photon's mass and kinetic coupling parameter $\epsilon$. Since the $\epsilon$ is a factor only interacting with the cross section, the kinematic performance of the final states largely relies on the dark photon's mass. Thus, 25 mass points are scanned with the fix $\epsilon = 1$, shown in the Table~\ref{tab:signal-production-summary}. The normalised cross section variation versus dark photon's mass is shown in Figure~\ref{fig:cross-section}. $ \sigma_{DM} $ scale quite large due to the effect form factor in the high mass region. In low mass region, the cross section is changed by $ \frac{1}{m_{A'}} $.

\begin{table}[!ht]
    \centering
    \caption{Summary table of generated signal samples.}
    \begin{tabular}{|l|l|l|l|l|l|l|l|l|l|l|l|l|l|}
    \hline
        Run ID & 1 & 2 & 3 & 4 & 5 & 6 & 7 & 8 & 9 & 10 & 11 & 12 & 13 \\ \hline
        A' Mass (MeV) & 1 & 2 & 3 & 4 & 5 & 6 & 7 & 8 & 9 & 10 & 20 & 30 & 40  \\ \hline
        Run ID & 14 & 15 & 16 & 17 & 18 & 19 & 20 & 21 & 22 & 23 & 24 & 25 & ~ \\ \hline
        A' Mass (MeV) & 50 & 60 & 70 & 80 & 90 & 100 & 200 & 500 & 700 & 1000 & 1500 & 2000 & ~ \\ \hline
    \end{tabular}
    \label{tab:signal-production-summary}
\end{table}

\begin{figure}[htbp]
    \centering
    \includegraphics[width=8.cm]{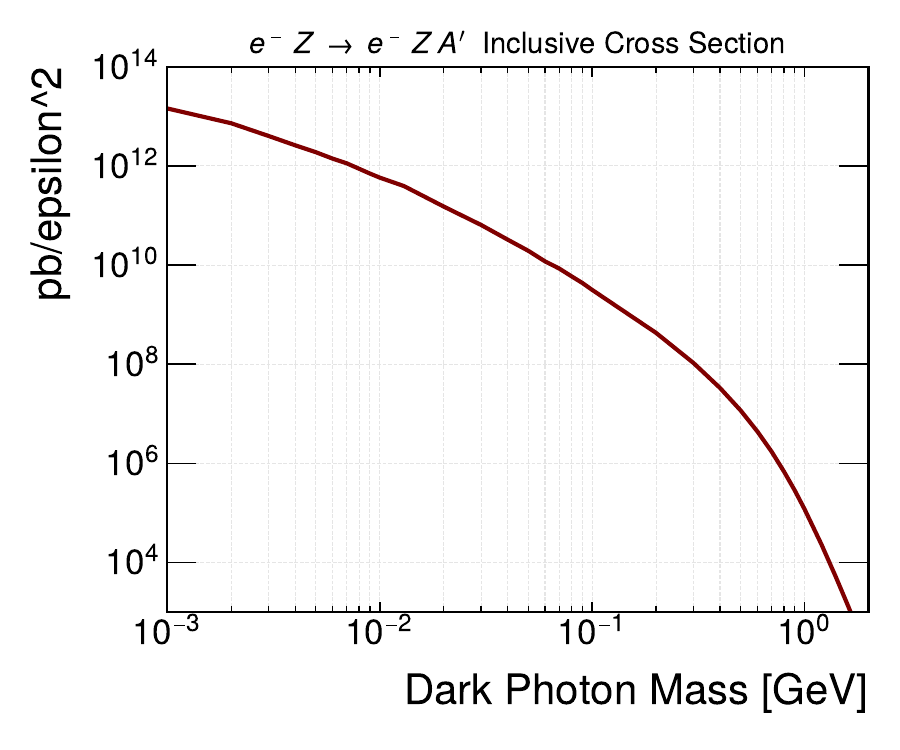}
    \caption{Inclusive cross section of dark photon bremsstrahlung from electron interacting with tungtsen target, which is produced by Calchep and is normalized to $\epsilon=1$, with 8\ GeV beam energy.}
    \label{fig:cross-section}
\end{figure}

As described in Sec~\ref{sec:target Sys}, ~\ref{sec:ECALIntroduction} and~\ref{sec:HCALIntroduction}, most of the incident electrons go through the target without any interaction. With relative rates $3.25\times 10^{-6}$ and $5.10\times 10^{-7}$, the electron-nuclear reaction can be produced in ECAL and target, respectively. Around 6.7$\%$ electrons can produce bremsstrahlung photons. This photon can induce photon-nuclear and $\gamma \to \mu \mu$ processes. These two processes can similarly occur in the two parts ECAL and target, where the former has relative rate $2.31\times 10^{-4}$ and $1.37\times 10^{-4}$, respectively, while the latter has ralative rate $1.63\times 10^{-6}$ and $1.50\times 10^{-8}$, respectively.  Figure~\ref{fig:flowbackground} summaries the relative rates of these backgrounds. Due to the small relative rate, the neutrino production reactions can be ignored for now. 

In general, the rare processes are denoted as EN\_target, EN\_ECAL, PN\_target, PN\_ECAL, GMM\_target, and GMM\_ECAL respectively according to their process type and occurrence location. One can tell EN stands for electron-nuclear, PN stands for photon-nuclear, and GMM stands for $\gamma \to \mu \mu$, respectively. In order to obtain higher statistics of these rare processes' sample, biasing was implemented in the sample production as mentioned in Sec~\ref{sec:simulation} by requesting the process of interest to have a much larger factor, which can suppress the rest processes to be negligible in the sample. Besides of these rare process samples, an inclusive background sample are also generated. All the background samples used in the sensitivity study are illustrated in Table~\ref{tab:bkg-production-summary}. 

\begin{figure}[htb]
\centering
\includegraphics
  [width=0.6\hsize]
  {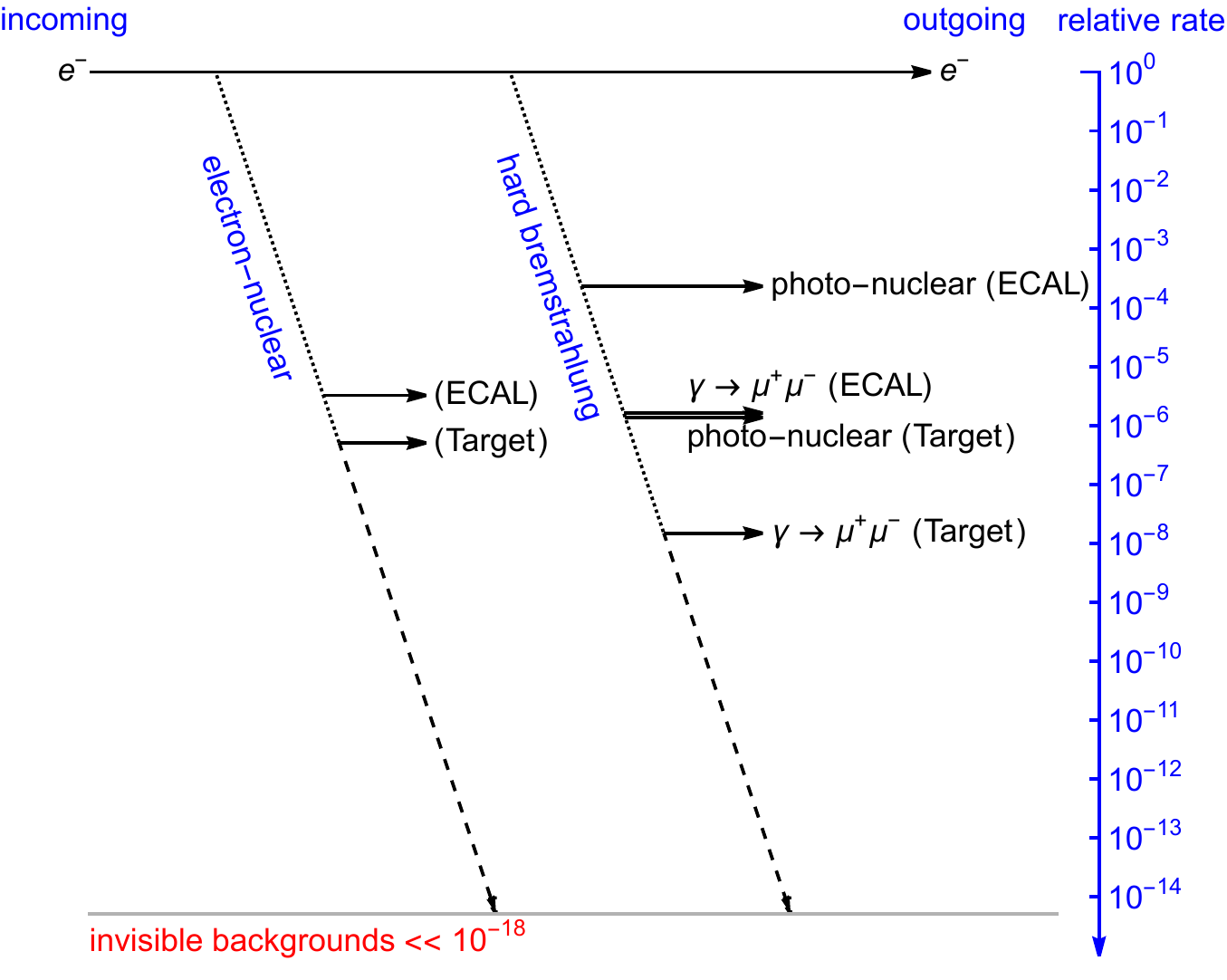}
\caption{Flow and  relative rates of background processes. "ECAL" and "target" refer to the locations where the processes occur.}
\label{fig:flowbackground}
\end{figure}

\begin{table}[!ht]
    \centering
    \caption{Summary table of generated background samples.}
    \begin{tabular}{|l|l|l|l|l|l|l|l|}
    \hline
    Type&Inclusive & EN\_Target & EN\_ECAL & PN\_Target & PN\_ECAL & GMM\_Target & GMM\_ECAL \\
   % \hline
   % Relative rate&$2.31\times 10^{-4}$&$2.31\times 10^{-4}$&$2.31\times 10^{-4}$&$2.31\times 10^{-4}$&$2.31\times 10^{-4}$&$2.31\times 10^{-4}$&$2.31\times 10^{-4}$ \\
    \hline
    Generate Events&$2.5\times 10^{9}$&$1\times 10^{8}$&$1\times 10^{7}$&$1\times 10^{7}$&$1\times 10^{8}$&$1\times 10^{7}$&$1\times 10^{7}$\\
    \hline
    effective EOT&$2.5\times 10^{9}$&$1.6\times 10^{12}$&$1.8\times 10^{12}$&$4.0\times 10^{12}$&$4.4\times 10^{11}$&$4.3\times 10^{14}$& $6.0\times 10^{12}$\\
    \hline
    \end{tabular}
    \label{tab:bkg-production-summary}
\end{table}

The signal performance compared to the inclusive background is shown in Figure~\ref{fig:sig_kinmetic}. The character of the dark photon bremsstrahlung, comparing the backgrounds represented in the grey lines, is the single electron track with large $p_T$ and low energy. They have large recoil angles resulting in large transverse separation on the surface of the detector due to the magnetic field. It distinctly shows the relationship that the larger $m_{A^{\prime}}$ would have more obvious detectable characters.

\begin{figure}[htbp!]
    \centering
    \includegraphics[width=0.35\linewidth]{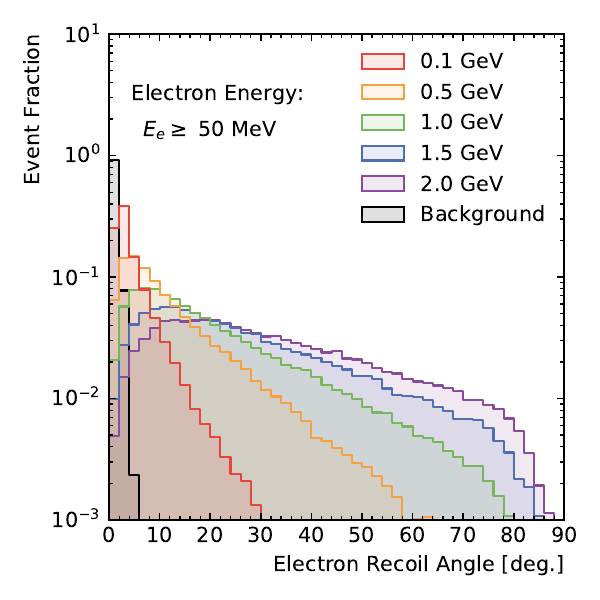}
    \includegraphics[width=0.35\linewidth]{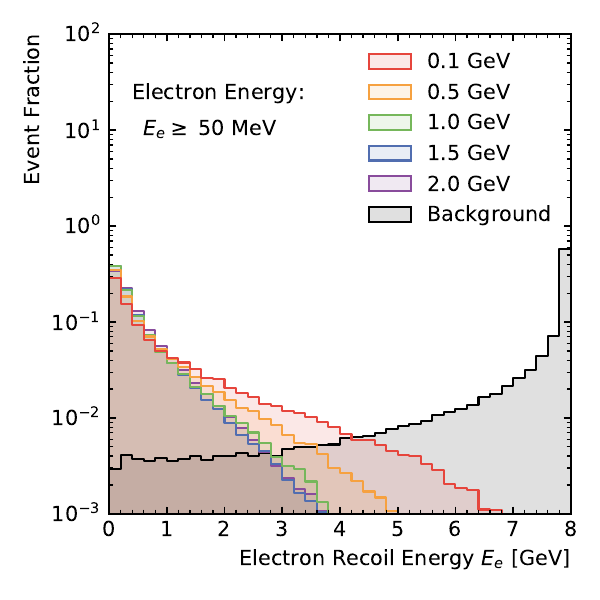}
    \includegraphics[width=0.35\linewidth]{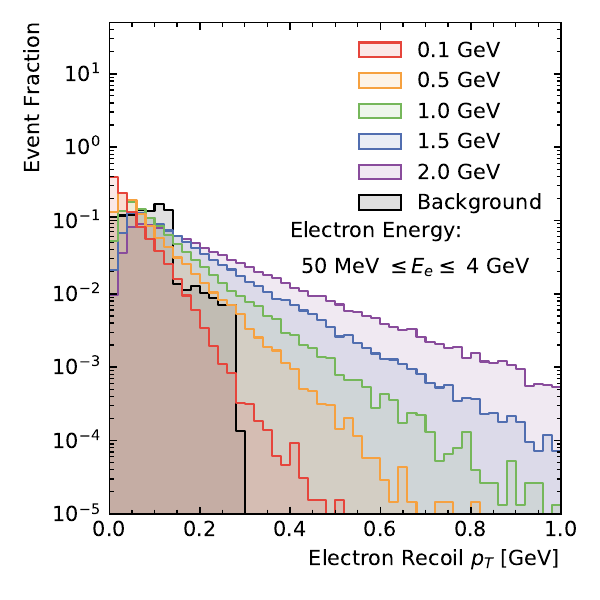}
    \includegraphics[width=0.35\linewidth]{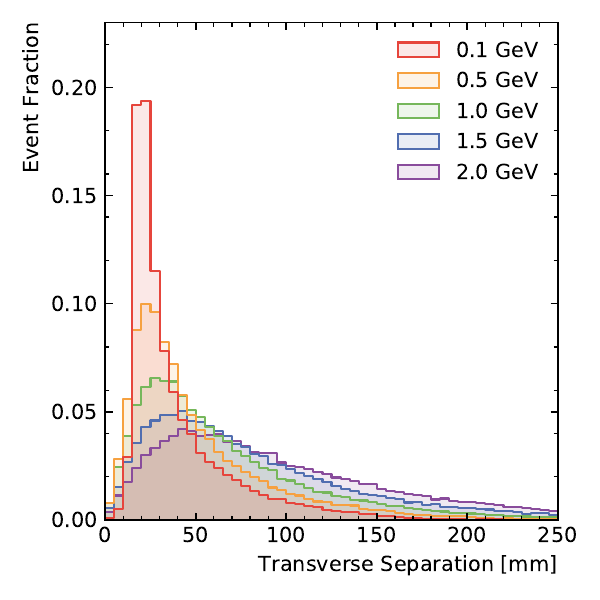}
    \caption{The signal kinetics distribution simulated with Geant4, comparing to the backgrounds.}
    \label{fig:sig_kinmetic}
\end{figure}

For background, some of them can be easily rejected using single variables, the remaining ones require a combination of information from three sub-detectors for rejection. Tracking system could powerfully suppress background by requesting only one reconstructed track in tagging tracker and recoil tracker, and by rejecting events with small difference between the momentum of the tagged track and the recoil track (missing momentum). Figure~\ref{fig:missingP} shows the distribution of missing momentum. For signal processes, large momentum was transferred to the dark photon. Most of signal events will fall into the large missing momentum region. But background events have small missing momentum, most of these events will fall into a low missing momentum region. 

\begin{figure}[htb]
\centering
\includegraphics
  [width=0.6\hsize]{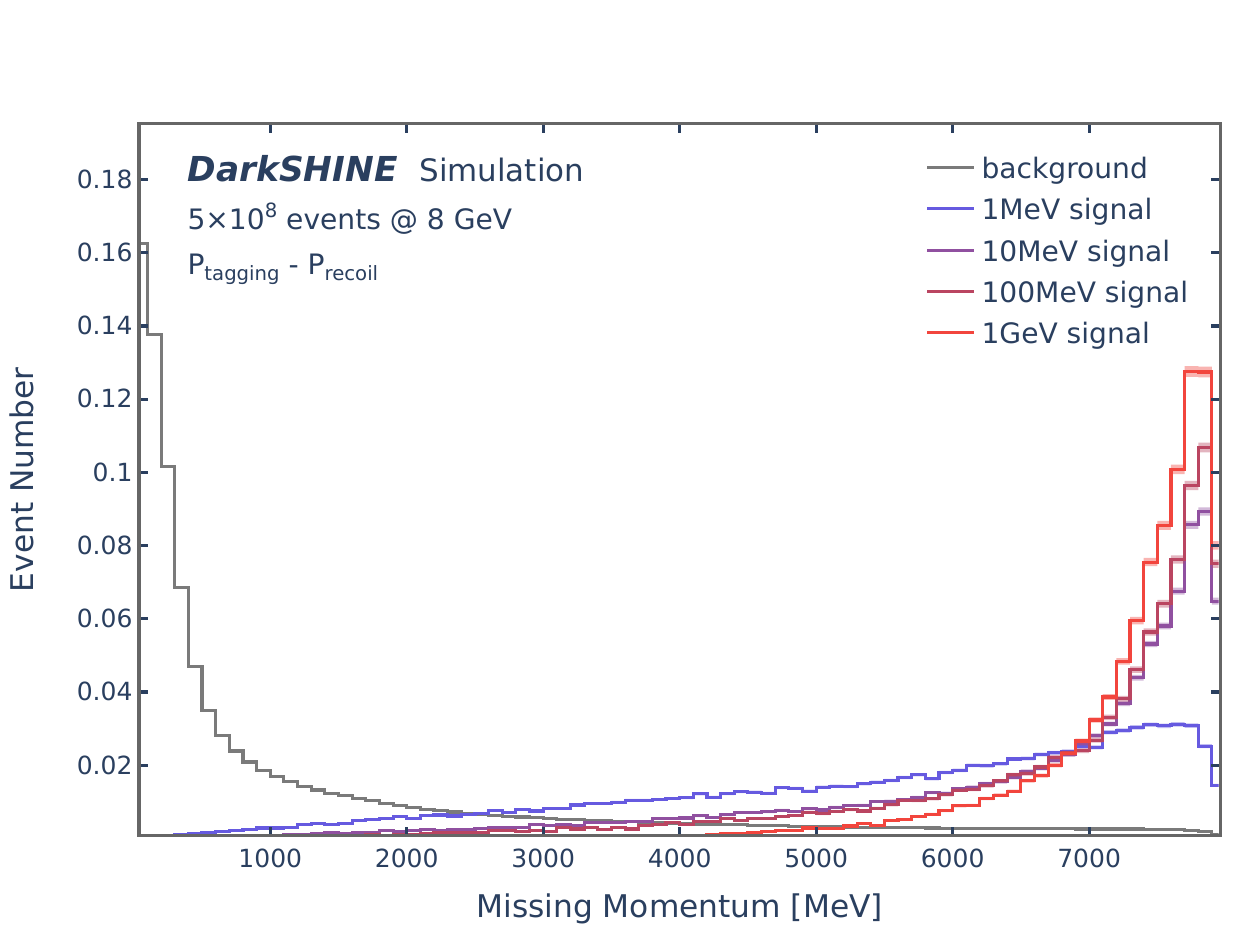}
\caption{Missing momentum: difference between tagging track momentum and recoil track momentum.}
\label{fig:missingP}
\end{figure}

The case of event survived from the tracker-based cuts is either that the electron loses a significant portion of its momentum by the dark photon, that it loses a significant portion by the photon, or that the electron reacts with the nucleus at the target and loses energy by the product. In addition to dark photons, these products suppose to deposit energy in the ECAL and HCAL. Figure~\ref{fig:SEmap} shows the HCAL energy vs. ECAL energy 2-dimensional map for $m_A^{\prime}=10$~MeV and $m_A^{\prime}=1$~GeV signal, respectively. Signal region should be defined in the bottom left corner of energy map in Figure~\ref{fig:SEmap}. It is far away from the energy distribution of inclusive background shows in Figure~\ref{fig:2DHCALECAL_Inc}. 

The signal box enriched with candidates dark photon generation process is optimized by the cuts from the reconstruction in the sub-detectors. The cuts are optimized to keep the highest signal efficiency with the smallest background yields in overall.
The events selection criteria are shown below:
\begin{itemize}
    \item number of the reconstructed track should be only 1.
    \item difference between the tagging track momentum and the recoil track momentum of the reconstructed electrons should be larger than 4 GeV
    \item the total energy deposit in the ECAL should be lower than 2.5 GeV
    \item the total energy detected in HCAL should be lower than 0.1 GeV
    \item maximum cell energy in HCAL should be less than 2 MeV
\end{itemize}
 Figure~\ref{fig:signalEff} illustrates the signal efficiency of the signals with different $m_{A^{\prime}}$ after the cuts mentioned above. The signal acceptance is over 60\% overall. For $m_{A^{\prime}}=1$~MeV signal point, missing momentum is more sensitive. The signal efficiency drops to 45\% with current missing momentum cut. For high $m_{A^{\prime}}$ signal points, the HCAL and ECAL cuts are too tight that the signal efficiencies drop to 55\% while The cut efficiencies are quite stable in the range between 100~MeV and 1~GeV.  

\begin{figure}[htbp]
    \centering
    \includegraphics[width=0.6\hsize]{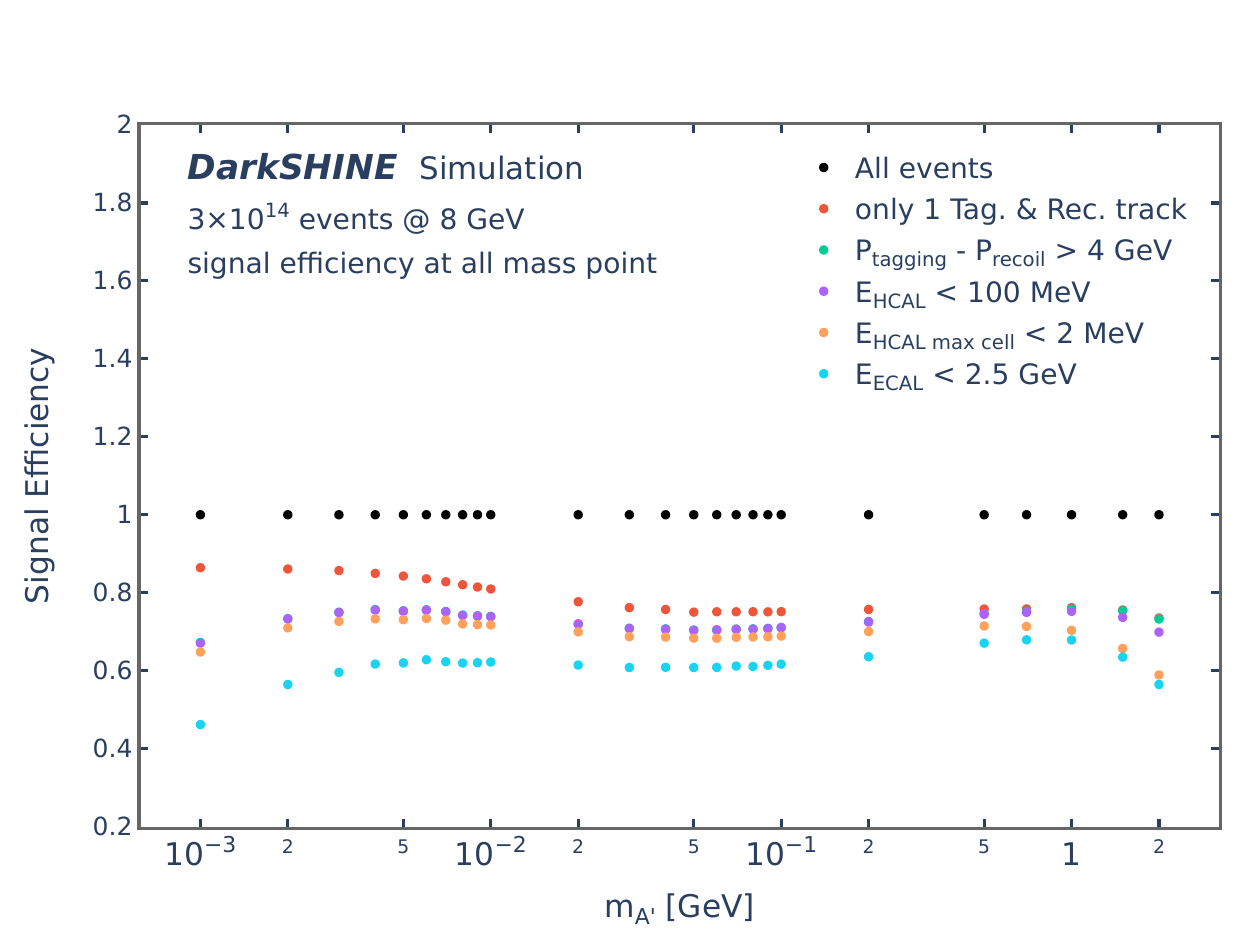}
    \caption{Signal efficiency.}
    \label{fig:signalEff}
\end{figure}

\begin{figure}[h]
\centering
\includegraphics[width=0.45\linewidth]{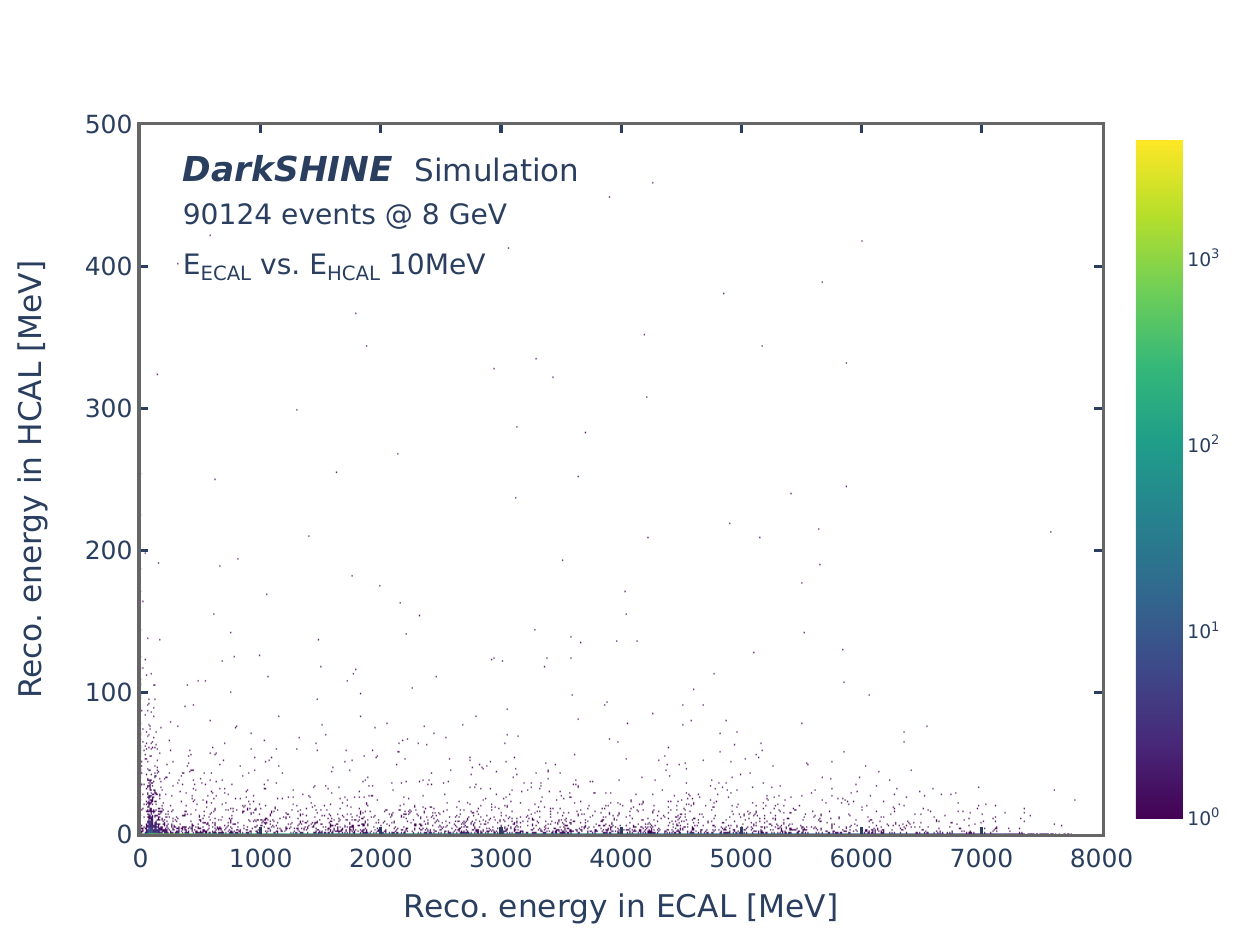}
\includegraphics[width=0.45\linewidth]{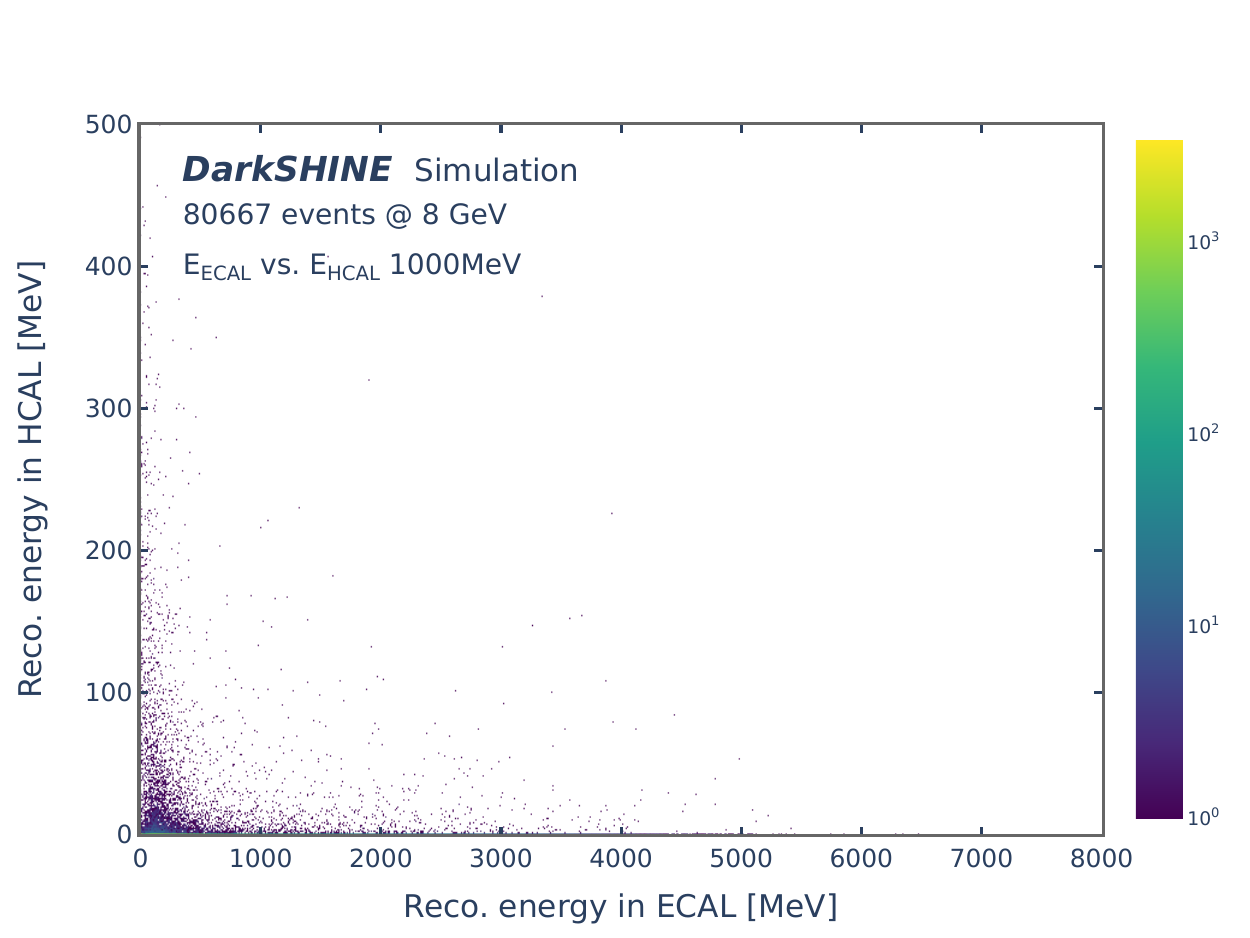}
\caption{\label{fig:SEmap}~Left: 2-dimensional deposit energy map in HCAL and ECAL with $m_{A^{\prime}}=10$~MeV. Right: 2-dimensional deposit energy map in HCAL and ECAL with $m_{A^{\prime}}=1$~GeV.}
\end{figure}

After signal region selections, no background survives. But as shown in Table~\ref{tab:bkg-production-summary}, only GMM\_Target sample has $4.3\times 10^{14}$ EOT, which is larger than  $3\times 10^{14}$ EOT. Due to lack of statistics, an extrapolation method~\cite{Chen:2022liu} is used to estimate background yields. In the end, 0.015 background yield per $3\times 10^{14}$ is derived.

%\subfile{SignalBackground/SignalEfficiency}
%\subfile{SignalBackground/BackgroundRejection}

\section{Expected Sensitivity}

In this section, we explore the sensitivity of Dark SHINE for $3\times10^{14}$ EOT, based on the preceding discussions on background rejection, signal efficiency, and dark photon production
cross-sections. Given the relatively low background yield estimated in Section~\ref{sec:SiganlAndBackground}, we assume that all observed events are background, following a Poisson distribution. Therefore, the upper limit on the signal times acceptance efficiency at 90\% confidence level (CL) is given by

\begin{equation}
\label{eq:upperlimit}
s_{\text{up}} \times \epsilon_{\text{sig}} = \frac{1}{2} F^{-1}_{\chi^2}(1 - \alpha; 2(n_{\text{obs}} + 1)) - b
\end{equation}

%\[
%s_{\text{up}} \times \epsilon_{\text{sig}} = \frac{1}{2} F^{-1}_{\chi^2}(1 - \alpha; 2(n_{\text{obs}} + 1)) - b,
%\]

where $F^{-1}_{\chi^2}$ is the inverse of the cumulative distribution function of the $\chi^2$ distribution, $\epsilon_{\text{sig}}$ is the acceptance efficiency for a given $m_{A'}$, and $n_{\text{obs}} = b$ is the background yield (considered as a constant). The relationship between the kinetic mixing parameter ($\epsilon$) and the expected signal yield ($N_{\text{sig}}$) is defined by equation~\ref{eq:signalYields}. Therefore, the upper limits on $\epsilon^2$ are given by

\begin{equation}
\label{eq:upperlimitEpsilon}
\epsilon^2 = \frac{s_{\text{up}}}{\epsilon_{\text{sig}} \times \sigma_{A'} \times 0.1X_0 \times L \times \frac{N_A}{M_W} \times 10^{-36}}
\end{equation}

%\[
%\epsilon^2 = \frac{s_{\text{up}}}{\epsilon_{\text{sig}} \times \sigma_{A'} \times 0.1X_0 \times L \times \frac{N_A}{M_W} \times 10^{-36}}.
%\]

Based on the sum of extrapolated background yields from each rare process, we ultimately obtained 0.015 background events, as described in Section~\ref{sec:SiganlAndBackground}. Additionally, the signal efficiencies were studied in the same Section. Finally, we obtain the 90\% C.L. sensitivity illustrated by the red line in the Figure~\ref{fig:limitsCurve}. Additionally, the projected sensitivity is separately estimated for $3\times10^{14}$EOTs, $9\times10^{14}$EOTs, $1.5\times10^{15}$EOTs and $10^{16}$EOTs, as shown in the colored dashed curves. The results suggest that the proposed analysis strategy above could improve sensitivity by nearly two orders of magnitude over current experiments. 

\begin{figure}[htbp]
    \centering
    \includegraphics[width=8.cm]{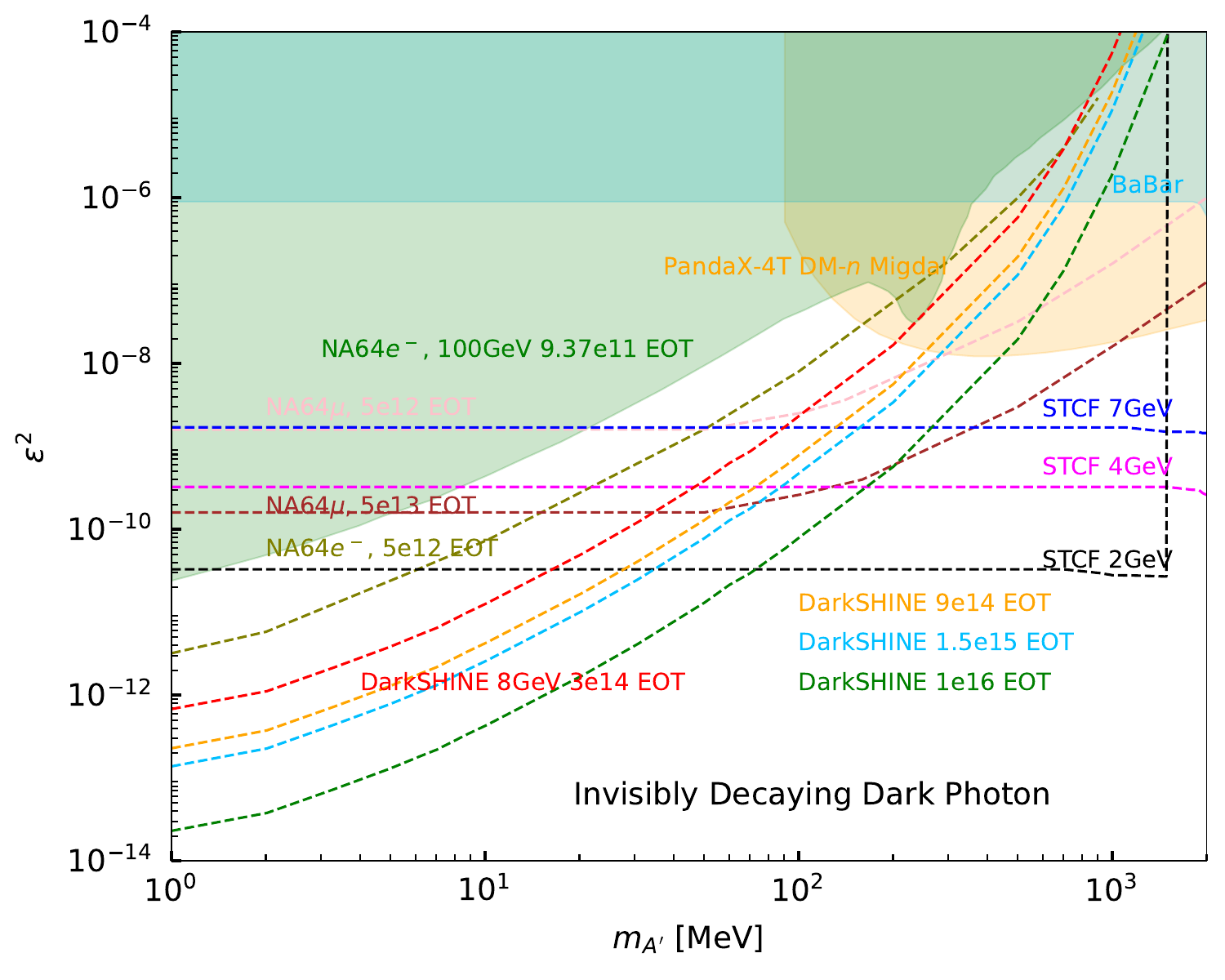}
    \caption{The DarkSHINE expected 90$\%$ C.L. exclusion limits in ($\epsilon^2$, $m_{A'}$) plane, estimated with $3\times10^{14}$ EOTs (red), $9\times10^{14}$EOTs(orange), $1.5 \times 10^{15}$ EOTs (blue), and $10^{16}$ EOTs (green). The existing constraints on $\epsilon^2$ from the NA64~\cite{NA64:2023wbi}, PandX-4T~\cite{PandaX:2023xgl} and BaBar~\cite{BaBar:2017tiz} experiments are shown as a reference. In addition, the STCF sensitivity curves shown in the plot are computed assuming 30 $ab^{-1}$ at $\sqrt{s} = 7$ GeV, $\sqrt{s} = 4$ GeV, and $\sqrt{s} = 2$ GeV, respectively~\cite{Zhang:2019wnz}. The expected curves from the future NA64 experiments are also shown in the plot.}
    \label{fig:limitsCurve}
\end{figure}

To investigate thermal relic dark matter, Figure~\ref{fig:dimensionlessLimitsCurve} presents the projected sensitivity of the dimensionless interaction strength $y=\epsilon^{2}\alpha_{D}(m_{\chi} / m_{A'})^{4}$ as a function of DM mass $m_{\chi}$. This results is assumed that the mass of the dark photon $m_{A'}$ is three times as large as the DM mass $m_{\chi}$ and that the coupling constant $\alpha_{D}$ between the dark photon $A'$ and the DM $\chi$ is equal to 0.5. Three benchmark thermal targets (elastic and inelastic scalar, Majorana fermion, and pseudo-Dirac fermion) are shown as solid lines. The filled regions represent existing and projected experimental constraints~\cite{Andreev:2021fzd,BaBar:2017tiz,deNiverville:2011it,Batell:2009di,Batell:2014mga,MiniBooNE:2017nqe,PandaX:2023xgl,Essig:2012yx}. The results indicate that DarkSHINE’s sensitivity could probe thermal relic dark matter in the MeV range. The result can also be compared with the LDMX R$\&$D experiment~\cite{Akesson:2022vza}. For $3 \times 10^{14}$ EOTs, DarkSHINE has comparable sensitivity with LDMX Phase 1 ($4 \times 10^{14}$ EOTs) but is slightly better at the high-mass range due to the higher incident electron energy. For $1 \times 10^{16}$ EOTs, DarkSHINE demonstrates better sensitivity in the low-mass region, attributed to the superior energy resolution of its LYSO crystals, while LDMX excels in the high-mass region with better spatial resolution from its Si-W calorimeter. Both experiments have similar objectives and will complement each other by providing cross-checks in the discovery phase space.

\begin{figure}[htbp]
    \centering
    \includegraphics[width=8.cm]{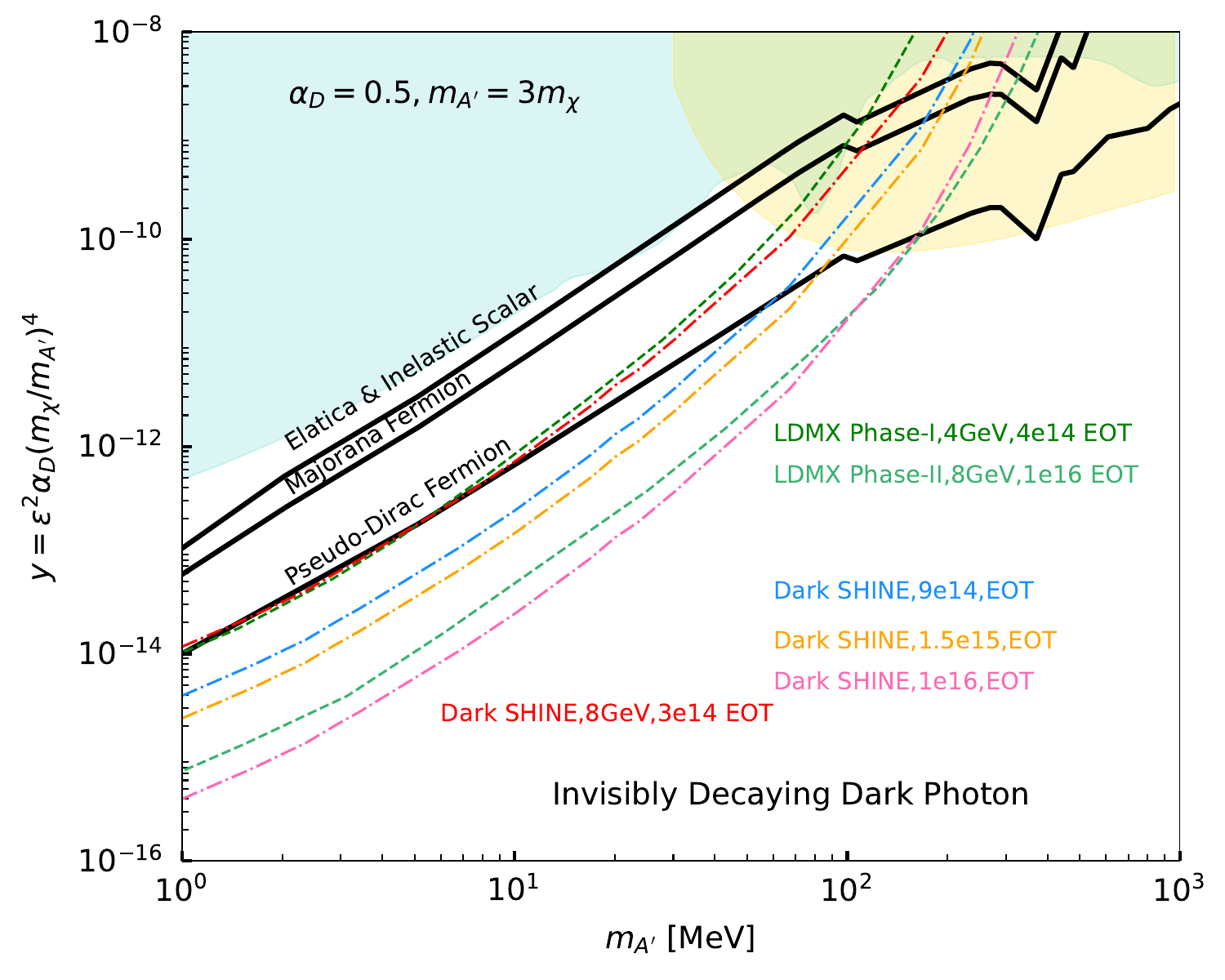}
    \caption{The DarkSHINE expected 90$\%$ C.L. exclusion limits in ($y$, $m_{\chi}$) plane obtained for $m_{A'} = 3m_{\chi}$ and $\alpha_{D} = 0.5$, with $3\times10^{14}$ EOTs (red), $9\times10^{14}$EOTs(blue), $1.5 \times 10^{15}$ EOTs (orange), and $10^{16}$ EOTs (pink).The turquoise region represents existing limits obtained in Ref.~\cite{NA64:2023wbi} from the results of the LSND~\cite{deNiverville:2011it,Batell:2009di},E137~\cite{Batell:2014mga}, BaBar~\cite{BaBar:2017tiz}, MiniBooNE~\cite{MiniBooNE:2017nqe}, COHERENT~\cite{COHERENT:2021pvd} and direct detection~\cite{Essig:2012yx} experiments. The latest result from PandX-4T experiment~\cite{Andreev:2021fzd} is also shown as the origin region in this plot. The projected limit from the LDMX experiment with $4 \times 10^{14} $EOTs with 4 GeV beam energy (Phase 1) and $10^{16}$ EOTs with 8 GeV beam energy (Phase 2) is shown in the plot as the green dashed curves~\cite{Akesson:2022vza}. The favored parameters for the observed relic DM density for the scalar, Majorana, and pseudo-Dirac of light thermal DM are shown as the solid curves~\cite{NA64:2017vtt}. } 
    \label{fig:dimensionlessLimitsCurve}
\end{figure}

\end{chapter}

\begin{chapter}{Conclusion}

In this report, we present our first baseline design for the future DarkSHINE experiment aiming to search for Dark Photon invisible decay signals through dark bremsstrahlung with electron-on-Tungsten-target setup. With the great potential of high repetition rate single electron beam to be deployed and delivered by the SHINE facility, DarkSHINE experiment will provide an unique opportunity to hunt for dark photon mediating the new physics forces to bridge SM particles and BSM DM candidate particles. To maximize the searched sensitivity, this report shows the overall design of the detector system consisting of the silicon strip trackers (two subsystems of tagging tracker + recoil tracker) to measure precisely the incident/recoiled electron trajectories and momenta, the electromagnetic calorimeter designed to reconstruct precisely the full energy deposition of recoiled electrons, the hadronic calorimeter providing extra background vetoing power against muons and neutral hadrons originating from the SM electron-nulcear/photon-nuclear interaction processes. Each sub-detector system are thoroughly presented with global setup integration, geometry description, particle detection response and expected performance, minimal unit/module design and tests, and moreover the readout electronics preliminary designs.
According to the baseline design, the dark photon invisible decay search sensitivities are studied with the corresponding full detector simulation. Signal and background characteristics are carefully simulated and studied so as to conclude a robust estimate. At the end, we estimate the background yield as 0.015 w.r.t. $3\times10^{14}$ electrons on target statitics. Furthermore, a 90\% confidence level exclusion limit is set on the kinetic mixing parameter $\varepsilon^{2}$ as a function of dark photon mass. To show also the potential of such experiment with future upgrade of beamline repetition rate and inclined statistics, different scenarios of $3\times 10^{14}$, $9\times 10^{14}$, $1.5\times 10^{15}$, and $10^{16}$, are also presented, which are referring to one, three, and five years of runs with the corresponding estimated statistics. By comparing with the latest international experiments/future experiments, a promising competitiveness of such experiment is expected in terms of searched sensitivity of dark photon invisible decays.

\end{chapter}

\begin{chapter}{Future Plan}

The baseline design report is meant to elaborate the present design, expected physics sensitivities and moreover have plenty of room to incorporate future works, not only to realize the present design as demonstrated larger scale prototypes but also to carry out further optimizations and more physics potential explorations.

\section{Tracker}
\label{sec:trackerfuture}

The current DarkSHINE simulations and prospective studies are based on the assumption of a single electron interacting with a fixed target, and the detector design has been optimized for this condition. Looking ahead, we plan to explore reconstructions involving multiple electrons interacting with the target, and to optimize the detector for this purpose. In the recoil region, vertex reconstruction can significantly suppress background processes from secondary interactions. The DarkSHINE experiment will also be sensitive to the visible decay signature of dark photons if vertex reconstruction can employed.

\section{ECAL}
\label{sec:ECALFuture}

As the central detector of DarkSHINE, the ECAL must be capable of accurately measuring the energy of recoil electrons and bremsstrahlung photons, while minimizing the leakage of electromagnetic showers into the HCAL. Such leakage could result in the dark photon signal being vetoed by the HCAL and also cause electrons from the inclusive background to deposit less energy in the ECAL, leading to misidentification as signal. To maximize the detector's sensitivity to dark photons while minimizing background, we investigated the causes of energy leakage into HCAL. The primary reason is that particles can pass through gaps between the crystals, penetrating the ECAL and reaching the HCAL. Although this occurrence is rare, it directly interferes with our ability to distinguish the dark photon signal. We can reduce or even eliminate such occurrences by optimizing the geometric arrangement of the ECAL. 

\begin{figure}[h]
\centering
    \subfigure[]{
    \includegraphics[width=0.6\textwidth]{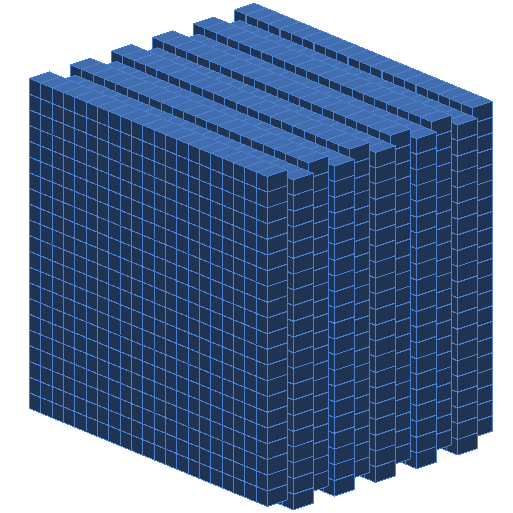}}
    \subfigure[]{
    \includegraphics[width=0.45\textwidth]{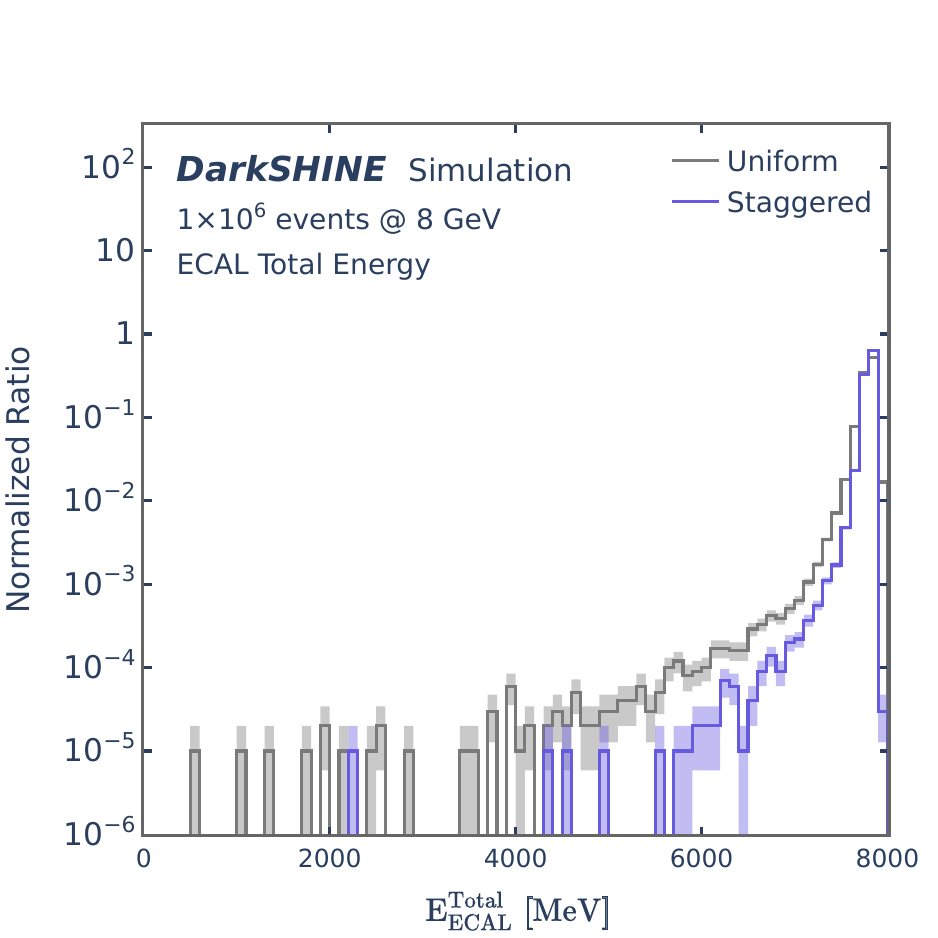}}
    \subfigure[]{
    \includegraphics[width=0.45\textwidth]{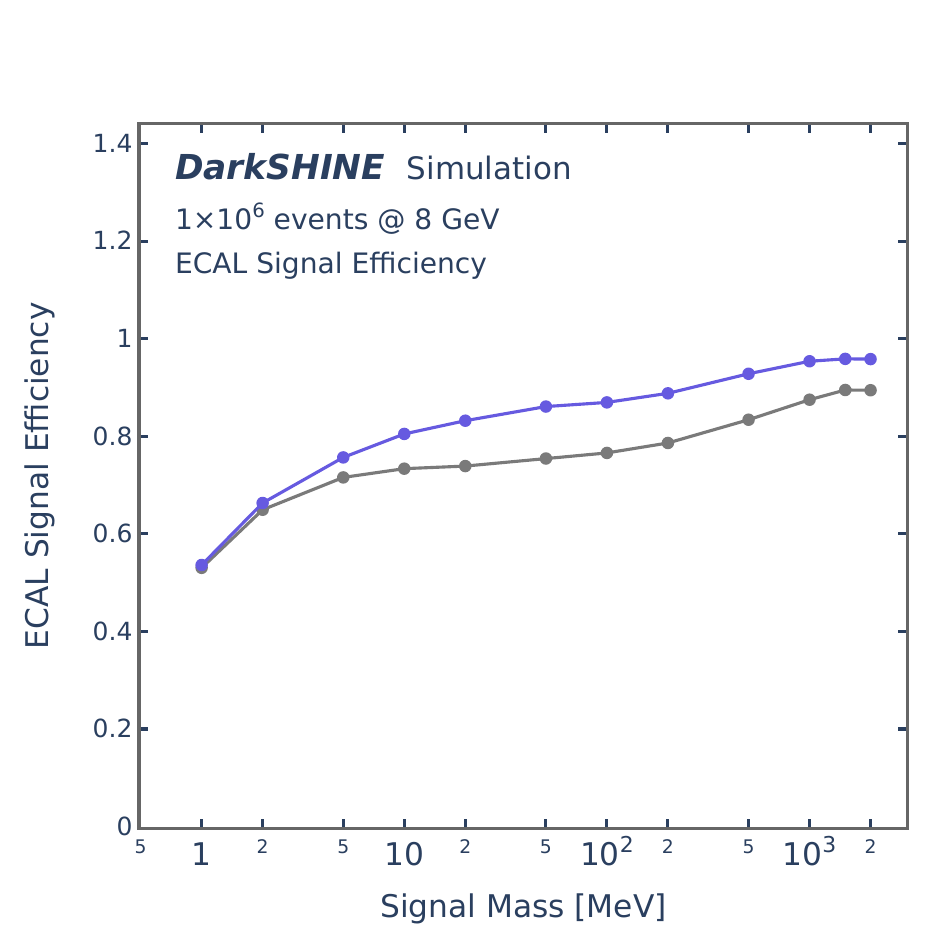}}
\caption{\label{fig:ECALStaggered}~(a) ECAL with staggered structure, where the placement of crystals in successive layers is shifted by half the transverse dimension of a crystal. (b) Deposited energy of 8 GeV electrons in ECAL with uniform layout and staggered layout. (c) Signal efficiencies in of ECAL with  uniform layout and staggered layout.}
\end{figure}

Figure~\ref{fig:ECALStaggered} (a) shows a staggered layout as a potential geometric configuration. In each layer, the crystals are still positioned in a uniform 21$\times$21 square pattern, which is same as the current design. However, to prevent particles from traversing the gaps, the placement of crystals in successive layers is shifted by half the transverse dimension of a crystal. This staggered structure can effectively reduce energy leakage in the ECAL. Figures~\ref{fig:ECALStaggered} (b) and (c) respectively show a comparison of the energy distribution of the inclusive background and the signal efficiency between the current uniform layout of the ECAL and the staggered layout of the ECAL. 

Compared to the uniform ECAL, the staggered ECAL allows fewer background events to enter the signal region of ECAL, where the ECAL energy is less than 2.5 GeV. In Figure~\ref{fig:ECALStaggered} (b), the ratio of background events entering the ECAL signal region to the total number of events is 1 in 100,000 for the staggered ECAL, with the only event being a hard muon event. In contrast, the uniform ECAL exhibits a ratio of 22 per 100,000, which is nore than one order of magnitude higher than the staggered ECAL. Additionally, Figure~\ref{fig:ECALStaggered} (c) shows that the staggered ECAL has a higher signal efficiency.

In the future, to achieve higher sensitivity to dark photons in DarkSHINE experiment, we will conduct further studies on the performance of the ECAL and continue to optimize its geometric structure.

\section{HCAL}
\label{sec:HCALFuture}

Being the heaviest sub-detector on the DarkSHINE detector, the weight of HCAL may be constrained by the loading capacity of the future experimental hall floor and significantly influences the mechanical structure and design of the entire apparatus. Consequently, consider in advance the effect of different sizes on HCAL performance in the case that the size of the whole machine needs to be limited.

A study is addressed to understand the veto inefficiency difference among several size choices from 1 $\times$ 1 m$^2$ to 4 $\times$ 4 m$^2$, as shown in figure~\ref{fig:horizontal_size}. It provides critical insights into the relationship between the HCAL's transverse size and its veto efficiency. The Y-axis represents the veto inefficiency, each curve corresponds to a specific size choice with same interaction length of materials on z-direction, and the x-axis denotes the incident particle energy. Preliminary findings suggest that variations in transverse dimensions significantly affect the HCAL's ability to veto background events effectively.

\iffalse
The HCAL is comprised of layers of plastic scintillators and iron absorbers, with the sensitive layer consisting of plastic scintillator strips. The main HCAL module has a transverse size of 1.5 $\times$ 1.5 m$^2$ and is divided into four horizontal modules measuring 75 $\times$ 75 cm$^2$ each. This design ensures that the length of an individual plastic scintillator is not excessively long. The scintillation strip size on the main hcal is 75 $\times$ 5 $\times$ 1 cm$^3$, and each layer on each module contains 15 scintillators. The scintillator strips in the two adjacent sensitive layers are oriented differently, along the x or y direction, respectively The main HCAL consists of alternating layers of plastic scintillator and iron, with a total length of 249~cm in the z direction. There are 89 layers of plastic scintillator, each with a thickness of 1~cm, followed by 70 layers of iron also measuring 1cm in thickness. The final 18 layers of iron have a thickness of 5~cm. It is worth noting that the total length exceeds ten times the interaction length.
\fi

\begin{figure}[htb]
\centering
\includegraphics
  [width=0.6\hsize]{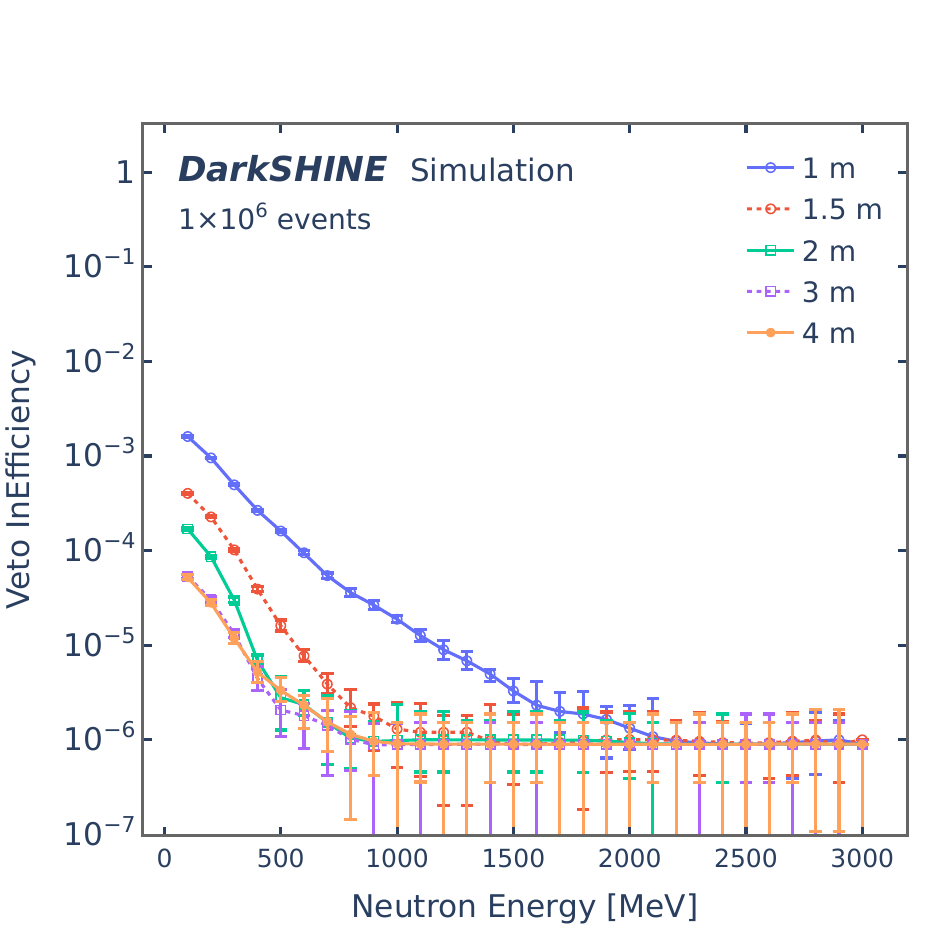}
\caption{Veto inefficiency as a function of different incident neutron energies. Larger size HCAL is showing better veto power as expected due to its capability of acceptance compared with smaller size HCAL designs. }
\label{fig:horizontal_size}
\end{figure}

In conclusion, these designs demonstrate equivalent veto power for high-energy neutrons in the energy range of 2–3~GeV, while the performance of 1~m $\times$ 1~m design deteriorates significantly between 1–2~GeV compared to the design above 1.5~m. For low-energy neutrons, larger area configurations offer enhanced performance. The veto inefficiency of low-energy neutrons in the 1.5~m $\times$ 1.5~m design was already less than 10$^{-3}$, which satisfied the specified requirement. Furthermore, the disparity in the low-energy neutron veto between the 1.5 and 1~m designs was more pronounced than that between the 4 and 1.5~m designs. Given the constraints of the SHINE facility on the supporting structure and weight, a 1.5~m $\times$ 1.5~m design has potential to be selected to ensure sufficient interaction length while minimizing weight. 

The optimization of HCAL geometric parameters could be achieved by considering a smaller size for future designs. The dimensions of the primary HCAL module directly impact veto inefficiency and are closely tied to overall weight. To maintain the same veto power while reducing material usage, the acceptance range should be taken into account. A potential solution is to explore a hybrid geometry where an additional layer of HCAL surrounds the ECAL. Furthermore, optimizing the absorber thickness in the main HCAL can effectively address sensitivity requirements for both high-energy and low-energy target particles. %As illustrated in Fig~\ref{fig:horizontal_size}, 

\section{More Physics Opportunities}
DarkSHINE experiment is presently aiming for Dark Photon invisible decay search, which drives the benchmark design and optimizations. However, the physics potentials can be further explored and enlarged towards the future studies on more diversified physics goals. Selective ideas are summarized as follows:
\begin{itemize}
    \item With the planned further design optimization and reconstruction studies as mentioned in Section~\ref{sec:trackerfuture}, Dark Photon visible decay search would have a better opportunity to pursue and a thorough study on the expected sensitivities will have to be performed.
    \item Given the electron beam delivered by the SHINE facility, a positron beam design is also being considered. Therefore, to explore the expected sensitivity of Dark Photon invisible decay search with positron-on-target experiment setup will be an orthogonal approach to provide complementary searches in addition to the present electron-on-target design.
    \item More exotic phenomena can be explored with the present and future optimized designs: to explore the Axion portal alongside the Dark Photon vector portal; to search for anomalous muonium new physics via the SM background process of $\gamma \to \mu\mu$ in Dark Photon searches; to look into more specific scenarios of Dark Photon phenomena such as time varying massive Dark Photons, other long-lived signatures, etc.
    \item Make use of DarkSHINE experiment setup to measure the photon-nuclear, electron-nuclear, neutrino-nuclear processes in SM interactions, and to test relevant phenomenological models examining potential anomalies beyond SM.
    \item Check and study the impact of repetition rates and beam energies on the searched Dark Photon sensitivities so as to better incorportate the SHINE facility readiness and strategic plannings.
\end{itemize}

\end{chapter}

\chapter*{Acknowledgement}
\addcontentsline{toc}{chapter}{Acknowledgement}
This work is supported by National Key R\&D Program of China (Grant No.: 2023YFA1606904 and 2023YFA1606900), National Natural Science Foundation of China (Grant No.: 12150006), and Shanghai Pilot Program for Basic Research—Shanghai Jiao Tong University (Grant No.: 21TQ1400209). The authors thank to Prof. Xiao-Gang He and Prof. Shao-Feng Ge for the theoretical discussions on the signal modeling, to Prof. Zhi Liu, Prof. Dong Wang, Prof. Weishi Wan, Prof. Dao Xiang and Prof. Meng Zhang for the technical and strategic discussions on the SHINE facility beamline design and scientific prospects. The authors also thank for the support from Key Laboratory for Particle Astrophysics and Cosmology (KLPPAC-MoE), Shanghai Key Laboratory for Particle Physics and Cosmology (SKLPPC).
\bibliographystyle{apsrev4-1}
\bibliography{references}

\begin{thebibliography}{58}%
\makeatletter
\providecommand \@ifxundefined [1]{%
 \@ifx{#1\undefined}
}%
\providecommand \@ifnum [1]{%
 \ifnum #1\expandafter \@firstoftwo
 \else \expandafter \@secondoftwo
 \fi
}%
\providecommand \@ifx [1]{%
 \ifx #1\expandafter \@firstoftwo
 \else \expandafter \@secondoftwo
 \fi
}%
\providecommand \natexlab [1]{#1}%
\providecommand \enquote  [1]{``#1''}%
\providecommand \bibnamefont  [1]{#1}%
\providecommand \bibfnamefont [1]{#1}%
\providecommand \citenamefont [1]{#1}%
\providecommand \href@noop [0]{\@secondoftwo}%
\providecommand \href [0]{\begingroup \@sanitize@url \@href}%
\providecommand \@href[1]{\@@startlink{#1}\@@href}%
\providecommand \@@href[1]{\endgroup#1\@@endlink}%
\providecommand \@sanitize@url [0]{\catcode `\\12\catcode `\$12\catcode `\&12\catcode `\#12\catcode `\^12\catcode `\_12\catcode `\%12\relax}%
\providecommand \@@startlink[1]{}%
\providecommand \@@endlink[0]{}%
\providecommand \url  [0]{\begingroup\@sanitize@url \@url }%
\providecommand \@url [1]{\endgroup\@href {#1}{\urlprefix }}%
\providecommand \urlprefix  [0]{URL }%
\providecommand \Eprint [0]{\href }%
\providecommand \doibase [0]{http://dx.doi.org/}%
\providecommand \selectlanguage [0]{\@gobble}%
\providecommand \bibinfo  [0]{\@secondoftwo}%
\providecommand \bibfield  [0]{\@secondoftwo}%
\providecommand \translation [1]{[#1]}%
\providecommand \BibitemOpen [0]{}%
\providecommand \bibitemStop [0]{}%
\providecommand \bibitemNoStop [0]{.\EOS\space}%
\providecommand \EOS [0]{\spacefactor3000\relax}%
\providecommand \BibitemShut  [1]{\csname bibitem#1\endcsname}%
\let\auto@bib@innerbib\@empty
%</preamble>
\bibitem [{\citenamefont {Zhang}\ \emph {et~al.}(2019{\natexlab{a}})\citenamefont {Zhang}, \citenamefont {Zhang}, \citenamefont {Song}, \citenamefont {Pan}, \citenamefont {Niu},\ and\ \citenamefont {Li}}]{bib:DP-ColSensi}%
  \BibitemOpen
  \bibfield  {author} {\bibinfo {author} {\bibfnamefont {Y.}~\bibnamefont {Zhang}}, \bibinfo {author} {\bibfnamefont {W.-T.}\ \bibnamefont {Zhang}}, \bibinfo {author} {\bibfnamefont {M.}~\bibnamefont {Song}}, \bibinfo {author} {\bibfnamefont {X.-A.}\ \bibnamefont {Pan}}, \bibinfo {author} {\bibfnamefont {Z.-M.}\ \bibnamefont {Niu}}, \ and\ \bibinfo {author} {\bibfnamefont {G.}~\bibnamefont {Li}},\ }\href {\doibase 10.1103/PhysRevD.100.115016} {\bibfield  {journal} {\bibinfo  {journal} {Phys. Rev. D}\ }\textbf {\bibinfo {volume} {100}},\ \bibinfo {pages} {115016} (\bibinfo {year} {2019}{\natexlab{a}})}\BibitemShut {NoStop}%
\bibitem [{\citenamefont {Banerjee}\ \emph {et~al.}(2019)\citenamefont {Banerjee}, \citenamefont {Burtsev}, \citenamefont {Chumakov}, \citenamefont {Cooke}, \citenamefont {Crivelli}, \citenamefont {Depero}, \citenamefont {Dermenev}, \citenamefont {Donskov}, \citenamefont {Dusaev}, \citenamefont {Enik}, \citenamefont {Charitonidis}, \citenamefont {Feshchenko}, \citenamefont {Frolov}, \citenamefont {Gardikiotis}, \citenamefont {Gerassimov}, \citenamefont {Gninenko}, \citenamefont {H\"osgen}, \citenamefont {Jeckel}, \citenamefont {Karneyeu}, \citenamefont {Kekelidze}, \citenamefont {Ketzer}, \citenamefont {Kirpichnikov}, \citenamefont {Kirsanov}, \citenamefont {Konorov}, \citenamefont {Kovalenko}, \citenamefont {Kramarenko}, \citenamefont {Kravchuk}, \citenamefont {Krasnikov}, \citenamefont {Kuleshov}, \citenamefont {Lyubovitskij}, \citenamefont {Lysan}, \citenamefont {Matveev}, \citenamefont {Mikhailov}, \citenamefont {Molina~Bueno}, \citenamefont {Peshekhonov}, \citenamefont {Polyakov}, \citenamefont {Radics},
  \citenamefont {Rojas}, \citenamefont {Rubbia}, \citenamefont {Samoylenko}, \citenamefont {Shchukin}, \citenamefont {Tikhomirov}, \citenamefont {Tlisova}, \citenamefont {Tlisov}, \citenamefont {Toropin}, \citenamefont {Trifonov}, \citenamefont {Vasilishin}, \citenamefont {Vasquez~Arenas}, \citenamefont {Volkov}, \citenamefont {Volkov},\ and\ \citenamefont {Ulloa}}]{bib:NA64}%
  \BibitemOpen
  \bibfield  {author} {\bibinfo {author} {\bibfnamefont {D.}~\bibnamefont {Banerjee}}, \bibinfo {author} {\bibfnamefont {V.~E.}\ \bibnamefont {Burtsev}}, \bibinfo {author} {\bibfnamefont {A.~G.}\ \bibnamefont {Chumakov}}, \bibinfo {author} {\bibfnamefont {D.}~\bibnamefont {Cooke}}, \bibinfo {author} {\bibfnamefont {P.}~\bibnamefont {Crivelli}}, \bibinfo {author} {\bibfnamefont {E.}~\bibnamefont {Depero}}, \bibinfo {author} {\bibfnamefont {A.~V.}\ \bibnamefont {Dermenev}}, \bibinfo {author} {\bibfnamefont {S.~V.}\ \bibnamefont {Donskov}}, \bibinfo {author} {\bibfnamefont {R.~R.}\ \bibnamefont {Dusaev}}, \bibinfo {author} {\bibfnamefont {T.}~\bibnamefont {Enik}}, \bibinfo {author} {\bibfnamefont {N.}~\bibnamefont {Charitonidis}}, \bibinfo {author} {\bibfnamefont {A.}~\bibnamefont {Feshchenko}}, \bibinfo {author} {\bibfnamefont {V.~N.}\ \bibnamefont {Frolov}}, \bibinfo {author} {\bibfnamefont {A.}~\bibnamefont {Gardikiotis}}, \bibinfo {author} {\bibfnamefont {S.~G.}\ \bibnamefont {Gerassimov}}, \bibinfo {author}
  {\bibfnamefont {S.~N.}\ \bibnamefont {Gninenko}}, \bibinfo {author} {\bibfnamefont {M.}~\bibnamefont {H\"osgen}}, \bibinfo {author} {\bibfnamefont {M.}~\bibnamefont {Jeckel}}, \bibinfo {author} {\bibfnamefont {A.~E.}\ \bibnamefont {Karneyeu}}, \bibinfo {author} {\bibfnamefont {G.}~\bibnamefont {Kekelidze}}, \bibinfo {author} {\bibfnamefont {B.}~\bibnamefont {Ketzer}}, \bibinfo {author} {\bibfnamefont {D.~V.}\ \bibnamefont {Kirpichnikov}}, \bibinfo {author} {\bibfnamefont {M.~M.}\ \bibnamefont {Kirsanov}}, \bibinfo {author} {\bibfnamefont {I.~V.}\ \bibnamefont {Konorov}}, \bibinfo {author} {\bibfnamefont {S.~G.}\ \bibnamefont {Kovalenko}}, \bibinfo {author} {\bibfnamefont {V.~A.}\ \bibnamefont {Kramarenko}}, \bibinfo {author} {\bibfnamefont {L.~V.}\ \bibnamefont {Kravchuk}}, \bibinfo {author} {\bibfnamefont {N.~V.}\ \bibnamefont {Krasnikov}}, \bibinfo {author} {\bibfnamefont {S.~V.}\ \bibnamefont {Kuleshov}}, \bibinfo {author} {\bibfnamefont {V.~E.}\ \bibnamefont {Lyubovitskij}}, \bibinfo {author}
  {\bibfnamefont {V.}~\bibnamefont {Lysan}}, \bibinfo {author} {\bibfnamefont {V.~A.}\ \bibnamefont {Matveev}}, \bibinfo {author} {\bibfnamefont {Y.~V.}\ \bibnamefont {Mikhailov}}, \bibinfo {author} {\bibfnamefont {L.}~\bibnamefont {Molina~Bueno}}, \bibinfo {author} {\bibfnamefont {D.~V.}\ \bibnamefont {Peshekhonov}}, \bibinfo {author} {\bibfnamefont {V.~A.}\ \bibnamefont {Polyakov}}, \bibinfo {author} {\bibfnamefont {B.}~\bibnamefont {Radics}}, \bibinfo {author} {\bibfnamefont {R.}~\bibnamefont {Rojas}}, \bibinfo {author} {\bibfnamefont {A.}~\bibnamefont {Rubbia}}, \bibinfo {author} {\bibfnamefont {V.~D.}\ \bibnamefont {Samoylenko}}, \bibinfo {author} {\bibfnamefont {D.}~\bibnamefont {Shchukin}}, \bibinfo {author} {\bibfnamefont {V.~O.}\ \bibnamefont {Tikhomirov}}, \bibinfo {author} {\bibfnamefont {I.}~\bibnamefont {Tlisova}}, \bibinfo {author} {\bibfnamefont {D.~A.}\ \bibnamefont {Tlisov}}, \bibinfo {author} {\bibfnamefont {A.~N.}\ \bibnamefont {Toropin}}, \bibinfo {author} {\bibfnamefont {A.~Y.}\
  \bibnamefont {Trifonov}}, \bibinfo {author} {\bibfnamefont {B.~I.}\ \bibnamefont {Vasilishin}}, \bibinfo {author} {\bibfnamefont {G.}~\bibnamefont {Vasquez~Arenas}}, \bibinfo {author} {\bibfnamefont {P.~V.}\ \bibnamefont {Volkov}}, \bibinfo {author} {\bibfnamefont {V.~Y.}\ \bibnamefont {Volkov}}, \ and\ \bibinfo {author} {\bibfnamefont {P.}~\bibnamefont {Ulloa}} (\bibinfo {collaboration} {NA64 Collaboration}),\ }\href {\doibase 10.1103/PhysRevLett.123.121801} {\bibfield  {journal} {\bibinfo  {journal} {Phys. Rev. Lett.}\ }\textbf {\bibinfo {volume} {123}},\ \bibinfo {pages} {121801} (\bibinfo {year} {2019})}\BibitemShut {NoStop}%
\bibitem [{\citenamefont {Åkesson}\ \emph {et~al.}(2018)\citenamefont {Åkesson}, \citenamefont {Berlin}, \citenamefont {Blinov}, \citenamefont {Colegrove}, \citenamefont {Collura}, \citenamefont {Dutta}, \citenamefont {Echenard}, \citenamefont {Hiltbrand}, \citenamefont {Hitlin}, \citenamefont {Incandela}, \citenamefont {Jaros}, \citenamefont {Johnson}, \citenamefont {Krnjaic}, \citenamefont {Mans}, \citenamefont {Maruyama}, \citenamefont {McCormick}, \citenamefont {Moreno}, \citenamefont {Nelson}, \citenamefont {Niendorf}, \citenamefont {Petersen}, \citenamefont {Pöttgen}, \citenamefont {Schuster}, \citenamefont {Toro}, \citenamefont {Tran},\ and\ \citenamefont {Whitbeck}}]{bib:LDMX}%
  \BibitemOpen
  \bibfield  {author} {\bibinfo {author} {\bibfnamefont {T.}~\bibnamefont {Åkesson}}, \bibinfo {author} {\bibfnamefont {A.}~\bibnamefont {Berlin}}, \bibinfo {author} {\bibfnamefont {N.}~\bibnamefont {Blinov}}, \bibinfo {author} {\bibfnamefont {O.}~\bibnamefont {Colegrove}}, \bibinfo {author} {\bibfnamefont {G.}~\bibnamefont {Collura}}, \bibinfo {author} {\bibfnamefont {V.}~\bibnamefont {Dutta}}, \bibinfo {author} {\bibfnamefont {B.}~\bibnamefont {Echenard}}, \bibinfo {author} {\bibfnamefont {J.}~\bibnamefont {Hiltbrand}}, \bibinfo {author} {\bibfnamefont {D.~G.}\ \bibnamefont {Hitlin}}, \bibinfo {author} {\bibfnamefont {J.}~\bibnamefont {Incandela}}, \bibinfo {author} {\bibfnamefont {J.}~\bibnamefont {Jaros}}, \bibinfo {author} {\bibfnamefont {R.}~\bibnamefont {Johnson}}, \bibinfo {author} {\bibfnamefont {G.}~\bibnamefont {Krnjaic}}, \bibinfo {author} {\bibfnamefont {J.}~\bibnamefont {Mans}}, \bibinfo {author} {\bibfnamefont {T.}~\bibnamefont {Maruyama}}, \bibinfo {author} {\bibfnamefont {J.}~\bibnamefont
  {McCormick}}, \bibinfo {author} {\bibfnamefont {O.}~\bibnamefont {Moreno}}, \bibinfo {author} {\bibfnamefont {T.}~\bibnamefont {Nelson}}, \bibinfo {author} {\bibfnamefont {G.}~\bibnamefont {Niendorf}}, \bibinfo {author} {\bibfnamefont {R.}~\bibnamefont {Petersen}}, \bibinfo {author} {\bibfnamefont {R.}~\bibnamefont {Pöttgen}}, \bibinfo {author} {\bibfnamefont {P.}~\bibnamefont {Schuster}}, \bibinfo {author} {\bibfnamefont {N.}~\bibnamefont {Toro}}, \bibinfo {author} {\bibfnamefont {N.}~\bibnamefont {Tran}}, \ and\ \bibinfo {author} {\bibfnamefont {A.}~\bibnamefont {Whitbeck}},\ }\href {https://arxiv.org/abs/1808.05219} {\enquote {\bibinfo {title} {Light dark matter experiment (ldmx)},}\ } (\bibinfo {year} {2018}),\ \Eprint {http://arxiv.org/abs/1808.05219} {arXiv:1808.05219 [hep-ex]} \BibitemShut {NoStop}%
\bibitem [{\citenamefont {et~al}(2022{\natexlab{a}})}]{bib:DarkQuest}%
  \BibitemOpen
  \bibfield  {author} {\bibinfo {author} {\bibfnamefont {A.~A.}\ \bibnamefont {et~al}} (\bibinfo {collaboration} {DarkQuest}),\ }\href@noop {} {\  (\bibinfo {year} {2022}{\natexlab{a}})},\ \Eprint {http://arxiv.org/abs/2203.08322} {arXiv:2203.08322 [hep-ex]} \BibitemShut {NoStop}%
\bibitem [{\citenamefont {Pachal}()}]{bib:DarkLight}%
  \BibitemOpen
  \bibfield  {author} {\bibinfo {author} {\bibfnamefont {K.}~\bibnamefont {Pachal}},\ }\href@noop {} {}\bibinfo {howpublished} {\url{https://indico-tdli.sjtu.edu.cn/event/1130/contributions/5825/attachments/2480/3765/DarkLight-MEPA-2022.pdf}}\BibitemShut {NoStop}%
\bibitem [{\citenamefont {et~al}(2022{\natexlab{b}})}]{bib:DarkMESA}%
  \BibitemOpen
  \bibfield  {author} {\bibinfo {author} {\bibfnamefont {S.~P.}\ \bibnamefont {et~al}} (\bibinfo {collaboration} {DarkMESA}),\ }\href {\doibase 10.1051/epjconf/202430305006} {\bibfield  {journal} {\bibinfo  {journal} {EPJ Web of Conferences}\ }\textbf {\bibinfo {volume} {303}},\ \bibinfo {pages} {05006} (\bibinfo {year} {2022}{\natexlab{b}})}\BibitemShut {NoStop}%
\bibitem [{\citenamefont {Khaw}()}]{bib:KSKhaw-ExpComp}%
  \BibitemOpen
  \bibfield  {author} {\bibinfo {author} {\bibfnamefont {K.~S.}\ \bibnamefont {Khaw}},\ }\href@noop {} {}\bibinfo {howpublished} {\url{https://indico-tdli.sjtu.edu.cn/event/192/contributions/678/attachments/341/597/DarkPhotonMeeting_20200529.pdf}}\BibitemShut {NoStop}%
\bibitem [{\citenamefont {Andreev}\ \emph {et~al.}(2024)\citenamefont {Andreev} \emph {et~al.}}]{NA64:2024nwj}%
  \BibitemOpen
  \bibfield  {author} {\bibinfo {author} {\bibfnamefont {Y.~M.}\ \bibnamefont {Andreev}} \emph {et~al.} (\bibinfo {collaboration} {NA64}),\ }\href@noop {} {\  (\bibinfo {year} {2024})},\ \Eprint {http://arxiv.org/abs/2409.10128} {arXiv:2409.10128 [hep-ex]} \BibitemShut {NoStop}%
\bibitem [{\citenamefont {Fabbrichesi}\ \emph {et~al.}(2021)\citenamefont {Fabbrichesi}, \citenamefont {Gabrielli},\ and\ \citenamefont {Lanfranchi}}]{Fabbrichesi:2020wbt}%
  \BibitemOpen
  \bibfield  {author} {\bibinfo {author} {\bibfnamefont {M.}~\bibnamefont {Fabbrichesi}}, \bibinfo {author} {\bibfnamefont {E.}~\bibnamefont {Gabrielli}}, \ and\ \bibinfo {author} {\bibfnamefont {G.}~\bibnamefont {Lanfranchi}},\ }\href {\doibase 10.1007/978-3-030-62519-1} {\emph {\bibinfo {title} {The Physics of the Dark Photon: A Primer}}}\ (\bibinfo  {publisher} {Springer International Publishing},\ \bibinfo {year} {2021})\BibitemShut {NoStop}%
\bibitem [{\citenamefont {Z.~Zhao}\ and\ \citenamefont {Yin}(2018)}]{Zhao:2018lcl}%
  \BibitemOpen
  \bibfield  {author} {\bibinfo {author} {\bibfnamefont {Z.~H.~Y.}\ \bibnamefont {Z.~Zhao}, \bibfnamefont {D.~Wang}}\ and\ \bibinfo {author} {\bibfnamefont {L.}~\bibnamefont {Yin}},\ }\href {https://doi.org/10.18429/JACoW-FEL2017-MOP055} {\enquote {\bibinfo {title} {Sclf: An 8-gev cw scrf linac-based x-ray fel facility in shanghai},}\ } (\bibinfo {year} {2018})\BibitemShut {NoStop}%
\bibitem [{\citenamefont {Nosochkov}\ \emph {et~al.}(2017)\citenamefont {Nosochkov}, \citenamefont {Beukers}, \citenamefont {Fry}, \citenamefont {Hast}, \citenamefont {Markiewicz}, \citenamefont {Nelson}, \citenamefont {Phinney}, \citenamefont {Raubenheimer}, \citenamefont {Schuster},\ and\ \citenamefont {Toro}}]{Nosochkov:2017xoc}%
  \BibitemOpen
  \bibfield  {author} {\bibinfo {author} {\bibfnamefont {Y.}~\bibnamefont {Nosochkov}}, \bibinfo {author} {\bibfnamefont {T.}~\bibnamefont {Beukers}}, \bibinfo {author} {\bibfnamefont {A.}~\bibnamefont {Fry}}, \bibinfo {author} {\bibfnamefont {C.}~\bibnamefont {Hast}}, \bibinfo {author} {\bibfnamefont {T.}~\bibnamefont {Markiewicz}}, \bibinfo {author} {\bibfnamefont {T.}~\bibnamefont {Nelson}}, \bibinfo {author} {\bibfnamefont {N.}~\bibnamefont {Phinney}}, \bibinfo {author} {\bibfnamefont {T.}~\bibnamefont {Raubenheimer}}, \bibinfo {author} {\bibfnamefont {P.}~\bibnamefont {Schuster}}, \ and\ \bibinfo {author} {\bibfnamefont {N.}~\bibnamefont {Toro}},\ }in\ \href {\doibase 10.18429/JACoW-IPAC2017-TUPIK121} {\emph {\bibinfo {booktitle} {{8th International Particle Accelerator Conference}}}}\ (\bibinfo {year} {2017})\BibitemShut {NoStop}%
\bibitem [{\citenamefont {Kimble}\ \emph {et~al.}(2002)\citenamefont {Kimble}, \citenamefont {Chou},\ and\ \citenamefont {Chai}}]{1239590}%
  \BibitemOpen
  \bibfield  {author} {\bibinfo {author} {\bibfnamefont {T.}~\bibnamefont {Kimble}}, \bibinfo {author} {\bibfnamefont {M.}~\bibnamefont {Chou}}, \ and\ \bibinfo {author} {\bibfnamefont {B.}~\bibnamefont {Chai}},\ }in\ \href {\doibase 10.1109/NSSMIC.2002.1239590} {\emph {\bibinfo {booktitle} {2002 IEEE Nuclear Science Symposium Conference Record}}},\ Vol.~\bibinfo {volume} {3}\ (\bibinfo {year} {2002})\ pp.\ \bibinfo {pages} {1434--1437 vol.3}\BibitemShut {NoStop}%
\bibitem [{\citenamefont {Liu}\ \emph {et~al.}(2023)\citenamefont {Liu}, \citenamefont {Li}, \citenamefont {Zhang}, \citenamefont {Sun}, \citenamefont {Fan}, \citenamefont {Liang}, \citenamefont {Wang}, \citenamefont {Zhao},\ and\ \citenamefont {Liu}}]{Liu:2023iip}%
  \BibitemOpen
  \bibfield  {author} {\bibinfo {author} {\bibfnamefont {K.}~\bibnamefont {Liu}}, \bibinfo {author} {\bibfnamefont {M.}~\bibnamefont {Li}}, \bibinfo {author} {\bibfnamefont {J.}~\bibnamefont {Zhang}}, \bibinfo {author} {\bibfnamefont {W.}~\bibnamefont {Sun}}, \bibinfo {author} {\bibfnamefont {Y.}~\bibnamefont {Fan}}, \bibinfo {author} {\bibfnamefont {Z.}~\bibnamefont {Liang}}, \bibinfo {author} {\bibfnamefont {Y.}~\bibnamefont {Wang}}, \bibinfo {author} {\bibfnamefont {M.}~\bibnamefont {Zhao}}, \ and\ \bibinfo {author} {\bibfnamefont {K.}~\bibnamefont {Liu}} (\bibinfo {collaboration} {DarkSHINE}),\ }\href@noop {} {\  (\bibinfo {year} {2023})},\ \Eprint {http://arxiv.org/abs/2310.13926} {arXiv:2310.13926 [physics.ins-det]} \BibitemShut {NoStop}%
\bibitem [{\citenamefont {Missio}(2024)}]{Missio:2024hqu}%
  \BibitemOpen
  \bibfield  {author} {\bibinfo {author} {\bibfnamefont {M.}~\bibnamefont {Missio}} (\bibinfo {collaboration} {ATLAS HGTD}),\ }\href {\doibase 10.1088/1748-0221/19/04/C04008} {\bibfield  {journal} {\bibinfo  {journal} {JINST}\ }\textbf {\bibinfo {volume} {19}},\ \bibinfo {pages} {C04008} (\bibinfo {year} {2024})}\BibitemShut {NoStop}%
\bibitem [{\citenamefont {Agapopoulou}\ \emph {et~al.}(2023)\citenamefont {Agapopoulou} \emph {et~al.}}]{Agapopoulou:2023jsd}%
  \BibitemOpen
  \bibfield  {author} {\bibinfo {author} {\bibfnamefont {C.}~\bibnamefont {Agapopoulou}} \emph {et~al.},\ }\href {\doibase 10.1088/1748-0221/18/08/P08019} {\bibfield  {journal} {\bibinfo  {journal} {JINST}\ }\textbf {\bibinfo {volume} {18}},\ \bibinfo {pages} {P08019} (\bibinfo {year} {2023})},\ \Eprint {http://arxiv.org/abs/2306.08949} {arXiv:2306.08949 [physics.ins-det]} \BibitemShut {NoStop}%
\bibitem [{\citenamefont {Workman}\ \emph {et~al.}(2022)\citenamefont {Workman} \emph {et~al.}}]{Workman:2022ynf}%
  \BibitemOpen
  \bibfield  {author} {\bibinfo {author} {\bibfnamefont {R.~L.}\ \bibnamefont {Workman}} \emph {et~al.} (\bibinfo {collaboration} {Particle Data Group}),\ }\href {\doibase 10.1093/ptep/ptac097} {\bibfield  {journal} {\bibinfo  {journal} {PTEP}\ }\textbf {\bibinfo {volume} {2022}},\ \bibinfo {pages} {083C01} (\bibinfo {year} {2022})}\BibitemShut {NoStop}%
\bibitem [{\citenamefont {Izaguirre}\ \emph {et~al.}(2015)\citenamefont {Izaguirre}, \citenamefont {Krnjaic}, \citenamefont {Schuster},\ and\ \citenamefont {Toro}}]{PhysRevD.91.094026}%
  \BibitemOpen
  \bibfield  {author} {\bibinfo {author} {\bibfnamefont {E.}~\bibnamefont {Izaguirre}}, \bibinfo {author} {\bibfnamefont {G.}~\bibnamefont {Krnjaic}}, \bibinfo {author} {\bibfnamefont {P.}~\bibnamefont {Schuster}}, \ and\ \bibinfo {author} {\bibfnamefont {N.}~\bibnamefont {Toro}},\ }\href {\doibase 10.1103/PhysRevD.91.094026} {\bibfield  {journal} {\bibinfo  {journal} {Phys. Rev. D}\ }\textbf {\bibinfo {volume} {91}},\ \bibinfo {pages} {094026} (\bibinfo {year} {2015})},\ \Eprint {http://arxiv.org/abs/1411.1404} {arXiv:1411.1404 [hep-ex]} \BibitemShut {NoStop}%
\bibitem [{\citenamefont {Klanner}(2019)}]{KLANNER201936}%
  \BibitemOpen
  \bibfield  {author} {\bibinfo {author} {\bibfnamefont {R.}~\bibnamefont {Klanner}},\ }\href {\doibase 10.1016/j.nima.2018.11.083} {\bibfield  {journal} {\bibinfo  {journal} {Nucl. Instrum. Meth. A}\ }\textbf {\bibinfo {volume} {926}},\ \bibinfo {pages} {36} (\bibinfo {year} {2019})},\ \Eprint {http://arxiv.org/abs/1809.04346} {arXiv:1809.04346 [physics.ins-det]} \BibitemShut {NoStop}%
\bibitem [{\citenamefont {Simon}(2019)}]{SIMON201985}%
  \BibitemOpen
  \bibfield  {author} {\bibinfo {author} {\bibfnamefont {F.}~\bibnamefont {Simon}},\ }\href {\doibase 10.1016/j.nima.2018.11.042} {\bibfield  {journal} {\bibinfo  {journal} {Nucl. Instrum. Meth. A}\ }\textbf {\bibinfo {volume} {926}},\ \bibinfo {pages} {85} (\bibinfo {year} {2019})},\ \Eprint {http://arxiv.org/abs/1811.03877} {arXiv:1811.03877 [physics.ins-det]} \BibitemShut {NoStop}%
\bibitem [{\citenamefont {Guo}\ \emph {et~al.}(2024)\citenamefont {Guo}, \citenamefont {Li}, \citenamefont {Liu}, \citenamefont {Liu}, \citenamefont {Tan}, \citenamefont {Tang}, \citenamefont {Wu}, \citenamefont {Yang}, \citenamefont {Zhao}, \citenamefont {Zhi},\ and\ \citenamefont {Zhou}}]{guo2024}%
  \BibitemOpen
  \bibfield  {author} {\bibinfo {author} {\bibfnamefont {Y.}~\bibnamefont {Guo}}, \bibinfo {author} {\bibfnamefont {S.}~\bibnamefont {Li}}, \bibinfo {author} {\bibfnamefont {K.}~\bibnamefont {Liu}}, \bibinfo {author} {\bibfnamefont {Y.}~\bibnamefont {Liu}}, \bibinfo {author} {\bibfnamefont {Y.}~\bibnamefont {Tan}}, \bibinfo {author} {\bibfnamefont {J.}~\bibnamefont {Tang}}, \bibinfo {author} {\bibfnamefont {W.}~\bibnamefont {Wu}}, \bibinfo {author} {\bibfnamefont {H.}~\bibnamefont {Yang}}, \bibinfo {author} {\bibfnamefont {Z.}~\bibnamefont {Zhao}}, \bibinfo {author} {\bibfnamefont {W.}~\bibnamefont {Zhi}}, \ and\ \bibinfo {author} {\bibfnamefont {Z.}~\bibnamefont {Zhou}},\ }\href@noop {} {\enquote {\bibinfo {title} {Design of high-speed readout electronics for the darkshine electromagnetic calorimeter},}\ } (\bibinfo {year} {2024}),\ \Eprint {http://arxiv.org/abs/2407.20723} {arXiv:2407.20723 [physics.ins-det]} \BibitemShut {NoStop}%
\bibitem [{\citenamefont {Agostinelli}\ \emph {et~al.}(2003{\natexlab{a}})\citenamefont {Agostinelli} \emph {et~al.}}]{GEANT4:2002zbu}%
  \BibitemOpen
  \bibfield  {author} {\bibinfo {author} {\bibfnamefont {S.}~\bibnamefont {Agostinelli}} \emph {et~al.} (\bibinfo {collaboration} {GEANT4}),\ }\href {\doibase 10.1016/S0168-9002(03)01368-8} {\bibfield  {journal} {\bibinfo  {journal} {Nucl. Instrum. Meth. A}\ }\textbf {\bibinfo {volume} {506}},\ \bibinfo {pages} {250} (\bibinfo {year} {2003}{\natexlab{a}})}\BibitemShut {NoStop}%
\bibitem [{\citenamefont {Chen}\ \emph {et~al.}(2023)\citenamefont {Chen} \emph {et~al.}}]{Chen:2022liu}%
  \BibitemOpen
  \bibfield  {author} {\bibinfo {author} {\bibfnamefont {J.}~\bibnamefont {Chen}} \emph {et~al.},\ }\href {\doibase 10.1007/s11433-022-1983-8} {\bibfield  {journal} {\bibinfo  {journal} {Sci. China Phys. Mech. Astron.}\ }\textbf {\bibinfo {volume} {66}},\ \bibinfo {pages} {211062} (\bibinfo {year} {2023})}\BibitemShut {NoStop}%
\bibitem [{\citenamefont {Zhao}\ \emph {et~al.}(2024{\natexlab{a}})\citenamefont {Zhao}, \citenamefont {Liu}, \citenamefont {Chen}, \citenamefont {Chen}, \citenamefont {Chen}, \citenamefont {Chen}, \citenamefont {Fu}, \citenamefont {Guo}, \citenamefont {Khaw}, \citenamefont {Li}, \citenamefont {Li}, \citenamefont {Liu}, \citenamefont {Liu}, \citenamefont {Song}, \citenamefont {Sun}, \citenamefont {Tang}, \citenamefont {Wang}, \citenamefont {Wang}, \citenamefont {Wu}, \citenamefont {Yang}, \citenamefont {Lin}, \citenamefont {Yuan}, \citenamefont {Zhang}, \citenamefont {Zhang}, \citenamefont {Zhou}, \citenamefont {Zhu},\ and\ \citenamefont {Zhu}}]{zhao2024DSECAL}%
  \BibitemOpen
  \bibfield  {author} {\bibinfo {author} {\bibfnamefont {Z.}~\bibnamefont {Zhao}}, \bibinfo {author} {\bibfnamefont {Q.}~\bibnamefont {Liu}}, \bibinfo {author} {\bibfnamefont {J.}~\bibnamefont {Chen}}, \bibinfo {author} {\bibfnamefont {J.}~\bibnamefont {Chen}}, \bibinfo {author} {\bibfnamefont {J.}~\bibnamefont {Chen}}, \bibinfo {author} {\bibfnamefont {X.}~\bibnamefont {Chen}}, \bibinfo {author} {\bibfnamefont {C.}~\bibnamefont {Fu}}, \bibinfo {author} {\bibfnamefont {J.}~\bibnamefont {Guo}}, \bibinfo {author} {\bibfnamefont {K.~S.}\ \bibnamefont {Khaw}}, \bibinfo {author} {\bibfnamefont {L.}~\bibnamefont {Li}}, \bibinfo {author} {\bibfnamefont {S.}~\bibnamefont {Li}}, \bibinfo {author} {\bibfnamefont {D.}~\bibnamefont {Liu}}, \bibinfo {author} {\bibfnamefont {K.}~\bibnamefont {Liu}}, \bibinfo {author} {\bibfnamefont {S.}~\bibnamefont {Song}}, \bibinfo {author} {\bibfnamefont {T.}~\bibnamefont {Sun}}, \bibinfo {author} {\bibfnamefont {J.}~\bibnamefont {Tang}}, \bibinfo {author} {\bibfnamefont
  {Y.}~\bibnamefont {Wang}}, \bibinfo {author} {\bibfnamefont {Z.}~\bibnamefont {Wang}}, \bibinfo {author} {\bibfnamefont {W.}~\bibnamefont {Wu}}, \bibinfo {author} {\bibfnamefont {H.}~\bibnamefont {Yang}}, \bibinfo {author} {\bibfnamefont {Y.}~\bibnamefont {Lin}}, \bibinfo {author} {\bibfnamefont {R.}~\bibnamefont {Yuan}}, \bibinfo {author} {\bibfnamefont {Y.}~\bibnamefont {Zhang}}, \bibinfo {author} {\bibfnamefont {Y.}~\bibnamefont {Zhang}}, \bibinfo {author} {\bibfnamefont {B.}~\bibnamefont {Zhou}}, \bibinfo {author} {\bibfnamefont {X.}~\bibnamefont {Zhu}}, \ and\ \bibinfo {author} {\bibfnamefont {Y.}~\bibnamefont {Zhu}},\ }\href {https://arxiv.org/abs/2407.17800} {\enquote {\bibinfo {title} {Design of a lyso crystal electromagnetic calorimeter for darkshine experiment},}\ } (\bibinfo {year} {2024}{\natexlab{a}}),\ \Eprint {http://arxiv.org/abs/2407.17800} {arXiv:2407.17800 [physics.ins-det]} \BibitemShut {NoStop}%
\bibitem [{\citenamefont {Zhu}\ \emph {et~al.}(2019)\citenamefont {Zhu} \emph {et~al.}}]{Zhu:2019ihr}%
  \BibitemOpen
  \bibfield  {author} {\bibinfo {author} {\bibfnamefont {R.-Y.}\ \bibnamefont {Zhu}} \emph {et~al.},\ }\href {\doibase 10.1088/1742-6596/1162/1/012022} {\bibfield  {journal} {\bibinfo  {journal} {J. Phys. Conf. Ser.}\ }\textbf {\bibinfo {volume} {1162}},\ \bibinfo {pages} {012022} (\bibinfo {year} {2019})}\BibitemShut {NoStop}%
\bibitem [{\citenamefont {Ulyanov}\ \emph {et~al.}(2020)\citenamefont {Ulyanov}, \citenamefont {Murphy}, \citenamefont {Mangan}, \citenamefont {Gupta}, \citenamefont {Hajdas}, \citenamefont {De~Faoite}, \citenamefont {Shortt}, \citenamefont {Hanlon},\ and\ \citenamefont {Mcbreen}}]{ULYANOV2020164203}%
  \BibitemOpen
  \bibfield  {author} {\bibinfo {author} {\bibfnamefont {A.}~\bibnamefont {Ulyanov}}, \bibinfo {author} {\bibfnamefont {D.}~\bibnamefont {Murphy}}, \bibinfo {author} {\bibfnamefont {J.}~\bibnamefont {Mangan}}, \bibinfo {author} {\bibfnamefont {V.}~\bibnamefont {Gupta}}, \bibinfo {author} {\bibfnamefont {W.}~\bibnamefont {Hajdas}}, \bibinfo {author} {\bibfnamefont {D.}~\bibnamefont {De~Faoite}}, \bibinfo {author} {\bibfnamefont {B.}~\bibnamefont {Shortt}}, \bibinfo {author} {\bibfnamefont {L.}~\bibnamefont {Hanlon}}, \ and\ \bibinfo {author} {\bibfnamefont {S.}~\bibnamefont {Mcbreen}},\ }\href {\doibase 10.1016/j.nima.2020.164203} {\  (\bibinfo {year} {2020}),\ 10.1016/j.nima.2020.164203},\ \Eprint {http://arxiv.org/abs/2007.10919} {arXiv:2007.10919 [physics.ins-det]} \BibitemShut {NoStop}%
\bibitem [{\citenamefont {Sanchez~Majos}\ \emph {et~al.}(2009)\citenamefont {Sanchez~Majos} \emph {et~al.}}]{SANCHEZMAJOS2009506}%
  \BibitemOpen
  \bibfield  {author} {\bibinfo {author} {\bibfnamefont {S.}~\bibnamefont {Sanchez~Majos}} \emph {et~al.},\ }\href {\doibase 10.1016/j.nima.2009.01.176} {\bibfield  {journal} {\bibinfo  {journal} {Nucl. Instrum. Meth. A}\ }\textbf {\bibinfo {volume} {602}},\ \bibinfo {pages} {506} (\bibinfo {year} {2009})}\BibitemShut {NoStop}%
\bibitem [{\citenamefont {Preghenella}\ \emph {et~al.}(2023)\citenamefont {Preghenella} \emph {et~al.}}]{PREGHENELLA2023167661}%
  \BibitemOpen
  \bibfield  {author} {\bibinfo {author} {\bibfnamefont {R.}~\bibnamefont {Preghenella}} \emph {et~al.},\ }\href {\doibase https://doi.org/10.1016/j.nima.2022.167661} {\bibfield  {journal} {\bibinfo  {journal} {Nuclear Instruments and Methods in Physics Research Section A: Accelerators, Spectrometers, Detectors and Associated Equipment}\ }\textbf {\bibinfo {volume} {1046}},\ \bibinfo {pages} {167661} (\bibinfo {year} {2023})}\BibitemShut {NoStop}%
\bibitem [{\citenamefont {HAHAMATSU}({\natexlab{a}})}]{S14160-3010PS}%
  \BibitemOpen
  \bibfield  {author} {\bibinfo {author} {\bibnamefont {HAHAMATSU}},\ }\href@noop {} {\enquote {\bibinfo {title} {S14160-3010ps},}\ }\bibinfo {howpublished} {\url{https://www.hamamatsu.com.cn/cn/zh-cn/product/optical-sensors/mppc/mppc_mppc-array/S14160-3010PS.html}} ({\natexlab{a}})\BibitemShut {NoStop}%
\bibitem [{\citenamefont {Laboratory}()}]{EQR06}%
  \BibitemOpen
  \bibfield  {author} {\bibinfo {author} {\bibfnamefont {N.~D.}\ \bibnamefont {Laboratory}},\ }\href@noop {} {\enquote {\bibinfo {title} {Eqr06 11-3030d-s},}\ }\bibinfo {howpublished} {\url{http://www.ndl-sipm.net/PDF/Datasheet-EQR06.pdf}}\BibitemShut {NoStop}%
\bibitem [{\citenamefont {HAHAMATSU}({\natexlab{b}})}]{S13360-6025PE}%
  \BibitemOpen
  \bibfield  {author} {\bibinfo {author} {\bibnamefont {HAHAMATSU}},\ }\href@noop {} {\enquote {\bibinfo {title} {S13360-6025pe},}\ }\bibinfo {howpublished} {\url{https://www.hamamatsu.com.cn/cn/zh-cn/product/optical-sensors/mppc/mppc_mppc-array/S13360-6025PE.html}} ({\natexlab{b}})\BibitemShut {NoStop}%
\bibitem [{\citenamefont {Zhao}\ \emph {et~al.}(2024{\natexlab{b}})\citenamefont {Zhao}, \citenamefont {Qi}, \citenamefont {Li},\ and\ \citenamefont {Liu}}]{zhao2024dynamic}%
  \BibitemOpen
  \bibfield  {author} {\bibinfo {author} {\bibfnamefont {Z.}~\bibnamefont {Zhao}}, \bibinfo {author} {\bibfnamefont {B.}~\bibnamefont {Qi}}, \bibinfo {author} {\bibfnamefont {S.}~\bibnamefont {Li}}, \ and\ \bibinfo {author} {\bibfnamefont {Y.}~\bibnamefont {Liu}},\ }\href {https://arxiv.org/abs/2407.17794} {\enquote {\bibinfo {title} {Dynamic range of sipms with high pixel densities},}\ } (\bibinfo {year} {2024}{\natexlab{b}}),\ \Eprint {http://arxiv.org/abs/2407.17794} {arXiv:2407.17794 [physics.ins-det]} \BibitemShut {NoStop}%
\bibitem [{\citenamefont {Tsang}\ \emph {et~al.}(2016)\citenamefont {Tsang}, \citenamefont {Rao}, \citenamefont {Stoll},\ and\ \citenamefont {Woody}}]{Tsang:2016cmc}%
  \BibitemOpen
  \bibfield  {author} {\bibinfo {author} {\bibfnamefont {T.}~\bibnamefont {Tsang}}, \bibinfo {author} {\bibfnamefont {T.}~\bibnamefont {Rao}}, \bibinfo {author} {\bibfnamefont {S.}~\bibnamefont {Stoll}}, \ and\ \bibinfo {author} {\bibfnamefont {C.}~\bibnamefont {Woody}},\ }\href {\doibase 10.1088/1748-0221/11/12/P12002} {\bibfield  {journal} {\bibinfo  {journal} {JINST}\ }\textbf {\bibinfo {volume} {11}},\ \bibinfo {pages} {P12002} (\bibinfo {year} {2016})}\BibitemShut {NoStop}%
\bibitem [{\citenamefont {Cordelli}\ \emph {et~al.}(2021)\citenamefont {Cordelli}, \citenamefont {Diociaiuti}, \citenamefont {Ferrari}, \citenamefont {Miscetti}, \citenamefont {M\"uller}, \citenamefont {Pezzullo},\ and\ \citenamefont {Sarra}}]{Cordelli:2018kgh}%
  \BibitemOpen
  \bibfield  {author} {\bibinfo {author} {\bibfnamefont {M.}~\bibnamefont {Cordelli}}, \bibinfo {author} {\bibfnamefont {E.}~\bibnamefont {Diociaiuti}}, \bibinfo {author} {\bibfnamefont {A.}~\bibnamefont {Ferrari}}, \bibinfo {author} {\bibfnamefont {S.}~\bibnamefont {Miscetti}}, \bibinfo {author} {\bibfnamefont {S.}~\bibnamefont {M\"uller}}, \bibinfo {author} {\bibfnamefont {G.}~\bibnamefont {Pezzullo}}, \ and\ \bibinfo {author} {\bibfnamefont {I.}~\bibnamefont {Sarra}},\ }\href {\doibase 10.1088/1748-0221/16/12/T12012} {\bibfield  {journal} {\bibinfo  {journal} {JINST}\ }\textbf {\bibinfo {volume} {16}},\ \bibinfo {pages} {T12012} (\bibinfo {year} {2021})},\ \Eprint {http://arxiv.org/abs/1804.09792} {arXiv:1804.09792 [physics.ins-det]} \BibitemShut {NoStop}%
\bibitem [{\citenamefont {Gu}\ \emph {et~al.}(2023)\citenamefont {Gu}, \citenamefont {Liu}, \citenamefont {Sun}, \citenamefont {Xu}, \citenamefont {Zhang}, \citenamefont {An}, \citenamefont {Gong}, \citenamefont {Li}, \citenamefont {Wen}, \citenamefont {Xiong}, \citenamefont {Zhang}, \citenamefont {Wang},\ and\ \citenamefont {Qu}}]{GU2023168381}%
  \BibitemOpen
  \bibfield  {author} {\bibinfo {author} {\bibfnamefont {F.}~\bibnamefont {Gu}}, \bibinfo {author} {\bibfnamefont {Y.}~\bibnamefont {Liu}}, \bibinfo {author} {\bibfnamefont {X.}~\bibnamefont {Sun}}, \bibinfo {author} {\bibfnamefont {Y.}~\bibnamefont {Xu}}, \bibinfo {author} {\bibfnamefont {D.}~\bibnamefont {Zhang}}, \bibinfo {author} {\bibfnamefont {Z.}~\bibnamefont {An}}, \bibinfo {author} {\bibfnamefont {K.}~\bibnamefont {Gong}}, \bibinfo {author} {\bibfnamefont {X.}~\bibnamefont {Li}}, \bibinfo {author} {\bibfnamefont {X.}~\bibnamefont {Wen}}, \bibinfo {author} {\bibfnamefont {S.}~\bibnamefont {Xiong}}, \bibinfo {author} {\bibfnamefont {F.}~\bibnamefont {Zhang}}, \bibinfo {author} {\bibfnamefont {C.}~\bibnamefont {Wang}}, \ and\ \bibinfo {author} {\bibfnamefont {G.}~\bibnamefont {Qu}},\ }\href {\doibase https://doi.org/10.1016/j.nima.2023.168381} {\bibfield  {journal} {\bibinfo  {journal} {Nuclear Instruments and Methods in Physics Research Section A: Accelerators, Spectrometers, Detectors and Associated
  Equipment}\ }\textbf {\bibinfo {volume} {1053}},\ \bibinfo {pages} {168381} (\bibinfo {year} {2023})}\BibitemShut {NoStop}%
\bibitem [{\citenamefont {Kalinnikov}\ \emph {et~al.}(2023)\citenamefont {Kalinnikov}, \citenamefont {Velicheva},\ and\ \citenamefont {Rozhdestvensky}}]{Kalinnikov:2023coj}%
  \BibitemOpen
  \bibfield  {author} {\bibinfo {author} {\bibfnamefont {V.}~\bibnamefont {Kalinnikov}}, \bibinfo {author} {\bibfnamefont {E.}~\bibnamefont {Velicheva}}, \ and\ \bibinfo {author} {\bibfnamefont {A.}~\bibnamefont {Rozhdestvensky}},\ }\href {\doibase 10.1134/S1547477123050412} {\bibfield  {journal} {\bibinfo  {journal} {Phys. Part. Nucl. Lett.}\ }\textbf {\bibinfo {volume} {20}},\ \bibinfo {pages} {995} (\bibinfo {year} {2023})}\BibitemShut {NoStop}%
\bibitem [{\citenamefont {H.}\ \emph {et~al.}(2018)\citenamefont {H.} \emph {et~al.}}]{LYSOIR}%
  \BibitemOpen
  \bibfield  {author} {\bibinfo {author} {\bibfnamefont {A.-S.}\ \bibnamefont {H.}} \emph {et~al.},\ }\href {\doibase 10.1016/S0168-9002(03)01368-8} {\bibfield  {journal} {\bibinfo  {journal} {Sci Rep.}\ }\textbf {\bibinfo {volume} {8(1)}},\ \bibinfo {pages} {17310} (\bibinfo {year} {2018})}\BibitemShut {NoStop}%
\bibitem [{\citenamefont {He}\ \emph {et~al.}(2021)\citenamefont {He} \emph {et~al.}}]{He:2021sbc}%
  \BibitemOpen
  \bibfield  {author} {\bibinfo {author} {\bibfnamefont {C.}~\bibnamefont {He}} \emph {et~al.},\ }\href {\doibase 10.1088/1748-0221/16/12/T12015} {\bibfield  {journal} {\bibinfo  {journal} {JINST}\ }\textbf {\bibinfo {volume} {16}},\ \bibinfo {pages} {T12015} (\bibinfo {year} {2021})},\ \Eprint {http://arxiv.org/abs/2108.11804} {arXiv:2108.11804 [physics.ins-det]} \BibitemShut {NoStop}%
\bibitem [{\citenamefont {Luo}\ \emph {et~al.}(2023)\citenamefont {Luo} \emph {et~al.}}]{Luo:2023inu}%
  \BibitemOpen
  \bibfield  {author} {\bibinfo {author} {\bibfnamefont {G.}~\bibnamefont {Luo}} \emph {et~al.},\ }\href {\doibase 10.1007/s41365-023-01263-7} {\bibfield  {journal} {\bibinfo  {journal} {Nucl. Sci. Tech.}\ }\textbf {\bibinfo {volume} {34}},\ \bibinfo {pages} {99} (\bibinfo {year} {2023})},\ \Eprint {http://arxiv.org/abs/2302.12669} {arXiv:2302.12669 [physics.ins-det]} \BibitemShut {NoStop}%
\bibitem [{\citenamefont {HAHAMATSU}({\natexlab{c}})}]{S14160-3015PS}%
  \BibitemOpen
  \bibfield  {author} {\bibinfo {author} {\bibnamefont {HAHAMATSU}},\ }\href@noop {} {\enquote {\bibinfo {title} {S14160-3015ps},}\ }\bibinfo {howpublished} {\url{https://www.hamamatsu.com.cn/cn/zh-cn/product/optical-sensors/mppc/mppc_mppc-array/S14160-3015PS.html}} ({\natexlab{c}})\BibitemShut {NoStop}%
\bibitem [{\citenamefont {HAHAMATSU}({\natexlab{d}})}]{S13360-3025CS}%
  \BibitemOpen
  \bibfield  {author} {\bibinfo {author} {\bibnamefont {HAHAMATSU}},\ }\href@noop {} {\enquote {\bibinfo {title} {S13360-3025cs},}\ }\bibinfo {howpublished} {\url{https://www.hamamatsu.com.cn/cn/zh-cn/product/optical-sensors/mppc/mppc_mppc-array/S13360-3025CS.html}} ({\natexlab{d}})\BibitemShut {NoStop}%
\bibitem [{\citenamefont {HAHAMATSU}({\natexlab{e}})}]{S13360-3050CS}%
  \BibitemOpen
  \bibfield  {author} {\bibinfo {author} {\bibnamefont {HAHAMATSU}},\ }\href@noop {} {\enquote {\bibinfo {title} {S13360-3050cs},}\ }\bibinfo {howpublished} {\url{https://www.hamamatsu.com.cn/cn/zh-cn/product/optical-sensors/mppc/mppc_mppc-array/S13360-3050CS.html}} ({\natexlab{e}})\BibitemShut {NoStop}%
\bibitem [{\citenamefont {Brun}\ \emph {et~al.}(2019)\citenamefont {Brun}, \citenamefont {Rademakers}, \citenamefont {Canal}, \citenamefont {Naumann}, \citenamefont {Couet}, \citenamefont {Moneta}, \citenamefont {Vassilev}, \citenamefont {Linev}, \citenamefont {Piparo}, \citenamefont {GANIS}, \citenamefont {Bellenot}, \citenamefont {Guiraud}, \citenamefont {Amadio}, \citenamefont {wverkerke}, \citenamefont {Mato}, \citenamefont {TimurP}, \citenamefont {Tadel}, \citenamefont {wlav}, \citenamefont {Tejedor}, \citenamefont {Blomer}, \citenamefont {Gheata}, \citenamefont {Hageboeck}, \citenamefont {Roiser}, \citenamefont {marsupial}, \citenamefont {Wunsch}, \citenamefont {Shadura}, \citenamefont {Bose}, \citenamefont {CristinaCristescu}, \citenamefont {Valls},\ and\ \citenamefont {Isemann}}]{rene_brun_2019_3895860}%
  \BibitemOpen
  \bibfield  {author} {\bibinfo {author} {\bibfnamefont {R.}~\bibnamefont {Brun}}, \bibinfo {author} {\bibfnamefont {F.}~\bibnamefont {Rademakers}}, \bibinfo {author} {\bibfnamefont {P.}~\bibnamefont {Canal}}, \bibinfo {author} {\bibfnamefont {A.}~\bibnamefont {Naumann}}, \bibinfo {author} {\bibfnamefont {O.}~\bibnamefont {Couet}}, \bibinfo {author} {\bibfnamefont {L.}~\bibnamefont {Moneta}}, \bibinfo {author} {\bibfnamefont {V.}~\bibnamefont {Vassilev}}, \bibinfo {author} {\bibfnamefont {S.}~\bibnamefont {Linev}}, \bibinfo {author} {\bibfnamefont {D.}~\bibnamefont {Piparo}}, \bibinfo {author} {\bibfnamefont {G.}~\bibnamefont {GANIS}}, \bibinfo {author} {\bibfnamefont {B.}~\bibnamefont {Bellenot}}, \bibinfo {author} {\bibfnamefont {E.}~\bibnamefont {Guiraud}}, \bibinfo {author} {\bibfnamefont {G.}~\bibnamefont {Amadio}}, \bibinfo {author} {\bibnamefont {wverkerke}}, \bibinfo {author} {\bibfnamefont {P.}~\bibnamefont {Mato}}, \bibinfo {author} {\bibnamefont {TimurP}}, \bibinfo {author} {\bibfnamefont
  {M.}~\bibnamefont {Tadel}}, \bibinfo {author} {\bibnamefont {wlav}}, \bibinfo {author} {\bibfnamefont {E.}~\bibnamefont {Tejedor}}, \bibinfo {author} {\bibfnamefont {J.}~\bibnamefont {Blomer}}, \bibinfo {author} {\bibfnamefont {A.}~\bibnamefont {Gheata}}, \bibinfo {author} {\bibfnamefont {S.}~\bibnamefont {Hageboeck}}, \bibinfo {author} {\bibfnamefont {S.}~\bibnamefont {Roiser}}, \bibinfo {author} {\bibnamefont {marsupial}}, \bibinfo {author} {\bibfnamefont {S.}~\bibnamefont {Wunsch}}, \bibinfo {author} {\bibfnamefont {O.}~\bibnamefont {Shadura}}, \bibinfo {author} {\bibfnamefont {A.}~\bibnamefont {Bose}}, \bibinfo {author} {\bibnamefont {CristinaCristescu}}, \bibinfo {author} {\bibfnamefont {X.}~\bibnamefont {Valls}}, \ and\ \bibinfo {author} {\bibfnamefont {R.}~\bibnamefont {Isemann}},\ }\href {\doibase 10.5281/zenodo.3895860} {\enquote {\bibinfo {title} {root-project/root: v6.18/02},}\ } (\bibinfo {year} {2019})\BibitemShut {NoStop}%
\bibitem [{\citenamefont {Agostinelli}\ \emph {et~al.}(2003{\natexlab{b}})\citenamefont {Agostinelli} \emph {et~al.}}]{Agostinelli:2002hh}%
  \BibitemOpen
  \bibfield  {author} {\bibinfo {author} {\bibfnamefont {S.}~\bibnamefont {Agostinelli}} \emph {et~al.} (\bibinfo {collaboration} {GEANT4}),\ }\href {\doibase 10.1016/S0168-9002(03)01368-8} {\bibfield  {journal} {\bibinfo  {journal} {Nucl. Instrum. Meth. A}\ }\textbf {\bibinfo {volume} {506}},\ \bibinfo {pages} {250} (\bibinfo {year} {2003}{\natexlab{b}})}\BibitemShut {NoStop}%
\bibitem [{\citenamefont {Belyaev}\ \emph {et~al.}(2013{\natexlab{a}})\citenamefont {Belyaev}, \citenamefont {Christensen},\ and\ \citenamefont {Pukhov}}]{Belyaev:2012qa}%
  \BibitemOpen
  \bibfield  {author} {\bibinfo {author} {\bibfnamefont {A.}~\bibnamefont {Belyaev}}, \bibinfo {author} {\bibfnamefont {N.~D.}\ \bibnamefont {Christensen}}, \ and\ \bibinfo {author} {\bibfnamefont {A.}~\bibnamefont {Pukhov}},\ }\href {\doibase 10.1016/j.cpc.2013.01.014} {\bibfield  {journal} {\bibinfo  {journal} {Comput. Phys. Commun.}\ }\textbf {\bibinfo {volume} {184}},\ \bibinfo {pages} {1729} (\bibinfo {year} {2013}{\natexlab{a}})},\ \Eprint {http://arxiv.org/abs/1207.6082} {arXiv:1207.6082 [hep-ph]} \BibitemShut {NoStop}%
\bibitem [{\citenamefont {Belyaev}\ \emph {et~al.}(2013{\natexlab{b}})\citenamefont {Belyaev}, \citenamefont {Christensen},\ and\ \citenamefont {Pukhov}}]{Belyaev_2013}%
  \BibitemOpen
  \bibfield  {author} {\bibinfo {author} {\bibfnamefont {A.}~\bibnamefont {Belyaev}}, \bibinfo {author} {\bibfnamefont {N.~D.}\ \bibnamefont {Christensen}}, \ and\ \bibinfo {author} {\bibfnamefont {A.}~\bibnamefont {Pukhov}},\ }\href {\doibase 10.1016/j.cpc.2013.01.014} {\bibfield  {journal} {\bibinfo  {journal} {Computer Physics Communications}\ }\textbf {\bibinfo {volume} {184}},\ \bibinfo {pages} {1729–1769} (\bibinfo {year} {2013}{\natexlab{b}})}\BibitemShut {NoStop}%
\bibitem [{\citenamefont {Andreev}\ \emph {et~al.}(2023)\citenamefont {Andreev} \emph {et~al.}}]{NA64:2023wbi}%
  \BibitemOpen
  \bibfield  {author} {\bibinfo {author} {\bibfnamefont {Y.~M.}\ \bibnamefont {Andreev}} \emph {et~al.} (\bibinfo {collaboration} {NA64}),\ }\href {\doibase 10.1103/PhysRevLett.131.161801} {\bibfield  {journal} {\bibinfo  {journal} {Phys. Rev. Lett.}\ }\textbf {\bibinfo {volume} {131}},\ \bibinfo {pages} {161801} (\bibinfo {year} {2023})},\ \Eprint {http://arxiv.org/abs/2307.02404} {arXiv:2307.02404 [hep-ex]} \BibitemShut {NoStop}%
\bibitem [{\citenamefont {Huang}\ \emph {et~al.}(2023)\citenamefont {Huang} \emph {et~al.}}]{PandaX:2023xgl}%
  \BibitemOpen
  \bibfield  {author} {\bibinfo {author} {\bibfnamefont {D.}~\bibnamefont {Huang}} \emph {et~al.} (\bibinfo {collaboration} {PandaX}),\ }\href {\doibase 10.1103/PhysRevLett.131.191002} {\bibfield  {journal} {\bibinfo  {journal} {Phys. Rev. Lett.}\ }\textbf {\bibinfo {volume} {131}},\ \bibinfo {pages} {191002} (\bibinfo {year} {2023})},\ \Eprint {http://arxiv.org/abs/2308.01540} {arXiv:2308.01540 [hep-ex]} \BibitemShut {NoStop}%
\bibitem [{\citenamefont {Lees}\ \emph {et~al.}(2017)\citenamefont {Lees} \emph {et~al.}}]{BaBar:2017tiz}%
  \BibitemOpen
  \bibfield  {author} {\bibinfo {author} {\bibfnamefont {J.~P.}\ \bibnamefont {Lees}} \emph {et~al.} (\bibinfo {collaboration} {BaBar}),\ }\href {\doibase 10.1103/PhysRevLett.119.131804} {\bibfield  {journal} {\bibinfo  {journal} {Phys. Rev. Lett.}\ }\textbf {\bibinfo {volume} {119}},\ \bibinfo {pages} {131804} (\bibinfo {year} {2017})},\ \Eprint {http://arxiv.org/abs/1702.03327} {arXiv:1702.03327 [hep-ex]} \BibitemShut {NoStop}%
\bibitem [{\citenamefont {Zhang}\ \emph {et~al.}(2019{\natexlab{b}})\citenamefont {Zhang}, \citenamefont {Zhang}, \citenamefont {Song}, \citenamefont {Pan}, \citenamefont {Niu},\ and\ \citenamefont {Li}}]{Zhang:2019wnz}%
  \BibitemOpen
  \bibfield  {author} {\bibinfo {author} {\bibfnamefont {Y.}~\bibnamefont {Zhang}}, \bibinfo {author} {\bibfnamefont {W.-T.}\ \bibnamefont {Zhang}}, \bibinfo {author} {\bibfnamefont {M.}~\bibnamefont {Song}}, \bibinfo {author} {\bibfnamefont {X.-A.}\ \bibnamefont {Pan}}, \bibinfo {author} {\bibfnamefont {Z.-M.}\ \bibnamefont {Niu}}, \ and\ \bibinfo {author} {\bibfnamefont {G.}~\bibnamefont {Li}},\ }\href {\doibase 10.1103/PhysRevD.100.115016} {\bibfield  {journal} {\bibinfo  {journal} {Phys. Rev. D}\ }\textbf {\bibinfo {volume} {100}},\ \bibinfo {pages} {115016} (\bibinfo {year} {2019}{\natexlab{b}})},\ \Eprint {http://arxiv.org/abs/1907.07046} {arXiv:1907.07046 [hep-ph]} \BibitemShut {NoStop}%
\bibitem [{\citenamefont {Andreev}\ \emph {et~al.}(2021)\citenamefont {Andreev} \emph {et~al.}}]{Andreev:2021fzd}%
  \BibitemOpen
  \bibfield  {author} {\bibinfo {author} {\bibfnamefont {Y.~M.}\ \bibnamefont {Andreev}} \emph {et~al.},\ }\href {\doibase 10.1103/PhysRevD.104.L091701} {\bibfield  {journal} {\bibinfo  {journal} {Phys. Rev. D}\ }\textbf {\bibinfo {volume} {104}},\ \bibinfo {pages} {L091701} (\bibinfo {year} {2021})},\ \Eprint {http://arxiv.org/abs/2108.04195} {arXiv:2108.04195 [hep-ex]} \BibitemShut {NoStop}%
\bibitem [{\citenamefont {deNiverville}\ \emph {et~al.}(2011)\citenamefont {deNiverville}, \citenamefont {Pospelov},\ and\ \citenamefont {Ritz}}]{deNiverville:2011it}%
  \BibitemOpen
  \bibfield  {author} {\bibinfo {author} {\bibfnamefont {P.}~\bibnamefont {deNiverville}}, \bibinfo {author} {\bibfnamefont {M.}~\bibnamefont {Pospelov}}, \ and\ \bibinfo {author} {\bibfnamefont {A.}~\bibnamefont {Ritz}},\ }\href {\doibase 10.1103/PhysRevD.84.075020} {\bibfield  {journal} {\bibinfo  {journal} {Phys. Rev. D}\ }\textbf {\bibinfo {volume} {84}},\ \bibinfo {pages} {075020} (\bibinfo {year} {2011})},\ \Eprint {http://arxiv.org/abs/1107.4580} {arXiv:1107.4580 [hep-ph]} \BibitemShut {NoStop}%
\bibitem [{\citenamefont {Batell}\ \emph {et~al.}(2009)\citenamefont {Batell}, \citenamefont {Pospelov},\ and\ \citenamefont {Ritz}}]{Batell:2009di}%
  \BibitemOpen
  \bibfield  {author} {\bibinfo {author} {\bibfnamefont {B.}~\bibnamefont {Batell}}, \bibinfo {author} {\bibfnamefont {M.}~\bibnamefont {Pospelov}}, \ and\ \bibinfo {author} {\bibfnamefont {A.}~\bibnamefont {Ritz}},\ }\href {\doibase 10.1103/PhysRevD.80.095024} {\bibfield  {journal} {\bibinfo  {journal} {Phys. Rev. D}\ }\textbf {\bibinfo {volume} {80}},\ \bibinfo {pages} {095024} (\bibinfo {year} {2009})},\ \Eprint {http://arxiv.org/abs/0906.5614} {arXiv:0906.5614 [hep-ph]} \BibitemShut {NoStop}%
\bibitem [{\citenamefont {Batell}\ \emph {et~al.}(2014)\citenamefont {Batell}, \citenamefont {Essig},\ and\ \citenamefont {Surujon}}]{Batell:2014mga}%
  \BibitemOpen
  \bibfield  {author} {\bibinfo {author} {\bibfnamefont {B.}~\bibnamefont {Batell}}, \bibinfo {author} {\bibfnamefont {R.}~\bibnamefont {Essig}}, \ and\ \bibinfo {author} {\bibfnamefont {Z.}~\bibnamefont {Surujon}},\ }\href {\doibase 10.1103/PhysRevLett.113.171802} {\bibfield  {journal} {\bibinfo  {journal} {Phys. Rev. Lett.}\ }\textbf {\bibinfo {volume} {113}},\ \bibinfo {pages} {171802} (\bibinfo {year} {2014})},\ \Eprint {http://arxiv.org/abs/1406.2698} {arXiv:1406.2698 [hep-ph]} \BibitemShut {NoStop}%
\bibitem [{\citenamefont {Aguilar-Arevalo}\ \emph {et~al.}(2017)\citenamefont {Aguilar-Arevalo} \emph {et~al.}}]{MiniBooNE:2017nqe}%
  \BibitemOpen
  \bibfield  {author} {\bibinfo {author} {\bibfnamefont {A.~A.}\ \bibnamefont {Aguilar-Arevalo}} \emph {et~al.} (\bibinfo {collaboration} {MiniBooNE}),\ }\href {\doibase 10.1103/PhysRevLett.118.221803} {\bibfield  {journal} {\bibinfo  {journal} {Phys. Rev. Lett.}\ }\textbf {\bibinfo {volume} {118}},\ \bibinfo {pages} {221803} (\bibinfo {year} {2017})},\ \Eprint {http://arxiv.org/abs/1702.02688} {arXiv:1702.02688 [hep-ex]} \BibitemShut {NoStop}%
\bibitem [{\citenamefont {Essig}\ \emph {et~al.}(2012)\citenamefont {Essig}, \citenamefont {Manalaysay}, \citenamefont {Mardon}, \citenamefont {Sorensen},\ and\ \citenamefont {Volansky}}]{Essig:2012yx}%
  \BibitemOpen
  \bibfield  {author} {\bibinfo {author} {\bibfnamefont {R.}~\bibnamefont {Essig}}, \bibinfo {author} {\bibfnamefont {A.}~\bibnamefont {Manalaysay}}, \bibinfo {author} {\bibfnamefont {J.}~\bibnamefont {Mardon}}, \bibinfo {author} {\bibfnamefont {P.}~\bibnamefont {Sorensen}}, \ and\ \bibinfo {author} {\bibfnamefont {T.}~\bibnamefont {Volansky}},\ }\href {\doibase 10.1103/PhysRevLett.109.021301} {\bibfield  {journal} {\bibinfo  {journal} {Phys. Rev. Lett.}\ }\textbf {\bibinfo {volume} {109}},\ \bibinfo {pages} {021301} (\bibinfo {year} {2012})},\ \Eprint {http://arxiv.org/abs/1206.2644} {arXiv:1206.2644 [astro-ph.CO]} \BibitemShut {NoStop}%
\bibitem [{\citenamefont {\r{A}kesson}\ \emph {et~al.}(2022)\citenamefont {\r{A}kesson} \emph {et~al.}}]{Akesson:2022vza}%
  \BibitemOpen
  \bibfield  {author} {\bibinfo {author} {\bibfnamefont {T.}~\bibnamefont {\r{A}kesson}} \emph {et~al.},\ }in\ \href@noop {} {\emph {\bibinfo {booktitle} {{Snowmass 2021}}}}\ (\bibinfo {year} {2022})\ \Eprint {http://arxiv.org/abs/2203.08192} {arXiv:2203.08192 [hep-ex]} \BibitemShut {NoStop}%
\bibitem [{\citenamefont {Akimov}\ \emph {et~al.}(2023)\citenamefont {Akimov} \emph {et~al.}}]{COHERENT:2021pvd}%
  \BibitemOpen
  \bibfield  {author} {\bibinfo {author} {\bibfnamefont {D.}~\bibnamefont {Akimov}} \emph {et~al.} (\bibinfo {collaboration} {COHERENT}),\ }\href {\doibase 10.1103/PhysRevLett.130.051803} {\bibfield  {journal} {\bibinfo  {journal} {Phys. Rev. Lett.}\ }\textbf {\bibinfo {volume} {130}},\ \bibinfo {pages} {051803} (\bibinfo {year} {2023})},\ \Eprint {http://arxiv.org/abs/2110.11453} {arXiv:2110.11453 [hep-ex]} \BibitemShut {NoStop}%
\bibitem [{\citenamefont {Banerjee}\ \emph {et~al.}(2018)\citenamefont {Banerjee} \emph {et~al.}}]{NA64:2017vtt}%
  \BibitemOpen
  \bibfield  {author} {\bibinfo {author} {\bibfnamefont {D.}~\bibnamefont {Banerjee}} \emph {et~al.} (\bibinfo {collaboration} {NA64}),\ }\href {\doibase 10.1103/PhysRevD.97.072002} {\bibfield  {journal} {\bibinfo  {journal} {Phys. Rev. D}\ }\textbf {\bibinfo {volume} {97}},\ \bibinfo {pages} {072002} (\bibinfo {year} {2018})},\ \Eprint {http://arxiv.org/abs/1710.00971} {arXiv:1710.00971 [hep-ex]} \BibitemShut {NoStop}%
\end{thebibliography}%

\end{document}